\documentclass[a4paper]{book}
\usepackage{cooltooltips}
\usepackage{xspace}

\newcounter{counter}

\newif\ifcom

\newif\ifcoms

\comtrue
\comstrue

\usepackage{marginnote} 
\usepackage{mathtools}
\usepackage{amssymb}
\usepackage[noend]{algorithm2e}
\usepackage[nottoc,notlot,notlof]{tocbibind}
\usepackage{url}
\usepackage{makeidx}
\usepackage{color}
\usepackage{multirow}
\usepackage{listings}
\usepackage{acronym}
\usepackage{multibib}
\usepackage{hhline}
\usepackage{setspace}
\usepackage[flushleft]{threeparttable}
\usepackage{array}
\usepackage{fancyhdr}
\usepackage{tikz}
\usepackage{tikz-uml}
\usetikzlibrary{calc}

\let\tmp\oddsidemargin
\let\oddsidemargin\evensidemargin
\let\evensidemargin\tmp
\reversemarginpar

\newcommand{\changefont}{%
    \fontsize{9}{11}\selectfont
}

\newcolumntype{L}{>{\centering\arraybackslash}m{1.5cm}}

\newcites{ltex}{Bibliography}
\newcites{web}{Website Bibliography}

\acrodef{tls}[TLS]{Transport Layer Security}
\acrodef{pki}[PKI]{Public Key Infrastructure}
\acrodef{hsts}[HSTS]{HTTP Strict Transport Security}
\acrodef{pat}[PAT]{Process Analysis Toolkit}
\acrodef{uml}[UML]{Unified Modelling Language}
\acrodef{csp}[CSP]{Communicating Sequential Processes}
\acrodef{ltl}[LTL]{Linear Temporal Logic}
\acrodef{ltl}[LTL]{Linear Temporal Logic}
\acrodef{ca}[CA]{Certification Authority}
\acrodef{seb}[SEB]{Safe Exam Browser}
\acrodef{mitm}[MITM]{Man-In-The-Middle}

\SetAlCapSkip{1em}

\newcommand{\ie}{i.e.,\xspace}
\newcommand{\eg}{e.g.,\xspace}
\newcommand{\tls}{TLS\xspace}

\newcommand{\UMLcheck}[1]{\mathit{#1}}
\newcommand{\UMLactivity}[1]{\mathsf{#1}}
\newcommand{\LTLprop}[1]{\mathtt{#1}}

\newcommand\mybox[2][]{\tikz[overlay]\node[fill=blue!20,inner sep=2pt, anchor=text, rectangle, rounded corners=1mm,#1] {#2};\phantom{#2}}
\newcommand\myboxer[2][]{\tikz[overlay]\node[fill=blue!20,inner sep=9pt, anchor=text, rectangle, rounded corners=1mm,#1] {#2};\phantom{#2}}

\newcommand*{\Ldot}{\raisebox{-0.45ex}{\scalebox{2}{$\cdot$}}}

\newcommand{\zerova}{
\begin{array}{c}{\begin{array}{c@{}c}{\blacksquare}&{\square}\end{array}}\\[-1ex]
{\begin{array}{c@{}c}{\square}&{\blacksquare}\end{array}}
 \end{array}}
 
\newcommand{\zerovb}{
\begin{array}{c}{\begin{array}{c@{}c}{\square}&{\blacksquare}\end{array}}\\[-1ex]
{\begin{array}{c@{}c}{\blacksquare}&{\square}\end{array}}
 \end{array}}

\newcommand{\onev}{
\begin{array}{c}{\begin{array}{c@{}c}{\blacksquare}&{\blacksquare}\end{array}}\\[-1ex]
{\begin{array}{c@{}c}{\blacksquare}&{\blacksquare}\end{array}}
 \end{array}}

\newcommand{\WTWO}{{WATA II}\xspace}
\newcommand{\WWW}{{WATA III}\xspace}

\def \etal{{\it et al.}~}

\newcommand{\attacker}{attacker\xspace}
\newcommand{\dishonest}{corrupted\xspace}

\newcommand{\appi}{applied $\pi$-}
\newcommand{\Appi}{Applied $\pi$-}
\newcommand{\inmessage}[2]{\mathit{in}(#1,#2)}
\newcommand{\outmessage}[2]{\mathit{out}(#1, #2)}
\newcommand{\freev}[1]{\mathit{fv}(#1)}
\newcommand{\boundv}[1]{\mathit{bv}(#1)}
\newcommand{\freen}[1]{\mathit{fn}(#1)}
\newcommand{\boundn}[1]{\mathit{bn}(#1)}

\newcommand{\pk}[1]{pk(#1)} 
 
\newcommand{\enc}[2]{enc(#1,#2)} 
\newcommand{\dec}[2]{dec(#1,#2)}

\newcommand{\id}{\mathit{id}}
\newcommand{\idcand}{\mathit{id\_c}}
\newcommand{\ans}{\mathit{ans}} 
\newcommand{\mrk}{\mathit{mark}} 
\newcommand{\ques}{\mathit{ques}} 
\newcommand{\idform}{\mathit{id\_test}}
\newcommand{\idexam}{\mathit{id\_e}}
\newcommand{\sign}{\mathit{Sign}}
\newcommand{\pid}{\mathit{pid}} 

\newcommand{\testUR}{\mathit{testUR}}

\newcommand{\holeprocess}[2]{#1_{#2}}
\newcommand{\phaseend}[2]{#1 \lvert_{#2}}
\newcommand{\EP}{\mathit{EP}}
\newcommand{\reg}[1]{\mathit{\texttt{registered}}\langle #1 \rangle} 
\newcommand{\fil}[4]{\mathit{\texttt{submitted}}\langle#1, #2, #3, #4 \rangle} 
\newcommand{\collect}[4]{\mathit{\texttt{collected}}\langle #1, #2, #3, #4 \rangle} 
\newcommand{\distribute}[5]{\mathit{\texttt{distributed}}\langle #1, #2, #3, #4, #5 \rangle} 
\newcommand{\marking}[5]{\mathit{\texttt{marked}}\langle #1, #2, #3, #4, #5 \rangle} 
\newcommand{\notify}[2]{\mathit{\texttt{notified}}\langle #1, #2 \rangle}

\newcommand{\regtt}[1]{\mathit{\mathtt{registered}}\langle #1 \rangle} 
\newcommand{\filtt}[4]{\mathit{\mathtt{submitted}}\langle#1, #2, #3, #4 \rangle} 
\newcommand{\collecttt}[4]{\mathit{\mathtt{collected}}\langle #1, #2, #3, #4 \rangle} 
\newcommand{\distributett}[5]{\mathit{\mathtt{distributed}}\langle #1, #2, #3, #4, #5 \rangle} 
\newcommand{\markingtt}[5]{\mathit{\mathtt{marked}}\langle #1, #2, #3, #4, #5 \rangle} 
\newcommand{\notifytt}[2]{\mathit{\mathtt{notified}}\langle #1, #2 \rangle} 
\newcommand{\requestedtt}[2]{\mathit{\mathtt{requested}}\langle #1, #2 \rangle} 
\newcommand{\storedtt}[2]{\mathit{\mathtt{stored}}\langle #1, #2 \rangle}

\newcommand{\OKa}[1]{\mathit{\texttt{OK}}\langle #1 \rangle} 
\newcommand{\OKd}[2]{\mathit{\texttt{OK}}\langle #1, #2 \rangle} 
\newcommand{\OKb}[3]{\mathit{\texttt{OK}}\langle #1, #2, #3 \rangle} 
\newcommand{\OKc}[4]{\mathit{\texttt{OK}}\langle #1, #2, #3, #4 \rangle} 
\newcommand{\generated}[1]{\mathit{\texttt{generated}}\langle #1 \rangle} 
\newcommand{\accepted}[3]{\mathit{\texttt{accepted}}\langle #1, #2, #3 \rangle} 
\newcommand{\markinga}[3]{\mathit{\texttt{marked}}\langle #1, #2, #3 \rangle} 
\newcommand{\correct}[3]{\mathit{\texttt{correct}}\langle #1, #2, #3 \rangle} 
\newcommand{\assigned}[4]{\mathit{\texttt{assigned}}\langle #1, #2, #3, #4 \rangle} 
\newcommand{\markingb}[4]{\mathit{\texttt{marked}}\langle #1, #2, #3, #4 \rangle} 
\newcommand{\markingc}[1]{\mathit{\texttt{marked}}\langle #1 \rangle} 
\newcommand{\assigneda}[2]{\mathit{\texttt{assigned}}\langle #1, #2 \rangle} 
\newcommand{\assignedb}[3]{\mathit{\texttt{assigned}}\langle #1, #2, #3 \rangle}

\newcommand{\CAu}{Test Origin Authentication\xspace}
 
\newcommand{\cautho}{Candidate Authorisation\xspace} 
\newcommand{\aau}{Answer Authenticity\xspace}

\newcommand{\mau}{Mark Authenticity\xspace}  
\newcommand{\ta}{Test Authenticity\xspace}  

\newcommand{\am}{Anonymous Marking\xspace} 
 
\newcommand{\mpr}{Mark Privacy\xspace} 
\newcommand{\man}{Mark Anonymity\xspace}

\newcommand{\dr}{Dispute Resolution\xspace} 

\newcommand{\have}{\checkmark}
\newcommand{\havenot}{\times}

\newcommand{\Huszti}{Huszti-Peth\H{o}\xspace}
\newcommand{\MIX}{\mathit{M}}
\newcommand{\ID}{\mathit{ID}}
\newcommand{\PK}{\mathit{PK}}
\newcommand{\SK}{\mathit{SK}}
\newcommand{\SSK}{\mathit{SSK}}
\newcommand{\SPK}{\mathit{SPK}}
\newcommand{\mi}[1]{\mathit{#1}}
\newcommand{\encr}[2]{\{#1\}_{{#2}}}

\newcommand{\ea}{exam authority\xspace}

\newcommand{\rr}{\overline{r}}
\newcommand{\BB}{bulletin board\xspace}
\newcommand{\set}[1]{\left\{#1\right\}}
\newcommand{\setremark}[1]{\langle{#1}\rangle}
\newcommand{\ppi}{\overline{\pi}}

\newcommand{\WW}{{WATA IV}\xspace}
\newcommand{\share}{\mathit{share}}

\newcommand{\data}{\mathit{data}}
\newcommand{\paper}{\mathit{paper}}
\newcommand{\transp}{\mathit{transp}}

\newcommand\FramedBox[3]{%
  \setlength\fboxsep{0pt}
  \fbox{\parbox[t][#1][c]{#2}{#3}}}

\newcommand{\remark}{{Remark!}\xspace}
\DeclareRobustCommand\picalc[0]{applied $\pi$-Calculus}

\newcommand{\examtest}{test\xspace}
\newcommand{\examtests}{tests\xspace}
\newcommand{\fun}[1]{\mathtt{#1}}

\newcommand{\red}[1]{#1}
\newcommand{\old}[1]{}
\newcommand{\evidences}{data\xspace}
\newcommand{\evidence}{data\xspace}
\newcommand{\type}[1]{\mathcal{#1}}
\newcommand{\bool}[0]{\fun{bool}}

\newcommand{\questionValidity}[0]{Question Validity}
\newcommand{\questionValidityFun}[0]{QV}
\newcommand{\etIntegrity}[0]{Test Integrity}
\newcommand{\etIntegrityFun}[0]{ETI}
\newcommand{\etMarkedness}[0]{Test Markedness}
\newcommand{\etMarkednessFun}[0]{ETM}
\newcommand{\markingCorrectness}[0]{Marking Correctness}
\newcommand{\markingCorrectnessFun}[0]{MC}
\newcommand{\markIntegrity}[0]{Mark Integrity}
\newcommand{\markIntegrityFun}[0]{MI}
\newcommand{\markNotificationIntegrity}[0]{Mark Notification Integrity}
\newcommand{\markNotificationIntegrityFun}[0]{MNI}

\newcommand{\registration}[0]{Registration}
\newcommand{\registrationFun}[0]{R}
\newcommand{\etIntegrityUniversal}[0]{Test Integrity}
\newcommand{\etIntegrityUniversalFun}[0]{ETI}
\newcommand{\etMarkednessUniversal}[0]{Test Markedness}
\newcommand{\etMarkednessUniversalFun}[0]{ETM}
\newcommand{\markingCorrectnessUniversal}[0]{Marking Correctness}
\newcommand{\markingCorrectnessUniversalFun}[0]{MC}
\newcommand{\markIntegrityUniversal}[0]{Mark Integrity}
\newcommand{\markIntegrityUniversalFun}[0]{MI}

\newcommand{\individual}{I.V.\xspace}
\newcommand{\universal}{U.V.\xspace}
\newcommand{\individualFun}{{IV}}
\newcommand{\universalFun}{{UV}}

\newcommand{\questionValidityAdj}[0]{question validity verifiable}
\newcommand{\etIntegrityAdj}[0]{test integrity verifiable}
\newcommand{\etMarkednessAdj}[0]{test markedness verifiable}
\newcommand{\markingCorrectnessAdj}[0]{marking correctness verifiable}
\newcommand{\markIntegrityAdj}[0]{mark integrity verifiable}
\newcommand{\markNotificationIntegrityAdj}[0]{mark notification integrity verifiable}
\newcommand{\registrationAdj}[0]{registration verifiable}
\newcommand{\etIntegrityUniversalAdj}[0]{test integrity universally verifiable}
\newcommand{\etMarkednessUniversalAdj}[0]{test markedness universally verifiable}
\newcommand{\markingCorrectnessUniversalAdj}[0]{marking correctness universally verifiable}
\newcommand{\markIntegrityUniversalAdj}[0]{mark integrity universally verifiable}

\newcommand{\haveall}{\checkmark({\text{all}})}

\newcommand{\suchthat}{such that\xspace}

\let\ab\allowbreak

\newtheorem{prop}{Definition}
\newtheorem{asse}{Assertion}
\newtheorem{rec}{\bf Recommendation}

\makeindex

\begin{document}

\lstset{
  basicstyle=\ttfamily,
  columns=fullflexible,
  keepspaces=true,
  frame=single, 
  emph={choice, collected, registered, submitted, distributed, marked, notified, OK, KO, accepted, assigned, requested, stored, Eguilty, Cguilty},
  emphstyle=\bfseries\itshape
}

\title{Design and Analysis of Secure Exam Protocols}
\author{Rosario Giustolisi}
\date{}
\frontmatter




    \begin{center}
        \vspace*{-3cm}

        \includegraphics[width=0.2\textwidth]{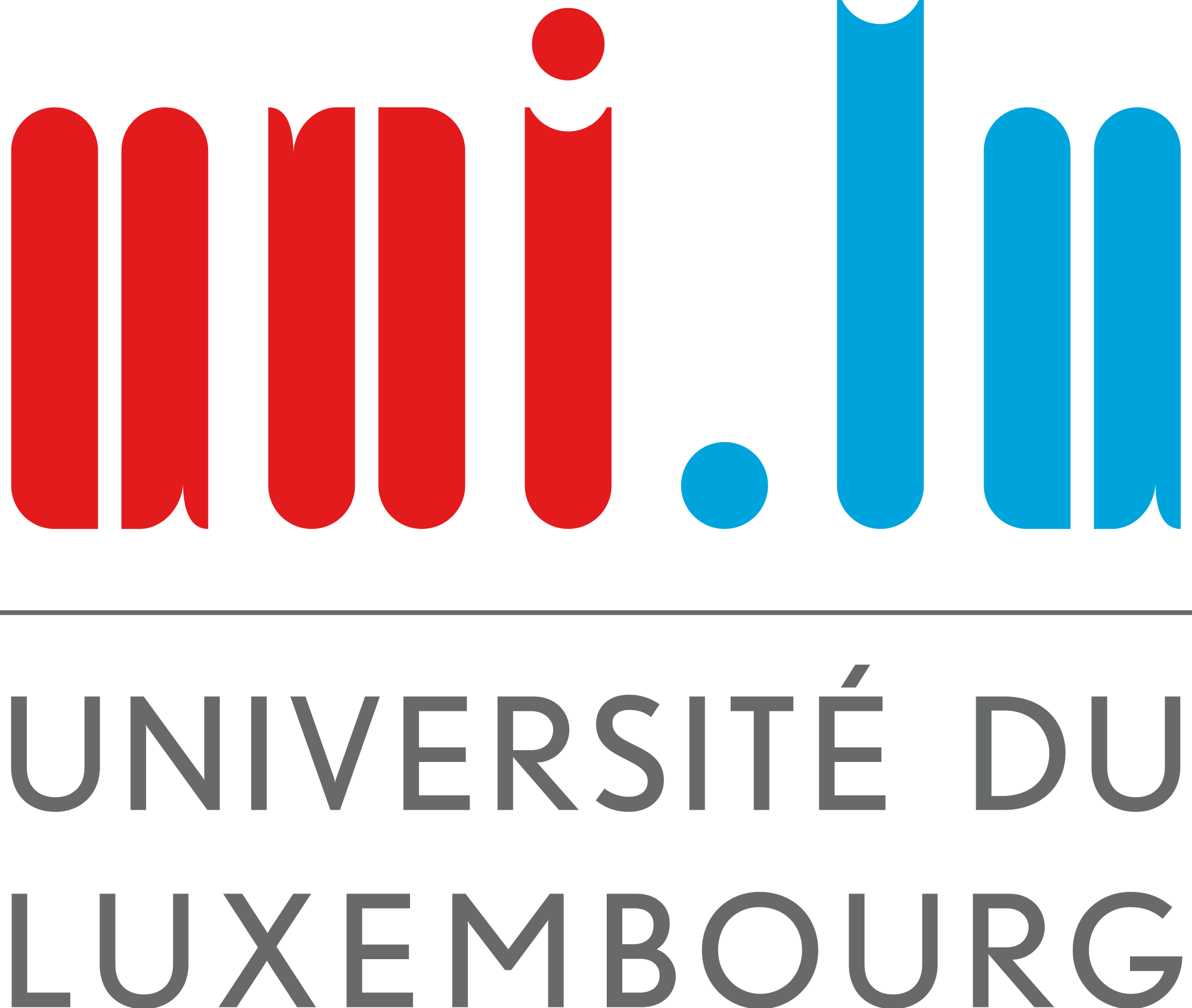}
        
        \vspace{0.7cm}
        {\fontfamily{cmss}\selectfont 
       PhD-FSTC-2015-47

        \vspace{0.2cm}

        The Faculty of Sciences, Technology and Communication

        \vspace{1cm}

{\LARGE        DISSERTATION}

        \vspace{0.8cm}

{\large        Defense held on 26/10/2015 in Luxembourg

  \vspace{0.4cm}

        to obtain the degree of}

  \vspace{0.5cm}

{\LARGE    DOCTEUR DE L'UNIVERSIT\'E DU


    LUXEMBOURG

  \vspace{0.5cm}

EN INFORMATIQUE}

  \vspace{0.5cm}

{\Large    by}

  \vspace{0.5cm}

{\LARGE Rosario GIUSTOLISI}


Born on 16 October 1983 in Giarre (Italy)

  \vspace{1cm}    

\huge \textsc{Design and Analysis of}

  \vspace{0.25cm}    

\textsc{Secure Exam Protocols}}

    \end{center}

  \vspace{0.5cm}    

\noindent\textbf{Dissertation defense committee}\\
{\setstretch{0.8}


\noindent Dr Peter Y.A. Ryan, dissertation supervisor\\
\emph{Professor, Universit\'e du Luxembourg}\\

\noindent Dr Gabriele Lenzini, vice-chairman\\
\emph{Universit\'e du Luxembourg}\\

\noindent Dr Sjouke Mauw, chairman\\
\emph{Professor, Universit\'e du Luxembourg}\\

\noindent Dr Steve Schneider\\
\emph{Professor, University of Surrey}\\

\noindent Dr Luca Vigan\`o\\
\emph{Professor, King's College London}

}

\pagestyle{empty}
\chapter*{Abstract} \pagestyle{plain} \setcounter{page}{1}

Except for the traditional threat that candidates may want to cheat, exams have historically not been seen like a serious security problem. That threat is routinely thwarted by having invigilators ensure that candidates do not misbehave during testing. However, as recent exam scandals confirm, also invigilators and exam authorities may have interest in frauds, hence they may pose security threats as well. 
Moreover, new security issues arise from the recent use of computers, which can facilitate the exam experience for example by allowing candidates to register from home.
Thus, exams must be designed with the care normally devoted to security protocols. 

This dissertation studies exam protocol security and provides an in-depth understanding that can be also useful for the study of the security of similar systems, such as public tenders, personnel selections, project reviews, and conference management systems. It introduces an unambiguous terminology that leads to the specification of a taxonomy of various exam types, depending on the level of computer assistance.
It then establishes a theoretical framework for the formal analysis of exams. The framework defines several authentication, privacy, and verifiability requirements that modern exams should meet, and enables the security of exam protocols to be studied.
Using the framework, we formally analyse traditional, computer-assisted, and Internet-based exam protocols. We find some security issues and propose modifications to partially achieve the desired requirements.

This dissertation also designs three exam protocols that guarantee a wide set of security requirements.
It introduces a novel protocol for Internet-based exams to thwart a malicious exam authority with minimal trust assumptions.
Then, it proposes secure protocols suitable for both computer-assisted and traditional pen-and-paper exams. A combination of oblivious transfer and visual cryptography schemes allows us to overcome the constraint of face-to-face testing and to remove the need of a trusted third party. Moreover, the protocols ensure accountability as they support the identification of the principal that is responsible for their failure. We evaluate the security of our protocols by a formal analysis in ProVerif.

Finally, this dissertation looks at exams as carried out through a modern browser, Safe Exam Browser (SEB). It was specifically designed to carry out Internet-based exams securely, and we confirm it immune to the security issues of certificate validation.
Using UML and CSP, we advance a formal analysis of its requirements that are not only logically conditioned on the technology but also on user actions. 
By extending this analysis onto other browsers, we state general best-practice recommendations to browser vendors.

\chapter*{Acknowledgements}

Pursuing a Ph.D. is a journey that involves many people. Here I want to acknowledge the people without whom this journey would not be The Journey.

First, I would like to express my gratitude to Gabriele Lenzini for his continuous supervision. He has devoted much of his time to help and support me in this journey. He has patiently taught me how to evaluate my work and how clearly formulate the results.
I am thankful to Peter Ryan for accepting me in his research group and for his valuable advice during our conversations. I am fascinated of his rare ability to quickly grasp the essence of any mathematical problem.
 I am particular grateful to Giampaolo Bella for his precious support and advice. He has started me on research and is always up to taught me how to develop my research skills further. Moreover, he is a great motivator. 

I would like to thank my Ph.D. examiners Sjouke Mauw, Steve Schneider, and Luca Vigan\`o for their interest and valuable expertise in giving some advice useful for  my dissertation. I am thankful to the Doctoral School of Computer Science and Computer Engineering for the financial support provided to attend conferences and summer schools. 

I am particular grateful to Jannik Dreier, Ali Kassem, and Pascal Lafourcade for the fruitful collaboration on the formal specification of exam requirements. I would like to thank Andrea Huszti for the interesting discussions about the security of exam protocols and efforts on how improve them. I also enjoyed discussions with people I met in scientific conferences and summer schools.
In particular, the annual Workshop on Security Frameworks has been a great source of feedback and ideas, and I benefited from discussions with people I met there  (Giampaolo Bella, Denis Butin, Gianpiero Costantino, Gabriele Lenzini, Giuseppe Patan\`e, Salvatore Riccobene, \ldots). 

I had great time with my fellow Ph.D. students (Arash, Massimo, Afonso, Jean-Louis, Miguel, Marjan, Dayana, Masoud, Jun) at ApSIA, and never felt too far from home thanks to my Italian office neighbours (Claudio and Vincenzo).
I was privileged to live at the R\'esidence des Dominicaines and to have a lot of friends living there.

I would like to thank my family: Daniela and Leo for taking care of everything during my stay abroad; 
my parents Franco e Pina for their unconditional love and support they have been giving to me, and for have allowed me to be the person I am today; my girlfriend Silvia for standing by me, for having patiently shared successes and failures, and for always believing in me.

Finally, I would like to thank you for reading this dissertation.

 \clearpage \pagestyle{empty}

\tableofcontents \pagestyle{plain} \clearpage \pagestyle{empty}
\listoffigures \pagestyle{plain}
\listoftables \clearpage  \pagestyle{empty}

\renewcommand{\chaptermark}[1]{\markboth{\thechapter.\space#1}{}} 
\mainmatter
\pagestyle{fancyplain}
\chapter{Introduction}\label{chap:introduction}
The Oxford English Dictionary defines an \emph{exam} as \emph{the process of testing the knowledge or ability of pupils, or of candidates for office, degrees, etc.}
Although the definition of exam is pivoted on individuals, exams assume a key role in fostering meritocracy in modern societies. 
On the individual side, exams are important as they help people understand their skills and knowledge in a particular subject.
On the societal side, exams are indispensable to select most appropriate people through an objective evaluation. \index{objective evaluation} \index{meritocracy}

The use of exams is widespread, with various examples derivable from the education sector (\eg admissions, coursework,  and final qualifications) as well as from the work sector (\eg recruitment, progression, and professional qualifications).
France is one of the countries that extensively use exams, some of which are very competitive: the ``Concours G\'en\'eral'' is the most prestigious academic exam, which is taken by about 15,000 French students, and typically bears a success rate of less than 2\% \citeweb{concours}; the exam to enter medicine studies attracts every year more than 50,000 candidates with a success rate of less than 15\% \citeweb{medicine}.
Exam meritocracy is one of the guiding principles in Singapore and is deemed to have contributed to the rapid growth of the country \citeltex{Gopinathan07}. 
High populated countries, such as China and India, resort on tough exams for the recruitment of various administrative officers. In particular, China has an old tradition in administering civil service exams, which dates back to the Han dynasty (206 BC - AD 220).
It is worth noting that the idea to promote people based on exams came to West only in the 17th century, when the British Empire began to hire employees using competitive exams \index{competitive exams} to eliminate favouritism and corruption \citeltex{KER09}. \index{corruption}

Exams fulfil the goal of testing knowledge and abilities of candidates only in absence of misbehaviour of the involved parties, and normally employ different methods to face possible threats. For example, threats ascribed to candidate cheating are normally mitigated by invigilation \index{invigilation} and anti-plagiarism methods.
Moreover, an exam requires people to follow many procedures to make sure that as few things as possible go wrong. For example, to ensure that only eligible candidates \index{eligible candidates} attend the exam, the invigilator checks their identity documents before allowing them to take the exam. To guarantee fair marking, each candidate may use a special number that replaces her name on the test so that the examiner can mark the test while ignoring its author. The author of a test is revealed only after the examiner marked the tests. 

The procedures outlined above help people's confidence that everything works as expected, provided that people can observe the procedures or trust the exam system. The latter means that people have to trust that the candidate's ID is properly checked, that special numbers are correctly employed, and that authorities do not reveal the author of a test before the examiner marks it. People have to trust authorities.

However, authorities and examiners may be corrupted, and so they may commit misconducts that are hard to eradicate. In the scandal known as Atlanta Cheating \citeweb{CheatingUSA2013}, about 35 people among school administrators, educators, and superintendents manipulated ranks and scores with the goal of gaining more \index{exam scandal}
school governmental funds.
In early 2014, the BBC revealed a fraud on the UK visa system, in which the invigilator read out all the correct answers during the exam of English proficiency \citeweb{cheatingets}. More recently, a medical school admission exam scandal in India has turned into thousands of arrests \citeweb{bbcindia}. The police revealed that candidates hired impersonators to take the written exam, and examiners gave higher marks to colluded candidates. \index{written exam}
The U.S. Navy disclosed cheating on the written exam that concerns the use of nuclear reactors that power carriers and subs \citeweb{usnavy2014}. It was found that questions and answers were illegally taken from a Navy computer since 2007 \citeweb{usnavyfox}.

In fact, computers have been increasingly introduced in the main procedures of exam systems. They can assist generation of questions \index{generation of questions} or automatic marking \index{automatic marking} procedures; also, they can be employed for remote registration, remote notification, and even remote testing, \index{remote registration} \index{remote notification} \index{remote testing} in which candidates can take the exam from home. For example, the most popular Massive Online Open Courses (MOOC), the education platforms that offer courses of study over the Internet, allow remote testing \citeweb{courseraexam, bbcmooc}. Such exams provide a formal recognition of the  candidates' achievements, and some universities already consider MOOC exams eligible for university credits \citeweb{mooc}. \index{MOOC} 

The use of computers simplifies certain tasks occurring during an exam, but does not necessarily make the exam more secure. For example, the registration of a candidate for the exam and the notification of the mark via the Internet should be at least as secure as they would be face-to-face.
Hence, we shall unfold the argument that an exam must be designed and analysed as carefully as security protocols normally are. 

This dissertation draws its main motivation from the observation that exams raise more challenging security and privacy issues than one may think. This is due to at least two main reasons. One is that threats may come from any of the roles playing in an exam. In particular, candidates and authorities may be corrupted to various extents. Exams then begin to look more balanced in terms of threats or benefits their participants pose or seek. 
The other reason is that there is no clear understanding of what the relevant security requirements for exams are.
We observe that the growth in the use of exam protocols has neither been followed nor preceded by a rigorous definition
and analysis of their security. Such absence may lead to the design and the practical adoption of insecure exam protocols, and makes people less confident
of exam trustworthiness.

These concerns are relevant also for other domains, such as voting. Significant parts of the related work in this dissertation highlight similarities and differences between exams and voting. Intuitively, both domains share similar security and privacy requirements. For example, only the answers originated by eligible candidates \index{eligible candidates} should be marked in an exam. Similarly, only ballots cast by eligible voters should be recorded in a voting system. However, a closer inspection reveals a number of different security concerns between the two domains. For example, the link between an answer and its author should be preserved through the phases of an exam. Conversely, unlinkability \index{unlinkability} between voter and vote is a desired property of a voting system. 
In a nutshell, a fundamental difference is that while fair elections aim to bring \emph{democracy}, fair exams aim to bring \emph{meritocracy} to societies. We observe that democracy is not meritocracy: in democracy, the selection of candidates is based on (people's) choices; in meritocracy, the selection of candidates is based on (the examiner's) assessment of their merit. \index{assessment}
Thus, understanding similarities and differences between exams and voting becomes an additional motivation of this manuscript. \index{meritocracy}




\section{Aims and Objectives}

This work aims to study the relevant security requirements for exam protocols and to design and analyse exam protocols that meet the stated security requirements. 
We intend to achieve those aims through four objectives.
\begin{enumerate}
\item {\bf To identify the relevant security requirements for exam protocols.} 
This is a fundamental objective as it provides the basics for further research and determines the meaning of \emph{secure} exam protocol. It requires the specification of a coherent terminology for exams including their phases and threat model. The desired outcome consists of a set of authentication, privacy, and verifiability requirements. \index{secure exam protocol}
\item {\bf To develop a formal framework for the specification of security requirements and the analysis of exam protocols.} 
This is a crucial objective that provides a rigorous and formal description of the security requirements for exams. It requires choosing a specific formalism in which the security requirements identified in objective 1 can be expressed. The desired outcome is to achieve a flexible framework that is suitable for the modelling and analysis of exams.
\item {\bf To design new secure protocols for different types of exams.} 
This objective consists in proposing novel exam protocols that meet the security requirements according to the restrictions of the different exam types, which depend on the level of computer assistance and span from traditional to Internet-based exams. It requires combining secure cryptographic schemes to guarantee the often contrasting requirements. The desired outcome is a number of protocols that provide the same level of security though they belong to different exam types. \index{Internet-based exams}
\item {\bf To explore novel security aspects of the critical components of exam protocols.} 
This objective is to expand the formal analysis of exams by considering also the user. In particular, it concerns the analysis of one of the components that interacts with the user most. As we shall see later, this component is the browser. This objective requires choosing a suitable approach that includes the user in the formulation of security requirements and in the analysis. The desired outcome is to understand how user's choices may influence the security of exam protocols.
\end{enumerate}

\section{Contributions}


\begin{figure}
\begin{center}
\begin{tikzpicture}
\node at (1,4.5)   {\myboxer[fill=blue!20]{Objective 1}};
\node[text width=9.2cm, fill=gray!20, rounded corners=1mm] at (7,4.5) {A clear description of the general building blocks of all exams, and the definition of the security requirements};

\node at (1,3)   {\myboxer[fill=blue!20]{Objective 2}};
\node[text width=9.2cm,fill=gray!20, rounded corners=1mm] at (7,3) {{Two formal frameworks for the security analysis of exam protocols}};

\node at (1,1.5)   {\myboxer[fill=blue!20]{Objective 3}};
\node[text width=9.2cm,fill=gray!20, rounded corners=1mm] at (7,1.5) {{Three  new  exam protocols  that guarantee several security requirements}};

\node at (1,0)   {\myboxer[fill=blue!20]{Objective 4}};
\node[text width=9.2cm,fill=gray!20, rounded corners=1mm] at (7,0) {{A socio-technical formal  analysis of browsers for Internet-based exams}};
\end{tikzpicture}
\caption{Overview of the contributions} 
\label{fig:contributions}
\end{center}
\end{figure}

This dissertation addresses the four objectives outlined in the previous section. It advances the state of the art in the design and analysis of secure exam protocols with at least four original contributions as reported in Figure \ref{fig:contributions}.  \index{secure exam protocol}

The first contribution is a clear specification of the general building blocks of all exams, in terms of tasks, roles, phases, and threats. This paves the way for the definition of the security requirements for exams and facilitates the description of exam protocols.

The second contribution consists of two formal frameworks for the security analysis of exam protocols. The frameworks enable the study of five authentication, five privacy, and eleven verifiability requirements for exam protocols, and support the specification of additional requirements. One framework formalises authentication and privacy requirements in the \appi calculus \citeltex{AF01}, while the other specifies verifiability requirements in a more abstract model \index{abstract model}. Both frameworks are validated in ProVerif \citeltex{proverif} on traditional, computer-assisted, and Internet-based exam protocols. We find that some protocols are flawed and propose modifications on their designs.

The third contribution consists of three new exam protocols that guarantee a set of security requirements. The first protocol is for Internet-based exams \index{Internet-based exams} and meets authentication, privacy, and verifiability requirements with minimal reliance on trusted parties. It distributes the trust across the servers that compose an \emph{exponentiation mixnet}. \index{exponentiation mixnet} The second protocol is for computer-assisted exams \index{computer-assisted exams} with face-to-face testing, \index{face-to-face testing} and meets the requirements by means of lightweight participation of a trusted third party (TTP). It exploits the use of signatures and visual cryptography to ensure authentication and privacy in the presence of corrupted authorities and candidates. The last protocol eliminates the need of the TTP by combining oblivious transfer and visual cryptography schemes. It still ensures the same security requirements as the previous protocol's and provides accountability without relying on a TTP.
The proposed exam protocols are formally analysed in ProVerif.

The last contribution is the formal analysis of six modern browsers for Internet-based exams. \index{Internet-based exams} This analysis tackles yet another aspect of the problem of secure exams as it concerns the human understanding of remote authentication, and in particular of certificate validation in browsers. Browsers are critical components of an exam as they are the main application users normally interact with. The list of analysed browsers includes Firefox, Chrome, Safari, Internet Explorer, Opera Mini, and  Secure Exam Browser (SEB), a kiosk browser \index{kiosk browser} that enforces the security of remote testing \index{remote testing} in Internet-based exams. Using UML Activity Diagrams and the CSP process algebra \citeltex{Hoare}, we provide a systematic approach to the security analysis of certificate validation in different scenarios, all considering user interactions. Parts of this contribution are the Linear Temporal Logic specification of five novel socio-technical requirements, each binding elements like TLS session, certificates and user choices. We find that each browser implements certificate validation differently, and some of them pose more security risks to users than others do. We propose four recommendations to improve the security of browsers' certificate validations.

The results of this research offer the basis for the design and analysis of secure protocols for traditional, computer-assisted, and Internet-based exams. 

\section{Outline}
This dissertation is structured in eight chapters. Most of the contents of the dissertation have been published as joint work with different co-authors in conference papers or submitted to journal articles. In the following, we outline the contents of each chapter.

\begin{description} 
\item
\textbf{Chapter \ref{chap:terminology}: Terminology.} This chapter introduces the basics of an exam. It begins with the description of the tasks that occur during an exam. In particular, it observes that  levels of detail and abstraction of an exam specification constraint the number of tasks. The chapter continues discussing the possible roles of an exam, possibly played by one or more principals. It identifies the typical phases of an exam and the basic threats coming from the main exam roles. The chapter concludes with a taxonomy that classifies exams by types and categories. 

The contents of this chapter have not been published. However, a paper containing an earlier version of the chapter was accepted for publication in a conference but not included in the proceedings.

\item \textbf{Chapter \ref{chap:formal}: Formalising Authentication and Privacy.} This chapter contains the formal definitions of authentication and privacy requirements for exams. It discusses different techniques to model security requirements and presents tools for the automatic analysis of security protocols. Then, it describes the \appi calculus, on which we rely to build the framework. The framework consists of the  formal model of an exam, five authentication, and five privacy requirements. It is validated via the ProVerif analysis of the \Huszti exam protocol. The chapter continues with the results of the analysis and concludes with a proposal that enhances the security of the \Huszti protocol.

Most of the contents of this chapter are based on different publications. In particular, the informal definitions of authentication and privacy requirements are based on a joint work with Giampaolo Bella and Gabriele Lenzini \cite{crisis13}. The formal framework and the analysis of the \Huszti protocol have been co-authored with Jannik Dreier, Ali Kassem, Pascal Lafourcade, Gabriele Lenzini, and Peter Y.A. Ryan \cite{DGK+14}. An extended version of this work that considers the fixes for the \Huszti protocol has been submitted to a journal.
The formalisation of the authentication requirement of Candidate Authorisation and its verification on the different protocols discussed in this dissertation are unpublished work. \index{Candidate Authorisation}

\item \textbf{Chapter \ref{chap:verifiability}: Formalising Verifiability.} This chapter proposes an abstract framework in which an exam protocol and its verifiability requirements are formalised via a set-theoretic approach. In this chapter, we distinguish the notions of \emph{individual verifiability} as verifiability from the point of view of the candidate, and \emph{universal verifiability} as verifiability from the point of view of an
external auditor. We propose six individual and five universal verifiability requirements.

This chapter is based on joint work with Jannik Dreier, Ali Kassem, Pascal Lafourcade, and Gabriele Lenzini \cite{DGK+15}.

\item \textbf{Chapter \ref{chap:remark}: The \remark~Internet-based Exam Protocol.} This chapter details \emph{Remark!}, a new protocol for Internet-based exams. It discusses the cryptographic building blocks on which \remark~is based, with a particular focus on the exponentiation mixnet. \index{exponentiation mixnet} The chapter continues with the description of the protocol and the formal analysis in ProVerif of authentication, privacy, and verifiability requirements. Notably, it discusses how to map the abstract definitions of verifiability in ProVerif. The chapter concludes with some security considerations of Remark!. \index{Internet-based exams}

The paper that proposes \remark~has been co-authored with Gabriele Lenzini and Peter Y.A. Ryan \cite{GLR14}. The formal analysis of \remark~is based on the papers that appears in \cite{DGK+14} and \cite{DGK+15}. The manual induction proofs that support the formal analysis of universal verifiability in \remark~are based on \cite{XtreportClear}.

\item \textbf{Chapter \ref{chap:wata}: Computer-assisted Exam Protocols.} This chapter focuses on \emph{WATA}, a family of computer-assisted exams each employing some level of computer assistance though keeping face-to-face testing. \index{computer-assisted exams} \index{face-to-face testing}
This chapter first introduces the protocol versions of the existing WATA II and III software and reviews their security. Then, it details \emph{WATA IV}, a novel exam protocol that meets more security requirements than the previous ones with less reliance on a TTP. WATA IV is then redesigned to meet the same security requirements without the need of any TTP. The chapter presents a detailed description of the enhanced version and a formal analysis in ProVerif, including the formalisation of an accountability requirement (Dispute Resolution). It concludes with a brief review of the protocols seen throughout the chapter.

This chapter is based on joint work with Giampaolo Bella, Gabriele Lenzini and Peter Y.A. Ryan.
The protocol versions of the WATA software and their informal analyses are unpublished work. WATA IV has been published in a conference paper \cite{BGL14}. The enhanced version and its formal analysis are based on \cite{BGL+15}.

\item \textbf{Chapter \ref{chap:socio}: Formal Analysis of Certificate Validation in SEB and  Modern Browsers.} This chapter considers exams taking place via modern browsers and focuses on 
the socio-technical analysis of certificate validation.
It clarifies the basics of certificate validation  and highlights the user involvement in the process. Using UML Activity Diagrams, it describes the certificate validation of six different browsers, considering both classic and private browsing modes. This chapter then introduces five socio-technical requirements in Linear Temporal Logic and outlines the translation of the UML diagrams to CSP\# \citeltex{SunLDP09}. It discusses the output of a model checker analysis, and concludes with four recommendations to browser vendors. 

This chapter is based on joint work with Giampaolo Bella and Gabriele Lenzini. A preliminary version appears in \cite{BGL13}, while the full treatment is based on \cite{BGL13pst}. The analysis of SEB, Safari, private browsing modes, and the interleaving with classic browsing are based on a version submitted to a journal.

\item \textbf{Chapter \ref{chap:conclusion}: Conclusions.} This chapter discusses the research presented throughout the dissertation, outlines future work, and  concludes the presentation. 
\nocite{FGH+13, BCG+14}
\end{description}


\chapter{Terminology}\label{chap:terminology}

This chapter introduces the reader to the general building blocks of all exams. In consequence, describing a specific exam becomes easier, at the sole price of further expanding or specifying these general concepts.  We view an exam as a security protocol that involves various tasks defining roles played by various principals through various phases. Hence, \emph{exam} or \emph{exam protocol} are used interchangeably. With a security take, an exam is expected to withstand a threat model meeting a number of security requirements.

\paragraph{Outline of the chapter.} Section \ref{sec:tasks} discusses the levels of detail and abstraction to characterise tasks. 
Section \ref{sec:roles} introduces possible roles for exams. 
Section \ref{sec:principals} outlines the principals that play the exam roles.
Section \ref{sec:phases} identifies the typical phases of an exam.
Section \ref{sec:threat} details the potential security threats associated with the exam roles.
Section \ref{sec:taxonomy} classifies exams by type and category, and concludes the chapter.

\section{Tasks}\label{sec:tasks}

A number of \emph{tasks} may occur during an exam, such as generating the set of questions, building the tests and marking them. 
%
We observe that the number of tasks that can be identified may change over two possible dimensions.

One is the required \emph{level of detail} for the specification of the exam protocol, establishing whether a task should be explicitly mentioned or not. For example, classical security protocols often prescribed the task of \emph{using} a nonce in a message, yet omitting the task of \emph{fetching} it. Experience teaches us that a specification should make very clear (in)security assumptions about the protocol environment, otherwise the analysis may yield debatable findings. In this vein, Needham stated that the public-key Needham-Schr\"oder protocol considered the attacker as an outsider but never made this explicit \citeltex{Needham02a}, a threat model that would remove the opportunity for Lowe's attack.

Another dimension is the \emph{level of abstraction} for the specification, establishing whether a task should be expanded into sub-tasks or not. For example, the task of fetching a nonce may be expanded into accessing a random number generator, running it, and receiving its output securely, and these may be further expanded in turn.

From the security analysis standpoint, the levels of detail and of abstraction must be chosen with care in order to limit the necessary assumptions to realism.

\section{Roles}\label{sec:roles}
A \emph{role} is a set of principals who perform a specific set, possibly of cardinality one, of tasks. During exams, an obvious role is the \emph{candidate} role, of taking the exam to get a mark that may accord the candidate a goal --- such as obtaining a qualification, passing a periodical academic assessment, or being selected through a public examination. Of course, the candidate role could be specified, if needed, at a lower level of abstraction, and examples can be derived from actual protocol specifications. Other possible roles, also called \emph{authority roles}, are as follows. \index{assessment}

\begin{itemize}

 \item The \emph{registrar} role, of checking the eligibility of candidates who wish to take an exam, and of populating a list of registered candidates accordingly.

 \item The \emph{question committee} role, of building the tests and passing them to invigilators, and of forming the test answers in case of multiple-choice tests and passing them to the examiners.
 
 \item The \emph{moderator} role, of liaising with the question committee to independently ensure that the tests conform to pre-existing quality standards, such as readability and appropriateness.
 
 \item The \emph{invigilator} role, of distributing tests to candidates, of checking candidates' identities, of following candidates while they take their test preventing them from misbehaving.
 
 \item The \emph{collector} role, of collecting the tests from the candidates at the end of the exam time, and distributing the test answers to the examiners.
 
 \item The \emph{examiner} role, of reading the test answers and of producing adequate marks for them.
  
 \item The \emph{recorder} role, of keeping records of what candidates received what marks at the exam.

 \item The \emph{notifier} role, of informing the candidates of the marks that their respective tests received, and of storing this information with some recorders.
 
 \item The \emph{observer} role, of watching everything.
 
\end{itemize}

It can be seen that each authority role clearly indicates its set of tasks, demonstrating the levels of detail and of abstraction that we advocate. If an exam features an additional task, then this could either extend an existing role or form a new role. Also, in order to meet the security requirements, an exam may allow two or more roles to merge into one, or may prescribe splitting a role into two or more. For example, the role of question committee can be split into exam convener, question setter, and question reviewer, as practised in some universities.
Typically, candidate and authority roles cannot be merged (and still meet the security requirements), except with exams such as MOOC \citeltex{mooctr}, where homeworks are peer-reviewed, namely candidates mark each other. \index{MOOC}

\section{Principals}\label{sec:principals}

Exams see the participation of a number of \emph{principals}, each playing one or more of the roles defined above. Principals may change depending on the specific exam, and various examples can be made. 

At university, the candidate role is played by students, while the roles of invigilator, collector and examiner are sometimes played by a single lecturer. At an extreme, there are only two principals, with a student as a candidate and a lecturer playing all authority roles.
Today, ProctorU \citeweb{proctoru} invigilates via webcams the candidates who take exams from home. At public examinations, the Police could take the invigilator role.

Principals are not necessarily humans. They may as well be pieces of software playing various roles, such as invigilator, by filming candidates during testing, or examiner, by marking multiple-choice tests mechanically.

Irrespectively or whether they are human or not, it must be assumed that principals may act somewhat maliciously. We anticipate that such assumptions are fundamental when considering corrupted principals in the formal definition of our security requirements. In fact, they will define a threat model, as we shall see below.

\section{Phases}\label{sec:phases}
We identify the four main phases that typically take place sequentially during an exam. They will be further detailed and specified by actual exams in the sequel of this work.

\begin{itemize}

\item At \emph{preparation}, certain authorities, typically registrars, file a new exam, and check the standard eligibility criteria, such as correct payment of fees and adequate previous qualifications, of candidates who wish to take the exam. Only those who satisfy the criteria get successfully registered, and the authorities ultimately produce a list of candidates registered for the exam. Similarly, the authorities might produce a list of eligible examiners. \index{eligible examiners}
Most importantly, this phase includes the preparation of tests and all the relevant material for the subsequent phase. For instance, creation of questions, printing of tests, and generation of pseudonyms to anoymise tests are tasks accomplished in this phase.


\item At \emph{testing}, each registered candidate gets a test containing a number of questions, which were previously built by authorities, normally question committee and moderators. Other authorities, typically invigilators, watch candidates through this phase. Each candidate answers their test, and may have to complete it with their personal details. The candidate then submits their test answers to an authority, such as an invigilator.

\item At \emph{marking}, the test answers of all candidates reach the examiner authority for evaluation. More precisely, the authority reads the test answers and evaluates their adherence to the required knowledge, then forming a mark, chosen on a given scale, for each test. Some real-world example scales are: pass/fail, A to E, 60\% to 100\% and 18 to 30. With multiple-choice tests, the examiner authority could be a computer program.

\item At \emph{notification}, an authority, commonly a notifier, gives each candidate the mark for the test answers the candidate submitted; either beforehand or afterwards, the notifier also stores this information with a recorder, commonly a server equipped with a DBMS.

\end{itemize}

Any subset of phases may either take place \emph{on site}, with candidates meeting the authorities face-to-face, or \emph{off site} otherwise. Regulated by the specific application requirements, these features will shape up the exam experience. 

Moreover, any subset of these phases may take place either traditionally, namely by pen and paper, or on computer. For example, we observed above that marking can be easily computer assisted with multiple-choice tests; similarly, notification could take place via dedicated workstations installed in the exam site.

The specification of the exam phases \index{exam phase} clarifies the terminology that is used coherently throughout this dissertation, but it may be useful to the reader if we point out that various synonyms are used in the literature. For example, ``registration'' or ``setup'' may refer to the preparation phase; ``examination'' is often taken to indicate  the testing phase; ``evaluation''  or ``grading'' may indicate the marking phase; ``exam'', ``examination'', or  ``assessment'' may sometimes even refer to the full sequence of phases. \index{assessment}

\section{Threats}\label{sec:threat}
A number of threats could be envisaged against exams, and some basic threats are enumerated here.

Threats may derive from each task. For example, even the task of printing may invite the principal who performs it to alter the printout or not print at all. We assume each principal to be rational in the sense that the principal does not misbehave unless there is a clear benefit for them or, in case of \emph{collusion}, for another principal.

We conveniently define threats on a per-role basis.
Therefore, augmenting a role with additional tasks would require extending the role-specific threats; adding new roles formed of new tasks would require extending the threats over the role; merging roles would yield the union of the threats of the original roles; splitting a role would partition its threats enabling each formed role to pose the threats deriving from its tasks. 


We define some basic threats coming from the preeminent roles used in the sequel of this dissertation: the candidate, the authority, and the observer roles. A specific protocol shall 
customise the list of the threats according its roles.

The threat model is the standard Dolev-Yao \citeltex{DY83} over the roles (or their portions) that are impersonated by computer programs. Additionally, the roles (or their portions) that are impersonated by humans pose the threats detailed below.

The corrupted candidate performs any tasks in order to:
\begin{itemize}
 \item register for an exam without being eligible;
 \item register on behalf of someone else;
 \item answer their test with knowledge obtained by cheating;
 \item get a higher mark than what the examiner assigns to their test.
\end{itemize}
These threats are significant. Eligibility criteria may be stringent, such as payment of fees and restricted participation to certain exam dates, hence the interest in misbehaviour. Candidates' attempts at cheating, for example by consulting books, and at getting a better mark than theirs are well known. In particular, the latter may see a candidate send someone else, more knowledgeable than them, to sit for the exam on the candidate's behalf, or see a candidate  swap their mark with that of another candidate known to be very knowledgeable. Thus candidates may collude each other to achieve their goal. It can be anticipated that these threats demand effective authorisation, authentication, invigilation \index{invigilation} and marking procedures.


The corrupted authority performs any tasks in order to:
\begin{itemize}
 \item assign an unfair mark to a specific candidate, namely to over-mark or under-mark her, or assign no mark at all.
\end{itemize}
This is the fundamental threat authorities may pose to candidates. This threat may hinder students' spontaneity during University lectures. This threat is well known in academic conferences, and is partially addressed with blind reviews. More seriously it has also brought corruption \index{corruption} into public competitions. Arguably, it calls for anonymous marking, verifiability, and accountability requirements.

The observer performs any tasks in order to:
\begin{itemize}
 \item gather any private information.
\end{itemize}
Observers may be allowed to watch parts of the exam, typically the candidates while they take their tests, to raise public acceptance of the regularity of procedures. They may, however, have malicious intentions and seek out any form of private information such as candidates' questions and marks. If the exam is somewhat assisted by computers or by the Internet, then this threat becomes a digital one.


\section{Taxonomy}\label{sec:taxonomy}
Having seen the phases of an exam, we can specify the dictionary definition of the word ``exam'' conveniently for our purposes.
\begin{prop}[Exam]\label{def:exam}
An \emph{exam} is a formal test taken to show candidate's knowledge of a subject.
It comprises the four sequential phases \emph{preparation}, \emph{testing}, \emph{marking} and \emph{notification}.
\end{prop}

Definition \ref{def:exam} only specifies the main functional requirement of an exam, that candidates take the exam. It purposely omits additional functional requirements that may depend on the application scenario, such as that candidates be allowed to register from home. 

\subsection{Exam Types}

Exams can be classified in various types according to the following definitions.
\begin{prop}[Computer-assisted exam]\label{def:eexam}
A \emph{computer-assisted exam} is an exam such that at least one of its phases receives some level of assistance from computers or Information Technology.
\end{prop} \index{computer-assisted exams}

At first glance, Definition \ref{def:eexam} appears to be too wide to the extent that every exam is a computer-assisted exam. 
For example, an exam that only requires computers to edit the questions could be classified as computer-assisted. However, as observed in Section \ref{sec:tasks}, the levels of detail and abstraction for the specification of the protocol should be considered to find the correct classification of the exam. Thus, if the way the questions are edited is not explicitly mentioned in the description of the  protocol, the exam should not be classified as computer-assisted.
A qualifying example of a computer-assisted exam is to allow candidates to register from home, but then continuing traditionally, namely fully on paper and without the use of computers.

\begin{prop}[Traditional exam or non-computer-assisted exam]\label{def:texam} \index{traditional exam}
A \phantom{sa} \mbox{\emph{traditional exam}} is an exam such that none of its phases receives any level of assistance from computers or Information Technology in general. A traditional exam is also said a \emph{non-computer-assisted exam} to indicate the absence of computer assistance.
\end{prop}

Definitions \ref{def:eexam} and \ref{def:texam} insist that exams can be partitioned between computer-assisted and traditional exams depending on computer assistance.
%
%
%
%
Having seen the main partition within exams, various types of e-exams can be defined. For brevity, it is useful to refer to the acronyms in Table \ref{table:a}.

\begin{table} [ht]
\centering
\begin{tabular}{r|l}
NCA & non-computer-assisted\\
CA & computer-assisted\\
CB & computer-based\\
IA & Internet-assisted\\
IB & Internet-based
\end{tabular}
\caption{Acronyms for exam types} 
\label{table:a}
\end{table}

\begin{prop}[Computer-based exam]\label{def:cbexam}
A \emph{computer-based exam} is an exam whose testing phase takes place fully on computer.
\end{prop}
We decide to pivot Definition \ref{def:eexam} around the testing phase because an exam is often somewhat simplistically understood as that phase alone in practice. 
The definition insists that the testing phase of a CB exam takes place fully on computer, ruling out exams where questions are given orally or are written on a board, which would only be CA exams. Clearly, a CB exam also is a CA exam but not vice versa.

\begin{prop}[Internet-assisted exam]\label{def:iaexam}
An \emph{Internet-assisted exam} is an \phantom{a} exam such that at least one of its phases receives some level of assistance from the Internet.
\end{prop}

It is logical that Definition \ref{def:eexam} and Definition \ref{def:iaexam} have the same structure, and similar considerations about levels of detail and abstraction apply here.
Definition \ref{def:iaexam} requires some use of the Internet in some phases. For example, an exam that only relies on the Internet to notify the candidates of their marks would be an IA exam. Clearly, an IA exam also is a CA exam. An IA exam may also be, but not necessarily, a CB exam. From a set-theory standpoint, if $\cal CB$ is the set of all CB exams, and $\cal IA$ is the set of all IA exams, it follows that $\cal CB$ and $\cal IA$ intersect but do not coincide.

\begin{prop}[Internet-based exam]\label{def:ibexam}
An \emph{Internet-based exam} is an exam whose testing phase takes place fully over the Internet.
\end{prop}

The formulation of Definition \ref{def:ibexam} closely maps one of Definition \ref{def:cbexam}. Clearly, an IB exam is also an IA exam; an IB exam must also be a CB exam because the testing phase could not happen fully over the Internet without happening fully over some computer (or similar devices such as smartphones). 

Also, Definition \ref{def:iaexam} and Definition \ref{def:ibexam} purposely omit the specification of the venue where the exam phases happen, whether locally, at the hosting institution's premises, or remotely from the candidate's place. For example, even an IB exam could happen locally. \index{exam phase}

\begin{table}
\begin{center}
\begin{tabular}{r|l}
$\cal NCA$ & set of all NCA exams\\
$\cal CA$ & set of all CA exams\\
$\cal CB$ & set of all CB exams\\
$\cal IA$ & set of all IA exams\\
$\cal IB$ & set of all IB exams
\end{tabular}
\end{center}
\caption{Sets of exam types}\label{table:examsets}
\end{table}

\begin{figure} [ht]
\centering
\includegraphics[scale=1.2]{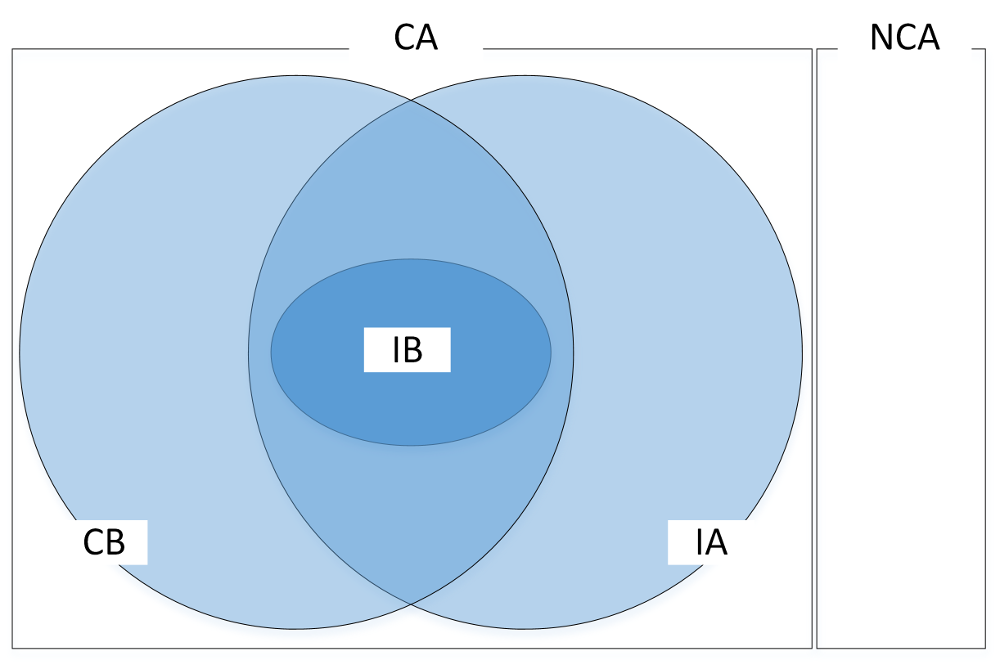}
\caption{A set-theory representation of exams} 
\label{fig:examtypes}
\end{figure}

Having defined the various types of exams, the main building blocks for our taxonomy of exams are available. Still, for each exam type, it is useful to define the set of all exams of that type, as Table \ref{table:examsets} does in a self-explaining form. Now, the exam taxonomy can be introduced using a set-theory notation; it is in Figure \ref{fig:examtypes}, and demonstrates the relations discussed above between the various exam types. 

As anticipated above, all roles span uniformly across the entire taxonomy because principals can either be human or not. Other objects may have to be re-interpreted depending on the exam type. For example, a test consists of some paper (showing questions and, later, answers) for all exams in $\cal NCA$, of some file(s) for all exams in $\cal CB$, of either some paper or some file(s) for all exams in ${\cal CA} \setminus {\cal CB}$.

The taxonomy in Figure \ref{fig:examtypes} seems to bear potential to capture more general scenarios than just exams, and precisely those commonly addressed as \emph{collaborative working environment} \citeltex{FJ95}. However, one would need to map the exam phases identified above to the main phases of other scenarios, and this exceeds the aims of the present research.

\subsection{Exam Categories}\label{sec:categories}

Exams can be categorised as \emph{written exams}, \emph{oral exams}, or \emph{performance exams}, as outlined below. Written exams are usually preferred when the number of candidates is high. Oral exams are considered the most effective practice to assess knowledge, but examiners tend to mark candidates less objectively compared to written exams. Performance exams are impractical when many candidates need to be evaluated. \index{written exam} \index{oral exam} \index{performance exam}

Our taxonomy was built to reflect written exams, where testing takes place in writing, namely with questions and answers given using various combinations of reading and (hand- or type-)writing on paper, boards, screens, etc.
However, 
our taxonomy can be stretched out to reflect also oral and performance exams as explained in the following, although this exceeds the scope of this dissertation. 

Oral exams see questions and answers given orally, either synchronously by interviews or asynchronously by some proxying. Synchronous exams can be represented by our taxonomy, as one can easily realise by looking at Figure \ref{fig:examtypes} with synchronous oral exams in mind. For example, a synchronous oral exam in $\cal IB$ may have the testing phase via videoconferencing; even one in the set ${\cal CB} \setminus {\cal IA}$ makes perfect sense if recording the interview is required.

Also asynchronous oral exams comply with our taxonomy. Through the use of techniques of speech processing or audio sampling, asynchronous oral exams could have the testing phase via voice chats. Also in this case exams in ${\cal CB} \setminus {\cal IA}$ are valuable, by yielding the full chat history of the testing. Envisaging an asynchronous oral exam that is NCA requires admitting the role of proxy to be played by a human without any computer assistance. If this is deemed impractical or implausible, then the taxonomy could be easily pruned of the NCA part, only for the category of asynchronous oral exams.

Performance exams differ from oral exams because the former require a candidate to actually perform an activity, rather than answer questions orally. The same techniques envisaged for synchronous and asynchronous oral exams (\eg videoconferencing, audio sampling, speech processing) hold for performance exams as well as the considerations that regard the exam types. 

There may also be \emph{hybrid exams}, for instance where testing combines written and oral means: a lecturer could dictate the questions for candidates to answer in writing. Although it might be less popular than written or oral exams, we argue that this category too could be represented with our taxonomy, perhaps with minor adjustments.


\chapter{Formalising Authentication and Privacy}\label{chap:formal}

%

Security protocols are distributed algorithms that use cryptography to achieve some security goals.
The design of such protocols can be a difficult task that might lead to serious security flaws.
The literature is full of security  protocols and standards that have been demonstrated to be bugged \citeltex{CJS+07,HM05,BGW01}, a trend that is unlikely to change \citeltex{ACC+08,AMR+14}. 
Formal approaches for the analysis of security protocol have been successfully used to  discover security flaws, such as the famous one in the Needham-Schroeder Public Key protocol \citeltex{Lowe96}, and eventually helped in fixing the protocols with some level of guarantee. Experience shows that formal approaches are important also in the design phase of protocols as they force designers to have a deep understanding of their models and what they aim to achieve. With this belief, we intend to formulate a formal framework for the design and analysis of secure exam protocols.

Similar to other systems like voting and auction, exams have not been designed with security in mind.
This is problematic since the recent growth in use of exam protocols has not been followed, nor preceded, by a rigorous understanding and analysis of their security. 
Although there are recent proposals for exam protocols with a security-by-design approach \citeltex{huszti10,DBLP:conf/IEEEares/Castella-RocaHD06,DBLP:conf/itcc/Herrera-JoancomartiPC04,BCC+11}, no formal analyses have been conducted against these proposals. Since almost all the existing exam protocols normally assume trusted authorities, only a small set of requirements, namely the ones concerning authentication of candidates, are usually considered in the analysis. As already noted in the previous chapter, exam authorities can be corrupted as well as can candidates,  and exam protocols should consider a larger set of requirements including privacy.

We identify and formalise a number of authentication and privacy requirements for exams. Although we find them highly desirable out of personal experience and discussions with colleagues, the list of requirements is not meant to be universal or exhaustive. It means that certain exam protocols might demand additional requirements. However, we consider our set of requirements to be 
\emph{fundamental} as similar requirements can be found in other independent works \citeltex{EW05,FurnellOKSBGR98}, still only informally.

From the formalisation of the requirements, we build a framework for the authentication and privacy analysis of exam protocols. The framework can be used to analyse different types of exams as corroborated by the three protocols analysed in this dissertation. The requirements are specified in the \appi calculus~\citeltex{AF01}, a process calculus that extends the $\pi$-calculus with support for a wide variety of cryptographic properties.

\paragraph*{Outline of the chapter.}
Section~\ref{sec:related} examines different formal approaches for the analysis of security protocols.
Section~\ref{sec:picalculus} outlines the basic constituents of the \appi-calculus.
Section~\ref{sec:exams} introduces the formal framework by specifying the formal model of an exam.
Section~\ref{sec:requirements}  continues the description of the framework by specifying authentication and privacy requirements for exam in the \appi-calculus. 
Section~\ref{sec:H} contains the description and analysis of the \Huszti~\citeltex{huszti10} exam protocol, which validates the framework. Moreover, the section discusses findings and proposes fixes to the protocol.
Finally, Section~\ref{sec:conclusion} concludes the chapter.

\section{Related Work}\label{sec:related}
Security protocols have been  historically analysed with two different approaches, one based on symbolic model, and one based on computational complexity theory.
Symbolic analysis methods for protocol analysis find their root in the seminal works of Needham and Schroeder \citeltex{NS78} and  Dolev and Yao \citeltex{DY83}, which assume perfect cryptography and an unbounded active attacker who controls the entire network. Formal logic and automated tools based on the symbolic model have been used successfully to analyse security protocols.

Methods based on computational complexity theory have been initially developed by Goldwasser and Micali \citeltex{GM84}. Analysis in the computational model usually see the attacker as a polynomial probabilistic Turing machine. Such analysis is deemed to be more realistic because avoids the perfect cryptography assumption, thus provides more insights about vulnerabilities of security protocols: an attack in the symbolic model leads to an attack in the computational model, while the contrary is not true in general. However, methods based on computational complexity theory are harder to mechanise (only recently mechanised tools have been proposed to assist manual proofs \citeltex{Blanchet08,BGZ09}) and are not suitable for automation. Proofs are mostly manual and difficult as they involve to reason about probability and computational complexity, hence prone to human errors. We thus choose to develop the formal framework for exam using the symbolic model. 

\paragraph{The symbolic model.} Several symbolic techniques have been proposed over the last 25 years. Merritt \citeltex{Merritt83} proposed to describe protocols by rewrite rules. Burrows \etal \citeltex{BAN90} introduced the so-called \emph{BAN logic} to reason about authentication goals. Meadows \citeltex{Meadows96} proposed a language based on events for specifying security protocols, and introduced the \emph{NRL Analyzer}, a model checker capable to verify authentication and secrecy goals. Ryan and Schneider \citeltex{Ryan:MAS} pioneered the idea of using process algebras for the analysis of security protocols. In Hoare's CSP \citeltex{Hoare, Schneider97} the protocol's principals are naturally modelled as processes, while security goals are modelled as reachability properties. Paulson \citeltex{Paulson98} introduced the \emph{Inductive Approach} wherein principals are modelled as a set of rules that inductively define an unbounded set of traces. A protocol guarantees a security goal if the goal inductively holds for all possible traces. Thayer \citeltex{THG+99} proposed the strand spaces approach, which features an intuitive way to reason about traces generated by the security protocols: a strand is a sequence of events in which a protocol's principal may participate. Abadi and Gordon \citeltex{AG99} introduced the spi-calculus, a process algebra that extends the $\pi$-calculus \citeltex{MPW92} with explicit representation of cryptographic operations. The next chapter extensively discusses the \appi-calculus \citeltex{AF01}, which extends further the spi-calculus with a richer algebra for the modelling of cryptographic primitives.

\paragraph{Security requirements.} The key requirements at  the heart of information security are authentication and privacy. The notion of authentication has found different flavours in the literature. Gollmann \citeltex{Gollmann} argued that capturing the notion of authentication is difficult. Lowe \citeltex{Lowe97} taxomised authentication in \emph{Aliveness}, \emph{Weak agreement}, \emph{Non-injective agreement}, and \emph{Injective agreement}. Despite the different interpretations, most of the approaches agree that authentication is a correspondence property: if the principal \emph{A} accepts a message from \emph{B}, then the principal \emph{B} has actually sent that message to \emph{A}. Typically, authentication can be captured by introducing  \emph{events} into the specification of protocol roles. An event explicitly signals that a principal has completed part of a run of the protocol, and what data has used in that run. 
The placement of events into the protocol, the corresponding data, and the relationship between events can thus capture a precise authentication goal.

Several definitions for privacy have been proposed in the literature, such as secrecy, anonymity, unlinkability, and untraceability. Dolev and Yao \citeltex{DY83} specified secrecy as a reachability \index{unlinkability} property, meaning that a secret is not made available or disclosed to the attacker. Schneider and Sidiropoulos \citeltex{SS96} formalised anonymity in CSP as the impossibility for an attacker to link a principal with a message. Deursen \etal \citeltex{DMR08} clarified that unlinkability considers whether links can be established between sender
and receiver principals, while untraceability considers whether different communications can be attributed to the same principal. Ryan and Schneider \citeltex{RS01} observed that the notion of non-interference \citeltex{GM82} and bisimulation can be used to express security requirements. On this vein, several formal definitions of privacy have been proposed as equivalence-based properties. For example, \emph{observational equivalence} states that an observer cannot distinguish any difference between two processes, although they might perform different computations on different data.

\paragraph{Tools.} There are several tools that support the automatic analysis of authentication and privacy. FDR \citeltex{Roscoe} is the most popular model checker for CSP, and contributed to the discovery of Lowe's attack. AVISPA \citeltex{AVISPA} combines four techniques to analyse reachability properties (authentication and secrecy), and recently a security flaw in Google implementation of SAML 2.0 Single Sign-On Authentication was discovered \citeltex{ACC+08} using that tool. ProVerif \citeltex{proverif} is an automatic protocol analyser that can prove reachability and equivalence-based properties. The input language of ProVerif is the \appi-calculus, which the tool automatically translates to Horn clauses. ProVerif proved to be one of the automatic analysers with the best performances \citeltex{CLN09}. Moreover, it allows user-defined equational theories that extend security models with algebraic properties in order to weaken the perfect cryptography assumption.
We thus choose formalise our security requirements in  the \appi-calculus and analyse the exam protocols with ProVerif.



\paragraph{Comparison with voting and auctions.} Few  works~\citeltex{EW05,FurnellOKSBGR98} list
 a number of security requirements for exams, still only informally.
Similar domains such as voting and auction have seen significant advances during the last few years, mostly in formal definitions of privacy requirements. Novel voting protocols \citeltex{PAV06, Adida08} have been formally analysed for a family of privacy requirements, such as ballot privacy, receipt-freeness, and coercion-resistance \citeltex{DBLP:conf/icc/DreierLL12,DBLP:conf/csfw/BackesHM08,DKR-jcs08}, while auction protocols have been analysed for privacy and fairness requirements  ~\citeltex{Dong10,DBLP:conf/post/DreierLL13,DBLP:conf/ccs/DreierJL13}.
It can be observed that only a few of security requirement definitions proposed in voting and auction domains are similar to ones we introduce for exams.  For example,
answers originated by eligible candidates \index{eligible candidates} should be marked in an exam. In the same way, only ballots cast by eligible voters should be recorded in a voting system, and only offers submitted by eligible bidders should be considered in an auction. 

The requirement of mark privacy for exams is intuitively close to the definitions of ballot privacy for voting and losing bid privacy for auctions.
However, it can be noticed a subtle difference: voting usually requires ballot privacy also towards the voting authority, while a mark eventually needs to be associated to the candidate usually by means of an exam authority.

Other requirements have fundamental differences. In exams, the association between an answer and its author should be preserved --- even in the presence of colluding candidates. Conversely, 
vote authorship is not a requirement for voting, in fact unlinkability between voter and vote is a desired property.  \index{unlinkability}
A peculiar  requirement for exam is to keep the questions secret until the testing phase. In voting, the list of candidates is public, while in auctions the list of goods is normally known to bidders.
Moreover, exams may require anonymous marking, namely answers are marked while ignoring their authors. This signifies a sort of fixed-term anonymity since each mark eventually needs to be assigned to the corresponding candidate.

\section{The \Appi calculus}\label{sec:picalculus}
The \appi calculus~\citeltex{AF01} is a formal language
for the description and analysis of security protocols, in which 
principals are represented as processes.
Its syntax consists of \emph{names}, \emph{variables}, and \emph{signatures}. The latter are function symbols each with an arity. Names represent channels and data, while function symbols represent cryptographic primitives such as encryption, decryption, digital signature, and hash functions.
Function symbols applied to names and variables generate \emph{terms}.
Tuples of arity $l$, such as $n_1,\ldots,n_l$, can be abbreviated in $\tilde{n}$.

\paragraph{Equational theories.} Whereas the $\pi$-calculus supports only a fixed set of cryptographic primitives, the \appi calculus allows one to model user-defined primitives by means of \emph{equational theories}.
An equational theory $E$ describes the equations that hold on terms
built from the signature. Terms are related by an equivalence relation $=$ induced by $E$. For instance, the equation $\dec{\enc{m}{\pk{k}}}{k} = m$  
models an asymmetric encryption scheme. The term $m$ is the message, the term $k$
is the secret key, the function $pk(k)$ models the corresponding public key, the term $enc$ models the encryption function, and the term $dec$ models the decryption function.

\paragraph{Processes.} The grammar for \emph{plain} processes is outlined in  Figure~\ref{plain processes}.
The null process $0$ does nothing; the process $P|Q$ is the parallel composition of processes $P$ and $Q$; the process $!P$ behaves as an unbounded number of copies of processes $P$
running in parallel; the process $\nu n.P$ generates a new \emph{private} name $n$, then
behaves like $P$; the conditional process `if $m=m'$ then $P$ else $Q$' behaves like the process $P$ if $m=m'$ and like the process $Q$ otherwise. For brevity, one can omit sub-term `else $Q$' when $Q$ is 0; the process $\inmessage{u}{x}.P$ awaits for an input from channel $u$, then behaves as the process $P$ with the received message replacing the variable $x$; Finally, the process $\outmessage{u}{m}.P$ outputs the message $m$ on the channel $u$, then behaves as the process $P$. For brevity, one can omit $.P$ when the process $P$ is 0.

\begin{figure}[htb]
\centering
\begin{tabular}{l l}
$P, Q, R :: =$ &  plain processes\\
0 &   null process\\ 
$P|Q$ & parallel composition\\
$!P$ & replication\\
$\nu n.P$ &name restriction (new)\\
if $m=m'$ then $P$ else $Q$~~ & conditional\\
$\inmessage{u}{x}.P$ & message input\\
$\outmessage{u}{m}.P$ & message output
\end{tabular}
\caption{The grammar for plain processes in the \appi calculus}\label{plain processes} 
\end{figure}

The grammar for \emph{extended} processes is outlined in  Figure~\ref{extended processes}. 
Extended processes model the knowledge exposed to the attacker.
An \emph{active substitution} $\{^m/_x\}$ is a process that replaces the variable $x$ with the term $m$.
We refer to a substitution also with  $\sigma$. We use $m \sigma$ to refer to the result of applying $\sigma$ to $m$. 
The sets $\freev{A}$,
$\boundv{A}$, $\freen{A}$ and $\boundn{A}$ respectively include free variables, bound variables,
free names, and bound names of the process $A$. An extended process is \emph{closed}
if all variables are bound or defined by an active substitution. 

\begin{figure}
\centering
\begin{tabular}{l l}
$A, B, C :: =$~~~~~ &  extended processes\\
P &   plain process\\ 
$A|B$ & parallel composition\\
$!P$ & replication\\
$\nu n.A$ &name restriction \\
$\nu x.A$ & variable restriction \\
$\{^m/_x\}$ & active substitution
\end{tabular}
\caption{The grammar for extended processes in the \appi calculus}\label{extended processes}
\end{figure}

The definition of \emph{corrupted process} \citeltex{DKR-csfw06} is useful to model corrupted principals who actively collaborate with the attacker. 

The definition outlined below specifies how to transform a process into a corrupted process. This transformation is based on two channels $c_1$ and $c_2$, which the process uses to receive and send data to the attacker.

\begin{prop}{\bf(Corrupted process $P^{c_1, c_2}$)}\label{def:cp} Let $P$ be a plain process and $c_1$, $c_2$ be two channel names such that $c_1,c_2 \notin \freen{P} \cup \boundn{P}$. The \emph{corrupted process} $P^{c_1,c_2}$ is defined as follows:
\begin{itemize}
\item $0^{c_1, c_2} ~ \hat{=} ~ 0$, 
 \item $(P|Q)^{c_1, c_2} ~ \hat{=} ~ P^{c_1, c_2}|Q^{c_1, c_2}$, 
 \item $(!P)^{c_1, c_2} ~ \hat{=} ~ !P^{c_1, c_2}$, 
 \item $(\nu n.P)^{c_1, c_2} ~ \hat{=} ~ \nu n.\outmessage{c_1}{n}.P^{c_1, c_2}$ if $n$ is a name of base type, otherwise $(\nu n.P)^{c_1, c_2} ~ \hat{=} ~ \nu n.P^{c_1, c_2}$, 
 \item $(\mbox{if } m=m' \mbox{ then } P \mbox{ else }  Q)^{c_1, c_2} ~ \hat{=} ~ \inmessage{c_2}{x}.\mbox{if } x=true \mbox{ then } P^{c_1, c_2} \mbox{ else }$ $Q^{c_1, c_2}$ where 
 $x$ is a fresh variable and true is a constant, 
 \item $(\inmessage{u}{x}.P)^{c_1, c_2} ~ \hat{=} ~  \inmessage{u}{x}.\outmessage{c_1}{x}.P^{c_1, c_2}$ if $x$ is a variable of base type, otherwise
 $(\inmessage{u}{x}.P)^{c_1, c_2} ~ \hat{=} ~  \inmessage{u}{x}.P^{c_1, c_2}$, 
  \item $(\outmessage{u}{m}.P)^{c_1, c_2} ~ \hat{=} ~  \inmessage{c_2}{x}.\outmessage{u}{x}.P^{c_1, c_2}$, where $x$ is a fresh variable. 
\end{itemize}
\end{prop}

A \emph{frame} is an extended process built from 0 and active substitutions of the form
$\{^m/_x \}$ by parallel composition and restriction. We use  $\Phi$ and $\Psi$ to range over frames. The domain $dom(\Phi)$ of a frame  $\Phi$
is the set of the variables for which $\Phi$ defines a substitution. Every extended process $A$ can be mapped to a frame $\Phi(A)$ 
by replacing every plain process in  $A$ with 0.  The frame $\Phi(A)$ can be seen as a representation of the knowledge of the process to its environment. 

Finally, a \emph{context} is an extended process $C$ with a hole, written $C$[\_]. It can be used to represent the environment in which the process is run.


\subsubsection*{Reachability and Correspondence properties} In the \appi-calculus, secrecy can be modelled as a reachability property. The secrecy of a term $m$ is preserved if an attacker, defined as an arbitrary process, cannot construct $m$ from any run of the protocol.
The definitions of \emph{name distinct}, and \emph{reachability-based secrecy} \citeltex{RS11}  models secrecy. A name-distinct process signifies that the names mentioned in a term appear unambiguously in the process either as free or bound names. The definition of reachability-based secrecy says that an attacker cannot build a process $A$ that can output the secret term $m$.

\begin{prop}{\bf(name-distinct for $\tilde{m}$)} A plain process $P$ is \emph{name-distinct for} a set of names $\tilde{m}$ if $\tilde{m} \cap \freen{P} \cap \boundn{P} =  \emptyset$ and for each name $n \in \tilde{m} \cap \boundn{P}$ there is exactly one restriction $\nu n$ in $P$. \end{prop}

\begin{prop}{\bf(Reachability-based secrecy)} A plain process $P$ that is name-distinct for the names mentioned in the term $m$ preserve \emph{reachability-based secrecy} if there is no plain process $A$ such that $(\freen{A} \cup \boundn{A}) \cap \boundn{P}=\emptyset$ and $P | A$ can output $m$. \end{prop}

In the \appi-calculus, authentication can be defined using \emph{correspondence assertions} \citeltex{WL93}. An event \texttt{e} is a message emitted into a special channel that is not under the  control of the attacker. Events may contain arguments $M_1,...M_n$, which are never revealed to the attacker. Events do not change the behaviour of the process in which they are located, but normally flag important steps in the execution of the protocol. 
To model correspondence assertions, we annotate processes with events such as $\mathtt{e}\langle M_1,...M_n \rangle$ and reason about the relationships ($\leadsto$) between events and their arguments in the form \emph{``if an event $\mathtt{e}\langle M_1,...M_n \rangle$ has been executed, then event $\mathtt{e'}\langle N_1,...N_n \rangle$ has been previously executed''}, which is formalised as the following definition.

\begin{prop}{\bf(Correspondence assertion)} A \emph{correspondence assertion} is a formula of the form $\mathtt{e}\langle M_1,...M_i \rangle \leadsto   \mathtt{e'}\langle N_1,...N_j \rangle$. \end{prop}

By adding the keyword \emph{inj}, it is possible to model an injective correspondence assertion, which signifies that \emph{``if an event $\mathtt{e}\langle M_1,...M_n \rangle$ has been executed, then a distinct earlier occurrence of event $\mathtt{e'}\langle N_1,...N_n \rangle$ has been previously executed''}.

\begin{prop}{\bf(Injective correspondence assertion)} An \emph{injective correspondence assertion} is a formula of the form $\mathtt{e}\langle M_1,...M_i \rangle \leadsto \mathit{inj}~   \mathtt{e'}\langle N_1,...N_j \rangle$. \end{prop}

Authentication is only one of the requirement that can be modelled by correspondence assertions. Correspondence assertions can, for instance, capture also verifiability requirements, as we shall see in chapter \ref{chap:verifiability}.


\subsubsection*{Observational Equivalence } The notion of observation equivalence can  capture privacy requirements. Informally, two processes are observational equivalent if an observer cannot distinguish the processes despite they might handle different data or perform different computations.
To formalise observational equivalence, we first introduce the notion of \emph{internal reduction} ($\to$), which captures the evolution of a process with respect to communication and conditionals as:

\begin{center}
\begin{itemize}
\item $\outmessage{c}{x}.P | \inmessage{c}{x}.Q \to P|Q$;
\item if $n=n$ then $P$ else $Q \to P$;
\item if $l=m$ then $P$ else $Q \to Q$, {\small where $L$ and $m$ are not equivalent}.
\end{itemize}
\end{center}

\begin{prop}{\bf(Observational Equivalence)}\label{def:oe} \emph{Observational equivalence} ($\approx$) is the   largest symmetric relation $\mathcal{R}$ on extended processes such that $A~\mathcal{R}~B$ implies: 
\begin{enumerate}
\item if $A\to^{\ast} C[\outmessage{c}{M}.P]$, then $B\to^* C[\outmessage{c}{M}.P]$;
\item if $A \rightarrow^{\ast} A'$, then $B \rightarrow^{\ast} B'$ and $A'~\mathcal{R}~B'$ for some $B'$;
\item $C[A]~\mathcal{R}~C[B]$ for all context $C[\_]$.
\end{enumerate}
\end{prop}

The relation $\rightarrow^{\ast}$ expresses the transitive and reflexive closure of the relation $\rightarrow$.
Definition \ref{def:oe} says that two processes $A$ and $B$ are observational equivalent if: 1. the process $A$ evolves to a process that can output on channel $c$, also $B$ can evolve to a similar process; 2. 
if $A$ evolves to some process $A'$, also $B$ can evolve to some process $B'$, and $A'$ and $B'$ are observational equivalent; 3. for all contexts, $C[A]$ and $C[B]$ are observational equivalent.

The definition of observational equivalence is impracticable because requires the quantification over contexts. To avoid quantification over contexts, we first introduce the definitions of \emph{equality of terms} and \emph{static equivalence}. The latter captures the static part of observational equivalence as it only examines the current state of the processes. Then, we formalise \emph{labelled bisimilarity}, which captures the dynamic behaviour of the processes and is equivalent to observational equivalence. In fact, Abadi and Fournet \citeltex{AF01} proved that observational equivalence and labelled bisimilarity coincide.

\begin{prop}{\bf(Equality of Terms)} Two terms $m$ and $m'$ are equal in
the frame $\Phi$, written $(m=m')\Phi$, if 
$\Phi \equiv \nu \tilde{n}.\sigma$, $m\sigma = m' \sigma$ and 
$\{\tilde{n}\} \cap (\freen{m} \cup \freen{m'}) = \emptyset$, 
for some names $\tilde{n}$ and some substitution $\sigma$.  \end{prop}

\begin{prop}{\bf(Static Equivalence)} 
Two closed frames $\Phi$ and $\Psi$ are \emph{statically equivalent}, written
$\Phi \approx_s \Psi$, if $dom(\Phi) = dom(\Psi)$, and for all terms $m$
and $m'$ we have that $(m=m')\Phi$ if and only if $(m=m')\Psi$. Two
extended processes $A$ and $B$ are statically equivalent, written
$A\approx_s B$ if their frames are statically equivalent.
\end{prop}
In fact, two processes are statically equivalent if all their previous operations gave the same results so that they cannot be distinguished from the messages they exchange with the environment.

\begin{prop}{\bf(Labelled Bisimilarity)}
\emph{Labelled bisimilarity} ($\approx_l$) is the $ $ largest symmetric relation $\mathcal{R}$ on extended processes, such that $A~\mathcal{R}~B$ implies:
\begin{enumerate}
\item $A \approx_s B$
\item if $A \rightarrow^{\ast} A'$, then $B \rightarrow^{\ast} B'$ and $A'~ \mathcal{R} ~B'$ for some $B'$
\item  if $A \overset{\alpha}{\rightarrow} A'$, $\freev{\alpha} \subseteq dom(A)$ and $\boundn{\alpha} \cap \freen{B} = \emptyset$, 
then $B \rightarrow^{\ast} \overset{\alpha}{\rightarrow} \rightarrow^{\ast} B'$ and $A' \mathcal{R} B'$ for some $B'$.
\end{enumerate}
\end{prop}

The relation $A\rightarrow^{\alpha}A'$ defines a labelled semantics that avoids the quantification over the contexts. It signifies that  $A$ can evolve to $A'$ using a labelled transition.
This happens when $A$ performs an input or an output due to some interaction with the environment, and $\alpha$ is the label standing for the involved action.

\section{Modelling Exams}\label{sec:exams}
The roles of an exam can be modelled as processes in the \appi calculus. These processes communicate via public or private channels, and can create fresh random values, which can serve as key or nonce, for example. 
Processes can perform tests and cryptographic operations, which are functions on terms with respect to an equational theory describing some algebraic properties.

The threat model of an exam protocol consists of a Dolev-Yao attacker who has full control of the network, namely of the public channels. The attacker can also inject messages of his choice into the public channels, and exploit the algebraic properties of cryptographic primitives due to an equational theory. Moreover, the ability of the attacker can be extended with corrupted principals according to Definition \ref{def:cp}.
However, the attacker has no control of private channels, which are normally used to model out-of-band communications between processes. The attacker cannot even know if any communication happens over private channels. 
Thus, he can eavesdrop, drop, substitute, duplicate, and delay messages that principals send one another over public channels. 

Having seen the basic constituents of the \appi-calculus, we can provide the definition of \emph{exam protocol} according the calculus.

\begin{prop}{\bf(Exam protocol)}\label{def:ep}
  An \emph{exam protocol} is a tuple $(C, E, Q, K,$ $ A_1,\ldots, A_l,
  \tilde{n}_p)$, where $C$ is the process executed by the candidates, $E$ is the process executed by the examiners, $Q$ is the process executed by the question committee, $K$ is the process executed by the collector, $A_1,\ldots, A_l$ are the processes executed by the remaining authorities, and $\tilde{n}_p$ is the set of private channel names. 
\end{prop} 

Note that we make explicit the examiner $E$, the question committee $Q$, and the collector $K$  among the authority processes although this is not strictly necessary. However, it turns out to be convenient for the formalisation of our security requirements.

All the principals playing the candidate role execute the same process $C$. However, each principal is instantiated with 
different variable values, \eg keys, identities, and answers. Similarly, each principal playing any one of the authority role (\eg examiner, question committee, collector, etc.) executes the respective  processes  with different values. 

\begin{prop}\label{def:epm}{\bf(Exam instance)}
An \emph{exam instance} of an exam protocol given by the tuple  $(C,  A_1,\ldots, A_l, \tilde{n})$ is a closed process \\ $\mathit{EP}=\nu \tilde{n}.
(C\sigma_{id_1}\sigma_{a_1} | \ldots
|$ $C\sigma_{id_j}\sigma_{a_j} | 
$ $E\sigma_{id'_1}\sigma_{m_1}| \ldots $
$|$ $E\sigma_{id'_k}\sigma_{m_k}|$ $Q \sigma_q |$ $K\sigma_{\mathit{test}} |$ $ A_{1} | \ldots
|A_{l})$, where 
\begin{itemize}
\item $\tilde{n}$ is the set of all restricted names, including
the private channels; 
\item $C\sigma_{id_i}\sigma_{a_i}$'s are
the processes run by the candidates, where the substitutions $\sigma_{id_i}$ and
$\sigma_{a_i}$ specify the identity and the answers associated with the $i^{th}$ candidate; 
\item $E\sigma_{id'_i}\sigma_{m_i}$'s are the processes run by the
examiner authorities, where the substitution $\sigma_{id'_i}$ and $\sigma_{m_i}$ specify the
identity and the mark associated with the $i^{th}$ examiner; 
\item $Q\sigma_q$ is the process run by the question committee authority, where the
substitution $\sigma_q$ specifies the exam questions; 
\item $K\sigma_{test}$ is the process run by the collector authority, where
the substitution $\sigma_{\mathit{test}}$ associates a test with an examiner for marking; 
\item $A_i$'s are the processes run by the remaining exam authorities.
\end{itemize}
    \end{prop}
Definitions \ref{def:ep}~and~\ref{def:epm} capture the levels of detail and abstraction advocated in chapter~\ref{chap:terminology}. The instance of an exam protocol can be customised by making processes $A_i$ explicit. For example, the exam instance can be expanded with processes that model a mixnet for the generation of test pseudonyms (see chapter \ref{chap:remark}), or a bulletin board that publishes the results of an exam (see chapter \ref{chap:wata}). 

As we shall see later, Definition \ref{def:epm} allows us to specify a considerable number of security requirements and is suitable for different types of exam protocols.
Moreover, it equally supports either machine or human examiners as principals that mark answers.


\section{Security Requirements}\label{sec:requirements}
We identify and formalise a set of fundamental \emph{authentication} and \emph{privacy} requirements in the \appi calculus. This set is not meant to be comprehensive, but it includes the basic security requirements that an exam protocol is normally expected to guarantee as corroborated in the literature~\citeltex{EW05,crisis13,FurnellOKSBGR98}. However, our set can be extended with additional security requirements. 

We introduce five authentication and five privacy requirements. The authentication requirements capture the associations between the candidate's identity, the answer, and the mark
being preserved through all the exam phases. When authentication holds there is no loss, no injection, and in general no manipulation of the exam tests from preparation to notification. \index{exam phase}
The five privacy requirements aim to capture  secrecy of marks, and anonymity of tests and examiners. 


\subsection{Authentication}
To model authentication requirements as correspondence properties, it is necessary to define a number of relevant events.
Events normally need to agree with some arguments to capture authentication. Thus, we  introduce the terms that serve as arguments in our events as follows.
\begin{itemize}
\item $\idcand$ refers to the identity of the candidate; 
\item $\ques$ denotes the question(s) of the test;
\item $\ans$ denotes the answer of a test; 
\item $\mrk$ denotes the mark assigned to the test;
\item $\idexam$ refers to the identity of the examiner;
\item $\idform$ refers to the identifier of the test.
\end{itemize}
The terms outlined above intuitively relates to the substitutions introduced in Definition~\ref{def:epm}. Their definitions are abstract enough to capture different exams. For example, the term $\idform$ may coincide with the identity of the candidate if the exam requires no blind marking, or may be a pseudonym to if the exam requires anonymous marking.

We define a list of six events that allow to specify five fundamental authentication requirements for exams. We stress that the list can be further extended to accommodate any additional requirements.

\begin{itemize}
\item $\reg{\idcand}$ means that the authority considers the candidate $\idcand$ registered for the exam. The event is inserted into the process of authority at the location where the registration of the candidate $\idcand$ concludes.
\item $\fil{\idcand}{\ques}{\ans}{\idform}$ means that the candidate $\idcand$ considers the test $\idform$, which consists of question $\ques$ and answer $\ans$, submitted for the exam. The event is inserted into the process of the candidate at the location where the test is sent to the collector.
\item $\collect{\idcand}{\ques}{\ans}{\idform}$ means that the collector accepts the test $\idform$, which originates from the candidate $\idcand$. The event is inserted into the process of the collector at the location where the test is considered as accepted.
\item $\distribute{\idcand}{\ques}{\ans}{\idform}{\idexam}$ means that the collector considers the test $\idform$, which originates from the candidate $\idcand$, associated with  the examiner $\idexam$ for marking.  The event is inserted into the process of the collector at the location where the test is distributed to the examiner.
\item $\marking{\ques}{\ans}{\mrk}{\idform}{\idexam}$ means that the examiner $\idexam$ considers the test $\idform$, which consists of question $\ques$ and answer $\ans$, evaluated with $\mrk$. 
The event is inserted into the process of the examiner at the location where the test is marked.
\item $\notify{\idcand}{\mrk}$: means that the candidate $\idcand$ accepts the mark $\mrk$. The event is inserted into the process of the candidate at the location where the mark is considered as accepted.
\end{itemize}

These events mark important steps of an exam protocol, and some can be associated with the phases of an exam. The event \texttt{registered} normally concludes the preparation phase, while \texttt{collected} concludes the testing phase. The event \texttt{distributed} begins the marking phase, which the event \texttt{marking} concludes. Finally, the event \texttt{notified} concludes the notification phase and the exam. 
Note that these events implicitly refer to the same exam session. However, one might want to parameterise all the events with a common term in order to distinguish among exam sessions.

The first authentication requirement we consider is \emph{\cautho}, which concerns preparation and testing. Informally, we want to capture the requirement that only registered candidates can take the exam. \index{Candidate Authorisation}
More specifically, the requirement says that if a candidate submits her test, then the candidate was correctly registered for the exam. This can be formalised as:

\begin{prop}[\bf \cautho] 
An exam protocol ensures \emph{\cautho} if for every exam process $\EP$
\begin{center}
 $\filtt{\idcand}{\ques}{\ans}{\idform}~\leadsto~\mathit{inj}\regtt{\idcand}$ 
\end{center}
on every execution trace.
\end{prop}
This requirement is modelled as injective correspondence assertion because the exam should consider only one submission per registered candidate. 

The second authentication requirement that we advance is \emph{\aau}, which concerns testing.
This requirement states that the collector should consider only answers that candidates actually submitted, and that the contents of each collected test are not modified after submission.
It says that a test must be bound to a candidate identity. A way to enforce this would be to only give a test to a candidate after she inserts in the test the same details that authenticated
her. This candidate becomes the test assignee. With exams that are not computer assisted, for example, an authority can check that the candidate writes down the right details on the test, or the authority can write them down personally. The requirement implies that two candidates will be unable to get tested on each other's questions, something that could be desirable if they found their respective questions too difficult.
Moreover, it should be considered only one test from each candidate, namely every time the collector process emits \texttt{collected}, there is a distinct earlier occurrence of the event \texttt{submitted} that satisfies the relationship between their arguments. This is enforced by the injective formula:

\begin{prop}[\bf \aau{}] \index{Answer Authenticity}
An exam protocol ensures $ $ \emph{\aau} if for every exam process $\EP$
\begin{center}
$\collecttt{\idcand}{\ques}{\ans}{\idform}~\leadsto~\mathit{inj}\filtt{\idcand}{\ques}{\ans}{\idform}$ 
\end{center}
on every execution trace.
\end{prop}

The third requirement is \emph{Test Origin Authentication} and concerns preparation and testing. \index{Test Origin Authentication}
Informally, it says that the collector should accept only tests that originate from registered candidates. 
This requirement should be modelled as an injective agreement to enforce that only one test from each registered candidate is actually collected. This can be formalised as:

\begin{prop}[\bf \CAu] 
$ $ An exam protocol ensures $ $ \emph{\CAu} if for every exam process $\EP$
\begin{center}
$\collecttt{\idcand}{\ques}{\ans}{\idform}~\leadsto~\mathit{inj}\regtt{\idcand}$ 
\end{center}
on every  execution trace.
\end{prop}

The forth authentication requirement is \emph{\ta} and concerns testing and marking. Since the collector distributes the tests possibly among different examiners, \index{Test Authenticity}
Test Authenticity insists that the examiner only marks the tests intended for him. Moreover, the contents of each test should not be modified  until after the tests are marked 
 by the examiner. This requirement can be modelled as injective agreement:

\begin{prop}[\bf \ta] 
An exam protocol ensures \emph{\ta} if for every exam process $\EP$
\begin{center}
$\markingtt{\ques}{\ans}{\mrk}{\idform}{\idexam}~\leadsto$\\
$\mathit{inj}\collecttt{\idcand}{\ques}{\ans}{\idform} \cup \mathit{inj}\distributett{\idcand}{\ques}{\ans}{\idform}{\idexam}$ 
\end{center}
on every  execution trace.
\end{prop}

The last requirement is \emph{Mark Authenticity}, which concerns marking and notification. It prescribes that the candidate should receive the mark assigned to her test by the examiner. Moreover, the examiner who evaluated the test should be the one chosen by the collector. In other words, if the candidate accepts the mark, then the examiner, which was appointed by the collector to evaluate the candidate's test, assigned that mark. 
This can be formalised as:

\begin{prop}[\bf \mau] \index{Mark Authenticity}
An exam protocol ensures \emph{\mau} if for every exam process $\EP$
\begin{center}
$\notifytt{\idcand}{\mrk}~\leadsto$\\
$\mathit{inj}\distributett{\idcand}{\ques}{\ans}{\idform}{\idexam} \cup
\mathit{inj}\markingtt{\ques}{\ans}{\mrk}{\idform}{\idexam}$
\end{center}
on every  execution trace.
\end{prop}

\begin{figure}
\begin{center}
\begin{tikzpicture}
\footnotesize
        \draw (0,0) node[inner sep=0.8pt, fill=black] (A) {};  
        \draw (1,0) node[circle, inner sep=0.8pt, fill=black, label={below:{\texttt{registered}}}] (B) {};  
        \draw (3,0) node[circle, inner sep=0.8pt, fill=black, label={below:{\texttt{submitted}}}] (C) {};  
        \draw (5,0) node[circle, inner sep=0.8pt, fill=black, label={below:{\texttt{collected}}}] (D) {};  
        \draw (7,0) node[circle, inner sep=0.8pt, fill=black, label={below:{\texttt{distributed}}}] (E) {};  
        \draw (9,0) node[circle, inner sep=0.8pt, fill=black, label={below:{\texttt{marked}}}] (F) {};  
        \draw (11,0) node[circle, inner sep=0.8pt, fill=black, label={below:{\texttt{notified}}}] (G) {};  
        \draw (12,0) node[inner sep=0.8pt, fill=black] (H) {};  
        \draw (A) to (H);   

        \draw[->] (B) to [bend left=45] (C);    
        \draw[->] (C) to [bend left=45] (D);    
        \draw[->] (E) to [bend left=45] (F);    
        \draw[->] (F) to [bend left=45] (G);    

        \draw[->] (B) to [bend left=45] (D);
        \draw[->] (D) to [bend left=45] (F);    
        \draw[->] (E) to [bend left=45] (G);

        \draw (1,2) node (b) {};  
        \draw (3,1.5) node (c) {};  
        \draw (5,2) node (d) {};  
        \draw (7,1.5) node (e) {};  
        \draw (9,1.2) node (f) {}; 
        \draw (9,1.4) node (F') {};   
        \draw (9,2) node (f') {};   
        \draw (11,1.5) node (g) {};  
        \draw [dashed] (B) to (b);   
        \draw [dashed] (C) to (c);   
        \draw [dashed] (D) to (d);   
        \draw [dashed] (E) to (e);   
        \draw [dashed] (F) to (f);   
        \draw [dashed] (F') to (f');   
        \draw [dashed] (G) to (g);   

        \node at (1,-0.7) {$A$};
        \node at (3,-0.7) {$C$};
        \node at (5,-0.7) {$K$};
        \node at (7,-0.7) {$K$};
        \node at (9,-0.7) {$E$};
        \node at (11,-0.7) {$C$};

\tiny
        \node [text width=1.5cm, align=center] at (2,1.3) {Candidate Authorisation};
        \node [text width=1.5cm, align=center] at (4,1.3) {Answer Authenticity};
        \node  at (7,1.8) {Test Authenticity};
        \node  at (9,1.3) {Mark Authenticity};
        \node  at (3,1.8) {Test Origin Authentication};

\end{tikzpicture}
\caption{A general view of authentication requirements for exams} 
\label{fig:authenticationrainbow}
\end{center}
\end{figure}
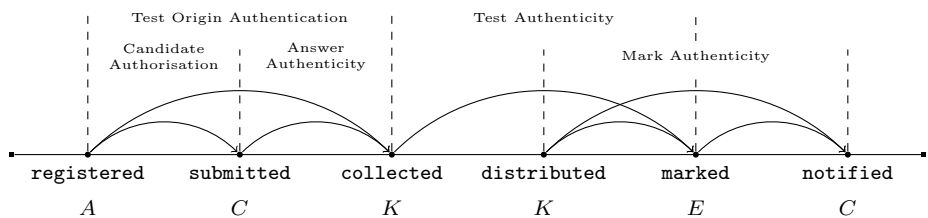

\paragraph{Remark.} It can be observed that the combination of these requirements produce novel requirements. If an exam  protocol guarantees {Candidate Authorisation} and {Answer Authenticity}, then the protocol also guarantees  {\CAu}, namely the tests submitted by registered candidates are actually collected. Conversely, a protocol that guarantees \CAu may guarantee neither \cautho nor \aau. \index{Candidate Authorisation} \index{Answer Authenticity}
If we consider a requirement in a certain phase of the exam, we cannot infer anything about other phases. For example, {Mark Authenticity} signifies that the candidate is notified with the mark delivered by the examiner on the test provided by the collector. However, the test provided by the collector may contain a different answer with respect to the answer the candidate submitted at testing. Only if the exam protocol also guarantees {Answer Authenticity} and {Test Authenticity}, then the contents of the tests are identical. Moreover, {Mark Authenticity} does not signify that a mark is computed correctly. \index{Test Authenticity}

In summary, an exam protocol that ensures all the requirements outlined above preserves the association between candidate identity, mark, and test, including question and answer, through all the phases of the exam. The relationships between authentication requirements with respect to exam run and principals are outlined in Figure \ref{fig:authenticationrainbow}. By looking at this figure, it can be seen that the stated  requirements produce an ordered sequence of events. It can be noted that there is no requirement that relates directly the events \texttt{collected} and \texttt{distributed}. We have not specified the requirement \emph{``the collector distributes the accepted tests''}. Such requirement is usually enforced by the sequential execution of the collector process, since both events belong to the same process. Moreover, it always holds if the common arguments of the two events are derived from the same source, for example, if the common arguments are build from the same message input. In general, this and other requirements get more interesting if events reflect tasks that are executed by different roles, according to the levels of detail and abstraction of the exam protocol. Hence, an exam protocol that specifies a huge number of tasks, might allow to express more events thus novel authentication requirements.



\subsection{Privacy} 

To model privacy requirements as equivalence properties, we use the definition of labelled bisimilarity ($\approx_l$), which was defined in Section \ref{sec:picalculus}.

We introduce two notations to make clear the requirements.
First, we denote with ``$\EP_I$[\_]'' the context of the process $\EP$ pruned of identities that appear in the set $I$.  For example, the process
$\nu \tilde{n}.( \_|\_|$ $C\sigma_{id_3}\sigma_{a_3} | \ldots
|C\sigma_{id_j}\sigma_{a_j} |$ $E\sigma_{id'_1}\sigma_{m_1}| \ldots $
$|E\sigma_{id'_k}\sigma_{m_k}| Q \sigma_q | K\sigma_{\mathit{dist}} | A_{1} | \ldots
|A_{l})$ can be concisely written as \ab $\EP_{\{id_1,id_2\}}$[\_].
Such compact notation is useful to specify and focus exactly on the processes concerned by the requirement. For example, we can write
$\holeprocess{\EP}{\{id_1,id_2\}}$[$C\sigma_{id_1}\sigma_{a_1} | C\sigma_{id_2}\sigma_{a_2}$] to reason about candidates $id_1$ and $id_2$ without repeating the entire exam instance.

Second, we denote with ``$\phaseend{\EP}{\mathtt{e}}$'' the process $\EP$ pruned of the code that follows the event $e$. For example, the process $\phaseend{\EP}{\mathtt{marked}}$ considers an exam instance that terminates at marking, namely after the event \texttt{marked} is emitted. This notation is useful to capture fixed-term requirements such as anonymous marking, which is intended to hold until after the marking, but is eventually falsified at notification when the mark is assigned to the candidate.

The first privacy requirement we consider is \emph{Question Indistinguishability}. This requirement says that the questions are not revealed until the testing phase begins. Thus, it is a fixed-term requirement that sees the exam process ending with the preparation phase. 
\begin{prop}[\bf Question Indistinguishability]  \index{Question Indistinguishability}
An exam protocol ensures \emph{Question Indistinguishability} if for any exam process $EP$ and any questions $q_1$
and $q_2$
\begin{center}
$\phaseend{\holeprocess{EP}{}[Q\sigma_{q_1}]}{\mathtt{registered}}~\approx_l
\phaseend{\holeprocess{EP}{}[Q\sigma_{q_2}]}{\mathtt{registered}}$
\end{center}
\end{prop}

Question Indistinguishability states that two processes with
different questions have to be observationally equivalent until after the preparation phase.  Note that this requirement is more stringent than reachability-based secrecy because the \attacker should not be able to distinguish whether the exam will use $q_1$ or $q_2$ despite he knows both the questions in advance.
For instance, the attacker cannot say whether the questions of the current exam are similar to the questions of the
previous exam, which are on attacker's knowledge. 
The analysis of Question Indistinguishability requires the question committee to be honest otherwise they could reveal the questions to the attacker making the requirement useless.  However, it is particularly interesting to consider a scenario with other corrupted roles. For example,  candidates might be interested to know the questions in advance. Such scenario can be explicitly captured by replacing honest candidates with corrupted ones using the definition of corrupted process, which is defined in Section \ref{sec:picalculus}.
Assuming candidate $\id_1$ being \dishonest, we obtain 
\begin{center}
$\phaseend{\holeprocess{EP}{\{\id_1\}}[(C\sigma_{\id_1}\sigma_{a_1})^{c_1,c_2} | 
Q\sigma_{q_1}]}{\mathtt{registered}}~\approx_l$ 
$\phaseend{\holeprocess{EP}{\{\id_1\}}[(C\sigma_{\id_1}\sigma_{a_1})^{c_1,c_2} 
| Q\sigma_{q_2}]}{\mathtt{registered}}$ 
\end{center}

The next requirement is \emph{Anonymous Marking}, which covers preparation, testing, and marking.
This requirement signifies that the examiner marks a test while ignoring its author, namely an anonymous test. It is a clear contribution to the fairness of the marking. As it stands, the requirement insists on anonymity only until the point that the examiner affixes a mark.
Anonymous Marking can be specified as  two exam instances in  which the processes of two candidates who swap their answers cannot be distinguished until after the end of the marking phase.

\begin{prop}[\bf Anonymous Marking]\label{am} \index{Anonymous Marking}
An exam protocol ensures $ $ \emph{Anonymous Marking} if any exam process $EP$,
any two candidates $\id_1$ and $\id_2$, and
any two answers $a_1$ and $a_2$
\begin{center}
$\phaseend{\holeprocess{EP}{\{\id_1,\id_2\}}[C\sigma_{id_1}\sigma_{a_1}|C\sigma_{id_2}\sigma_{a_2}]}{\mathtt{marked}}~
\approx_l
\phaseend{\holeprocess{EP}{\{\id_1,\id_2\}}[C\sigma_{\id_1}\sigma_{a_2}|C\sigma_{\id_2}\sigma_{a_1}]}{\mathtt{marked}}$
\end{center}
\end{prop}

In other words,  Anonymous Marking says that the
process where  candidate $\id_1$ submits $a_1$  and candidate $\id_2$ submits $a_2$ is indistinguishable to
 the process where candidate $\id_1$ submits $a_2$ and candidate $\id_2$ submits $a_1$. 
It prevents the attacker to obtain the identity of the candidate who submits a certain answer before the marking ends.
Candidate 

Similarly to Question Indistinguishability, it is interesting to consider corrupted principals in the analysis of Anonymous Marking, which means that nobody knows who submitted a test while this is being marked, except the official author of
the test. An implication is that test anonymity during marking will even resist collusion of the examiner with other authorities and candidates. Again, the definition of corrupted process can model corrupted examiners and authorities.
The definition can be also used to specify corrupted candidates, however the candidates $\id_1$ and $\id_2$ who submit two different answers have to be honest. This avoids the corner case in which all candidates but one reveal their answers to the attacker, who can easily associate the remaining answer with the honest candidate and thus trivially violate the requirement.

We now consider the requirement of \emph{Anonymous Examiner}, which concerns all the phases of an exam. In fact, an exam could require examiner anonymity forever to prevent bribing or coercion. Thus, the requirement of Anonymous Examiner says that no candidate knows which examiner marked their tests. This can be formalised as:

\begin{prop}[\bf Anonymous Examiner] \index{Anonymous Examiner}
  An exam protocol ensures \emph{Anonymous Examiner} if for any
  exam process $EP$, any two candidates $\id_1$ and $\id_2$, any two
  examiners $\id'_1$ and $\id'_2$, any two marks $m_1$ and $m_2$, and two associations  $\mathit{test_1}$ and $\mathit{test_2}$
\begin{center}
  $\holeprocess{EP}{\{\id_1, \id_2, \id'_1, \id'_2\}}~[C \sigma_{id_1}\sigma_{a_1} | C\sigma_{id_2}\sigma_{a_2} |
  E \sigma_{id'_1}\sigma_{m_1} | E\sigma_{id'_2}\sigma_{m_2} | K
  \sigma_{test_1}] ~\approx_l$
  $\holeprocess{EP}{\{\id_1, \id_2, \id'_1, \id'_2\}}$ 
$[C $ $\sigma_{id_1}$ $\sigma_{a_1} | C\sigma_{id_2}\sigma_{a_2} | 
  E \sigma_{id'_1}\sigma_{m_2} | E\sigma_{id'_2}\sigma_{m_1} | K
  \sigma_{test_2} ]$ 
\end{center}

where 
\begin{itemize}
\item[--] $\sigma_{test_1}$ associates the test  of candidate $\id_1$ to examiner $\id'_1$ and the test of
  candidate $\id_2$ to examiner $\id'_2$;
\item[--] $\sigma_{test_2}$ associates the test of candidate $\id_1$ to examiner $\id'_2$
  and the test of candidate $\id_2$ to examiner $\id'_1$.
\end{itemize}

\end{prop}

Thus, Anonymous Examiner states that a process in which the examiner
$\id'_1$ evaluates the test of candidate $\id_1$ while the examiner
$\id'_2$ evaluates the test of candidate $\id_2$ is indistinguishable to
the process in which the examiner $\id'_1$ evaluates the test of candidate $\id_2$ while
the examiner $\id'_2$ evaluates the test of candidate $\id_1$.  
Note that the two marks $\sigma_{m_1}$ and $\sigma_{m_2}$ are swapped on examiner processes to ensure that each test is evaluated with the same mark in both cases.
In the field of peer review systems, this requirement is known as \emph{blind review}. The requirement of \emph{double-bind review} instead refers to a peer review system that ensure both Anonymous Examiner and Anonymous Marking, namely anonymity is provided to both authors and examiners.
However, peer review systems usually assume that the collector knows which examiner evaluates a test, while other systems may not. To ensure a stronger version of Anonymous Marking it is possible to model corrupted collectors, candidates, and any other principal, provided that examiners $id_1'$ and $id_2'$ are honest. This would avoid the corner case in which an examiner reveals the mark to the attacker, a case that would trivially violate the requirement.

The requirement of \emph{Mark Privacy} concerns all phases of an exam. It states that the mark ultimately attributed to a candidate is treated as valuable personal information of the candidate’s. More specifically, no one learns the marks, besides the examiner, the concerned candidate, and the authority responsible for the notification. This means that the marks cannot be public.

\begin{prop}[\bf Mark Privacy]  \index{Mark Privacy}
An exam protocol ensures \emph{Mark Privacy} if for any exam 
process $EP$ and any two marks $m_{1}$ and $m_{2}$
\begin{center}
$  \holeprocess{EP}{\{id'\}}[E\sigma_{id'}\sigma_{m_1}] \approx_l 
\holeprocess{EP}{\{id'\}}[E$ $\sigma_{id'}\sigma_{m_2}].$
\end{center}
\end{prop}

The definition of Mark Privacy means that a process in which the examiner $id'$ assigns the mark $m_1$ to an answer  cannot be distinguished from a process in which the same examiner assigns a different mark $m_2$ to the same answer. 
This is a strict requirement because an exam protocol that guarantees Mark Privacy cannot publicly disclose the marks even if these cannot be associated with the corresponding candidates. In fact, the publication of the marks allows the attacker to distinguish the processes.
Again, it can be assumed that some candidates and examiners are corrupted, namely collaborate with the attacker to find out the marks of other candidates.
However, the examiner who assigns the different marks, the two candidates who submit the tests, and the authority in charge of the notification of the marks should be honest. Otherwise, any of these could violate the requirement by revealing the mark to the attacker.

Since Mark Privacy can be a requirement too strong to satisfy, we introduce a variant called \emph{Mark Anonymity}. This requirement states that no one learns the association between a mark and the corresponding candidate. Intuitively, an exam protocol that publishes the list of all marks might still ensure Mark Anonymity, but not Mark Privacy.  This is a common privacy requirement in scenarios like public competitions,  in which marks are published and associated with a list of pseudonyms for transparency. Mark Anonymity can be defined as follows:
\begin{prop}[\bf Mark Anonymity]  \index{Mark Anonymity}
An exam protocol ensures \emph{Mark Anony\-mity} if for any exam 
process $EP$, any two candidates $id_1$, $id_2$, any examiners $id'$, any two answers $a_1$, $a_2$, two substitutions $\sigma_{m_a}$ and $\sigma_{m_b}$ and an association $\mathit{test}$
\begin{center}
$  \holeprocess{EP}{\{id_1, id_2, id'\}}[C\sigma_{id_1}\sigma_{a_1} | C\sigma_{id_2}\sigma_{a_2} | E\sigma_{id'}\sigma_{m_a}| K\sigma_{test}] \approx_l 
\holeprocess{EP}{\{id_1, id_2, id'\}}[C\sigma_{id_1}\sigma_{a_1} | C\sigma_{id_2}\sigma_{a_2} | E\sigma_{id'}\sigma_{m_b} | K\sigma_{test}]$ 
\end{center}
where
\begin{itemize}
\item[--] $\sigma_{test}$ associates the tests of both candidates $\id_1$  and $\id_2$ to the examiner $\id'$;
\item[--] $\sigma_{m_a}$ attributes the mark $m_{1}$ to the answer $a_1$ and the mark $m_{2}$ to the answer $a_2$;
\item[--] $\sigma_{m_b}$ attributes the mark $m_{2}$ to the answer $a_1$ and the mark $m_{1}$ to the answer $a_2$.
\end{itemize}
\end{prop}

In other words, a process in which an examiner evaluates two  answers $a_1$ and $a_2$ respectively with $m_1$ and $m_2$ is indistinguishable for the attacker with a process in which the examiner evaluates the same answers but with swapped marks, namely the examiner marks $a_1$ and $a_2$ respectively with $m_2$ and $m_1$.
In doing so, the authority can make the list of marks public assuming the attacker cannot associate the marks to the candidates.
The analysis of Mark Anonymity requires the two concerned candidates, the examiner, and the notifier authority to be honest.
Otherwise, they can simply reveal the answer and the associated mark to allow the attacker to  distinguish the two case processes.
Other principals can be considered corrupted.
It can be noted that an exam protocol that guarantees \emph{Mark Privacy} also guarantees \emph{Mark Anonymity}. In fact, $\sigma_{m_a}$ and $\sigma_{m_b}$ defined in Mark Anonymity are special instances of $\sigma_{m_1}$ and $\sigma_{m_2}$ defined in Mark Privacy.

\section{The Huszti-Peth\H{o} Protocol}\label{sec:H}
We validate our formal framework with the analysis of the  \Huszti~\citeltex{huszti10} exam protocol.
To the best of our knowledge, this is the first exam protocol proposed in the literature that aims to guarantee authentication and privacy requirements in presence of corrupted candidates and exam authorities.

\begin{table}[]
\begin{center}
\footnotesize
\begin{tabular}{|c|p{5.2cm}|p{3.6cm}|}

\hline
\multicolumn{2}{|c| }{\bf Huszti and Peth\H{o}}  & \multicolumn{1}{c|}{\multirow{2}{*}{\bf This dissertation}} \\
\cline{1-2}
{\bf Requirement} & \multicolumn{1}{c|}{\bf Description} &  \\ \hline\hline
\multirow{4}{*}{Authenticity} & \emph{Only eligible students’ tests should be considered} & \multirow{2}{3.75cm}{Candidate Authorisation Answer Authenticity}\\
 \cline{2-3}
 & \emph{It should be verified whether the exam grade is proposed by a teacher} & \multirow{2}{*}{Mark Authenticity} \\ \hline\hline
\multirow{4}{*}{Anonymity} &  \emph{Teachers do not know which paper belonging to which student he is correcting} & \multirow{2}{*}{Anonymous Marking} \\ \cline{2-3}
 &  \emph{Students do not know who corrects their papers} & \multirow{2}{*}{Anonymous Examiner} \\ \hline\hline
\multirow{4}{*}{Secrecy}  & \emph{\mbox{Exam questions are kept secret} \phantom{fdaffadfdf fdfadsfd fda fd fdsa sdf}} & \multirow{2}{*}{Question Indistinguishability}  \\ \cline{2-3}
 & \emph{Only the corresponding student should know his mark} & \multirow{2}{*}{Mark Anonymity} \\ \hline\hline
\multirow{3}{*}{Robustness}  & \multirow{3}{5.5cm}{\emph{Exam questions can not be altered after submission}} & \multirow{3}{3cm}{Answer Authenticity Test Authenticity Mark Authenticity}\\ 
& & \\
& &\\
\hline
\multirow{2}{*}{Correctness}  & \emph{Students are not allowed to take the same exam more than once} & \multirow{2}{*}{Answer Authenticity} \\ \hline\hline
\multirow{3}{*}{Receipt}  & \multirow{3}{5.5cm}{\emph{Students are able to make sure of the successful submission}} & \multirow{3}{3cm}{Answer Authenticity Test Authenticity Mark Authenticity} \\ 
& &\\
& &\\
\hline
\end{tabular}
\caption{Comparison of Huszti and Peth\H{o}'s requirements with ones proposed in this dissertation.}
\label{tab:comparisonHP}
\end{center}
\end{table}

Huszti and Peth\H{o} informally analyse their protocol with respect to six security requirements. They state the requirements informally, and each requirement contains sub requirements. For example, the requirement of \emph{Secrecy} implicitly specifies two sub requirements as it states that \emph{``exam questions are kept secret''} and \emph{``only the corresponding student should know his mark''}. 
Table \ref{tab:comparisonHP} clarifies the sub requirements and shows how to map them to the formal requirements proposed in this dissertation. 
By looking at the table, it can be seen that combinations of our formal requirements capture any informal requirement defined by Huszti and Peth\H{o}. For example, Huszti and Peth\H{o}'s definition of  Robustness, which says that questions can not be altered after submission, is captured by the combination of Answer Authenticity, Test Authenticity, and Mark Authenticity. We anticipate that the results of  our analysis show the protocol guarantees only one of our formal requirements, but none of the six envisaged by Huszti and Peth\H{o}. 
\index{Answer Authenticity} \index{Test Authenticity} \index{Mark Authenticity}
The Huszti-Peth\H{o} exam protocol uses four cryptographic building blocks, namely ElGamal encryption \citeltex{elgamal}, zero-knowledge proof \citeltex{GMR85}, reusable anonymous return channel \citeltex{GJ03}, and a timed-release service based on Shamir's secret sharing \citeltex{DBLP:journals/cacm/Shamir79}. 

\subsubsection{ElGamal encryption} This cryptographic primitive for public-key cryptography consists of three algorithms of \emph{key generation}, \emph{encryption}, and \emph{decryption}. 
The key generation algorithm outputs the public key $PK=(G, q, g, h)$ and the secret key $SK=s$;  $G$ is a cyclic group of order $q$ with generator $g$; $s$ is a random value in $\mathbb{Z}_q^*$; and $h=g^s$.
The encryption algorithm takes in a message $m$ and a random value $k \in \mathbb{Z}_q^*$, and outputs the ciphertext $(g^k, m \cdot h^k)$, which is denoted with $\{m\}_{PK}$. 
The decryption algorithm takes as input the ciphertext $\{m\}_{PK}$ and the secret key $SK$, and outputs the message $m$. In fact, $\frac{h^{k}}{g^{ks}}=\frac{g^{sk}}{g^{ks}}=m$.
The ElGamal encryption primitive is semantically secure assuming the decisional Diffie–Hellman problem is intractable.

\subsubsection{Zero-knowledge proof} This cryptographic scheme allows a \emph{prover} to convince a \emph{verifier} that a given statement is true without revealing any extra information except the correctness of the statement. A zero-knowledge scheme must guarantee completeness, soundness, and zero knowledge. Completeness means that if the statement is true, then the verifier always accepts the statement.
Soundness means that if the statement is false, then the verifier always rejects the statement.
Zero knowledge means that the verifier cannot get any information apart from the fact that the statement is indeed true.

Zero-knowledge schemes can be categorised in non-interactive and interactive.
Interactive zero-knowledge proofs require prover and verifier to exchange at least two messages. The input of the prover is usually a challenge message sent by the verifier. In so doing, the proof is only valid for that challenge, and cannot replayed by the prover to someone else.  

Non-interactive  zero-knowledge proof schemes contain only the message sent by a prover to the verifier. They are simpler and more efficient than interactive schemes, hence more suitable for the inclusion in the design of cryptographic protocols. In the remainder, we only consider non-interactive schemes as zero-knowledge proof.

\subsubsection{Reusable anonymous return channel} This cryptographic scheme implements anonymous two-way conversations between a \emph{sender} and a \emph{receiver} by means of a \emph{mixer}.
The mixer is implemented by a re-encryption mix network that consists of a chain of mix servers. The servers take in messages from multiple senders, randomly shuffle them, and send them to the receivers. 
One goal of reusable anonymous return channel is to ensure the anonymity of the sender who send a message via the mixer.
Another goal is to allow the receiver to reply to the sender still guaranteeing the anonymity of the  sender.
Note that reusable anonymous return channel scheme aims to ensure anonymity, but not secrecy of the messages~\citeltex{GJ03}.
As we shall see, the \Huszti protocol resorts on this primitive for both message secrecy and sender anonymity.

Reusable anonymous return channel consists of five algorithms, namely \emph{setup}, \emph{submission of messages}, \emph{delivery of messages}, \emph{submission of reply}, and \emph{delivery of reply}.
The scheme assumes a primitive for digital signature, but the authors do not specify which one. However, digital signature primitives usually employ  public-key cryptography and consist of three algorithms of \emph{key generation}, \emph{signing}, and \emph{verification}. Key generation outputs the secret signing key $SSK$ and verification public key $SPK$. The signing algorithm takes in a message $m$ and the signing key $SSK$, and outputs the signature $\sign_{SSK}(m)$. The verification algorithm takes as input the signature $\sign_{SSK}(m)$ and the verification public key $SPK$, and returns  \texttt{true} if the signature is correct, namely the message $m$ was actually signed with the signing key $SSK$.       

The setup algorithm consists of mix servers jointly generating an ElGamal key pair ($\SK_{\MIX}$, $\PK_{\MIX}$),   and  signature keys  ($\SSK_{\MIX}$, $\SPK_{\MIX}$). Sender and receiver also generate respectively the ElGamal pairs ($\PK_{A}$, $\SK_{A}$) and ($\PK_{B}$, $\SK_{B}$).  The identities of sender and receiver are denoted respectively with the tags  $\ID_A$ and $\ID_B$. For example,
  email addresses can serve as identity tag.

The algorithm of the submission of messages  is run by the sender $A$. It allows $A$ to send an anonymous message $m$ to the receiver with tag $\ID_B$.
The sender generates the triplet  {\[
(\lbrace{\ID_A, \PK_{A}}\rbrace_{\PK_{\MIX}}, \lbrace{m}\rbrace_{PK_{\MIX}},
\lbrace{\ID_B, PK_{B}}\rbrace_{PK_{\MIX}})\]} and  two \emph{proofs of knowledge} of
$\lbrace{\ID_A, \PK_{A}}\rbrace$ and of $\lbrace{\ID_B, \PK_{B}}\rbrace$. 
Proofs of knowledge are similar to zero-knowledge schemes but guarantee only completeness and soundness.
In this case, they aim to avoid the attacker to decrypt the triplets by using the mixer as an oracle. The sender  outputs the triplet and the proofs.

The algorithm of  delivery of messages is run by the mixer.
It takes as input a batch of triplets and proofs sent from different senders. The mixer checks the proofs and then randomly shuffles the batch of triplets. Each triplet is extended with a checksum to ensure they are not separated during the shuffle.
The message $m$ is then re-encrypted with $\PK_{B}$, resulting in the ciphertext $\lbrace{m}\rbrace_{\PK_{B}}$.
The mixer signs the first element of the triplet in input $\{\ID_A, \PK_A\}_{PK_{\MIX}}$, and outputs the pair \[(\mathit{Sign_{SSK_{\MIX}}(\{\ID_A,
\PK_A\}_{PK_{\MIX}}),\lbrace{m}\rbrace_{\PK_{B}})}.\]

The receiver can reply an anonymous message with a new message $m'$ using the algorithm for submission of reply. 
The receiver takes in the pair $(\mathit{Sign_{SSK_{\MIX}}(\{\ID_A,
\PK_A\}_{PK_{\MIX}}),\lbrace{m}\rbrace_{\PK_{B}})}$, encrypts the message $m'$ with the public key of the mixnet $\PK_{\MIX}$ resulting in  ${m'}_{PK_{\MIX}}$, 
 and outputs the triplet \[(\lbrace{\ID_B, \PK_{B}}\rbrace_{\PK_{\MIX}},\mathit{\{m'\}_{PK_{\MIX}},\sign_{SSK_{\MIX}}(\{\ID_A,\PK_A\}_{PK_{\MIX}}))}\] 
and the proof of knowledge of $\lbrace{\ID_B, \PK_{B}}\rbrace$.  

The algorithm of delivery of reply is similar to one of delivery of messages. The only difference is that 
the input consists of triplet $(\lbrace{\ID_B, \PK_{B}}\rbrace_{\PK_{\MIX}},$ $\mathit{{m'}_{PK_{\MIX}},Sign_{SSK_{\MIX}}(\{\ID_A,\PK_A\}_{PK_{\MIX}}))}$ 
and proof of knowledge of $\lbrace{\ID_B,\PK_{B}}\rbrace$, thus the mixer verifies only one proof of knowledge.

\begin{figure}
    \centering
\FramedBox{2.8cm}{12.1cm}
{
\begin{enumerate}
\item  \emph{A}$\rightarrow$\emph{M}: $\lbrace{\ID_A, \PK_{A}}\rbrace_{\PK_{\MIX}}, \lbrace{m}\rbrace_{PK_{\MIX}},
\lbrace{\ID_B, PK_{B}}\rbrace_{PK_{\MIX}}$
\item  \emph{M}$\rightarrow$\emph{B}: $\mathit{Sign_{SSK_{\MIX}}(\{\ID_A,
\PK_A\}_{PK_{\MIX}}),\lbrace{m}\rbrace_{\PK_{B}}}$
\item  \emph{B}$\rightarrow$\emph{M}: $\lbrace{\ID_B, \PK_{B}}\rbrace_{\PK_{\MIX}},\mathit{\{m'\}_{PK_{\MIX}},Sign_{SSK_{\MIX}}(\{\ID_A,\PK_A\}_{PK_{\MIX}})}$

\end{enumerate}
}
    \caption{Reusable anonymous return channel in the Alice-Bob notation}
    \label{fig:rarc}
\end{figure}

A succinct description of reusable anonymous return channel in the Alice-Bob notation is provided in Figure \ref{fig:rarc}.

\subsubsection{Timed-release service} This service is based on threshold Shamir's secret sharing, a cryptographic primitive that ensures fixed-term secrecy. A secret is shared among $n$ servers and cannot be reconstructed unless some servers collaborate to reveal the secret.  
This service assumes the existence of a trusted third party, which in the Huszti-Peth\H{o} protocol is known as \emph{registry}. The registry knows the secret, bootstraps the servers, and serves as authority to provide absolute time reference to the servers. The Huszti-Peth\H{o} protocol uses the timed-release service to deanonymise the candidate's pseudonym for notification. We do not detail this service further because in our analysis we assume the servers to be trusted.

\subsection{Description}
The Huszti-Peth\H{o} protocol specifies six roles: exam authority (EA), registry, timed-release service (NET), question committee (COM), candidate (C), examiner (E).
The exam authority manages the entire exam process. Specifically, it generates the pseudonyms to anonymise candidates' tests and examiner's identities, collects the tests, distributes the tests to examiners, and notifies the marks to the candidates. The registry generates the necessary cryptographic keys for the other roles and bootstraps the timed-release service. The NET consists of the servers that implement the timed-release service, and contributes to generate and revoke the candidate pseudonyms using threshold Shamir's secret sharing.  The question committee generates the questions for the exam, the candidate takes the exam, and the examiner marks the tests.

The protocol assumes that no candidate reveals their private keys to other candidates, and that invigilators supervise candidates during the testing phase. All communications take place via reusable anonymous return channels.

\begin{figure}[!ht]
\begin{center}
\FramedBox{18.5cm}{12.1cm}
{\small
\texttt{~Preparation} 
\begin{enumerate}
\item $EA$ publishes $g$ and $h = g^s$ 
\item $\mi{COM} \rightarrow EA:$ 
 $\lbrace \sign_{\SSK_{\mi{COM}}}(question,time_{1}) \rbrace_{\PK_{\MIX}}$  
\item $EA$ checks $E$ eligibility, and calculates $\tilde{q}=PK_E^s$ \hfill //Examiner Registration 
\item $EA \rightarrow E: \tilde{q}, g_E$ 
\item $E$ calculates $q'=(\tilde{q})^{\alpha}$,
$t=(g_E)^{\alpha},$ and $q=t^{SK_E}$ 
\item $EA \leftrightarrow E:$ $ZKP_{eq}((q,q'),(g,h))$
\hfill  //$E$ pseudonym is $(t, q, q')$ 
\item $E \rightarrow  EA: t, q, q',subject$  
\item $EA$ checks $q^s=q'$ 
\item  $E \leftrightarrow EA: \mi{ZKP}_{\mi{sec}}(\SK_E)$  
\item $EA$ stores ZKP data plus $\lbrace{\ID_E,
  \PK_E}\rbrace_{\PK_{\MIX}}$ and $subject$ 
\item $EA$ checks $C$ eligibility, and calculates $\tilde{p}=(PK_C)^s$ \hfill //Candidate Registration
\item  $EA\rightarrow NET:  \tilde{p}, g_C $ 
\item  $NET$ calculates $p'=(\tilde{p})^{\Gamma}$, and
$r=(g_C)^{\Gamma}$, and stores $\mi{time}$ of notification, $\tilde{p}$, and
$g_C$. 
\item  $NET \rightarrow  C: r, p'$ 
\item  $C$ calculates $p=r^{SK_C}$  
\item  $EA \leftrightarrow C:
  \mi{ZKP}_{eq}((p,p'),(g,h))$  \hfill //$C$ pseudonym is $(r, p, p')$ 
\end{enumerate}
\texttt{~Testing} 
\begin{enumerate}
\setcounter{enumi}{16}
\item $C \rightarrow EA: r, p, p',subject$ 
\item  $EA$ checks $p^s=p'$ 
\item  $C \leftrightarrow EA: \mi{ZKP}_{\mi{sec}}(\SK_C)$  
\item  $EA \rightarrow  C:$ 
$\sign_{\SSK_{\mi{COM}}}(question),
    \mi{time}_{1}$  
\item  $C\rightarrow EA:
 r,p,\lbrace{\mi{answer}}\rbrace_{\PK_{\MIX}},\mi{time}_{2} $  
\item  $EA \rightarrow  C:$ 
$ \mi{Hash}(r,p,p',subject,\mi{trans}_{C},\mi{question}, \mi{time}_{1},
\mi{time}_{2}, \lbrace{\mi{answer}}\rbrace_{\PK_{\MIX}})$
\end{enumerate}
\texttt{~Marking} 
\begin{enumerate}
\setcounter{enumi}{22}
\item  $ EA \rightarrow  E:$ 
 $\lbrace{\mi{answer}}\rbrace_{\PK_{\MIX}}$ 
\item  $E \rightarrow EA:$   
$ \mi{mark}, \mi{Hash}(\mi{mark},\mi{answer}),
[\mi{Hash}(\mi{mark},\mi{answer})]^{\SK_E} , \mi{verzkp}
$ 
\\ 
\hfill where  $\mi{verzkp=}$
$ \mi{ZKP}_{eq}(\mi{Hash}(\mi{mark}, \mi{answer}),[\mi{Hash}(\mi{mark},\mi{answer})]^{\SK_E}),t,q$ 
\end{enumerate}
\texttt{~Notification} 
\begin{enumerate}
\setcounter{enumi}{25}
\item  $ EA \rightarrow NET:  p' $ \hfill  // Note that $r=(g_C)^{\Gamma}$, $p=(PK_C)^{\Gamma}$, $p'=(g_C)^{\Gamma s}$ 
\item  $NET$ calculates $p'=(\tilde{p})^{\Gamma}$ 
\item  $NET \rightarrow EA: \{p', \tilde{p}\}_{\PK_{EA}}$ 
\item  $EA$ stores  $\mi{mark}, \mi{Hash}(\mi{mark}, \mi{answer}), [\mi{Hash}(\mi{mark}, \mi{answer})]^{\SK_E} , \mi{verzkp}$  
\end{enumerate}
}

\caption{The \Huszti e-exam protocol in the Alice-Bob notation}
\label{fig:hp}
\end{center}
\end{figure}

The protocol originally sees three stages: \emph{registration}, \emph{exam}, and \emph{grading}. It can be observed that the exam stage begins with the exam authority checking the candidate's eligibility, and concludes with the examiner sending the mark to the exam authority. 
Thus, to match this structure with our phases, we map the registration stage to preparation,  the grading stage to notification, and we divide the exam stage in testing and marking. We provide a high-level description of the protocol, which is supported by a more detailed specification in the Alice-Bob notation in Figure \ref{fig:hp}.  

\subsubsection{Preparation} This phase concerns the registration of both candidate and examiner, and the generation of the pseudonyms.
The \ea publishes the public parameters to identify a new exam (step 1).  The question committee
 then signs and sends the questions, which are encrypted with the public key of the mixer implementing the reusable anonymous return channel (step 2).
The mixer will publish the questions only at time of testing ($\mathit{time_1}$).

The registration of the  examiner consists of creating a pseudonym, which is jointly generated by the \ea and the examiner.
The examiner verifies the correctness of the pseudonym by
using a zero-knowledge proof ($\mathit{ZKP_{eq}}$) on the equality of the discrete logarithms with the \ea (step 6). 
To enrol for an exam, the examiner sends pseudonym and subject to the \ea (step 9), and proves the knowledge of his secret key ($\mathit{ZKP_{sec}(\SK_E})$).
Note that the \ea knows that the examiner is eligible for the exam, but cannot learn the examiner identity since the communication takes place via reusable anonymous return channel. Thus, at the end of examiner registration,  the \ea stores the encrypted identity of the examiner ($\lbrace{\ID_E,
  \PK_E}\rbrace_{\PK_{\MIX}}$) (step 10), which the \ea will use to send the answer to the anonymous examiner at marking.

The registration of a candidate slightly differs from the registration of an examiner since the anonymity of the candidate eventually will be broken at notification, while the anonymity of  examiner may last forever.
The  pseudonym  of the candidate is jointly calculated by the \ea, the candidate, and also the NET.
The NET stores the secret values used for the generation of the candidate pseudonym (step 13), and will use the secret values at notification to allow the \ea to associate the candidate with the mark.
Similarly to the registration of examiner, the candidate can verify the correctness of
the pseudonym using a  zero-knowledge proof ($\mathit{ZKP_{eq}}$) of the equality of the discrete logarithms with the \ea being the prover (step 16).

\subsubsection{Testing} The candidate sends the pseudonym to the \ea (step 17) and proves the knowledge of the private key  ($\mathit{ZKP_{sec}(\SK_C)}$) (step 19). We stress that the protocol assumes the candidate does not share with other principals the private key.
The \ea checks whether the pseudonym is allowed for the exam (step 18), and then sends the questions signed by the question committee (step 20).
The candidate then sends the answer (step 21) encrypted with the public key of the mixer  ($\lbrace{\mi{answer}}\rbrace_{\PK_{\MIX}}$). At marking, the mixer will re-encrypt the answer with the public key of the examiner. Thus, the exam authority cannot learn the answer submitted by the candidate.
Testing concludes with the \ea sending to the candidate  a receipt (step 22), which consists of
the hash of all parameters seen by the \ea during testing, the
transcription of the zero-knowledge proof ($\mathit{trans_C}$), and the time when the answer was submitted ($\mathit{time_2}$).

\subsubsection{Marking} We recall that at preparation the \ea stored  the encrypted identities of the examiners, thus it can choose an examiner who is eligible for the exam to forwards the candidate's answer. The examiner assigns the answer with a mark (step 23), and sends it to the \ea with a  zero-knowledge proof ($\mi{verzkp}$), which proves the examiner actually marked the answer (step 24).

\subsubsection{Notification} When all the answers are marked, the NET
de-anonymises the pseudo\-nyms associated to the answers, so the \ea can link back the pseudonym with the corresponding candidate (steps 27-28). Finally,  the \ea stores the marks and the zero-knowledge proof provided at marking by the examiner (step 29).

\subsection{Formal Analysis of Reusable Anonymous Return \ab Channel}
\label{sec:analysisrarc}

Prior to verify the \Huszti protocol, we provide a formal analysis of reusa\-ble anonymous return channel. In particular, we verify whether the scheme ensures message secrecy and anonymity of sender and receiver as it is assumed in the \Huszti protocol.

\subsubsection{Model choices}
The equational theory in Table \ref{table:eqt_rarc} models the cryptographic primitives for reusable anonymous return channel. It includes models for ElGamal public-key encryption,  digital signatures, and zero-knowledge proof.
ElGamal encryption consists of two functions $\mi{encryption}$ and $\mi{decryption}$. A message encrypted with a public key
can only be decrypted using the corresponding secret key. Digital signature consists of two equations: the function $\mi{getmess}$ checks integrity as it returns a message embedded into a signature; the function $\mi{checksign}$ checks also for authenticity as it returns the message only if the function is provided with the correct  verification key. 
The theory for zero-knowledge proof that we use is inspired by Backes  \etal\citeltex{BMU08}, who model  zero-knowledge proof as two functions. The function $\mathit{zkp\_proof}$ models the proof that the prover builds to demonstrate the knowledge of the secret, which in the case of reusable anonymous return channel is the message encrypted with the public key of the mixer. The function $\mathit{zkpsec}$ models the verification of the proof by the verifier,  which in the case of reusable anonymous return channel is the mixer.  The function $\mathit{zkp\_proof}(\mathit{public, secret})$ takes as arguments 
 public and  secret parameters. In this case, the public parameter is the encryption of the message, and the private parameter is  the message. Note that the correct function can be constructed only by the prover who knows the private parameter.  The verification function  $\mathit{zkpsec}(\mathit{zkp\_proof}(\mathit{public,secret}),\mathit{verinfo})$
 takes as arguments the proof function and the verification parameter  $\mathit{verinfo}$. The verifier only accepts the proof if the
 relation between $\mathit{verinfo}$ and $\mathit{secret}$ is  satisfied.

\begin{figure}[h!]
\begin{center}
\begin{lstlisting}
(*----Sender----*)
let A (SKa: skey, PKb: pkey, PKmix: pkey, SPKmix: spkey) =
 out(c, (encrypt(pkey_to_bitstring(pk(SKa)), PKmix), 
         encrypt(secret_message, PKmix), 
         encrypt(pkey_to_bitstring(PKb), PKmix))).

(*----Receiver 1----*)
let B (SKb: skey, PKmix: pkey, SPKmix: spkey) =
 in (c, (c1: bitstring , s1: bitstring, cm1: bitstring)).

(*----Receiver 2-----------------------*)
let C (SKc: skey, PKmix: pkey, SPKmix: spkey) =
 in (c, (c1: bitstring , s1: bitstring, cm1: bitstring)).

(*----Mixer----*)
let MIX (SKmix: skey, SSKmix: sskey ) =
 !MIX1(SKmix, SSKmix) | !MIX2(SKmix, SSKmix).

(*----Mixer 1----*)
let MIX1 (SKmix: skey, SSKmix: sskey ) =
 in (c, (c1: bitstring, c2: bitstring, c3: bitstring, 
         p1:zkp, p2:zkp));
 let (xmsg: bitstring) = decrypt(c2, SKmix) in
 let (xdst: pkey) = bitstring_to_pkey(decrypt(c3, SKmix)) in
 if(checkproof(p1,c1) && checkproof(p2,c3)) then 
  (out(c, (c1, sign(c1, SSKmix), encrypt( xmsg, xdst)))).

(*----Mixer 2----*)
let MIX2 (SKmix: skey, SSKmix: sskey ) =
 in (c, (c1': bitstring, c2': bitstring, c3': bitstring, 
         p1':zkp, p2':bitstring));
 let (xmsg': bitstring) = decrypt(c2', SKmix) in
 let (xdst': pkey) = bitstring_to_pkey(decrypt(c3', SKmix)) in
 if(checkproof(p1',c1') && checksign(p2',spk(SSKmix)) = c3') then 
  (out(c, (c1', sign(c1', SSKmix), encrypt( xmsg', xdst')))).
\end{lstlisting}
\caption{The processes of sender, receivers, and mixer.}
\label{fig:rarc_proc}
\end{center}
\end{figure}

The ProVerif description of  sender, receivers, and mixer processes is outlined in Figure \ref{fig:rarc_proc}. 
We specify one sender and two receivers, and model a simpler version of reusable anonymous return channel without considering the submission of reply, namely we omit step 3 of Figure \ref{fig:rarc}. As we shall see later, this simpler version is sufficient to show successful attacks on message secrecy and sender anonymity.  

\begin{figure}
\begin{center}
\begin{lstlisting}
process
 new skA: skey; let pkA = pk(skA) in out (c, pkA);
 new skB: skey; let pkB = pk(skB) in out (c, pkB);
 new skC: skey; let pkC = pk(skC) in out (c, pkC);
 new skMIX: skey; let pkMIX = pk(skMIX) in out (c, pkMIX);
 new sskMIX: sskey; let spkMIX = spk(sskMIX) in out (c, spkMIX);
 (
	(A(skA, pkB, pkMIX, spkMIX)) | 
	(B(skB, pkMIX, spkMIX)) |
	(C(skC, pkMIX, spkMIX)) |
	(MIX(skMIX, sskMIX)) 
 )
\end{lstlisting}
\caption{The instance of sender, receiver, and mixer processes.}
\label{fig:rarc_inst}
\end{center}
\end{figure}

\begin{figure}
\begin{center}
\begin{lstlisting}
(*----Sender----*)
let A (SKa: skey, PKb: pkey, PKmix: pkey, SPKmix: spkey) =
 out(c, (encrypt(pkey_to_bitstring(pk(SKa)), PKmix), 
         encrypt(choice[secret_message1, secret_message2], PKmix), 
         encrypt(pkey_to_bitstring(PKb), PKmix))).
\end{lstlisting}
\caption{The instance of sender to analyse message secrecy.}
\label{fig:rarc_secrecy}
\end{center}
\end{figure}

An instance of reusable anonymous return channel is in Figure \ref{fig:rarc_inst}. We recall that submission of a message consists of the triplet {$(\lbrace{\ID_A, \PK_{A}}\rbrace_{\PK_{\MIX}},$ $ \lbrace{m}\rbrace_{PK_{\MIX}},\lbrace{\ID_B, PK_{B}}\rbrace_{PK_{\MIX}})$}.
We use the \texttt{choice} command in ProVerif to check both message secrecy and sender anonymity. This command allows us to  verify if the processes obtained by instantiating a variable with two different values are bisimilar.

We analyse secrecy of the message by checking whether the attacker can distinguish the two scenarios in which the sender outputs two different messages. Thus,  \texttt{choice} is applied in the second element of the triplet, and the process of the sender becomes as in Figure \ref{fig:rarc_secrecy}.

\begin{figure}
\begin{center}
\begin{lstlisting}
(*----Sender----*)
let A (SKa: skey, SKb:skey, PKb: pkey, PKc: pkey, 
       PKmix: pkey, SPKmix: spkey) =
 out(c, (encrypt(pkey_to_bitstring(pk(choice[SKa, SKb]), PKmix), 
         encrypt(secret_message, PKmix), 
         encrypt(pkey_to_bitstring(PKb), PKmix))).
\end{lstlisting}
\caption{The instance of sender to analyse anonymity.}
\label{fig:rarc_anonymity}
\end{center}
\end{figure}

We analyse anonymity of the sender by checking whether the attacker can say if the message is sent either to Receiver 1 or Receiver 2. In this case,  \texttt{choice}  is applied in the first element of the triplet. Figure \ref{fig:rarc_anonymity} shows the process of the sender to check anonymity.

\begin{figure}
    \centering
\FramedBox{2.8cm}{12.1cm}
{
\begin{enumerate}
\item  \emph{A}$\rightarrow$\emph{M}: $\lbrace{\ID_A, \PK_{A}}\rbrace_{\PK_{\MIX}}, \lbrace{m}\rbrace_{PK_{\MIX}},
\lbrace{\ID_B, PK_{B}}\rbrace_{PK_{\MIX}}$
\item  \emph{I}$\rightarrow$\emph{M}: $\lbrace{\ID_A, \PK_{A}}\rbrace_{\PK_{\MIX}}, \lbrace{\ID_A, \PK_{A}}\rbrace_{\PK_{\MIX}},
\lbrace{\ID_I, PK_{I}}\rbrace_{PK_{\MIX}}$
\item  \emph{M}$\rightarrow$\emph{I}: $\mathit{Sign_{SSK_{\MIX}}(\{\ID_A,
\PK_A\}_{PK_{\MIX}}),\lbrace{\ID_A, \PK_{A}}\rbrace_{\PK_{I}}}$
\end{enumerate}
}
    \caption{Attack trace on sender anonymity}
    \label{fig:rarc_attack}
\end{figure}

\begin{table}[]
\begin{center} 
\vspace{0.65cm}
\begin{tabular}{|c|r|}
\hline
{\bf Primitive} & \multicolumn{1}{c|}{{\bf Equation}}
\\ \hline
ElGamal encryption  &
$\begin{aligned}
\mi{decrypt}(\mi{encrypt}(m, pk(sk), r), sk)  =  m\phantom{ru}
\end{aligned}$
\\ \hline
Digital signature &
$\begin{aligned}
\mi{getmess}(\mi{sign}(m,ssk)) & = m\phantom{ru} \\
\mi{checksign}(\mi{sign}(m,ssk), \mi{spk}(ssk)) & = m\phantom{ue}
\end{aligned}$ 
\\ \hline
Zero-knowledge proof &
$\begin{aligned}
 \mi{zkpsec}(\mi{zkp}\_proof(\mi{encrypt}(m,pk(sk),r),(r,m)), \\
 \mi{encrypt}(m,pk(sk),r)) =  true
\end{aligned}$ 
\\ \hline
\end{tabular}
\normalsize
\caption{Equational theory to model reusable anonymous return channel}
\label{table:eqt_rarc}
\end{center}
\end{table}

\subsubsection*{Results} The results of the automatic analysis in ProVerif indicate that reusable anonymous return channel fails to guarantee both secrecy of messages and anonymity of sender and receiver identities. 
According the attack traces generated by ProVerif, both message secrecy and sender anonymity can be exploited using the same attack strategy.
The attacker can use the mixer as decryption oracle, letting the mixer reveal any of the plaintexts contained in the triplet. 
The zero-knowledge proofs required to avoid this very attack reveal to be insufficient. In fact, the attack traces provided by ProVerif show the attacker can input the mixer with valid zero-knowledge proofs.

In the following we detail the attack traces. The attacker chooses one of the three elements of the triplet. This choice depends on what the attacker wants to learn: if the target is the content of the message, the attacker chooses the second element; if the target is the identity of the sender, the attacker chooses the first element; if the target is the identity of the receiver, the attacker chooses the third element. Whatever the element of the triplet, the attacker submits this as a new message.  

Figure \ref{fig:rarc_attack} shows how the attacker can defeat sender anonymity. The attacker targets $\encr{ID_A,PK_A}{\PK_{\MIX}}$, which becomes the second element of the new triplet submitted by the attacker. 
Note that the attacker can leave the first element of the triplet and the zero-knowledge proof unchanged. The attacker replaces the third element of the triplet with a public key $PK_I$ for which the attacker knows the corresponding secret key $SK_I$. Thus, the attacker can also provide the necessary proof of knowledge of the plaintext contained in the third element.
The mixer then shuffles the input messages, and encrypts the message with the attacker public key. Since the
attacker knows the secret key $SK_I$, he can decrypt the message, which in this case is  $ID_A,PK_A$, namely the identity of the sender.

Since the attacker can substitute any of the elements of the triplet as a new message, reusable anonymous return channel can neither ensure secrecy of the messages nor anonymity of sender and receiver. It can be observed that the checksum meant to guarantee the integrity of the triplet is added after the submission of the triplet,  and is only used inside the mixer. Hence, the checksum does
not prevent the attacker from submitting a modified triplet. Unfortunately even adding the checksum before the submission of the triplet does not prevent the attack as the knowledge of the ciphertexts is sufficient to compute the checksum.

\paragraph{Remark.} Reusable anonymous return channel was originally designed to withstand a passive attacker that
however can statically corrupt parties~\citeltex{GJ03}. Our analysis in ProVerif considers an active attacker. We observe that a passive attacker is not realistic in exam,  where corrupted principals could actively try to cheat. However, an attacker who statically corrupts principals can still defeat reusable anonymous return channel. A corrupted principal can be instructed to send and receive messages via
the reusable anonymous return channel on behalf of the attacker.  The attacker still need to intercept those
messages before they enter the mixer, but this is possible with insecure networks such as the Internet.

\subsection{Formal Analysis of the \Huszti Protocol}
\label{sec:analysisH}
We now introduce the formal model of the \Huszti protocol and then the results of the analysis of authentication and privacy.

\subsubsection*{Model choices} The first model choice is about the channels. The \Huszti protocol assumes that all messages are exchanged using reusable anonymous return channel. In the previous section, we demonstrated that reusable anonymous return channel fails to guarantee both message secrecy and sender anonymity. We choose to model the \Huszti protocol with the \emph{ideal} implementation of reusable anonymous return channel, which  guarantees anonymity of senders and receivers. This can be implemented with ProVerif's  anonymous
channels.

\begin{table}
\begin{center} 
\begin{tabular}{|c|r|}
\hline
{\bf Primitive} & \multicolumn{1}{c|}{{\bf Equation}}
\\ \hline
ElGamal encryption  &
$\begin{aligned}
\mi{decrypt}(\mi{encrypt}(m, pk(sk), r), sk)  =  m\phantom{ue}
\end{aligned}$
\\ \hline
Digital signature &
$\begin{aligned}
\mi{getmess}(\mi{sign}(m,ssk)) & = m\phantom{ue} \\
\mi{checksign}(\mi{sign}(m,ssk), \mi{spk}(ssk)) & = m\phantom{ue}
\end{aligned}$ 
\\ \hline
Zero-knowledge proof &
$\begin{aligned}
\mi{zkpsec}(\mi{zkp}\_proof(\mi{exp}(\mi{exp}(g,e1),e2), e2),& \\
\mi{exp}(\mi{exp}(g,e1),e2)) & =  true
\end{aligned}$ 
\\ \hline
Diffie-Hellman exp. &
$\begin{aligned}
 \mi{exp}(\mi{exp}(\mi{exp}(g,x),y),z)  = \mi{exp}(\mi{exp}(\mi{exp}(g,y),z),x)
\end{aligned}$ 
\\ \hline
ZKP of discrete log. &
$\begin{aligned}
 \mi{checkproof}(\mi{xproof}( p, p',t, \mi{exp}(t,e), e),& \\
 p, p', t, \mi{exp}(t,e)) & = true
\end{aligned}$ 
\\ \hline
\end{tabular}
\normalsize
\caption{Equational theory to model the \Huszti protocol}
\label{table:eqt_hp}
\end{center}
\end{table}

The equational theory depicted in Table~\ref{table:eqt_hp} models the  cryptographic primitives used within the \Huszti protocol. It
includes the same models of ElGamal encryption, digital signatures, and zero-knowledge proofs defined for reusable anonymous return channels. 
In addition, we provide an equation to model the Diffie-Hellmann exponentiation. This model is limited because it just takes into account the equation needed for the protocol to work, and does not capture the full set of algebraic properties of Diffie-Hellmann exponentiation that an attacker may exploit to break the protocol. However, this has a limited influence on our analysis because, as we shall see later, the protocol ensures only one out ten security requirements, namely in most cases the attacker breaks the protocol though the simple model of Diffie-Hellmann exponentiation.

The equations for zero-knowledge proofs are customised according to the exponentiation operator. In particular, we support the model of zero-knowledge proof of the equality of discrete logarithms $\mathit{check\_proof}$ with tables in ProVerif. This approach is needed because ProVerif cannot deal with the associativity of multiple exponents. This approach is sound because it limits the attacker capability to generate fake zero-knowledge proofs, since the attacker cannot write and read ProVerif tables.
 
We assume that the same generator $g$ is used to generate the pseudonyms of candidates and examiners. 
This choice is sound because we distinguish the roles, and each principal is identified by its public
 key. We replace the candidate identity with the corresponding pseudonym inside the events to check authentication requirements. Note that the replacement is also sound because the equational theory preserves the bijective mapping between the keys that identify the
 candidates and the pseudonyms.

\begin{figure}
\begin{center}
\begin{lstlisting}
let EAi (SSKea: sskey,  SPKcom: spkey) =
(*Preparation*)
 new s: exponent;
 let h=exp(g, s) in
 let sh=sign(h,SSKea) in
 out(c, (h,sh));
 in(c, (quest: bitstring, squest: bitstring));
 (!Ereg(h,s) | !Creg(h,s,quest,squest, SPKcom, SSKea)).

let Ereg (h: bitstring, s: exponent)=
 (* Examiner Registration *)
 get keys (=g_e, xpk_e) in
 let q_tilde=exp(xpk_e, s) in
 out(c, (q_tilde, xpk_e)); 
 insert zkpeq(q_tilde, xpk_e);
 (* EA inserts q_tilde into a table to support the zkp with E *)
 in(c, (q: bitstring, q': bitstring));
 out(c, xproof(q,q',g, h, s));
 in (c, (t: bitstring, =q, =q'));
 if exp(q,s)=q' then
  in(c, zkp_sec_proof: bitstring);
  if zkpsec2(zkp_sec_proof,t, q)=true then 0.

let Creg (h: bitstring, s: exponent, quest: bitstring, 
          squest: bitstring,  SPKcom: spkey, sskea:sskey)=
 (* Candidate Registration *)
 get keys (=g_c, xpk_c) in
 let p_tilde=exp(xpk_c, s) in
 insert reg_cand(xpk_c, p_tilde, h);
 out(c, (p_tilde, g_c));
 (* EA inserts p_tilde into a table to support the zkp with C *)
 insert zkpeq(p_tilde, xpk_c); 
 in(c, (p: bitstring, p': bitstring));
 (* EA registered C with pseudonym 'p' to the exam 'h' *) 
 out(c, xproof(p,p',g, h, s));
 Exam(h,s,quest,squest, SPKcom,sskea).
\end{lstlisting}
\caption{The process of the exam authority that concerns preparation.}
\label{fig:hp_eap}
\end{center}
\end{figure}

\begin{figure}
\begin{center}
\begin{lstlisting}
let Exam (h: bitstring, s: exponent, quest: bitstring, 
          squest: bitstring, SPKcom: spkey, SSKea: sskey)=
(*Testing*)
 in (c, (r: bitstring,p: bitstring,p':bitstring));
 if exp(p,s)=p' then  
  in (c, zkp_sec_proof: bitstring); 
  if zkpsec2(zkp_sec_proof, r, p)=true then
   if quest=checksign(squest, SPKcom) then
    out(c, (quest, squest));
    in (c, (=r, =p, answer: bitstring));
    (* EA succesfully collects the pair ('quest','answer') *)
    (* from C with pseudonym 'p' for the exam 'h' *)
    event collected(p', h, quest, answer);
    out(c,  hash( (r, p, p', zkp_sec_proof, quest, answer)));

(*Marking*)
    (* EA chooses an E who registered for the exam 'h' *)
    get examinertable(t, q, =h, xzkp_sec_proof) in
    new eap: bitstring;
    (* EA assigns 'eap' to the pair ('quest','answer') *)
    (* submitted by C with pseudonym 'p' for the exam 'h',*)
    (* and distributes them to E with pseudonym 'q' *)
    event distributed(p',h, quest, answer, eap, q);
    out(c, (eap, answer)); 
    get zkpeq(hma,=spkeytobitstring(spk(SSKea))) in 
    in (c, (mark: bitstring, =hma, hma_e_enc: bitstring, 
    zkp_sec_hma: bitstring, =t, =q));
    if (checkproof(zkp_sec_hma, hma, hma_e_enc,t,q)=true) then
     out(c, p');
(*Notification*)
     in (c, (=p', p_tilde: bitstring));
     get reg_cand(xpk_c, =p_tilde, =h) in  
     insert marks(xpk_c,mark, hma, hma_e_enc, zkp_sec_hma, t,q).
\end{lstlisting}
\caption{The process of the exam authority that concerns testing, marking, and notification.}
\label{fig:hp_eatmn}
\end{center}
\end{figure}

The process of the exam authority is modelled as in Figure \ref{fig:hp_eap} and \ref{fig:hp_eatmn}. It is conveniently split into four sub processes. Three sub processes concern preparation and consist of the initialisation of the exam authority, the registration of the candidate, the registration of the examiner. The last sub process concerns the remaining phases, namely testing, marking, and notification.

\begin{figure}
\begin{center}
\begin{lstlisting}
let C (SKc: skey, SPKea: spkey, SPKcom: spkey) =
(*Preparation*)
 in (c, (h: bitstring, sh: bitstring)); 
 if h=checksign(sh, SPKea) then 
  (* The zkp of equivalence of discrete log is supported by *)
  (* the tables 'zkpeq' and 'zkpeqnet'. *)
  (* In doing so, C can verify that 'ea_tilde' and 'p_tilde' *)
  (* have been correctly generated resp. by EA and NET *)
  get zkpeq(ea_tilde, =exp(g,skey_to_exponent(SKc))) in  
  get zkpeqnet(=ea_tilde,r,p') in 
  let p=exp(r,skey_to_exponent(SKc)) in
  out(c, (p, p'));
  in (c, zproof: bitstring);
  if (checkproof(zproof, p, p',g,h)=true) then 
   let zkp_sec_c = zkp_proof2(r,p,skey_to_exponent(SKc)) in
   out(c, (r,p,p')); 
   out(c, zkp_sec_c);
   
(*Testing*)
   in (c, (quest: bitstring, squest: bitstring));
   if quest=checksign(squest, SPKcom) then
    new answ: bitstring;
    (* C submits 'answ' and 'quest' for the exam 'h' *)
    event submitted(p', h, quest, answ);
    out(c, (r, p, answ));
    in (c, receipt: bitstring);
    if (hash( (r,p,p', zkp_sec_c, quest, answ))=receipt) then
     
(*Notification*)
     get marks(=exp(g,skey_to_exponent(SKc)),mark, hma, hma_e_enc, 
     zkp_sec_hma, t,q) in  
     get zkpeq(=hma,=spkeytobitstring(SPKea)) in 
     if (hash((mark, answ)) = hma && checkproof(zkp_sec_hma, hma, 
      hma_e_enc,t,q)=true ) then
      (* C is notified with 'mark' for the exam 'h' *)
      event notified(p',h,mark).
\end{lstlisting}
\caption{The process of the candidate.}
\label{fig:hp_candidate}
\end{center}
\end{figure}

\begin{figure}
\begin{center}
\begin{lstlisting}
let E (SKe: skey,  SPKea: spkey, SPKcom: spkey) =
 (*Preparation*)
 in (c, (h: bitstring, sh: bitstring));
 if h=checksign(sh, SPKea) then
  new alpha: exponent;
  get zkpeq(q_tilde, =exp(g,skey_to_exponent(SKe))) in 
  let q'=exp(q_tilde, alpha) in
  let t=exp(g, alpha) in
  let q=exp(t, skey_to_exponent(SKe)) in
  out (c, (q, q'));
  in (c, zproof: bitstring);
  if (checkproof(zproof, q, q',g,h)=true) then
   let zkp_sec_e = zkp_proof2(t,q,skey_to_exponent(SKe)) in
   out(c, (t,q,q'));
   out(c, zkp_sec_e);
   insert examinertable (t, q, h, zkp_sec_e);
   
   (*Marking*)
   in (c, (quest: bitstring, squest: bitstring));
   if quest=checksign(squest, SPKcom) then
    in (c, (eap: bitstring, answer: bitstring));
    new mark: bitstring;
    event marked(quest, answer, mark, eap, q, h);
    let hma=hash( (mark, answer) ) in
    let hma_e= exp(hma, skey_to_exponent(SKe)) in
    insert zkpeq(hma, spkeytobitstring(SPKea));
    let zkp_sec_hma=xproof(hma,hma_e,t,q,skey_to_exponent(SKe)) in
    out(c, (mark, hma, hma_e, zkp_sec_hma, t, q)).
\end{lstlisting}
\caption{The process of the examiner.}
\label{fig:hp_examiner}
\end{center}
\end{figure}

\begin{figure}
\begin{center}
\begin{lstlisting}
let NET () =
(*Preparation*)
 in (c, (p_tilde: bitstring, =g_c));
 new ro: exponent;
 let p'=exp(p_tilde, ro) in
 let r=exp(g, ro) in
 (* The NET registered the candidate with pseudonym p' *)
 event registered(p');
 out(c,(p', r));
 (* The NET inserts 'p_tilde' into a table to support *)
 (* the zkp between EA (the prover) and C (the verifier) *)
 insert zkpeqnet(p_tilde,r,p');
 
(*Notification*)
 in (c, =p'); 
 out(c, (p', p_tilde)).
\end{lstlisting}
\caption{The process of the NET.}
\label{fig:hp_NET}
\end{center}
\end{figure}

\begin{figure}
\begin{center}
\begin{lstlisting}
process
 !(
 new sskEA: sskey; let spkEA = spk(sskEA) in out (c, spkEA);
 new sskCom: sskey; let spkCom = spk(sskCom) in out (c, spkCom);
 new question: bitstring;
 let squestion=sign(question, sskCom) in
 out(c, (question, squestion));

 !(new skC: skey; let pkC = exp(g,skey_to_exponent(skC)) in 
   out (c, pkC); insert keys(g_c, pkC); C(skC, spkEA, spkCom)
  )| 
 !(EAi(sskEA , spkCom))|
 !(new skE: skey; let pkE = exp(g,skey_to_exponent(skE)) 
   in out (c, pkE); insert keys(g_e, pkE); E(skE,  spkEA, spkCom)
  )| 
 !(NET())
 )
\end{lstlisting}
\caption{The exam process.}
\label{fig:hp_exam}
\end{center}
\end{figure}

The ProVerif model of the candidate is depicted in Figure \ref{fig:hp_candidate}, the  examiner process is in Figure \ref{fig:hp_examiner}, the NET process is in Figure  \ref{fig:hp_NET}, and the exam process is modelled as in Figure \ref{fig:hp_exam}. All the processes are augmented with the events that allow verifying the authentication requirements. 
To verify Question Indistinguishability we use the \texttt{noninterf} command of ProVerif, which checks that any two instances of the exam protocol that only differ in the value of the variable of questions are bisimilar.  To verify
Mark Privacy, Anonymous Examiner and Anonymous Marking we use the ProVerif command \texttt{choice[]}.  \index{Anonymous Marking} \index{Mark Privacy}
The full ProVerif code used to analyse the requirements of the \Huszti protocol is available on the Internet \citeweb{thesiscode}. 

\begin{table}
\begin{center}
{\begin{tabular}{|c|c|c|c|}
\hline
{\bf Requirement} & {\bf Result} & {\bf Time} \\ \hline
  {\cautho} &  $\have$&  26 s \\ \hline
  {\aau} &  $\havenot$ & 3 s  \\ \hline 
  {\CAu} &  $\havenot$ & 3 s  \\ \hline 
      {\ta} &  $\havenot$ & 33 s \\ \hline
    {Mark Authenticity} &  $\havenot$  &  52 s \\ \hline
     {Question Indistinguishability} &  $\havenot$ & $<$ 1 s\\ \hline
      {Anonymous Marking} &  $\havenot$ & 1h 58 m 33 s\\ \hline
       {Anonymous Examiner} &  $\havenot$ & 6h 37 m 33 s \\ \hline
       {Mark Privacy} &  $\havenot$ & 23 m 59 s  \\ \hline       
       {Mark Anonymity} &  $\havenot$ & 49 m 5 s \\ \hline
       \end{tabular}}
\caption{Summary of the analysis of the \Huszti protocol.}
\label{tab:resultshp}
\end{center}
\end{table}%

\subsubsection*{Results} The analysis in ProVerif shows that the \Huszti protocol only ensures \cautho as reported in Table \ref{tab:resultshp}.
Regarding \aau, ProVerif shows an attack trace that allows the \ea to accept a test that has not been submitted by a registered candidate.
In fact, the attacker can generate a fake pseudonym that allows him to take part in an exam for which the attacker did not register. This is possible because the \ea does not check whether the pseudonym has been actually created using the partial information provided by \index{Answer Authenticity}
the NET. The attacker generates a secret key $SK_I$, and calculates an associate pseudonym, which sends to the exam authority. The exam
authority successfully verifies the received data since the attacker knows $SK_I$, hence the exam authority accepts the test.
In other words, the \ea may collect a test whose pseudonym is replaced with one chosen by the attacker. The same attack trace violates \CAu because the attacker can generate a valid pseudonym for a candidate who did not register for the exam.  \index{Test Origin Authentication}

ProVerif finds a counterexample that invalidates \ta. The requirement cannot be achieved because there is no mechanism that allows \index{Test Authenticity}
the examiner to check whether the answers have been forwarded by the \ea. Although the answer is encrypted with the public key of the mixer, this does not guarantee that the exam authority actually sent the message, because anyone can submit any message to the mixer.

Regarding Mark Authenticity, ProVerif provides an attack trace in which the attacker can forward any answer to any \index{Mark Authenticity}
examiner, even if the answer was not collected by the exam authority. Moreover, no mechanism ensure that the notified mark originates from the \ea. In fact, the attacker can notify the candidate by himself with a mark of his choice.

The \Huszti exam protocol does not guarantee any privacy requirement.
Intuitively, all privacy requirements can be violated because reusable anonymous return channel does not guarantee anonymity.

However, even assuming anonymous channels, ProVerif shows an attack trace for each requirement. 
Question Indistinguishability does not hold because messages sent via reusable anonymous return channel are not secret, as our analysis \index{Question Indistinguishability}
demonstrates. Since the questions are sent through the anonymous channel, the attacker can still  obtain them.  

Anonymous Marking is violated since the attacker can check whether a candidate accepts the zero-knowledge proof, hence associates the candidate identity with the pseudonym, and then identifies the candidate's test.

Anonymous Examiner can be also violated because the attacker can track which examiner accepts the zero-knowledge proof when receiving \index{Anonymous Examiner}
the partial pseudonym, hence associates the answer to the examiner. 

Mark Privacy fails because the examiner sends the mark to the exam authority via reusable anonymous return channel, which does not ensure secrecy.  

Finally, ProVerif shows that the \Huszti protocol does not also ensure \emph{Mark Anonymity}. Since one can track which \index{Mark Anonymity}
pseudonym is assigned to the candidate, and the mark is not secret, the attacker can link the candidate to the assigned mark.

\begin{table}
\begin{center}
{\begin{tabular}{|c|c|c|c|}
\hline
{\bf Requirement} & {\bf Result} & {\bf Time} \\ \hline
  {\cautho} &  $\have$ &  3 s \\ \hline
  {\aau}&  $\havenot$& 2 s  \\ \hline
  {\CAu}&  $\have$& 2 s  \\ \hline
    {\ta} &  $\have$ & 3 s \\ \hline
    {Mark Authenticity} &  $\have$ & 4 s \\ \hline
       \end{tabular}}
\caption{Summary of the analysis of authentication of the modified \Huszti protocol.}
\label{tab:resultshpfix}
\end{center}
\end{table}%

\subsection{Fixing Authentication} We propose four modifications to the
\Huszti protocol in order to achieve most of authentication requirement. In
particular, we prove in ProVerif that the modified \Huszti protocol achieves \cautho, \CAu, \ta, and Mark Authenticity as in Table \ref{tab:resultshpfix}. We found no easy solution for \aau because the protocol sees no signatures for candidates, and reusable anonymous return channel does not guarantee authentication.
\index{Candidate Authorisation}
The first modification consists in the NET receiving the partial pseudonyms generated by the \ea via a secure channel instead via reusable anonymous return channel.
It can be observed that the \ea and the NET do not need to communicate anonymously via RARC, as the original protocol prescribes. Conversely, they need a secure channel to avoid 
that the attacker injects messages.  In doing so, the attacker cannot use the NET to generate fake pseudonyms.

Similarly, the second modification consists in the \ea receiving the eligible pseudonyms from NET via a  secure channel. Thus, the \ea generates zero-knowledge proofs of the equality of
discrete logarithm to eligible pseudonyms only. The \ea can also store the eligible pseudonyms, which can be checked at testing before the \ea accepts a test from a candidate. 

The third modification concerns marking and consists in the \ea signing the collected test prior to distribute it to the chosen examiner. In fact, it is required the examiner identity to be anonymous but not the \ea's. Thus, the examiner marks the test only if the signature can be correctly verified. In the original protocol, the examiner could not verify whether a test was sent by the \ea. 

The last modification concerns the test identifier the \ea affixes to the test before distributing it to the examiner, and consists in a modified receipt of candidate's submission. The \ea adds the test identifier to the receipt and signs it. The examiner also adds the test identifier into the receipt of marking, hence the candidates can verify whether they are notified with the correct marks. 
In the original protocol, the attacker could notify the candidate with any other examiner's mark because the candidate was unaware of the test identifier.

Table~\ref{tab:resultshp} summarises the results of the formal analysis of the \Huszti protocol assuming all principal being honest.
The reported times refer to ProVerif analyses over an Intel Core i7 3.0 GHz machine with 8 GB RAM.

It can be seen that the modified  \Huszti protocol guarantees four out five authentication requirements. Unfortunately, no easy solution can be envisaged for privacy. The design of the original protocol is heavily based on the assumption that reusable anonymous return channel guarantees secrecy, authentication, and anonymity. Since reusable anonymous return channel guarantees none  of these properties, the \Huszti protocol would need a complete redesign to achieve privacy.
  
\section{Conclusion}\label{sec:conclusion}
This chapter discusses a formal framework for the security analysis of exam protocols. Although many symbolic analysis methods have been proposed, our choice to model exams in the \appi calculus reveals to be interesting. We advance five authentication and five privacy requirements for exam, counting a total of ten novel requirements. The proposed framework is validated with the formal analysis of the 
\Huszti protocol, the first secure exam scheme proposed in the literature. The protocol succumbed to our analysis, though being quite complex.

It is found that the protocol guarantees  only one of the ten requirements. Authentication is compromised because of
inaccuracies in the protocol design. Privacy requirements are mostly violated because assumptions on reusable anonymous return channel.
It is demonstrated that an attacker can compromise message secrecy and  anonymity on reusable anonymous return channel.
  
This chapter also introduces a few modifications on the \Huszti protocol in order to 
guarantee most of the authentication requirements. 
A formal analysis in ProVerif confirms that the modified protocol ensures these requirements.
However, even with an ideal reusable anonymous return channel implementation that  ensures  anonymity, the \Huszti protocol does not ensure any privacy requirements. Thus, we think that the  protocol requires fundamental changes.

Generally speaking, the proposed formal framework brings exams into the attention of the security community.  Computer-based exams are becoming widespread, and it can be difficult to discover exam protocol vulnerabilities as they may be  exposed to unprecedented cheating attacks. 

This work poses the first research step in the formal understanding of exam protocols.
This is corroborated by the analysis of more protocols as discussed in chapters \ref{chap:remark} and \ref{chap:wata}.


\chapter{Formalising Verifiability}\label{chap:verifiability}
A fundamental requirement of exams is transparency. While traditional exams \index{traditional exam} can be normally observed through all the phases of the exam, computer-assisted exams may introduce opaqueness. In general, any operation performed by computers may not be observable, depending on the level of computer assistance. For example, computer markers may alter the evaluations of the tests, or a malware in the collector may modify or drop the submitted tests. Thus, transparency demands for \emph{verifiable} exams.  A verifiable exam can be checked for the presence or the absence of irregularities and provides evidences about fairness and correctness of marking.  \index{verifiable exam}
Moreover, verifiable exams foster public trust, as transparency can persuade the involved parties to comply with regulations.

To analyse whether an exam protocol is verifiable, it is necessary to clarify the relevant verifiability requirements, and then build a framework to check the protocol against these requirements.
In this chapter, we present a clear understanding of verifiability for exam protocols and propose a methodology to analyse their verifiability.
Differently from the framework advanced in chapter \ref{chap:formal}, we propose a formal framework based on multisets that abstract away from the applied $\pi$-calculus constraints and is suitable for both symbolic and computational analysis.
We also formalise eleven verifiability requirements for exams. Each requirement is pivoted on a \emph{verifiability-test}, and we state the conditions that a sound and complete verifiability-test has to satisfy.  Following a practice already explored in other domains~\citeltex{CohenFischer85,Benaloh94,Benaloh96,HS00}, we classify our verifiability requirements into individual and universal. 



\paragraph*{Outline of the chapter.}
Section~\ref{sec:relatedveri} discusses the related work about the formalisation of verifiability in other contexts.
Section~\ref{sec:modelveri} introduces the  constituents of the formal framework.
Section~\ref{sec:reqveri} contains the specification of eleven verifiability requirements for exams.
Section~\ref{sec:concveri} draws the conclusions and outlines the future work.


\section{Related Work}
\label{sec:relatedveri}
To the best of our knowledge, verifiability for exams has not been studied at the time of writing this manuscript. Few papers list informally a few security requirements. Castella-Roca \etal \citeltex{DBLP:conf/IEEEares/Castella-RocaHD06} discuss a secure exam management system, and informally define  authentication, privacy, correction and receipt fullness. Huszti and Peth\H{o}~\citeltex{huszti10}, whose protocol is analysed in chapter \ref{chap:formal}, extend the requirements with secrecy and robustness. None of the works outlined above addresses verifiability. 

However, verifiability has been studied in domains close to exams, such as
 voting and auctions, and different models and  requirements have been proposed from the beginning of 2010s \citeltex{DJL13,KuesTrudVogt10,Kremer10}. 
In voting, \emph{individual verifiability} signifies that voters can verify that  their votes have been 
handled correctly, namely ``cast as intended'', ``recorded as cast'', and ``counted as recorded'' \citeltex{Benaloh94,HS00}. 
The concept of \emph{universal verifiability} has been introduced to express
that auditors can verify the correctness of the tally using only
public information \citeltex{CohenFischer85,Benaloh94,Benaloh96}.
Kremer \etal\citeltex{Kremer10} formalised both individual and universal
verifiability in the \picalc{}. They also introduced the requirement of  
\emph{eligibility verifiability}, which  expresses  that
auditors can verify that each vote in the election result
was cast by a registered voter, and there is at most one vote per voter.
Smyth \etal\citeltex{Smyth10} used ProVerif to check verifiability in three voting protocols. They
express the requirements  as reachability properties.  
In the next chapter, we also analyse an exam protocol in ProVerif to validate our framework. However, our model and
definitions  are constrained neither to the \picalc{}~ nor to ProVerif.
The sound and complete verifiability-tests that we use to specify our requirements are inspired by the work of Dreier \etal\citeltex{DJL13}, who formalised verifiability for e-auction.

K\"usters \etal\citeltex{KuesTrudVogt10} studied
\emph{accountability}. This requirement says that, in presence of a protocol failure, one can identify the principal responsible for the failure. The notion of accountability is strongly related to verifiability as the latter's goal is to check the presence of protocol failures.
In their work, K\"usters \etal give symbolic and computational definitions of verifiability, which they argue to be a weak variant of accountability.
Differently from our approach, their framework has to be instantiated for each application by identifying relevant verifiability goals. 

Guts \etal\citeltex{GutsFournetNardelli09} defined \emph{auditability} as the quality of a protocol, which stores a sufficient  number of evidences, to convince an honest judge that specific properties are
satisfied. As we shall see later, auditability expresses the same concept of universal verifiability as defined in this dissertation: anyone, even an outsider without a private knowledge about the protocol execution, can verify that the system relies only on the available pieces of evidence.


\section{A More Abstract Model} \label{sec:modelveri}
In chapter \ref{chap:terminology}, we observed that any exam involves at least the candidate role plus other possible authority roles, and that the run of an exam can be represented as the sequential execution of preparation, testing, marking, and notification phases. From these observations, we build a formal framework for the analysis of verifiability requirements.
While in chapter \ref{chap:formal} we advanced a framework based on the applied $\pi$-calculus, hence supported by an operational semantics, here we propose an abstract framework for the analysis of verifiability that comes without such semantics.
On the one hand, a more abstract framework requires to map the model onto one that allows the analysis of the protocol. On the other hand, it allows for a wider choice of analysis methods, regardless whether they are based on the symbolic or computational model. In chapter \ref{chap:remark} we validate this framework by analysing verifiability in a novel exam protocol, using ProVerif as analysis method.

We view the abstract model \index{abstract model} of an exam consisting  of four sets, three relations, and one function. The four sets are of  candidate's identities $I$,  questions $Q$, answers $A$,  and marks $M$. 
The three relations \red{$\fun{Accepted}$,  $\fun{Marked}$, and  $\fun{Assigned}$}, link candidates, questions, answers, and marks along the four phases. A test consists of the pair  $(Q \times A)$ of questions and answers. The function $\fun{Correct}$ maps a mark to a test.
It is assumed that sets and relations are built from data logs such as registers or repositories.

\begin{prop}{\bf(Exam (abstract model))}
\label{examdef}
An exam $E$ is a tuple  $(I,Q,A,M, $ $\alpha)$ where $I$ of type $\type{I}$ is a set
of candidate identities, $Q$ of type $\type{Q}$ is a set of questions, $A$ of
type $\type{A}$ is a set of answers, $M$ of type $\type{M}$ is a set of marks,
and $\alpha$ is the set of the
following relations: 
\begin{itemize}
\item $\fun{Accepted} \subseteq I \times (Q \times A)$: the
  candidates' \examtests accepted by the collector authority;
\item $\fun{Marked} \subseteq I \times (Q \times A) \times M$: the
  marks given to the   candidates' \examtests;
\item $\fun{Assigned} \subseteq I \times M$: the marks assigned to the candidates;
\item $\fun{Correct}: (\type{Q} \times \type{A}) \rightarrow
  \type{M}$: the function used to mark a \examtest.
\end{itemize} 
\end{prop}

Definition~\ref{examdef} can be extended with two specific subsets:
\begin{itemize}
\item $I_r\subseteq I$ as the set of identities of candidates  who registered for the exam;
\item $Q_g \subseteq Q$ as the set of questions generated by the question committee.
\end{itemize} 

It can be noted that this approach can model exam executed either honestly or with frauds. For example, the set $I\setminus I_r$ contains the identities of the unregistered candidates who
took the exam. Similarly, the set  $Q\setminus Q_g$ contains the illegitimate questions administered at the exam.  An honest execution of an exam requires $(I\setminus I_r) = (Q\setminus Q_g) = \emptyset$

The function $\fun{Correct}$ models any objective mapping that assigns a mark to an
answer. This works well for multiple-choice tests, but it is
inappropriate for free-response tests. The evaluation of  a free-response question is
hardly objective: the ambiguities of natural language can lead to
subjective interpretations by the examiner. In this case, it is not possible to verify the correctness of the marking, whatever model is considered.
In other words, an exam protocol that does not allow a definition of  the function $\fun{Correct}$  cannot be checked for the correctness of the marking.



\section{Verifiability Requirements}
\label{sec:reqveri}
In this section we present eleven verifiability requirements for exams. Verifiability is a sort of meta-requirement as a protocol is verifiable with respect to specific properties.


To be verifiable an exam should be \emph{testable}, namely it should provide an executable procedure (\emph{verifiability-test}) that checks a specific property on the exam execution. For executable procedure we refer to the existence of a procedure that takes in some data and outputs true or false. Such procedure may be either explicitly provided by the protocol designer or found by the protocol analyst.
A verifiability requirement has the form $t(e) \Leftrightarrow c$, where the verifiability-test  $t(e)$ is a function from ${\cal E} \rightarrow \bool$, where ${\cal E}$ is the set of \evidences $e$, and $c$ is a predicate that models the specific property. The \evidences needed to run the verifiability-test are obtained from the available information about the  execution of the exam and from the private knowledge of the involved roles. It is assumed that the pieces of \evidences become available after the exam concludes and are not subject to further changes. 
The following definition resumes the notion of testable exam.  \index{testable exam}
\begin{prop}{\bf{(Testable exam)}}
An exam protocol is \emph{testable} if it provides a verifiable-test that checks a desired property.
\end{prop}

Being testable is not a sufficient condition for an exam to be verifiable because the verifiability-test should be sound and complete ($\Leftrightarrow$) for the specific property: the success of the verifiability-test is a \emph{sufficient} condition for $c$ to hold (soundness $\Rightarrow$), and the success of the verifiability-test  is a \emph{necessary} condition for $c$ to hold (completeness $\Leftarrow$).  
This is captured by the definition of \emph{verifiable exam}. \index{verifiable exam}
\begin{prop}{\bf{(Verifiable exam)}}
An exam protocol is \emph{verifiable} for a desired property if the exam is testable and the corresponding verifiable-test is \emph{sound} and \emph{complete}.
\end{prop}

With a security take, a verifiability-test should be sound in presence of an attacker and corrupted principals.
It means that when the test succeeds the property holds despite the presence of attacker and corrupted principals.
The verifiability-test should be complete to avoid trivialities: a verifiability-test that always returns false is sound but useless. 

Sound verifiability-tests cannot be complete if a corrupted principal is allowed to submit incorrect \evidence, since the verifiability-test would fail although the property holds. Thus, a verifiability-test should be complete in the sense that if all principals follow the protocol, then the verifiability-test must succeed.

A verifiability-test can be run by exam principals or outsiders, a distinction that leads to two notions of verifiability requirements: \emph{individual} and \emph{universal}.  
In the scenario of exam, we view individual verifiability as verifiability from the perspective of the candidate role. The candidate  can feed the verifiability-test with the private knowledge
acquired during the exam, namely the candidate's identity, the  test, the  mark, and the messages the candidate exchanged with the other principals through the exam.

We view universal verifiability as verifiability from the perspective of an external auditor or outsider. This role can  be played by auditors who 
acquire no private knowledge during the exam. The auditor typically has no tasks associated to an exam, thus he has no candidate's identity,  he has not seen the exam's questions, answered any of them, and he did not receive any mark. Besides, he has not interacted with any of the exam principals.
In short, the auditor runs the verifiability-tests only using the exam's available pieces of \evidence.  

\begin{table}
{\footnotesize\centering
\begin{tabular}{|c||p{4.1cm}|p{3.55cm}|}
\hline
{\bf Requirement} & \multicolumn{1}{c|}{{\bf Individual Verifiability}} & \multicolumn{1}{c|}{{\bf Universal Verifiability}} \\\hline
Registration &  &  
$\begin{array}{l}
\fun{\registrationFun{}_\universalFun{}}(e)
\Leftrightarrow \\
I_r \supseteq 
\set{i: (i,x) \in \fun{Accepted }}
\end{array}$
\\ \hline
Question Validity &
$\begin{array}{l}
\fun{\questionValidityFun{}_\individualFun{}}(i, q, a, m, p) 
\Leftrightarrow 
(q\in Q_g)
\end{array}$ & 
\\\hline
\red{Marking 
Correctness} & 
$\begin{array}{l}
\fun{\markingCorrectnessFun{}_\individualFun{}}(i, q, a, m, p)
\Leftrightarrow \\
(\fun{Correct}(q,a) = m) 
\end{array}$ 
& $\begin{array}{l}
\fun{\markingCorrectnessUniversalFun{}_\universalFun{}}(e) 
\Leftrightarrow \\
(\forall (i,x,m) \in \fun{Marked},\\
\fun{Correct}(x) = m
\end{array}$ 
\\\hline
Test Integrity & 
$\begin{array}{l}
\fun{\etIntegrityFun{}_\individualFun{}}(i, q, a, m, p)
\Leftrightarrow \\
\bigl((i,(q,a))\in\fun{Accepted } \\
\wedge \exists m':(i,(q,a),m') \in 
\fun{Marked}\bigr)
\end{array}$
& $\begin{array}{l} 
\fun{\etIntegrityUniversalFun{}_\universalFun{}}(e)
\Leftrightarrow 
\fun{Accepted} = \\
\set{(i,x): (i,x,m) \in
  \fun{Marked}} \end{array}$ \\\hline
Test Markedness & 
$\begin{array}{l}
\fun{\etMarkednessFun{}_\individualFun{}} (i, q, a, m, p)
\Leftrightarrow  \\ 
(\exists m' :(i,(q,a),m') \in \fun{Marked}))
\end{array}$ & 
$\begin{array}{l}
\fun{\etMarkednessUniversalFun{}_\universalFun{}}(e)
\Leftrightarrow 
\fun{Accepted} \supseteq 
\\ \set{(i,x): (i,x,m)
  \in \fun{Marked}}
  \end{array}$ \\\hline
Marking Integrity & 
$\begin{array}{l}
\fun{\markIntegrityFun{}_\individualFun{}}(i, q, a, m, p)
 \Leftrightarrow \\
 \exists m': \big((i,(q,a),m') \in \fun{Marked} \\ 
\wedge (i,m') \in \fun{Assigned}\big) \end{array}$
& 
$\begin{array}{l}
\fun{\markIntegrityUniversalFun{}_\universalFun{}}(e)
\Leftrightarrow 
\fun{Assigned} = \\
\set{(i,m): (i,x,m) \in
  \fun{Marked}}
  \end{array}$ \\\hline
\begin{tabular}{c}
Marking Notification \\ 
Integrity\end{tabular} & 
$\begin{array}{l}\fun{\markNotificationIntegrityFun{}_\individualFun{}}(i, q, a, m, p)
 \Leftrightarrow \\
 (i,m) \in \fun{Assigned}
\end{array}$ & \\ \hline
\end{tabular}
}

\caption{Individual and Universal Verifiability}
\label{iuv:table}
\end{table}

The list of proposed verifiability requirements is not meant to be exhaustive but aims to cover all the  phases of an exam.
The requirements concern the verifiability of candidate registration, the validity of questions, and the integrity of tests, marks, and notification. In the remainder of the section we detail the requirements, which are concisely listed in Table \ref{iuv:table}.
Generally speaking, an exam is fully verifiable,  if it ensures all the verifiability requirements.

\subsection{Individual Verifiability}
Individual verifiability allows the candidate to verify some aspects of the exam using the public data that is available from the execution of the exam plus the candidate's private knowledge.  
The candidate knows her identity $i$, the \examtest she submitted, which consists of question $q$ and answer $a$, and the notified mark $m$. 
The candidate also knows the perspective $p$ of the run of the exam. The perspective consists of the messages the candidate sent and received during the run of the exam.
Thus, the \evidence is a tuple $(i, q, a, m, p)$. Note that the candidate's perspective $p$ is not necessary to specify the predicate that models the properties to verify. In fact, the perspective never appears in the predicate $c$. However, the perspective may be necessary to implement the verifiability-test $t(e)$.

The six individual verifiability requirements concern the validity of the questions, the integrity of the submitted test, and the correctness and integrity of the mark notified to the candidate.

The first requirement is \emph{\questionValidity{}}, which signifies  that the candidate can check that she received the questions actually generated by the question committee.
The requirement is modelled by a verifiability-test that returns true if the questions $q$ received by the candidate belong to the set of the valid questions $Q_g$ generated by the question committee. This is formalised as follows:

\begin{prop}[\questionValidity{} \individual]  \index{Question Validity}
\label{QV}
Given an exam $E$ and a set of $ $ \mbox{verifiability-tests} $\beta$, then $(E,\beta)$ is \emph{\questionValidityAdj{}} 
if there is a $ $ verifiability-test $\fun{\questionValidityFun{}_\individualFun{}}: 
{\cal E}
\rightarrow \bool$ in $\beta$ 
\suchthat
\[\fun{\questionValidityFun{}_\individualFun{}}(i, q, a, m, p) 
\Leftrightarrow (q\in Q_g)\]
\end{prop}

The next requirement is \emph{\markingCorrectness{}}, which says that the candidate can verify that the mark she received is correctly computed on her \examtest. It can be formalised as:

\begin{prop}[\markingCorrectness{} \individual] \index{Marking Correctness}
\label{MC}
Given an exam $E$ and a set of verifiability-tests $\beta$, then $(E,\beta)$ is \emph{\markingCorrectnessAdj{}} if there is a verifiability-test $\fun{\markingCorrectnessFun{}_\individualFun{}}: 
{\cal E}
\rightarrow \bool$ in $\beta$
\suchthat
\[\fun{\markingCorrectnessFun{}_\individualFun{}}(i, q, a, m, p) \Leftrightarrow (\fun{Correct}(q,a) = m)\]
\end{prop}

A way to ensure  \markingCorrectness{} is to give the candidate access to the marking algorithm, so she can compute again
the mark  and compare it with the one she received. As we discussed in Section~\ref{sec:modelveri}, this makes perfect sense with multiple-choice tests, but it makes not in the case of free-response tests.
However, one can envisage other ways to convince the candidate she received the correct mark, provided the examiner follows the marking algorithm correctly. For example, the candidate could check that the integrity of her \examtest is preserved from submission until marking, and that the integrity of the mark is preserved from marking until notification. 
The remaining individual verifiability requirements cover these very checks.

The third requirement is \emph{\etIntegrity{}}, which  states that the candidate can check that her test is accepted and marked as she submitted it. It is formalised as follows:

\begin{prop}[\etIntegrity{} \individual] 
\label{ETI}
Given an exam $E$ and a set of verifi\-ability-tests $\beta$, then $(E,\beta)$ is
\emph{\etIntegrityAdj{}} if there is a verifiability-test $\fun{\etIntegrityFun{}_\individualFun{}}: {\cal E}
\rightarrow \bool$ in $\beta$ 
\suchthat
\[\fun{\etIntegrityFun{}_\individualFun{}}(i, q, a, m, p)
\Leftrightarrow \bigl((i,(q,a))\in\fun{Accepted} 
\wedge \exists m':(i,(q,a),m') \in 
\fun{Marked}\bigr)\]
\end{prop}

Since the verifiability-tests are  run after the conclusion of the exam, \etIntegrity{} cannot capture the scenario in which a test is modified before the marking and put back to its original version after marking. Such scenario can be detected by verifying \markingCorrectness{}. 

%

Another requirement that concerns the integrity of the \examtest is \emph{\etMarkedness{}}, which says that the candidate can check that the \examtest she submitted is marked without modification. It can be specified as follows:
  
\begin{prop}[\etMarkedness{} \individual] 
\label{ETM}
Given an exam $E$ and a set of $\phantom{a}  $ verifiability-tests $\beta$, then $(E,\beta)$ is
\emph{\etMarkednessAdj{}} if there is a $ $ verifiability-test $\fun{\etMarkednessFun{}_\individualFun{}}: {\cal E} \rightarrow \bool$ in $\beta$ \suchthat 
\[\fun{\etMarkednessFun{}_\individualFun{}} (i, q, a, m, p)
\Leftrightarrow (\exists m' :(i,(q,a),m') \in \fun{Marked})\]
\end{prop}

Note that the predicate of \etMarkedness{} coincides with the one of \etIntegrity{} pruned of ``$(i,(q,a))\in\fun{Accepted}$''. Thus, 
if $\fun{\etIntegrityFun{}_\individualFun{}}$ succeeds, then $\fun{\etMarkednessFun{}_\individualFun{}}$ also succeeds, namely
$\fun{\etIntegrityFun{}_\individualFun{}}(i, q, a, m, p) \Rightarrow \fun{\etMarkednessFun{}_\individualFun{}}(i, q, a, m, p)$.  However, if the  $\fun{\etIntegrityFun{}_\individualFun{}}$ fails, but   $\fun{\etMarkednessFun{}_\individualFun{}}$ succeeds, it follows that  the test of the  candidate is modified upon acceptance but put back to its original version before marking. 
This may be not a security issue for the candidate since her \examtest is  marked as submitted. However, the candidate can report this issue to the responsible authority for further investigation. 
Another scenario where an exam protocol may provide  $\fun{\etMarkednessFun{}_\individualFun{}}$ but not $\fun{\etIntegrityFun{}_\individualFun{}}$ is when  there is a lack of available data at the conclusion of the exam.

The next requirement is \emph{\markIntegrity{}}, which signifies that the candidate can verify that the mark attributed to her test is assigned to her without any modification. This requirement is formalised as follows:

\begin{prop}[\markIntegrity{} \individual]
\label{MI} 
Given an exam $E$ and a set of verifi\-ability-tests $\beta$, then $(E,\beta)$ is
\emph{\markIntegrityAdj{}} if there is a verifiability-test $\fun{\markIntegrityFun{}_\individualFun{}}: {\cal E} \rightarrow \bool$ in $\beta$ \suchthat 
\[\fun{\markIntegrityFun{}_\individualFun{}}(i, q, a, m, p)
\Leftrightarrow \exists m': \big((i,(q,a),m') \in \fun{Marked}  
\wedge (i,m') \in \fun{Assigned}\big)\]
\end{prop}

The last requirement is  \emph{\markNotificationIntegrity{}}, which says that the candidate can check she received the mark assigned to her. This requirement is formalised as:
\begin{prop}[\markNotificationIntegrity{} \individual] \label{MNI}
Given an exam $E$ and a set of verifiability-tests $\beta$, then $(E,\beta)$ is
\emph{\markNotificationIntegrityAdj{}} if there is a verifiability-test  $\fun{\markNotificationIntegrityFun{}_\individualFun{}}:  {\cal E} \rightarrow \bool$ in $\beta$ \suchthat 
\[\fun{\markNotificationIntegrityFun{}_\individualFun{}}(i, q, a, m, p)
\Leftrightarrow 
(i,m) \in \fun{Assigned}\]
\end{prop}

There is a subtle difference between the two last definitions. $\fun{\markIntegrityFun{}_\individualFun{}}$ can succeed despite the candidate is  notified with a mark that is different from the one assigned to her, while $\fun{\markNotificationIntegrityFun{}_\individualFun{}}$ cannot. Conversely, if $\fun{\markNotificationIntegrityFun{}_\individualFun{}}$ succeeds,  then $\fun{\markIntegrityFun{}_\individualFun{}}$ could fail if the examiner evaluated the test with a different mark.

\subsection{Universal Verifiability}
\label{uvp}
The definitions of universal verifiability are not pivoted around any exam role, but consider the viewpoint of an external auditor. The auditor runs the \phantom{~} \mbox{verifiability-tests} on the public \evidences available after an exam protocol run. Hence, the knowledge of the auditor 
consists of a general variable $e$ that contains the \evidences.

The five universal verifiability requirements concern the registration of the candidates and the integrity of the batch of tests from the submission until after the marking. The requirements of \questionValidity{} and of \markNotificationIntegrity{}, which are  definitions relevant for individual verifiability, are hard to capture in the context of universal verifiability. This is  because the external auditor has no knowledge of the questions nor of the marks received by the candidates, but only of public \evidences.

The first universal verifiability requirement we consider is \emph{\registration{}}, which says that an auditor can check that all accepted \examtests are submitted by registered candidates. Thus, the collector should have considered only tests that originated from eligible candidates. \index{eligible candidates} This requirement can be specified as:
\begin{prop}[\registration{} \universal] \label{UR} 
Given an exam $E$ and a set of verifi\-ability-tests $\beta$, then $(E,\beta)$ is
\emph{\registrationAdj{}} if there is a verifiability-test $\fun{\registrationFun{}_\universalFun{}}: {\cal E} \rightarrow \bool$ in $\beta$ \suchthat 
\[\fun{\registrationFun{}_\universalFun{}}(e) \Leftrightarrow I_r \supseteq \set{i: (i,x) \in \fun{Accepted}}\]
\end{prop}

Note that the superset symbol is preferred over strict equality since a candidate may register for an exam but may not show at testing. Thus, the collector may accept fewer tests than registered candidates. 

The next requirement is \emph{\markingCorrectnessUniversal{}}, which signifies that an auditor can check that all the marks attributed by the examiners to the tests are computed correctly. It is formalised as follows:

\begin{prop}[\markingCorrectnessUniversal{} \universal] 
\label{UMC}
Given an exam $E$ and a set of verifiability-tests $\beta$, then $(E,\beta)$ is
\emph{\markingCorrectnessUniversalAdj{}} if there is a verifiability-test $\fun{\markingCorrectnessUniversalFun{}_\universalFun{}}: {\cal E} \rightarrow \bool$ in 
$\beta$ \suchthat \index{Marking Correctness}
\[(\fun{\markingCorrectnessUniversalFun{}_\universalFun{}}(e)) \Leftrightarrow (\forall (i,x,m) \in \fun{Marked},~\fun{Correct}(x) = m))\]
\end{prop}

\markingCorrectnessUniversal{} makes the same arguments about free-response of tests we observed for \markingCorrectness{}.
However, for the sake of transparency, it can be assumed that exam authorities allow auditors to access the logs of the exam such that the auditors can inspect the marking process.

The third requirement is \emph{\etIntegrityUniversal{}}, which says that an auditor can verify that all and only accepted \examtests are marked without any modification.  It means that the auditor can be convinced that  no \examtest is modified, added, or deleted until the end of marking. This requirement is formalised as:

\begin{prop}[\etIntegrityUniversal{} \universal] 
\label{UETI}
Given an exam $E$ and a set of verifi\-ability-tests $\beta$, then $(E,\beta)$ is  \emph{\etIntegrityUniversalAdj{}} 
if there is a verifiability-test $\fun{\etIntegrityUniversalFun{}_\universalFun{}}:  {\cal E} \rightarrow \bool$ in $\beta$ \suchthat 
\[ (\fun{\etIntegrityUniversalFun{}_\universalFun{}}(e))
\Leftrightarrow (\fun{Accepted} = \set{(i,x): (i,x,m) \in
 \fun{Marked}})\]
\end{prop}

The equality symbol in the predicate specification enforces that at marking no test has been added or removed from the batch of accepted tests.

The next requirement is \emph{\etMarkednessUniversal{}}, which says that an auditor can check that only the accepted \examtests are marked without modification. It is formalised as follows:

\begin{prop}[\etMarkednessUniversal{} \universal] 
\label{UETM}
Given an exam $E$ and a set of $ $ \mbox{verifiability-tests} $\beta$, then $(E,\beta)$ is  
\emph{\etMarkednessUniversalAdj{}} if there exists a verifiability-test
$\fun{\etMarkednessUniversalFun{}_\universalFun{}}: {\cal E}
\rightarrow \bool$ in $\beta$ \suchthat 
\[(\fun{\etMarkednessUniversalFun{}_\universalFun{}}(e))
\Leftrightarrow (\fun{Accepted} \supseteq \set{(i,x): (i,x,m)
 \in \fun{Marked}})\]
\end{prop}

It can be noted that  \etMarkednessUniversal{} is a relaxed version of \etIntegrityUniversal{} because the predicate of the former definition does not require strict equality of the two multisets. Thus, if $\fun{\etIntegrityUniversalFun{}_\universalFun{}}$ fails but $\fun{\etMarkednessUniversalFun{}_\universalFun{}}$ succeeds, it follows that at least one accepted test has not been marked. This scenario however may not be a security problem.
For example, the rules of the exam may allow the candidate to drop the examination after testing. Conversely, the scenario in which an examiner marks a test that was not accepted  is normally considered a violation of the exam.

The last requirement we consider is \markIntegrityUniversal{}, which signifies that an auditor can check that all and only the marks associated to the \examtests are assigned to the corresponding candidates with no modifications. This is formalised as:

\begin{prop}[\markIntegrityUniversal{} \universal] \label{UMNI}
Given an exam $E$ and a set of verifi\-ability-tests $\beta$, then $(E,\beta)$ is  
\emph{\markIntegrityUniversalAdj{}} if there exists a verifiability-test $\fun{\markIntegrityUniversalFun{}_\universalFun{}}: {\cal E}
\rightarrow \bool$ in $\beta$ \suchthat
\[(\fun{\markIntegrityUniversalFun{}_\universalFun{}}(e) )
\Leftrightarrow (\fun{Assigned} = \set{(i,m): (i,x,m) \in
 \fun{Marked}})\]
\end{prop}

The equality symbol in the  specification of the predicate enforces that no pair of candidates and marks  have been added or removed from the batch of marked tests.

To conclude, it can be observed that the combination of registration, test integrity, and mark integrity universal verifiability enforces the verifiability from preparation to notification of an exam protocol.


\section{Conclusion}\label{sec:concveri}
This chapter advances verifiability requirements for exam protocols. The domain of exam has unique features that call for verifiability definitions that are different from ones proposed for voting and auctions. \index{abstract model}
The eleven requirements, which are classified in individual and universal verifiability categories,  are specified in a formal and abstract model that opens up opportunities for both symbolic and computational analysis. This model contrasts the frameworks proposed for the verifiability of voting and auction protocols,  as the latter usually focus on cryptographic protocols. Intuitively, the proposed model is sufficiently abstract to specify any type of exam, from traditional to Internet-based exams. Traditional exams usually provide evidence data via log-books and registers, while Internet-based exams implement the electronic versions, such as web bulletin-boards. \index{Internet-based exams} \index{traditional exam}

Individual verifiability definitions consider a candidate who can check if she got a valid set of questions, if her test was properly 
processed through the phases of the exam, and if her mark was correctly computed.
Universal verifiability definitions consider an external auditor who can check the correct execution of the exam with no private knowledge about the run of the exam. The auditor can verify if only registered candidates took the exam, if all tests were properly processed and marked, and if the marks were assigned correctly.

We validate the proposed model in the next chapter, in which  we introduce an Internet-based exam protocol. We analyse it for the eleven verifiability requirements, and show how to map the relations defined in the proposed abstract model in ProVerif. 



\chapter{The Remark! Internet-based Exam Protocol}\label{chap:remark}

In the last decade, computers have been extensively introduced in the design of critical systems such as voting, auctions, and exams. Computers are progressively becoming the main components of such systems providing the core tasks. In the context of exam, computers can be used for local tasks, such as generation of questions  and automatic marking, but also to support remote tasks. For example, remote registration and remote notification of candidates are must-have functionalities of exams that expect many candidates; remote testing, in which distant located candidates take the exam at their place, is the distinctive functionality of Internet-based exams. At an extreme, all the phases of an exam may take place remotely. \index{automatic marking} \index{remote registration} \index{remote notification} \index{remote testing} \index{Internet-based exams}

The use of computers exposes exams to new threats and requires changing the well-established procedures used in traditional exam. The design of secure exam protocols is further complicated by the conflicting interests that roles typically have in an exam. In fact, it may be hard to find a role who can play as TTP, as recent exam scandals confirm \citeweb{CheatingUSA2013, cheatingets, usnavy2014}. \index{exam scandal} \index{traditional exam}
In this chapter, we advance Remark!, a new Internet-based exam protocol that guarantees several authentication, privacy, and verifiability requirements without the need of a TTP.
The idea behind \remark~is to distribute the trust across the several servers that compose an \emph{exponentiation mixnet}. \index{exponentiation mixnet} As we shall see later, the mixnet generates the pseudonyms that allow the exam principals to encrypt and sign messages anonymously. Using ProVerif, we prove that \remark~ensures all the authentication and privacy requirements proposed in chapter \ref{chap:formal} with minimal reliance on trusted parties. Moreover, we demonstrate that \remark~provides the verifiability-tests listed in chapter \ref{chap:verifiability}, and discuss the necessary assumptions to make \remark~fully verifiable.

\subsubsection{Assumptions} Like any other security protocol, \remark~is not designed to withstand every possible threat. For instance, it cannot cope with plagiarism, but assumes appropriate invigilation \index{invigilation} during testing. Principals may still collude and communicate via subliminal channels, for instance by using steganography. Although it is hard to rule out completely such a threat, steganalysis techniques can be of some help here. Other countermeasures may be needed against collusion attacks that exploit covert channels. 
We thus specify seven assumptions conveniently for the goals of Remark!. In particular, we assume that:
\begin{enumerate}
\item Each principal holds a long-term public/private pair of keys.
\item The candidate holds a smart card in which the personal details of the candidate are visibly engraved. The smart card securely stores the candidate's private key, namely the private key cannot be extracted from the smart card.
\item The candidate is invigilated during testing to mitigate cheating. Invigilation for remote testing can be guaranteed with online invigilation software, such as  ProctorU~\citeweb{proctoru}.
\item The model answers are kept secret from the candidates until after testing. The examiners may be provided with the model answers at marking. \index{remote testing}
\item It is available an authenticated append-only bulletin board that guarantees everyone to see the same data, though write access might be restricted to appropriate entities \citeltex{BRT13}. An implementation of a bulletin board and its formal analysis has been proposed by Culnane and Schneider \citeltex{CS14}.
\item It is available an implementation of TLS channel that ensures integrity and confidentiality of messages.
\item At least one of the servers that compose the exponentiation mixnet is honest. \index{exponentiation mixnet}
\end{enumerate}

\paragraph*{Outline of the chapter.}
Section~\ref{sec:relatedremark} reviews a few proposals of secure protocols for Internet-based exams. \index{Internet-based exams}
Section~\ref{sec:exp} details exponentiation mixnet, a cryptographic scheme on which \remark~is based.
Section~\ref{sec:descriptionremark} describes \remark~according the four phases of an exam.
Section~\ref{sec:analysisremark} contains the formal analysis in ProVerif of ten authentication and privacy requirements
Section~\ref{sec:remark_ver} contains the formal analysis in ProVerif of eleven verifiability requirements.
Finally, Section~\ref{sec:conclusionremark} discusses future work and concludes the chapter.



\section{Related work} \label{sec:relatedremark} 
To the best of our knowledge only  few works propose protocols for Internet-based exams.
TOEFL \citeweb{toeflurl}, which is one of the major English-language test in the world,  has replaced its traditional exams \index{traditional exam} with Internet-based exams. Neither the specification nor the security requirements of the TOEFL protocol are publicly available. Moreover, its design probably includes a trusted exam authority that is in charge of the critical tasks of the exam. The same concerns apply for MOOCs, which  offer Internet-based exams that grant credits for many universities \citeweb{mooc}. \index{MOOC}
 Conversely, \remark~is designed to minimise the reliance on trusted parties.
\Huszti \citeltex{huszti10} advanced an Internet-based exam with few trust requirements on principals, but in chapter \ref{chap:formal} we have shown that the protocol has several security issues. In contrast, we prove that \remark~ensure all the security requirements. \index{Internet-based exams}

In the domain of Computer Supported Collaborative Working, Foley and Jacob \citeltex{FJ95} formalised confidentiality requirements and proposed an exam as case study.
Maffei \etal \citeltex{MPR13} implemented a course evaluation system that guarantees privacy using anonymous credential schemes without a trusted third party.
Similarly, Hohenberger \etal \citeltex{HMP+14} advanced \emph{ANONIZE}, a protocol for surveys that ensures authentication and privacy in presence of corrupted authorities. However, surveys have different security requirements than exams, for instance, surveys do not consider test authorship and fixed-term anonymity definitions. 

Some related protocols  have been proposed in the area of conference management systems. Kanav \etal \citeltex{KLP14} introduced \emph{CoCon}, a formally verified implementation of conference management system that guarantees confidentiality. Arapinis \etal \citeltex{ABR11} introduced and formally analysed \emph{ConfiChair}, a cryptographic protocol that addresses secrecy and privacy risks coming from a \emph{malicious-but-cautious} cloud. Their work has been recently extended to support any cloud-based system that assumes honest managers, such as public tender management and recruitment process \citeltex{ABR13}. In Remark!, a different attacker is considered since exam authorities, which are analogous to managers in cloud-based systems, can be corrupted.

\section{Exponentiation mixnet} \label{sec:exp} \index{exponentiation mixnet}
\remark~relies on ElGamal encryption, digital signature, and exponentiation mixnet.
In this section, we detail the rudiments of exponentiation mixnet, while the reader can refer to chapter \ref{chap:formal} for a brief description of ElGamal encryption and digital signature. 

The main functionality provided by exponentiation mixnet is to generate a \emph{pseudo} public key that allows the owner of the corresponding private key to encrypt and sign messages anonymously.
An exponentiation mixnet takes in a batch of public keys and outputs a set of new pseudo public keys. The scheme ensures that no one but the owner of the public/private key pair can link a public key of the original batch with any of the pseudo public keys. 
In contrast to  conventional re-encryption mixnet \citeltex{Chaum81} in which each term is independently re-encrypted, the peculiarity of  exponentiation mixnet is that each mix server re-encrypts the terms by a common exponent value. This idea appeared first in the work of Haenni and Spycher \citeltex{HS11}.

\begin{figure}
\begin{center}
\begin{tikzpicture}
\footnotesize
\node at (0.3,4) {$C_1$};
\node at (0.3,3) {$C_2$};
\node at (0.3,2) {$\vdots$};
\node at (0.3,1) {$C_n$};

\node at (1.3,4.5) {$\mathcal{BB}$};
\node at (1.3,4)  {$\PK_1$};
\node at (1.3,3)  {$\PK_2$};
\node at (1.3,2)  {$\vdots$};
\node at (1.3,1)  {$\PK_n$};
\node at (1.3,0)  {$g$};

\draw[thin,->] (1.7,2) -- (2,2);
\node at (2.4,2)   {\mybox[fill=blue!20]{$\mathit{mix_1}$}};
\draw[thin,->] (2.8,2) -- (3.1,2);
\draw[thin,-] (2.6,1.8) -- (2.6,0);
\draw[thin,-] (2.6,1.8) -- (2.6,0);
\draw[thin,-] (2.6,0) -- (4.8,0);

\node at (3.7,4.5) {$\mathcal{BB}$};
\node at (3.7,4)  {$\PK^{r_1}_{\pi_1(1)}$};
\node at (3.7,3)  {$\PK^{r_1}_{\pi_1(2)}$};
\node at (3.7,2)  {$\vdots$};
\node at (3.7,1)  {$\PK^{r_1}_{\pi_1(n)}$};
\node at (3.7,0.15)  {$g^{r_1}$};

\draw[thin,->] (4.3,2) -- (4.6,2);
\draw[thin,->] (4.8,0) -- (4.8,1.8);
\node at (5,2) {\mybox[fill=blue!20]{$\mathit{mix_2}$}};
\draw[thin,->] (5.4,2) -- (5.7,2);
\draw[thin,-] (5.2,1.8) -- (5.2,0);
\draw[thin,-] (5.2,0) -- (7.4,0);

\node at (6.5,4.5) {$\mathcal{BB}$};
\node at (6.5,4) {$\PK^{r_1r_2}_{\pi_2\circ\pi_1(1)}$};
\node at (6.5,3) {$\PK^{r_1r_2}_{\pi_2\circ\pi_1(2)}$};
\node at (6.5,2) {$\vdots$};
\node at (6.5,1) {$\PK^{r_1r_2}_{\pi_2\circ\pi_1(n)}$};
\node at (6.5,0.15) {$g^{r_1r_2}$};

\node at (7.65,4) {$\cdots$};
\node at (7.65,3) {$\cdots$};
\node at (7.65,2) {$\cdots$};
\node at (7.65,1) {$\cdots$};
\node at (7.65,0) {$\cdots$};

\draw[thin,->] (7.9,2) -- (8.18,2);
\draw[thin,-] (7.9,0) -- (8.4,0);
\draw[thin,->] (8.4,0) -- (8.4,1.8);
\node at (8.6,2) {\mybox[fill=blue!20]{$\mathit{mix_m}$}};
\draw[thin,->] (9,2) -- (9.3,2);

\node at (10,4.5) {$\mathcal{BB}$};
\node at (10,4) {$\PK^{\rr_m}_{\ppi_m(1)}$};
\node at (10,3) {$\PK^{\rr_m}_{\ppi_m(2)}$};
\node at (10,2) {$\vdots$};
\node at (10,1) {$\PK^{\rr_m}_{\ppi_m(n)}$};
\node at (10,0) {$g^{\rr_m}$};

\node at (11,4) {$=$};
\node at (11,3) {$=$};
\node at (11,1) {$=$};
\node at (11,0) {$=$};

\node at (11.5,4) {$\overline{\PK}_1$};
\node at (11.5,3) {$\overline{\PK}_2$};
\node at (11.5,1) {$\overline{\PK}_n$};
\node at (11.5,0) {$h_C$};

\draw (0.9, 4.3) to (0.9, -0.2);
\draw (1.7, 4.3) to (1.7, -0.2);
\draw (0.9, 4.3) to (1.7, 4.3);
\draw (0.9, -0.2) to (1.7, -0.2);

\draw (3.1, 4.3) to (3.1, 0.5);
\draw (4.3, 4.3) to (4.3, 0.5);
\draw (3.1, 4.3) to (4.3, 4.3);
\draw (3.1, 0.5) to (4.3, 0.5);

\draw (5.7, 4.3) to (5.7, 0.5);
\draw (7.3, 4.3) to (7.3, 0.5);
\draw (5.7, 4.3) to (7.3, 4.3);
\draw (5.7, 0.5) to (7.3, 0.5);

\draw (9.3, 4.3) to (9.3, -0.2);
\draw (10.65, 4.3) to (10.65, -0.2);
\draw (9.3, 4.3) to (10.65, 4.3);
\draw (9.3, -0.2) to (10.65, -0.2);

\end{tikzpicture}
\caption{The exponentiation mixnet scheme} 
\label{fig:expo}
\end{center}
\end{figure}
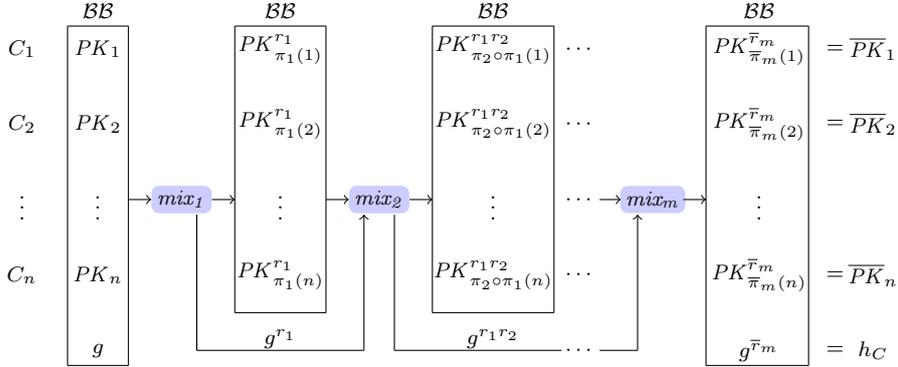

In the following, we detail the construction of an exponentiation mixnet, which is depicted in Figure  \ref{fig:expo}.
Let $g$ be a generator of a multiplicative subgroup $\mathbb{G}$ of order $q$.
Let us assume $n$ principals $\setremark{C_1,\ldots C_n}$, each have a pair of public/private keys $(\PK,\SK)$ such that  $\PK=g^{\SK}$.
Let us assume $m$ servers composing the exponentiation mixnet. 
The mix server $\mathit{mix}_1$ takes the batch of the public keys 
$\setremark{\PK_1,\ldots \PK_n}$, generates a fresh random $r_1 \in \{1, q-1\}$, and computes the batch of temporal pseudo public keys $\setremark{\PK_1^{r_1},\ldots \PK_n^{r_1}}$. Then, $\mathit{mix}_1$ 
signs and sends to the bulletin board the computed batch in a secret shuffled order, namely 
 the server posts $\setremark{\PK^{\rr_1}_{\ppi_1(1)},\ldots PK^{\rr_1}_{\ppi_1(n)}}$. Additionally, $\mathit{mix}_1$ 
posts a zero-knowledge proof of correctness and sends the new generator $g^{r_1}$ to the next server over a secure channel. Further servers repeat the steps above as required. 
The last server, $\mathit{mix}_m$, publishes the final batch of pseudo public key $\setremark{\PK^{\rr_m}_{{\ppi}_m(1)},\ldots \PK^{\rr_m}_{{\ppi}_m(n)}}$ and the final generator 
$g^{\rr_m}$, where $\overline{r}_m=\prod_{i=1}^m r_i$ and $\overline{\pi}_m=\pi_k\circ \cdots \circ \pi_1$.
Note that the intermediate  $g^{r_1},\ldots,g^{\rr_{m-1}}$ terms are not posted on the bulletin board. This prevents each principal to trace their intermediate pseudo public keys through the mixnet.
Although it is not clear whether such eventuality is an attack, it is normally considered  an undesired feature.
Each principal $C_i$ can find the corresponding pseudo public key using their private keys, since $g^{\rr_m\SK_i}=\PK^{\rr_m}_{{\ppi}_m(i)}$. 

\remark~makes use of exponentiation mixnet at preparation to create the pseudonyms for candidates and examiners. The mixnet is also required at notification to revoke the  pseudonyms of candidates. In so doing, each server $\mathit{mix}_i$ reveals its random value $r_i$, hence by revealing  all the values \index{exponentiation mixnet}
$\rr_m$ it is possible to link the pseudonyms to the identities of the candidates.

\section{Description} \label{sec:descriptionremark}
\label{protocol} 
\remark~has four roles: exam authority (EA), candidate (C), examiner (E), and mixnet (NET). The exam authority manages the exam and also plays the roles of collector and notifier.

\remark~relies on a \BB to publish  pseudonyms,  questions, tests, and marks.
As we discussed in the previous section, the bulletin board is also used in the exponentiation mixnet scheme. In the remainder, we assume that anyone can post messages on the bulletin board, even the attacker. Thus, we require each principal to sign their messages. However, if one assumes that the bulletin board has appropriate write access control mechanisms, namely it only publishes messages that originate from eligible principals, signatures may not be necessary.

\setlength{\abovecaptionskip}{25pt}
\setlength{\belowcaptionskip}{0pt}

\begin{figure}
{
\footnotesize
\setlength{\unitlength}{3.1cm}
\begin{picture}(3,4)
\put(0.25,4.88){\line(0,-1){1.6}}
\put(0.25,3.08){\line(0,-1){1.91}}
\put(0.25,0.98){\line(0,-1){1.11}}

\put(1.35,4.88){\line(0,-1){0.33}}
\put(1.35,4.25){\line(0,-1){0.35}}
\put(1.35,3.6){\line(0,-1){.8}}
\put(1.35,2.5){\line(0,-1){.55}}
\put(1.35,1.77){\line(0,-1){.17}}
\put(1.35,1.3){\line(0,-1){.1}}
\put(1.35,0.7){\line(0,-1){.18}}
\put(1.35,0.27){\line(0,-1){.38}}

\put(2.5,4.88){\line(0,-1){.7}}
\put(2.5,3.9){\line(0,-1){.37}}
\put(2.5,3.23){\line(0,-1){3.21}}

\put(3.75,4.88){\line(0,-1){1.6}}
\put(3.75,3.08){\line(0,-1){0.61}}
\put(3.75,2.29){\line(0,-1){2.43}}

\put(0.25,4.98){\makebox(0,0){Examiner}}
\put(3.75,4.98){\makebox(0,0){Candidate}}
\put(1.35,4.98){\makebox(0,0){Mixnet}}
\put(2.5,4.98){\makebox(0,0){Exam Authority}}

\thicklines
\multiput(0,4.83)(0.05,0){80}{\circle*{0.02}}

\put(0.01,4.23){\rotatebox[origin=l]{90}{\makebox{\texttt{Preparation}}}}

\put(.35 ,4.38){\framebox{
\parbox{6cm}{
\begin{tabular}{lll}
$\bar{r}_m=\prod\limits_{i=1}^m r_{i}$, & 
$\overline{{\PK}}_{C}=\PK_{C}^{\rr_m}$, &
$h_{C}=g^{\bar{r}_m}$
\end{tabular}
}}}

\put(.35 ,3.73){\framebox{
\parbox{6cm}{
\begin{tabular}{lll}
$\bar{r}'_m=\prod\limits_{i=1}^m r'_{i}$, &
$\overline{{\PK}}_{E}=\PK_{E}^{\bar{r}'_m}$, &
$h_{E}=g^{\bar{r}'_m}$
\end{tabular}
}}}

\put(1.35,4.04){\vector(1,0){1.2}}
\put(1.85,4.11){\makebox(0,0){$1:\sign_{SK_M}( \overline{{\PK}}_{C}, h_{C}) $}}
\put(2.57 ,4.01){\framebox{$\mathcal{BB}$}}

\put(1.35,3.38){\vector(1,0){1.2}}
\put(1.85,3.46){\makebox(0,0){$2:\sign_{SK_M}( \overline{{\PK}}_{E}, h_{E} )$}}
\put(2.57,3.35){\framebox{$\mathcal{BB}$}}

\put(2.94 ,3.15){\framebox{Check $\overline{{\PK}}_{C}=h_{C}^{\SK_C}$}}
\put(0 ,3.15){\framebox{Check $\overline{{\PK}}_{E}=h_{E}^{\SK_E}$}}

\multiput(0,2.85)(0.05,0){80}{\circle*{0.02}}
\put(0.01,2.43){\rotatebox[origin=l]{90}{\makebox{\texttt{Testing}}}}

\put(2.5,2.60){\vector(-1,0){1.72}}
\put(1.8,2.68){\makebox(0,0){$3:\lbrace \sign_{SK_A}(  \mathit{quest}, \overline{\PK}_{C}) \rbrace_{\overline{\PK}_{C}}$}}
\put(.5 ,2.58){\framebox{$\mathcal{BB}$}}

\put(3.75,2.09){\vector(-1,0){1.25}}
\put(2.75,2.36){\framebox{
\parbox{3.2cm}{
$T_C= \setremark{\mathit{quest},\mathit{ans},\overline{\PK}_C}$}}}

\put(3.15,2.16){\makebox(0,0){$4: \lbrace  \sign_{SK_C,h_{C}} (T_C)\rbrace_{\PK_{A}} $}}

\put(2.5,1.86){\vector(-1,0){1.72}}
\put(1.92,1.94){\makebox(0,0){$5:\lbrace  \sign_{SK_A}( H(T_C) \rbrace_{\overline{\PK}_C}$}}
\put(.5 ,1.83){\framebox{$\mathcal{BB}$}}

\multiput(0,1.66)(0.05,0){80}{\circle*{0.02}}

\put(0.01,1.25){\rotatebox[origin=l]{90}{\makebox{\texttt{Marking}}}}

\put(.5 ,1.38){\framebox{$\mathcal{BB}$}}
\put(2.5,1.38){\vector(-1,0){1.72}}
\put(1.86,1.47){\makebox(0,0){$6:\lbrace \sign_{SK_A}(\overline{\PK}_E, T_C)\rbrace_{\overline{\PK}_E}$}}

\put(0.25,0.76){\vector(1,0){2.25}}
\put(0,1.05){\framebox{$M_C= \setremark{ \sign_{SK_A}(\overline{\PK}_E, T_C), \mathit{mark}}$}}
\put(0.86,.83){\makebox(0,0){$7:\lbrace \sign_{SK_E,h_{E}}( M_C) \rbrace_{\PK_{A}}$}}

\multiput(0,0.58)(0.05,0){80}{\circle*{0.02}}
\put(0.01,-0.07){\rotatebox[origin=l]{90}{\makebox{\texttt{Notification}}}}

\put(2.5,0.35){\vector(-1,0){1.72}}
\put(1.9,0.43){\makebox(0,0){$8: \lbrace  \sign_{SK_E,h_{E}}(M_C) \rbrace_{\overline{\PK}_C}$}}
\put(.5 ,0.33){\framebox{$\mathcal{BB}$}}

\put(1.35,0.15){\vector(1,0){1.15}}
\put(1.5,0.2){\makebox(0,0){$9:\bar{r}_C$}}
\put(1.47,0.11){\makebox(0,0){\tiny $(TLS)$}}
\put(2.05 ,-0.08){\framebox{Register $ \PK_C, \mathit{mark}$}}
\end{picture}
\caption{The \remark~Internet-based exam protocol} 
\label{fig:remark}
}
\end{figure}
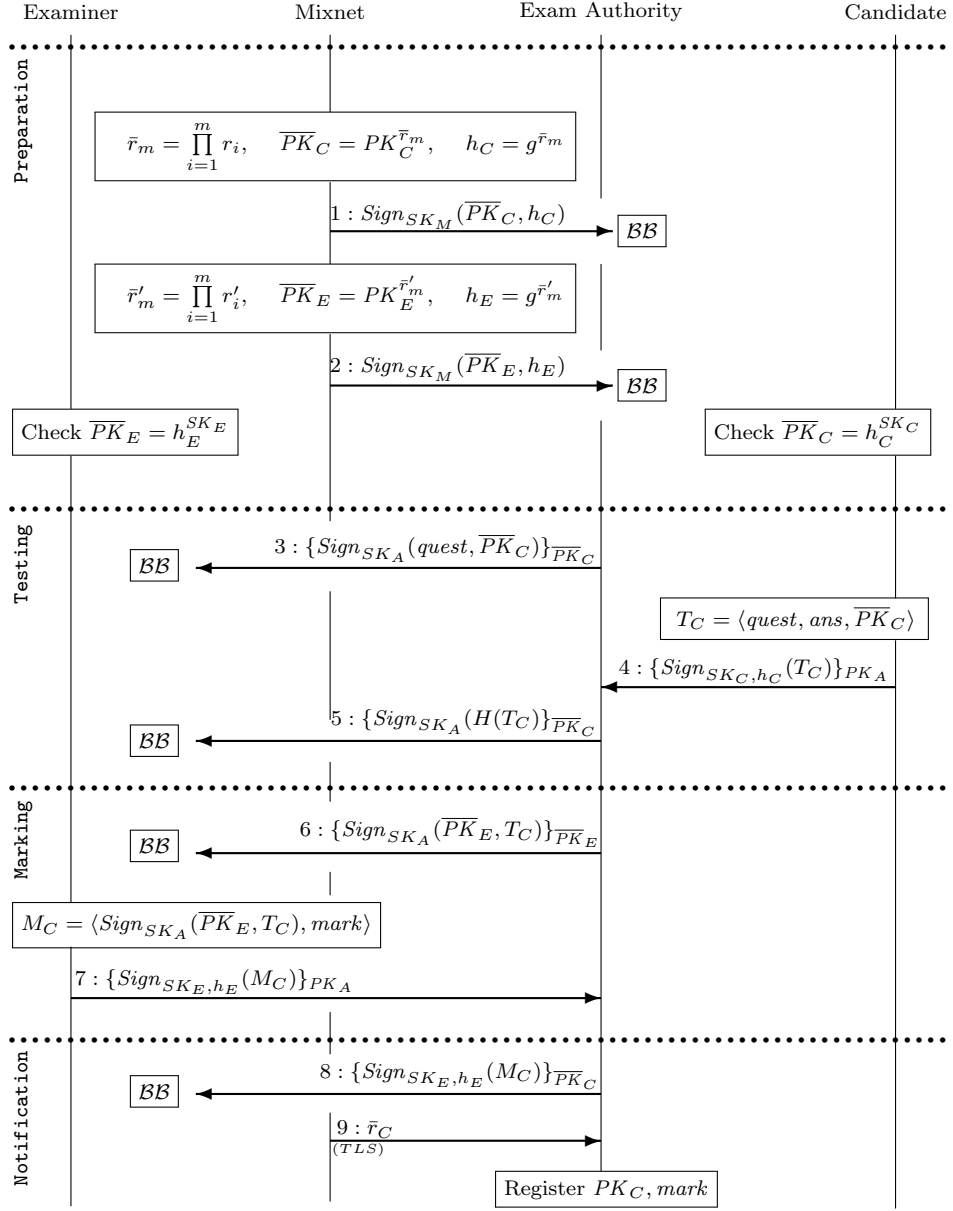
\setlength{\abovecaptionskip}{10pt}
\setlength{\belowcaptionskip}{0pt}

In the following, we detail \remark~according the phases of an exam. Figure \ref{fig:remark} illustrates the protocol's steps in form of a message sequence chart.

\subsubsection{Preparation}
The exponentiation mixnet generates the pseudonyms of candidate and examiner. \index{exponentiation mixnet}
The generation takes place in two independent runs: first the mixnet generates the pseudonym of candidates and then the pseudonym of examiners.
Such separation is necessary because only the identities of candidates should be revealed at notification.

The public key $\PK_C$ of an eligible candidate \index{eligible candidates} $C$ is processed by the exponentiation mixnet among the public keys of other candidates.
After the last mix server publishes the list of pseudonyms and the new generator $h_C := g^{\rr_m}$, the candidate identifies her pseudo public key $\overline{\PK}_C$ computing $h_C^{\SK_C}$. The pseudo public key from now on serves as the pseudonym for $C$.

After the pseudonyms of candidates have been published (step~1), the mixnet generates the pseudonyms for examiners in a similar way.
Since the mixnet generates the pseudonyms of examiners using different random values, a new generator $h_E$ is published at the end of the mix (step~2).

\subsubsection{Testing}
The \ea generates the questions, signs them with its private key $\SK_A$, and encrypts each question under a candidate pseudonym.  
We do not specify how the \ea generates the  questions in order to include different forms of questions (e.g., multiple choice, free-response, etc.) and assignments (e.g., single question, different questions for candidate, random permutations of a set of questions, etc.). In the remainder, with \emph{question} we actually refer to a list of questions possibly of size one. 

\remark~assumes that an invigilator authenticates the candidate by checking whether the personal details printed on the top of the smart card matches the candidate identity. For remote authentication, this procedure can be supported with tools such as ProctorU.    
Then, the \ea publishes the encrypted questions on the bulletin board (step~3).
After the candidate answers the test, she  appends the answer to her pseudonym and question, so the filled test consists of $T_C= \setremark{\mathit{ques}, 
\mathit{ans}, \overline{\PK}_C}$. Then, she signs the test $T_C$ with her private key $\SK_C$ using the generator $h_C$. Thus, the signature can be verified using the pseudonym of the candidate $\overline{\PK}_C$ with respect to $h_C$. The candidate then encrypts the signed test with the public key of the \ea $\PK_A$, and submits it (step~4).
The \ea decrypts the test, and then signs the hash of $T_C$ using its private key $\SK_A$. It then encrypts the signed hash under the corresponding candidate's pseudonym, that is, $\mathit{\{Sign_{SK_A}(H(T_C))\}_{\overline{\PK}_C}}$, and publishes such encryption as receipt (step~5). 

\subsubsection{Marking}
The \ea randomly chooses an eligible examiner \index{eligible examiners} pseudonym $\overline{\PK}_E$, and encrypts the signed test $T_C$ under the chosen examiner pseudonym (step~6). Note that the \ea does not know the real identity of the examiner. Moreover, it is possible to introduce a universally verifiable deterministic assignment of test to examiners. For example, encrypted tests and the examiner pseudonyms could be posted in two lexically ordered lists, and the \ea cyclically assigns a test to an examiner according the order. 

After the designated examiner marks the test,  he appends the mark to the signed test, thus generating the evaluation $\mathit{M_C= \setremark{ Sign_{SK_A}(\overline{PK}_E, T_C), mark}}$. The examiner then signs $M_C$ with his private key $\SK_E$ and the generator $h_E$, and encrypts the signed evaluation under the public key of the \ea $\PK_A$ (step~7).

\subsubsection{Notification}
The \ea receives from the examiner the encrypted evaluation, which it decrypts and re-encrypts under the corresponding candidate pseudonym $\overline{\PK}_C$. After the \ea publishes all the test evaluations (step~8), it asks the mixnet to reveal the random values $r$ used to generate the pseudonyms of the candidates (step~9). In so doing, the candidate anonymity is revoked, and the mark can finally be registered. Note that each candidate learns the corresponding mark before $\rr_m$ is revealed.



\section{Formal Analysis of Authentication and Privacy}\label{sec:analysisremark}
We analyse  \remark~in ProVerif. We consider the authentication and privacy requirements formally specified in chapter \ref{chap:formal}.
The requirements with short descriptions are recalled below: 
\begin{itemize}
\item Authentication
\begin{itemize}
\item \emph{Candidate Authorisation}, which says that only registered candidates can take the exam. \index{Candidate Authorisation}
\item \emph{Answer Authenticity}, which says that the collector considers only the answers that candidates actually submitted.
\item \emph{Test Origin Authentication}, which says that the collector accepts only tests that originate from registered candidates.
\item \emph{Test Authenticity}, which says that the examiner only marks the tests intended for him.
\item \emph{Mark Authenticity}, which says that the candidate receives the mark assigned to her test by the examiner chosen by the collector.
\end{itemize} 
\item Privacy
\begin{itemize}
\item \emph{Question Indistinguishability}, which says that the questions are not revealed until testing begins.
\item \emph{Anonymous Marking}, which says that the examiner marks a test while ignoring its author.
\item \emph{Anonymous Examiner}, which says that the candidate cannot learn which examiner marked her test.
\item \emph{Mark Privacy}, which says that no one learns the marks, besides the examiner, the concerned candidate, and the notifier.
\item \emph{Mark Anonymity}, which says that no one learns the association between a mark and the corresponding candidate.
\end{itemize}
\end{itemize}

\begin{table}
\begin{center} 
\begin{tabular}{|c|r|}
\hline
{\bf Primitive} & \multicolumn{1}{c|}{{\bf Equation}}
\\ \hline
ElGamal encryption  &
$\begin{aligned}
\mi{decrypt}(\mi{encrypt}(m, pk(k), r), k) & =  m \\
\mi{decrypt}(\mi{encrypt}(m,\mi{pseudo\_pub}(\mi{pk}(k), \\
rce),r),\mi{pseudo\_priv}(k,\mi{exp}(rce))) & = m \\
\mi{checkpseudo}(\mi{pseudo\_pub}(\mi{pk}(k),rce),  \\
\mi{pseudo\_priv}(k,\mi{exp}(rce))) & = \mi{true}\\
\end{aligned}$
\\ \hline
Digital signature &
$\begin{aligned}
\mi{getmess}(\mi{sign}(m,k)) & = m\phantom{ii'} \\
\mi{checksign}(\mi{sign}(m,k), \mi{spk}(k)) & = m \\
\mi{checksign}(\mi{sign}(m,\mi{pseudo\_priv}(k,  \\
\mi{exp}(rce))),\mi{pseudo\_pub}(pk(k),rce)) & = m
\end{aligned}$ 
\\ \hline
\end{tabular}
\normalsize
\caption{Equational theory to model Remark!}
\label{table:eqt-mscremark}
\end{center}
\end{table}

\subsubsection*{Model choices} 
We model the bulletin board as a public channel, and use the 
equational theory depicted in Table~\ref{table:eqt-mscremark}.  The theory consists of the standard equations for ElGamal encryption and digital signatures extended with novel equations that model pseudonyms as public keys. The pseudonym, which also serves as test identifier, can be generated using the function $\mathit{pseudo\_pub}$, which takes in a public key and a random exponent. In fact, this function models the main feature of exponentiation mixnet. The function $\mathit{pseudo\_priv}$ can be used by a principal to decrypt or sign anonymous messages. The function takes in the private key of the principal and the new generator published by the mixnet. The function $\mathit{checkpseudo}$ allows a principal to check whether a pseudonym is associated with the principal's private key. In practice,  \index{exponentiation mixnet}
principals use this function to identify their pseudonyms published on the bulletin board.

\begin{figure}
\begin{center}
\begin{lstlisting}
let EA (skA:skey, pkN:pkey, ques:bitstring) =
(*Preparation*)
 in(bbn, (pseudo_C:pkey, hc:bitstring, r: role, spseC:bitstring));
 if (pseudo_C, hc, r) = checksign(spseC, pkN) && r = C then 
  let sques = sign(ques, skA) in	
  let eques = encrypt( (ques,sques), pseudo_C) in 
  out(bba, eques);
  in(ch, eca:bitstring);
  let ((=ques, ans:bitstring, =pseudo_C), sca:bitstring) = 
  decrypt(eca, skA) in 
  if (ques, ans, pseudo_C) = checksign(sca, pseudo_C) then
   (* EA collects the test from C with pseudonym pseudo_C *)
   event collected(pseudo_C, ques, ans);
   let ca = (ques, ans, pseudo_C) in 
   let sca' = sign(ca, skA)  in 
   let eca' = encrypt((ca, sca'), pseudo_C) in 
   out(bba, eca');	

(* Marking *)
   in(bbn, (pseudo_E:pkey, he:bitstring, rolet:role, 
      spseE:bitstring));
   if (pseudo_E, he, rolet) = checksign(spseE, pkN) && 
      rolet = E then 
    let ca'' = (ques, ans, pseudo_C, pseudo_E) in 
    let sca'' = sign(ca'', skA)  in 
    let eca'' = encrypt((ca'', sca''), pseudo_E) in 
    (* EA distributed the test (pseudo_C,ques,ans) *)
    (* identified by pseudo_C (id_form = pseudo_C) to E pseudo_E*)
    event distributed(pseudo_C,ques,ans,pseudo_C,pseudo_E);
    out(bba, eca'');
    in(ch, ema:bitstring);
    let ((=ca'', =sca'', mark:bitstring), sma:bitstring) = 
    decrypt(ema, skA) in 
    if (ca'', sca'', mark) = checksign(sma, pseudo_E)  then 

(* Notification *)
     let ma = (ca'', sca'', mark) in 
     let ema' = encrypt((ma, sma), pseudo_C) in 
     out(bba, ema');
     (*Reveal ID*)
     in(ch, encsignetX: bitstring).
\end{lstlisting}
\caption{The process of the exam authority}
\label{fig:remark_ea}
\end{center}
\end{figure}

\begin{figure}
\begin{lstlisting}
let Cand (skC:skey, pkA:pkey, pkN:pkey, ans:bitstring) =
(*Preparation *)
 in(bbn, (pseudo_C:pkey, hc:bitstring, r: role, spseC:bitstring));
 if (pseudo_C, hc, r) = checksign(spseC, pkN) && r = C then 
  let priv_C = pseudo_priv(skC, hc) in 
  if checkpseudo(pseudo_C, priv_C) =true then 

(*Testing*)
  in(bba, eques:bitstring);
  let (ques: bitstring, sques:bitstring)=decrypt(eques, priv_C) in
  if ques=checksign(sques, pkA) then
   let ca = (ques, ans, pseudo_C) in 
   let sca = sign(ca,priv_C) in 
   let eca = encrypt((ca, sca), pkA) in 
   (* C with pseudo_C  submits his test (ques, ans) *)
   event submitted(pseudo_C, ques, ans);
   out(ch, eca);
   in(bba, eca':bitstring);
   let (=ca, sca':bitstring) = decrypt(eca', priv_C) in 
   if (ques, ans, pseudo_C) = checksign(sca', pkA) then 

(*Notification*)
    in(bba, ema':bitstring);
    in(bbn, (pseudo_E:pkey, he:bitstring, role_E: role,
       spseE:bitstring));
    if (pseudo_E, he, E) = checksign(spseE, pkN) then 
     let ((ca'': bitstring, sca'': bitstring, mark:bitstring), 
           sma:bitstring) = decrypt(ema', priv_C) in 
     if ca''=(ques, ans, pseudo_C, pseudo_E) && 
        ca''=checksign(sca'', pkA) then 
      if ((ques, ans, pseudo_C, pseudo_E), sca'', mark) = 
         checksign(sma, pseudo_E) then 
       (* C with pseudo_C is notified with "mark" *)
       event notified(pseudo_C, mark).
\end{lstlisting}
\vspace{-0.5cm}
\caption{The process of the candidate}
\label{fig:remark_can}
\end{figure}

\begin{figure}
\begin{lstlisting}
let NET (skN:skey, pkA:pkey, rc:bitstring) =
 in(ch, (R: role));
 get publickey(=R, rx, pkX) in
 let hx = exp(rx) in
 let pseudo_X = pseudo_pub(pkX,rx) in 
 let spseX = sign ((pseudo_X, hx, R), skN) in 
 out(bbn, ( pseudo_X, hx, R, spseX));
 (*Reveal rc*)
 let signetX = sign( (rc), skN) in
 let encsignetX = encrypt ( (rc, signetX), pkA) in
 out(ch, encsignetX).
\end{lstlisting}
\vspace{-0.5cm}
\caption{The process of the mixnet}
\label{fig:remark_mix}
\end{figure}

\begin{figure}
\begin{center}
\begin{lstlisting}
let Ex (skE:skey, pkA:pkey, pkN:pkey, mark:bitstring) =
(*Preparation *)
 in(bbn, (pseudo_E:pkey, he:bitstring, r: role, spseE:bitstring));
 if (pseudo_E, he, E) = checksign(spseE, pkN) then 
  let priv_E = pseudo_priv(skE, he) in 
  if checkpseudo(pseudo_E, priv_E) =true then 

(* Marking *)
   in(bba, eca'':bitstring);
   let ((ques:bitstring, ans:bitstring, pseudo_C:pkey, =pseudo_E),
       sca':bitstring) = decrypt(eca'', priv_E) in 
   if (ques, ans, pseudo_C,pseudo_E) = checksign(sca', pkA) then 
    let ca = (ques, ans, pseudo_C, pseudo_E) in 
    let ma:bitstring = (ca, sca', mark) in 
    let sma:bitstring =  sign(ma,priv_E) in 
    let ema = encrypt((ma, sma), pkA) in
    event marked(ques,ans,mark,pseudo_C,pseudo_E); 
    (* E with  pseudo_E marked the test (ques, ans) *)
    (* identified by pseudo_C with mark *)
    out(ch, ema).
\end{lstlisting}
\caption{The process of the examiner}
\label{fig:remark_ex}
\end{center}
\end{figure}

\begin{figure}
\begin{center}
\begin{lstlisting}
process
 !(
 (*Products of the secret exponent values of the servers *)
 (* (represented by the NET): rc for C and re for E *)
 new rc: bitstring;
 new re: bitstring;
 (*Assume one NET and one EA*)
 new skA: skey; let pkA = pk(skA) in out (ch, pkA); 
 new skN: skey; let pkN = pk(skN) in out (ch, pkN); 

 (!( NET(skN, pkA, rc))) | 
 (!( new ques:bitstring;  EA(skA, pkN, ques))) |
 (!( new skC: skey; let pkC = pk(skC) in out (ch, pkC);  
     new ans:bitstring; insert publickey(C, rc, pkC); 
     Cand(skC, pkA, pkN, ans))
   ) |
 (!( new skE: skey; let pkE = pk(skE) in out (ch, pkE); 
     new mark:bitstring; insert publickey(E, re, pkE); 
     Ex(skE, pkA, pkN, mark))
   )

 )
\end{lstlisting}

\caption{The exam process}
\label{fig:remark_ep}
\end{center}
\end{figure}

The process of the \ea is in Figure \ref{fig:remark_ea}, the process of the mixnet is in Figure \ref{fig:remark_mix}, the process of  the candidate is in Figure \ref{fig:remark_can}, and the process of examiner is in Figure \ref{fig:remark_ex}. The exam process is depicted in Figure \ref{fig:remark_ep}.
In each process we replace the identity of candidate  with the corresponding candidate's pseudonym inside the events. This choice is sound because the equational theory preserves the bijective mapping between keys and pseudonyms.

We analyse \remark~in ProVerif with the same approach used to verify the \Huszti protocol in chapter \ref{chap:formal}.
In particular, we use ProVerif's \texttt{noninterf} and  \texttt{choice[]} commands to verify the privacy requirements. 
The full ProVerif code is available on the Internet \citeweb{thesiscode}.

\subsubsection*{Results} 
Assuming an attacker in control of the network and  honest principals, ProVerif successfully proves all authentication and privacy requirements. Table \ref{tab:results_remark} reports the execution times over an Intel Core i7 3.0 GHz machine with 8 GB RAM.
Also assuming corrupted principals, ProVerif proves that \remark~ensures all the requirements. Table \ref{tab:results_remark} also reports the honest roles that are required for each requirement to hold. Note that we only model the processes  needed to specify the requirement. For example, the specification of Anonymous Marking requires two  candidates to be honest, otherwise they could just reveal their tests to the attacker, who would  trivially violate the protocol. However, all other candidates can be corrupted and collude with the attacker to violate the protocol.

Notably, \remark~ensures a stronger version of Anonymous Examiner since no one, even the exam authority, knows which examiner marks which test. It can be observed that  Mark Anonymity is not in Figure \ref{tab:results_remark}: since \remark~ensures Mark Privacy, it also guarantees Mark Anonymity. \index{Anonymous Examiner} \index{Mark Anonymity}


{
\begin{table}[]
\begin{center}
\begin{tabular}{|c|c|c|l|}
\hline
{\bf Requirement} & {\bf Result} & {\bf Time} & {\bf Honest roles} \\ \hline
  {\cautho} &  $\have$  &  1 s & (C, EA, NET) \\ \hline
  {\aau} &  $\have$  & 1 s & (E, EA, NET) \\ \hline 
  {\CAu} &  $\have$  & 1 s & (NET) \\ \hline 
      {\ta} &  $\have$ & 1 s & (E, EA, NET) \\ \hline
    {Mark Authenticity} &  $\have$  &  1 s & (E, EA, NET)\\ \hline
     {Question Indistinguishability} &  $\have$  &  1 s & (E, EA, NET) \\ \hline
      {Anonymous Marking} &  $\have$ & 1 s & (C, NET)\\ \hline
       {Anonymous Examiner} &  $\have$  & 1 s & (E, NET) \\ \hline
       {Mark Privacy} &  $\have$ & 3 m 39 s & (EA, NET) \\ \hline       
\end{tabular}
\caption{Summary of authentication and privacy analysis of Remark!}
\label{tab:results_remark}
\end{center}
\normalsize
\end{table}%

\paragraph{Remark.}
We report an issue on an early version of \remark~that witnesses how formal approaches contribute to achieve a deep understanding of the design models.
In the first draft of Remark!, the receipt of submission of a test $T_C$ consisted of the message $\lbrace{\mathit{Sign_{\SK_{A}}(T_C)}}\rbrace_{\overline{\PK}_C}$, that is, the \ea signs the test and posts the signed test encrypted with the candidate's pseudonym.
Moreover, the assignment of the test to the examiner consisted of the message  $\lbrace{\mathit{Sign_{\SK_{A}}(T_C)}}\rbrace_{\overline{\PK}_E}$, namely the signed test encrypted with an eligible examiner's pseudonym. The rest of the protocol was unchanged respect to the current version. \index{eligible examiners} \index{Test Authenticity}
With these two modifications ProVerif cannot prove Test Authenticity. In fact, the attack trace shows that a corrupted candidate can pick an examiner of her choice by re-encrypting the signed receipt received from the exam authority. It means that the candidate can influence the choice of the examiner who marks her test. 
Such attack could be avoided assuming  an access control mechanism that would not allow the candidate to post on the bulletin board.

However, the fixes implemented in the final version of \remark~shows that there is no need of access control mechanisms to secure the protocol. The first fix consists in signing the hash of the test as receipt. The second fix consists in making the pseudonym of the chosen examiner explicit. In doing so, the signature of the exam authority within the receipt cannot be used by a candidate to designate any examiner.


\section{Formal Analysis of Verifiability}\label{sec:remark_ver}

The ProVerif model proposed to check authentication and privacy in the previous section can be also used to analyse \remark~for verifiability.
The definitions of authentication and privacy introduced in chapter \ref{chap:formal} are formulated in the applied $\pi$-calculus and can be coded straight to ProVerif.
Conversely,  verifiability definitions are expressed in a more abstract model. Thus, it is necessary to map sets and relations specified in the verifiability model to \remark. We briefly recall the informal descriptions of the verifiability requirements specified in chapter \ref{chap:verifiability}. \index{abstract model}
\begin{itemize} 
\item Individual Verifiability: a candidate can check that
\begin{itemize}
\item \emph{Question Validity}: she received the questions actually generated by the question committee.
\item \emph{Marking Correctness}: the mark she received is correctly computed on her \examtest.
\item \emph{Test Integrity}:  her test is accepted and marked as she submitted it.
\item \emph{Test Markedness}: the \examtest she submitted is marked without modification.
\item \emph{Marking Integrity}: the mark attributed to her test is assigned to her without any modification.
\item \emph{Marking Notification}: she received the mark assigned to her.
\end{itemize}
\item Universal Verifiability: an auditor can check that
\begin{itemize}
\item \emph{Registration}: all accepted \examtests are submitted by registered candidates.
\item \emph{Marking Correctness}: all the marks attributed by the examiners to the tests are computed correctly.
\item \emph{Test Integrity}: all and only accepted \examtests are marked without any modification.
\item \emph{Test Markedness}: only the accepted \examtests are marked without modification.
\item \emph{Marking Integrity}: all and only the marks associated to the \examtests are assigned to the corresponding candidates with no modifications.
\end{itemize}
\end{itemize}

We recall that Definition~\ref{examdef}, which we introduced in chapter \ref{chap:verifiability}, considers the data sets  $I$, $Q$, $A$, $M$, and their elements $i$, $q$, $a$, $m$, which specify the candidate identities, the questions, the answers, and the marks respectively.
In Remark!, the set $I$ contains the candidate pseudonyms rather than the identities. In the previous section we argued this choice to be sound.
The sets $Q$, $A$, and $M$ contains the messages that correspond to questions, answers and marks generated by the protocol's principals, possibly manipulated by the attacker.

The relations $\fun{Accepted}$, $\fun{Marked}$, and $\fun{Assigned}$ are built from the posts that appear on the bulletin board.  The
tuples $(i,(q,a))$ of the relation $\fun{Accepted}$ consist of the receipts of submission that the \ea publishes on the bulletin board at the end of testing.
The tuples $(i,(q,a),m)$ of the relation $\fun{Marked}$ coincide with the tuples $(i,m)$  of $\fun{Assigned}$, and consist of the messages that the \ea publishes on the bulletin board at marking.
Precisely, the tuples $(i,(q,a),m)$ are generated from the marked \examtest signed by the examiner, that is, $Sign_{SK_E,h_{E}}( M_C)$. The tuples $(i,m)$ instead are built from the encryption of the marked \examtest generated by the \ea, that is, $ \lbrace  Sign_{SK_E,h_{E}}(M_C) \rbrace_{\overline{\PK}_C}$.
It can be observed that encryption under the candidate's pseudonym officially assigns the mark to the candidate.

Finally, the function $\fun{Correct}$, which is the algorithm used to mark the \examtests, can be modelled as a ProVerif table.

\subsection{Individual Verifiability} 
Individual verifiability definitions require the existence of verifiability-tests that candidates run to check the properties of the protocol. We show that \remark has the necessary verifiability-tests.
In ProVerif, we model the verifiability-tests as processes that emit the event \texttt{OK} when the verifiability-test succeeds, and emit the event \texttt{KO} when the verifiability-test fails.

We use correspondence assertions to prove soundness. ProVerif checks verifiability as a reachbility property. The verification strategy normally consists of checking that the event \texttt{OK} is always preceded by the event emitted in the part of the code where the predicate becomes satisfied.  In the ProVerif model of \remark~we assume an honest candidate principal who plays the role of the verifier. The other principals are usually corrupted, if not stated otherwise. The verifiability-test receives the data from the candidate via a private channel, and the remaining data posted on the bulletin board via public channels. This allows an attacker to manipulate the input data. Corrupted principals may collude with the attacker. 

We resort to unreachability of the event \texttt{KO} to prove completeness. In this case, the ProVerif model enforces only honest principals and prevents the attacker to manipulate the input data of the verifiability-tests. In fact, a complete verifiability-test must succeed if its input data is correct. 

In the following paragraphs, we specify the verifiability-tests for Remark!. We also discuss the conditions to achieve sound and complete verifiability-tests according to each individual verifiability requirement.

\subsubsection*{\questionValidity}   \index{Question Validity}
\remark~assumes that the \ea generates the questions at preparation and publishes them at testing. Thus, we model the \ea as an honest process in ProVerif, otherwise a corrupted \ea would publish questions that are different from the ones actually generated.

The verifiability-test \emph{testQV} , which is depicted in Figure \ref{fig:rqvtest}, receives the question \emph{eques} that is published on the bulletin board from a public channel. It also receives the candidate's question \emph{ques} and  private key \emph{priv\_C} from a private channel. 
The verifiability-test checks whether  the candidate actually received the question published by the \ea on the bulletin board. 

To prove soundness, we annotate the ProVerif process of the \ea with the event \texttt{generated} where the questions are generated.
Then, ProVerif checks if the verifiability-test emits the event \verb+OK+ only if the \ea actually generated the question received from the candidate, namely ProVerif checks the following correspondence assertion:
\begin{center}
 $\OKa{\ques}~\leadsto~\generated{ques}$ 
\end{center}

To prove completeness, ProVerif checks that the verifiability-test process does not emit the event \verb+KO+ when the input data is correct.
ProVerif confirms the verifiability-test is sound and complete, so we can conclude that \remark~is \questionValidityAdj.

\begin{figure}
\begin{center}
\begin{lstlisting}
let testQV(pkA: pkey, pch: channel) =
 in(bba, eques:bitstring);
 in(pch, (ques: bitstring, priv_C: skey));

 let (ques':bitstring, sques:bitstring) = 
     decrypt(eques, priv_C) in
 let (ques'':bitstring,p seudoC:bitstring) = 
     checksign(sques, pkA) in

 if ques'=ques && ques''=ques' then event OK 
 else event KO.
\end{lstlisting}
\caption{The Question Validity individual verifiability-test}
\label{fig:rqvtest}
\end{center}
\end{figure}

\subsubsection*{\markingCorrectness}  \index{Marking Correctness}
\remark~is designed to support different forms of questions (e.g., multiple choice, free-response, etc.), hence there is no universal marking algorithm that can be used to evaluate the answers. However, we can assume that the \ea publishes the table of evaluations that maps an answer to a mark after the exam concludes. We thus model the exam authority as an honest process in ProVerif to check \markingCorrectness.

The verifiability-test \emph{testMC}, which is in Figure \ref{fig:mctest}, receives the test ($\ques$, $\ans$) submitted by the candidate, and the mark \emph{mark} notified to her.  The verifiability-test checks if the mark reported on the table of evaluations and associated to the candidate's answer coincides with the mark received from the candidate.

To prove soundness, we annotate the ProVerif process of the candidate with the event \texttt{correct} where the candidate receives the mark at notification. ProVerif checks that if the verifiability-test emits the event \texttt{OK}, then a previous event \texttt{correct} was emitted. This is formalised as:
\begin{center}
 $\OKb{\ques}{\ans}{\mathit{mark}}~\leadsto~\correct{\ques}{\ans}{\mathit{mark}}$
\end{center}

ProVerif checks that the verifiability-test does not emit the event \texttt{KO} to prove completeness.

Thus, assuming an honest \ea that provides the table of evaluations at end of exam, \remark~is 
\markingCorrectnessAdj.

\begin{figure}
\begin{center}
\begin{lstlisting}
let testMC (pkA: pkey, priv_ch: channel) =
 in(priv_ch, (ques: bitstring, ans: bitstring, mark: bitstring));
 
 get correct_ans(=ques,=ans,mark':bitstring) in

 if mark'=mark then
  event OK
 else KO.
\end{lstlisting}
\caption{The \markingCorrectness~individual verifiability-test}
\label{fig:mctest}
\end{center}
\end{figure}

\begin{figure}
\begin{center}
\begin{lstlisting}
let testTI (pkA: pkey, priv_ch: channel) =
 in(priv_ch, (priv_C: skey, ques: bitstring, ans: bitstring, 
              pseudo_C: pkey));
 in(bbn, (pseudo_E:pkey, he:bitstring, rolet: role, 
          spseE:bitstring));
 in(bba, eca': bitstring);
 in(bba, ema':bitstring);

 (* If the message on the BB is signed by the authority, *)
 (* it is considered as part of the relation Accepted. *)
 let (ca: bitstring, sca':bitstring) = decrypt(eca', priv_C) in 
 let (ques': bitstring, ans': bitstring, pseudo_C': pkey) = 
     checksign(sca', pkA) in 
 (* If the message on the BB is signed by the examiner, *)
 (* it is considered as part of the relation Marked. *)
 let (((ques'': bitstring, ans'': bitstring, pseudo_C'': pkey), 
       sca1: bitstring, mark:bitstring), sma:bitstring) = 
     decrypt(ema', priv_C) in 
 let ((ques''': bitstring, ans''': bitstring, pseudo_C''': pkey),
      sca1': bitstring, mark':bitstring) = 
     checksign(sma, pseudo_E) in

 if ques'=ques && ans'=ans && pseudo_C'=pseudo_C && ques''=ques && 
    ans''=ans && pseudo_C''=pseudo_C && (ques', ans', pseudo_C') = 
    checksign(sca1,pkA) && ques'''=ques && ans'''=ans && 
    pseudo_C'''=pseudo_C && sca1'=sca1 then 
  event OK 
 else event KO.
\end{lstlisting}
\caption{The \etIntegrity~individual verifiability-test}
\label{fig:retitest}
\end{center}
\end{figure}

\subsubsection*{\bf \etIntegrity} 
The verifiability-test \emph{testTI} in Figure \ref{fig:retitest} takes in the test ($\ques$,$\ans$) submitted by the candidate via a private channel, and the receipt of submission \emph{eca'} and the notification \emph{ema'} published on the bulletin board via a public channel. The verifiability-test checks if candidate's submission, the receipt, and the notification contain the same question, answer, and pseudonym.

To prove soundness, we annotate the verifiability-test with the events \phantom{a} \verb+accepted+ and \verb+marked+ that map the corresponding relations.  In particular, the receipt of submission is part of the relation \verb+Accepted+ if it is signed by the exam authority and encrypted under the pseudonym of the candidate. Similarly, the notification is part of the relation \verb+Marked+ if it is signed by the examiner and encrypted under the pseudonym of the candidate. The requirement can be formalised with the following correspondence assertion:
\begin{center}
 $\OKb{\id}{\ques}{\ans}~\leadsto~\markinga{\id}{\ques}{\ans} \cup \accepted{\id}{\ques}{\ans}$ 
\end{center}

To prove completeness, ProVerif checks that the verifiability-test process does not emit the event KO when the input data is correct.

ProVerif shows that the verifiability-test for \etIntegrity~is sound and complete.
Note that a corrupted exam authority can publish two different receipts for the same test on the bulletin board. However, since the bulletin board is append-only, the candidate notices if the exam authority appends two different receipts for her submission because only the candidate knows the private key.

\subsubsection*{\etMarkedness} 
Since \remark~has a sound and complete verifiability-test for \etIntegrity, we can build from this a sound and complete verifiability-test for \etMarkedness. It is sufficient to not consider the receipt of submission as input, and just check whether the candidate's submission and the data obtained from notification contain the same question, answer, and pseudonym. The verifiability-test \emph{testTM} is depicted in Figure \ref{fig:markednesstest}.
To prove soundness, it is sufficient to prove the following correspondence assertion:
\begin{center}
 $\OKb{\id}{\ques}{\ans}~\leadsto~\markinga{\id}{\ques}{\ans}$
\end{center}

\begin{figure}
\begin{center}
\begin{lstlisting}
let testTM (pkA: pkey, priv_ch: channel) =
 in(priv_ch, (priv_C: skey,ques: bitstring, ans: bitstring, 
              pseudo_C: pkey));
 in(bbn, (pseudo_E:pkey, he:bitstring, rolet: role, 
          spseE:bitstring));
 in(bba, ema:bitstring);

 (* If the message on the BB is signed by the examiner, *)
 (* it is considered as part of the relation Marked. *)
 let (((ques': bitstring, ans': bitstring, pseudo_C': pkey), 
       sca1: bitstring, mark:bitstring), sma:bitstring) = 
     decrypt(ema, priv_C) in 
 let ((ques'': bitstring, ans'': bitstring, pseudo_C'': pkey), 
      sca1': bitstring, mark':bitstring) = 
     checksign(sma, pseudo_E) in
 
 if ques'=ques && ans'=ans && pseudo_C'=pseudo_C &&ques''= ques && 
    ans''=ans && pseudo_C''=pseudo_C && (ques', ans', pseudo_C') = 
    checksign(sca1,pkA) && sca1'=sca1 then 
  event OK
 else event KO.
\end{lstlisting}
\caption{The \etMarkedness~individual verifiability-test}
\label{fig:markednesstest}
\end{center}
\end{figure}

\begin{figure}
\begin{center}
\begin{lstlisting}
let testMI (pkA: pkey, priv_ch: channel) =
 in(priv_ch, (priv_C: skey,ques: bitstring, ans: bitstring, 
    pseudo_C: pkey));
 in(bbn, (pseudo_E:pkey, he:bitstring, rolet: role, 
    spseE:bitstring));
 in(bba, ema':bitstring);

 (* Assigned is the mark sent by the authority *)
 let (((ques': bitstring, ans': bitstring, pseudo_C': pkey), 
       sca': bitstring, mark:bitstring), sma:bitstring) = 
     decrypt(ema', priv_C) in 
 (* Marked are the marks signed by the examiner *)
 let ((ques'': bitstring, ans'': bitstring, pseudo_C'': pkey), 
      sca'': bitstring, mark': bitstring) = 
     checksign(sma, pseudo_E) in

 if ques'=ques && ans'=ans && pseudo_C'=pseudo_C && ques''=ques && 
    ans''=ans && pseudo_C''=pseudo_C && mark=mark' && 
    (ques', ans', pseudo_C')=checksign(sca',pkA) then
  event OK
 else KO.
\end{lstlisting}
\caption{The \markIntegrity~individual verifiability-test}
\label{fig:rmitest}
\end{center}
\end{figure}

\subsubsection*{\markIntegrity} 
The verifiability-test \emph{testMI} in Figure \ref{fig:rmitest} takes in the test ($\ques$,$\ans$) submitted by the candidate via a private channel,  and the notification \emph{ema'} published by the exam authority on the bulletin board. 
The verifiability-test checks if the test provided by the candidate and the notification on the bulletin board contain the same question, answer, and pseudonym, and if the examiner's signature on the mark is correct.

To check soundness in ProVerif, we annotate the verifiability-test with the events \verb+assigned+ and \verb+marked+ that map the corresponding relations. The data on the notification message is part of the relation \verb+Assigned+ if the data is signed by the exam authority and encrypted under the pseudonym of the candidate. This data is also part of the relation \verb+Marked+ if it also include the signature of the examiner. The correspondence assertion to check soundness is:
\begin{center}
 $\OKc{\id}{\ques}{\ans}{\mathit{mark}}~\leadsto~\markingb{\id}{\ques}{\ans}{\mathit{mark}} \cup \assigned{\id}{\ques}{\ans}{\mathit{mark}}$ 
\end{center}

We check completeness as usual, and ProVerif confirms that the verifiability-test for \markIntegrity~is sound and complete.

\begin{figure}
\begin{center}
\begin{lstlisting}
let testMNI (pkA: pkey, priv_ch: channel) =
 in(priv_ch, (priv_C: skey,mark: bitstring, pseudo_C: pkey, 
              ema': bitstring));
 in(bba, ema: bitstring);

 (* Assigned is the mark sent by the authority *)
 let (((ques:bitstring, ans: bitstring, pseudo_C':pkey), 
       sca: bitstring, mark':bitstring), sma:bitstring) =
     decrypt(ema, priv_C) in

 if (ques,ans,pseudo_C')=checksign(sca, pkA) && 
    pseudo_C'=pseudo_C &&  mark'=mark then
  event OK
 else event KO.
\end{lstlisting}
\caption{The \markNotificationIntegrity~individual verifiability-test}
\label{fig:rmnitest}
\end{center}
\end{figure}

\subsubsection*{\markNotificationIntegrity} 
The last individual verifiability definition concerns the check of the integrity of the notified mark. The verifiability-test \emph{testMNI} in Figure \ref{fig:rmnitest} is fed via a private channel with the mark $\mathit{mark}$ that the candidate received at notification. The verifiability-test \emph{testMNI} also takes in the  official notification \emph{ema'} published in the bulletin board via a public channel, and checks if the mark provided in the notification coincides with the one received from the candidate.

Similarly to \markIntegrity, we annotate the verifiability-test  \emph{testMNI} with the event \texttt{assigned} to prove soundness. ProVerif checks if the verifiability-test emits the event \texttt{OK} only if the mark notified to the candidate is the same officially assigned at the end of the exam. This is formalised as:
\begin{center}
 $\OKd{\id}{\mathit{mark}}~\leadsto~ \assigneda{\id}{\mathit{mark}}$ 
\end{center}

Also in this case  ProVerif checks that the verifiability-test process
does not emit the event KO when the input data is correct to prove completeness. ProVerif confirms
the verifiability-test is sound and complete, hence \remark~is  \markNotificationIntegrityAdj.

{
\begin{table}
\begin{center}
\begin{tabular}{|c|c|c|}
\hline
{\bf Requirement} & {\bf Soundness} & {\bf Completeness} \\ \hline
\questionValidity &
			$\have$ (EA) & $\haveall$ \\
			\hline \etIntegrity & $\have$ & $\haveall$ \\
			\hline \etMarkedness & $\have$ & $\haveall$ \\
			\hline \markingCorrectness & $\have$ (EA) & $\haveall$ \\
			\hline \markIntegrity & $\have$ & $\haveall$  \\ 
			\hline \markNotificationIntegrity & $\have$ & $\haveall$  \\
 \hline       
\end{tabular}
\caption{Summary of the analysis of \remark~for I.V. requirements}
\label{tab:riv}
\end{center}
\normalsize

\end{table}%

Table \ref{tab:riv} summarises the results of the individual verifiability analysis of \remark~and reports the roles required to be honest.

\subsection{Universal verifiability}
Also for the specification of universal verifiability definitions, we model verifi\-ability-tests as processes that emit the event \texttt{OK} when the verifiability-test succeeds, and emit the event \texttt{KO} when the verifiability-test fails.

In the case of universal verifiability an auditor runs the verifiability-tests, namely the auditor plays the role of the verifier. This  requires a different approach to prove soundness compared to the approach used for individual verifiability definitions, in which the candidate plays as verifier.
In fact, in the case of universal verifiability also the candidate can be corrupted, hence it can be hard to find a ProVerif process that can be annotated with events to check soundness via correspondence assertions.

The different approach consists of proving soundness of the verifiability-tests using unreachability of the event \texttt{KO}.
The underlying idea is that every time the verifiability-test succeeds, which means that it emits the event \texttt{OK},
we check if the decryption of the concerned ciphertext gives the expected plaintext. If not, the event \texttt{KO} is emitted, thus the verifiability-test is not sound. 

As we shall see later, it is however possible to prove the soundness of the  verifiability-test  for  Registration requirement using correspondence assertions. This is possible because the NET is assumed to be honest, hence the corresponding ProVerif process can be annotated with an event that is emitted when registration concludes.
We  always use unreachability of the event \texttt{KO} to prove completeness of the verifiability-tests.

\paragraph{Remark.} It can be noted that all messages posted by the \ea on the bulletin board are encrypted under the pseudonym of either 
the candidate or the examiner, hence no public data can be used as it is by the auditor. Candidates and examiners hold  long-term pairs of public/private keys, and it is implausible that they reveal their private keys for audit purposes. Since the auditor cannot decrypt a ciphertext message posted on the bulletin board, the auditor should be rather provided with the corresponding plaintext and pseudonym. In so doing, the auditor can encrypt the plaintext with the pseudonym and check if the encryption coincides with the same ciphertext message posted on the bulletin board.
Since \remark~uses ElGamal encryption, which is probabilistic, the auditor should be also provided with the random value used to encrypt a message.

In the following, we specify the data that the \ea should provide to the auditor after the exam concludes. 
\begin{itemize}
\item Registration:  the \ea reveals the signatures inside the receipts $\mathit{receipt}=\lbrace  Sign_{SK_A}( H(T_C) \rbrace_{\overline{\PK}_C}$ posted
on the bulletin board and the random values used to encrypt the receipts.
\item \markingCorrectnessUniversal: the \ea reveals the marked \examtests inside the evaluations $\mathit{sma}=\lbrace Sign_{SK_E,h_{E}}( M_C) \rbrace_{\PK_{A}}$, the random values used to encrypt the marked tests, and the table $\mathit{correct\_ans}$ that maps each mark to each answer. \index{Marking Correctness}
\item \etIntegrityUniversal: the \ea reveals the marked \examtests inside the evaluations, the random values used to encrypt the marked tests, plus the data disclosed for \registration.
\item \etMarkednessUniversal: the \ea reveals the same data disclosed for \etIntegrityUniversal.
\item \markIntegrityUniversal: the \ea reveals the examiners' signatures on the marked \examtests inside the evaluations, and the random values  used to encrypt the notifications $\mathit{notif}=\lbrace  Sign_{SK_E,h_{E}}(M_C) \rbrace_{\overline{\PK}_C}$  before posting them on the bulletin board.
\end{itemize}

We anticipate that it is not possible to automatically prove the  universal verifiability requirements in ProVerif. 
To prove such requirements it is needed to iterate over all candidates, but ProVerif does not support loops.
We thus prove the base case of each requirement automatically in ProVerif, in which it is considered only one accepted test or one assigned mark. Then, we provide manual induction proofs that generalise the ProVerif result to the general case with an arbitrary number of candidates.

\begin{figure}
\begin{center}
\begin{lstlisting}
let testUR(pkN, pkA, ch1,...,chn, bbn1,...,bbnm, bba1,...,bban)=
 in(bbn1, (pseudo_C1, hc, r, NET_sign1));
 ...
 in(bbnm, (pseudo_Cm, hc, r, NET_signm));

 in(bba1, receipt1);
 ...
 in(bban, receiptn);

 in(ch1, (rcoin1, EA_sign_rcpt1));
 ...
 in(chn, (rcoinn, EA_sign_rcptn));

 let (quest1, answ1, pseudo_C'1) = 
 checksign(EA_sign_rcpt1, pkA) in
 ...
 let (questn, answn, pseudo_C'n) = 
 checksign(EA_sign_rcptn, pkA) in    
 
 (* If the pseudonym on the BB is signed by the NET, *)
 (* it is considered as part of the relation Accepted. *)
 if (pseudo_C1, hc, r)=checksign(NET_sign1, pkN) && r=C && 
    pseudo_C1=pseudo_C'1 
    ||...||
    (pseudo_C1, hc, r)=checksign(NET_sign1, pkN) && r=C &&
    pseudo_C1=pseudo_C'n 
    &&...&&
    (pseudo_Cm, hc, r)=checksign(NET_signm, pkN) && r=C && 
    pseudo_Cm=pseudo_C'1    
    ||...||
    (pseudo_Cm, hc, r)=checksign(NET_signm, pkN) && r=C && 
    pseudo_Cm=pseudo_C'n  then
  if receipt1=int_encrypt(((quest1, answ1, pseudo_C'1), 
                          EA_sign_rcpt1), pseudo_C1, rcoin1)
     &&...&&
     receiptn=int_encrypt(((questn, answn, pseudo_C'n), 
                          EA_sign_rcptn), pseudo_Cm, rcoinn)
  then event OK
  else event KO
 else event KO.
\end{lstlisting}
\caption{The \registration~universal verifiability-test}
\label{fig:rutest}
\end{center}
\end{figure}

\subsubsection*{\registration}
The verifiability-test \emph{testUR}, which is depicted in Figure \ref{fig:rutest}, takes in from the bulletin board 
the pseudonyms of the candidates signed by the mixnet and the receipts of submissions generated by the \ea.
In so doing, the auditor can check that the \ea  accepted only tests signed with pseudonyms posted by the mixnet during preparation.

ProVerif proves that the verifiability-test is complete and sound for the base case, which considers one accepted test and an unbounded number of candidates. 
To prove the general case, namely for an unbounded number of accepted tests and candidates, it is necessary to show that 
\begin{equation*}
\testUR(E) = \mi{true} \Leftrightarrow 
 \set{i: (i,x) \in \fun{Accepted}} \subseteq I_r
\end{equation*}
holds for an exam execution $E$ that considers any size $n$ of the relation $\fun{Accepted}$ and any number $m$ of registered candidates. 

Let $\testUR_k(\cdot)$ be the verifiability-test applied to an exam execution that has $k$ accepted tests;
let $\testUR_k(\cdot) \rightarrow^{\ast} OK$ denote the verifiability-test that outputs  \verb+OK+ (true) 
after some steps;
let $E$ be an exam execution that has $m$ registered candidates and $n$ accepted tests;
let $E_j$ be a version of $E$ that only considers the $j^{th}$ accepted test, which is submitted by the candidate $i_j$. 
Since ProVerif proves that  the verifiability-test is complete and sound for one accepted test and any number of registered candidates, it follows that for soundness we have
$$\forall 1 \leq j \leq n: \testUR_1(E_j) \rightarrow^{\ast} OK \Rightarrow i_j \in I_r$$ and for completeness we have
$$\forall 1 \leq j \leq n: i_j \in I_r \Rightarrow \testUR_1(E_j) \rightarrow^{\ast} OK.$$
The verifiability-test $\testUR_n(E)$ checks if each of the accepted tests received on channels \verb+bba1+, \ldots, \verb+bban+ was submitted by one of the candidates given on  channels  \verb+bbn1+, \ldots, \verb+bbnn+. The verifiability-test $\forall 1 \leq j \leq n: \testUR_1(E_j)$ checks if the $j^{th}$ accepted test received on the channel \verb+bbaj+ was submitted by one of the candidates given on the channels \verb+bbn1+, \ldots, \verb+bbnn+. 
\paragraph{}

We have that
\begin{center}
$\testUR_n(E)\rightarrow^{\ast} OK$\\
$ \Downarrow$\\
$\forall 1 \leq j \leq n: \testUR_1(E_j) \rightarrow^{\ast} OK$\\
$ \Downarrow_{(by~ProVerif)}$\\
$\forall 1 \leq j \leq n: i_j \in I_r$\\
$ \Downarrow$\\
$\set{i: (i,x) \in \fun{Accepted}} \subseteq I_r$
\end{center}
Thus, the verifiability-test \emph{testUR} is sound also for the general case.

\begin{center}
$\set{i: (i,x) \in \fun{Accepted}} \subseteq I_r$\\
$ \Downarrow$\\
$\forall 1 \leq j \leq n: i_j \in I_r$\\
$ \Downarrow_{(by ~ProVerif)}$\\
$\forall 1 \leq j \leq n: \testUR_1(E_j) \rightarrow^{\ast} OK$\\
$ \Downarrow$\\
$\testUR_n(E)\rightarrow^{\ast} OK$
\end{center}
Also the verifiability-test \emph{testUR} is complete for the general case.

\subsubsection*{\markingCorrectnessUniversal} 
The verifiability-test \emph{testUMC}, which is depicted in Figure \ref{fig:umctest}, takes as input from the bulletin board  \index{Marking Correctness}
the pseudonym of the examiner signed by the mixnet, and the mark notifications signed by the examiner and published by the  \ea.
The auditor can obtain the evaluations generated by the examiner from the mark notifications. 
Then, the auditor checks if the mark assigned to the question of each test coincides with the mark associated to the same question on the table provided by the \ea. \remark~intuitively ensures this requirement only if the \ea is honest as it provides the table at the conclusion of the exam. For simplicity, we assume that one examiner marks all the tests.

ProVerif proves soundness and completeness of the verifiability-test assuming only one marked test, namely the relation $\fun{Marked}$ has only one entry.
To prove the general case, we should consider an unbounded number of marked test. Thus, it is necessary to show that
\begin{equation*}
\begin{split}
\mathit{testUMC}(E) = \mi{true} \Leftrightarrow & \forall (i,x,m) \in \fun{Marked},~\fun{Correct}(x) = m
\end{split}
\end{equation*}
holds for an exam execution $E$ that considers any size $n$ of the relation $\fun{Marked}$.

\begin{figure}
\begin{center}
\begin{lstlisting}
let testUMC (pkN, bbn, ch1,...,chn) =
 in(bbn, (pseudo_E, he, r, spseE));
 in(ch1, sma1);
 ...
 in(chn, sman);

 let ((ques1, ans1, pseudo_C1), sca1, mark1) = 
     checksign(sma1, pseudo_E) in
 ...
 let ((quesn, ansn, pseudo_Cn), scan, markn) = 
     checksign(sman, pseudo_E) in
 get correct_ans(ques'1,ans'1,=mark1) in
 ...
 get correct_ans(ques'n,ans'n,=markn) in

 if (pseudo_E, he, r) = checksign(spseE, pkN) && r = E then
  if (ques1=ques'1 && ans'1=ans1)
     &&...&&   
     (quesn=ques'n && ans'n=ansn)
  then event OK
  else event KO
 else KO.
\end{lstlisting}

\caption{The \markingCorrectnessUniversal~universal verifiability-test}
\label{fig:umctest}
\end{center}
\end{figure}

Let $\mathit{MC_k}(\cdot)$ be the verifiability-test applied to an exam execution that has $k$ marked tests;
let $\mathit{MC_k}(\cdot) \rightarrow^{\ast} OK$ denote the verifiability-test that outputs  \verb+OK+ (true)
after some steps;
let $E$ be an exam execution that has $n$ marked tests;
let $E_j$ be a version of $E$ that only considers the $j^{th}$ marked test, which was submitted by the candidate $i_j$ and evaluated with the mark $m_j$, namely $(i_j,(x_j),m_j) \in \fun{Marked}$.
Since ProVerif proves that  the verifiability-test is complete and sound for one marked test, it follows that for soundness we have
$$\forall 1 \leq j \leq n: \mathit{MC}_1(E_j) \rightarrow^{\ast} OK \Rightarrow \fun{Correct}(x_j) = m_j$$ and for completeness we have
$$\forall 1 \leq j \leq n: \fun{Correct}(x_j) = m_j \Rightarrow \mathit{MC}_1(E_j) \rightarrow^{\ast} OK.$$

The verifiability-test  $\mathit{MC}_n(E)$ obtains the mark evaluations from  channels \verb+ch1+, \ldots, \verb+chn+, and checks if 
 all the tests that are contained in evaluation  are marked correctly.
The verifiability-test $\forall 1 \leq j \leq n: \mathit{MC}_1(E_j)$ checks if the $j^{th}$ test, whose evaluation is obtained from the channel \verb+chj+, is  marked correctly. Thus, it follows that
\begin{center}
$\mathit{MC}_n(E)\rightarrow^{\ast} OK$\\
$ \Downarrow$\\
$\forall 1 \leq j \leq n: \mathit{MC}_1(E_j) \rightarrow^{\ast} OK$\\
$ \Downarrow_{(by~ProVerif)}$\\
$\forall 1 \leq j \leq n: \fun{Correct}(x_j) = m_j$\\
$ \Downarrow$\\
$\forall (i,(q,a),m) \in \fun{Marked},~\fun{Correct}(x) = m$
\end{center}
Thus, the verifiability-test \emph{testUMC} is sound also for the general case. 

\begin{center}
$\forall (i,x,m) \in \fun{Marked},~\fun{Correct}(x) = m$\\
$ \Downarrow$\\
$\forall 1 \leq j \leq n: \fun{Correct}(x_j) = m_j$\\
$ \Downarrow_{(by ~ProVerif)}$\\
$\forall 1 \leq j \leq n: \mathit{MC}_1(E_j) \rightarrow^{\ast} OK$\\
$ \Downarrow$\\
$\mathit{MC}_n(E)\rightarrow^{\ast} OK$
\end{center}
Also the verifiability-test  \emph{testUMC} is complete  for the general case. 

\begin{figure}
\begin{center}
\begin{lstlisting}
let testUTI(pkN, pkA, bba1,..., bban, bbn, ch1,...,chn)=
 in(bbn, (pseudo_E, he, re, spseE));
 in(ch1,((rcoin1, sca1, pseudo_C1),(rcoinA1, smaA1, pseudo_CA1)));
 ...
 in(chn,((rcoinn, scan, pseudo_Cn),(rcoinAn, smaAn, pseudo_CAn)));
 in(bba1, (receipt1, notif1));
 ...
 in(bban, (receiptn, notifn));

 let (quest1, answ1, pseudo_C'1) = checksign(sca1, pkA) in
 ...
 let (questn, answn, pseudo_C'n) = checksign(scan, pkA) in
 let ((quest'1, answ'1, pseudo_C''1), sca'1, mark1) =
     checksign(smaA1, pseudo_E) in
 ...
 let ((quest'n, answ'n, pseudo_C''n), sca'n, markn) = 
     checksign(smaAn, pseudo_E) in

 if (receipt1=int_encrypt(((quest1, answ1, pseudo_C'1), sca1),
                          pseudo_C1, rcoin1) && 
     notif1=int_encrypt((((quest'1, answ'1, pseudo_C''1), sca'1, 
                         mark1), smaA1), pseudo_CA1, rcoinA1) && 
     sca'1=sca1)
     &&...&&
    (receiptn=int_encrypt(((questn, answn, pseudo_C'n), scan),
                          pseudo_Cn, rcoinn)&& 
     notifn=int_encrypt((((quest'n, answ'n, pseudo_C''n), sca'n,
                         markn), smaAn), pseudo_CAn, rcoinAn) && 
     sca'n=scan)	    
 then 
  if (pseudo_C1=pseudo_CA1 && pseudo_CA1=pseudo_C'1 && 
      pseudo_C'1=pseudo_C''1 && quest1=quest'1 && answ1=answ'1)
     &&...&& 
     (pseudo_Cn=pseudo_CAn && pseudo_CAn=pseudo_C'n && 
      pseudo_C'n=pseudo_C''n && questn=quest'n && answn=answ'n)
  then event OK
  else KO    
 else KO.
\end{lstlisting}
\caption{The \etIntegrityUniversal~universal verifiability-test}
\label{fig:uetitest}
\end{center}
\end{figure}

\subsubsection*{\etIntegrityUniversal} 
The verifiability-test \emph{testUTI} (Figure \ref{fig:uetitest}) takes as input from the bulletin board 
the pseudonyms of the candidates signed by the mixnet and the receipts of submissions plus the mark notifications generated by the \ea. 
The auditor can obtain the evaluations generated by the examiner from the mark notifications. 
The verifiability-test then checks  if the submitted tests were marked without any modification.
Similarly to \markingCorrectnessUniversal, we assume that one examiner marks all the tests for simplicity.

ProVerif can prove that the verifiability-test is complete and sound when one accepted test and one marked test are considered. 
To prove the general case that considers an unbounded number of accepted tests, it is necessary to show that
\begin{equation*}
\begin{split}
\mathit{testUTI}(E) = \mi{true} \Leftrightarrow
\fun{Accepted} = 
\set{(i,x): (i,x,m) \in \fun{Marked}}
\end{split}
\end{equation*}
holds for an exam execution $E$ that considers any size of the relations $\fun{Accepted}$ and $\fun{Marked}$. 

It can be assumed that the size of the relation $\fun{Accepted}$ is equal to the relation  $\fun{Marked}$.
In fact, by looking at the bulletin board, the auditor can check that the number of the receipts of submissions coincides with the number of mark notifications. 

Let $\mathit{testUTI_k}(\cdot)$ be the \etIntegrityUniversal{} verifiability-test applied to an exam execution that has $k$ accepted tests  
and $k$ marked tests;
let $\mathit{testUTI_k}(\cdot) \rightarrow^{\ast} OK$ denote the verifiability-test that outputs  \verb+OK+ (true)
after some steps;
let $E$ be an exam execution that has $n$ accepted tests and $n$ marked tests;
let us assume that the tests are marked in the same order as they were accepted;
let $E_j$ be a version of $E$ that only considers the $j^{th}$ accepted test $x_j$ submitted 
by the candidate $i_j$, and the $j^{th}$ marked test $x'_j$ associated to the candidate $i'_j$.  

Since ProVerif proves that  the verifiability-test is complete and sound for one accepted test and one marked test, it follows that for soundness we have
$$\forall 1 \leq j \leq n: \mathit{testUTI_1}(E_j) \rightarrow^{\ast} OK \Rightarrow (i_j,(q_j, a_j)) = (i'_j,(q'_j, a'_j))$$ and for completeness 
we have
$$\forall 1 \leq j \leq n:  (i_j,(q_j, a_j)) = (i'_j,(q'_j, a'_j)) \Rightarrow \mathit{testUTI_1}(E_j) \rightarrow^{\ast} OK.$$

The verifiability-test  $\mathit{testUTI_n}(E)$ checks if each pair of accepted and marked tests obtained from channels \verb+bba1+, \ldots, \verb+bban+ has the same pseudonym, question, and answer.
Similarly, the verifiability-test  $\forall 1 \leq j \leq n: \mathit{testUTI_1}(E_j)$ checks if the $j^{th}$ accepted and marked tests obtained from the channel \verb+bbaj+ are identical. Thus, it follows that 
\begin{center}
$\mathit{testUTI_n}(E)\rightarrow^{\ast} OK$\\
$ \Downarrow$\\
$\forall 1 \leq j \leq n: \mathit{testUTI_1}(E_j) \rightarrow^{\ast} OK$\\
$ \Downarrow_{(by~ProVerif)}$\\
$\forall 1 \leq j \leq n: (i_j,x_j) = (i'_j,x'_j)$\\
$ \Downarrow$\\
$\fun{Accepted} = \set{(i,x): (i,x,m) \in \fun{Marked}}$
 \end{center}
Thus, the verifiability-test  \emph{testUTI} is sound also for the general case. 
\begin{center}
$\fun{Accepted} = \set{(i,x): (i,x,m) \in \fun{Marked}}$\\
$ \Downarrow$\\
$\forall 1 \leq j \leq n: (i_j,x_j) = (i'_j,x'_j)$\\
$ \Downarrow$\\
$\forall 1 \leq j \leq n: \mathit{testUTI_1}(E_j) \rightarrow^{\ast} OK$\\
$ \Downarrow$\\
$\mathit{testUTI}_n(E)\rightarrow^{\ast} OK$
 \end{center}
Also the verifiability-test  \emph{testUTI} is complete for the general case.

\subsubsection*{\etMarkednessUniversal} Since \remark~is \etIntegrityUniversalAdj, it is also \etMarkednessUniversalAdj.
The proof strategy is the same outlined above for \etIntegrityUniversal. However, it is not necessary to assume that the size of the relation $\fun{Accepted}$ is equal to the relation  $\fun{Marked}$, since \etMarkednessUniversal~does not require strict equality of the two multisets.

\begin{figure}
\begin{center}
\begin{lstlisting}
let testUMI (bbn, bba1,...,bban) =
 in(bbn, (pseudo_E, he, re, spseE));
 in(bba1, (notif1,rcoin1, sma1));
 ...
 in(bban, (notifn,rcoinn, sman)); 

 let ((quest1, answ1, pseudo_C1), sca'1, mark1) = 
     checksign(sma1, pseudo_E) in 
 ...
 let ((questn, answn, pseudo_Cn), sca'n, markn) = 
     checksign(sman, pseudo_E) in 
 
 if notif1=int_encrypt((((quest1, answ1, pseudo_C1),sca'1, mark1), 
                        sma1), pseudo_C1, rcoin1) 
    &&...&&                     
    notifn=int_encrypt((((questn, answn, pseudo_Cn),sca'n, markn),
                        sman), pseudo_Cn, rcoinn)                     
 then event OK
 else event KO.
\end{lstlisting}
\caption{The \markIntegrityUniversal~universal verifiability-test}
\label{fig:umitest}
\end{center}
\end{figure}

\subsubsection*{\markIntegrityUniversal} 
The verifiability-test \emph{testUMI}, which is in Figure \ref{fig:umitest}, is fed with mark notifications posted on the bulletin board by the \ea. The auditor obtains the evaluations generated by the examiner from the mark notifications, and checks if the marks that the \ea assigned to the candidates coincide with the marks that the examiner assigned to the candidates' tests.
Also for this requirement we assume one examiner who marks all the tests.

In the case that the relations $\fun{Assigned}$ and $\fun{Marked}$ contain each  one entry, ProVerif proves that the verifiability-test  \markIntegrityUniversal{} is complete and sound.
The general case, which considers an unbounded number of entries, consists on proving that
\begin{equation*}
\begin{split}
\mathit{testUMI}(E) = \mi{true} \Leftrightarrow \fun{Assigned} =  \set{(i,m): (i,x,m) \in \fun{Marked}}
\end{split}
\end{equation*}
holds for an exam execution $E$ that considers any size of the relations $\fun{Assigned}$ and $\fun{Marked}$. 

Similarly to \etIntegrityUniversal, it can be assumed that the size of the relation $\fun{Assigned}$ is equal to the relation  $\fun{Marked}$, as the auditor can check such equality by looking at the bulletin board.

Let $\mathit{testUMI_k}(\cdot)$ be the \markIntegrityUniversal{} verifiability-test applied to an exam execution that has $k$ marks assigned to the candidates  and $k$ marks associated to the tests;
let $\mathit{testUMI_k}(\cdot) \rightarrow^{\ast} OK$ denote the verifiability-test that outputs  \verb+OK+ (true)
after some steps;
let $E$ be an exam execution that has $n$ marks assigned to the candidates  and $n$ marks associated to the candidates' tests;
let us assume that the tests are assigned to the candidates in the same order as they were marked;
let $E_j$ be a version of $E$ that only considers  the $j^{th}$ mark $m_j$ assigned to the candidate $i_j$, and the $j^{th}$ mark $m'_j$ associated to the test of candidate $j'_j$. 

Since ProVerif proves that the verifiability-test is complete and sound for one entry, it follows that for soundness we have
$$\forall 1 \leq j \leq n: \mathit{testUMI}_1(E_j) \rightarrow^{\ast} OK \Rightarrow (i_j,m_j) = (i'_j, m'_j)$$ and for completeness we have
$$\forall 1 \leq j \leq n: (i_j,m_j) = (i'_j, m'_j) \Rightarrow \mathit{testUMI}_1(E_j) \rightarrow^{\ast} OK.$$

The verifiability-test $\mathit{testUMI}_n(E)$  receives from the channels  \verb+bba1+, \ldots, \verb+bban+ the notifications of the \ea, and checks if the  pseudonyms and marks obtained from the notifications coincide with the ones obtained from the evaluations of the examiner.
Similarly, the verifiability-test $\forall 1 \leq j \leq n: \mathit{testUMI}_1(E_j)$ checks if the $j^{th}$ 
 pseudonym and mark obtained from the  evaluation  and notification on channel \verb+bbaj+ are identical.
Thus, it follows that
\begin{center}
$\mathit{testUMI}_n(E)\rightarrow^{\ast} OK$\\
$ \Downarrow$\\
$\forall 1 \leq j \leq n: \mathit{testUMI}_1(E_j) \rightarrow^{\ast} OK$\\
$ \Downarrow_{(by~ProVerif)}$\\
$\forall 1 \leq j \leq n: (i_j,m_j) = (i'_j, m'_j)$\\
$ \Downarrow$\\
$\set{(i,m): (i,x,m) \in \fun{Marked}} = \fun{Assigned}$
\end{center}
Thus, the verifiability-test \emph{testUMI} is sound also for the general case.
\begin{center}
$\set{(i,m): (i,x,m) \in \fun{Marked}} = \fun{Assigned}$\\
$ \Downarrow$\\
$\forall 1 \leq j \leq n: (i_j,m_j) = (i'_j, m'_j)$\\
$ \Downarrow_{(by ~ProVerif)}$\\
$\forall 1 \leq j \leq n: \mathit{testUMI}_1(E_j) \rightarrow^{\ast} OK$\\
$ \Downarrow$\\
$\mathit{testUMI}_n(E)\rightarrow^{\ast} OK$
\end{center}
Thus, the verifiability-test \emph{testUMI} is complete also for the general case.

\begin{table}
\begin{center}
\begin{tabular}{|c|c|c|}
\hline
{\bf Requirement} & {\bf Soundness} & {\bf Completeness} \\ \hline
\registration & $\have$ & $\haveall$ \\\hline
\markingCorrectnessUniversal & $\have$ (EA)  & $\haveall$  \\\hline
\etIntegrityUniversal & $\have$ & $\haveall$ \\\hline
\etMarkednessUniversal & $\have$ & $\haveall$  \\\hline
\markIntegrityUniversal & $\have$ & $\haveall$  \\ \hline
\end{tabular}
\caption{Summary of the analysis of \remark~for U.V. requirements}
\label{tab:remark_uv}
\end{center}
\end{table}

Table \ref{tab:remark_uv} summarises the results of the universal verifiability analysis of \remark~and reports the roles required to be honest.


\section{Conclusion} \label{sec:conclusionremark}
This chapter presents Remark!, a protocol for Internet-based exam that guarantees authentication, privacy, and verifiability
with minimal trust assumptions. \remark~meets its requirements in most of the cases by assuming only one honest server among the servers that compose the exponentiation mixnet. \index{exponentiation mixnet} According to each requirement, \remark~can resist against collusion of candidate and exam authority (\eg Anonymous Examiner), exam authority and examiner (\eg Anonymous Marking), or candidate and examiner (\eg Question Indistinguishability) without the presence of a trusted third party. \index{Anonymous Marking}

A formal analysis in ProVerif confirms that \remark~ensures all the authentication and privacy requirements proposed in chapter \ref{chap:formal}. Notably, thanks to this formal analysis, we found and solved an issue on an earlier version of the protocol.

\remark~proves to be fully verifiable, according the individual and universal verifiability definitions proposed in chapter \ref{chap:verifiability}. ProVerif automatically proves all the individual verifiability requirements. Assuming an honest mixnet, all the requirements but Question Validity \index{Question Validity} and Marking Correctness \index{Marking Correctness} can be proved assuming corrupted candidates, examiners, and exam authority. Question Validity and Marking Correctness still require an honest exam authority.
Concerning the universal verifiability requirements, ProVerif cannot deal with the general cases, thus we completed the analysis with manual proofs. It turns out that \remark~ensures all the requirements but Marking Correctness assuming  corrupted candidates, examiners, and exam authority. Also in this case,  Marking Correctness can be proved assuming an honest exam authority. However, it is also assumed that the \ea provides the auditor with some additional data at the conclusion of the exam, since all the messages posted on the bulletin board are encrypted.


\chapter{Computer-assisted Exam Protocols}\label{chap:wata}

According to Definition \ref{def:eexam} proposed in chapter \ref{chap:terminology}, a protocol belongs to the category of computer-assisted exams if at least one if its phases receives some level of assistance from computers. We argued that the levels of detail and abstraction of the protocol specification determine whether a protocol belongs to traditional or to computer-assisted exams. \index{computer-assisted exams}

In this chapter, we focus on some specifications of computer-assisted exam protocols that share traditional testing, namely testing takes place by pen and paper. These protocols however have different  functional requirements and threat models: one considers local tasks, such as notification of marks,  and no TTP; some others consider remote tasks, such as remote registration, but assume TTP; one achieves remote tasks without TTP.  \index{remote registration} \index{remote notification}

In a way, \remark~already achieves remote registration and remote notification with minimal reliance on trusted parties. As \remark~belongs to the class of Internet-based exams, it requires candidate and exam authority to use computers at testing in order to sign and encrypt the tests. Therefore, testing cannot take place by pen and paper. \index{Internet-based exams}
Moreover, \remark~assumes at least one honest mix server. As we shall see later, we propose a  computer-assisted exam protocol that ensures the same authentication and privacy requirements of \remark~though relying neither on mixnet nor on TTP.

\index{remote testing}
Either testing is carried out traditionally or remotely is a key aspect for security. Remote testing is supported with computers, which intuitively introduce more security risks, but facilitate the design of the other remote phases. In contrast, traditional testing is less risky but complicates such remote design. The major security risks introduced by remote testing are due to remote invigilation and computer devices required at testing. Normally, it is better not to allow such devices at testing because they can promote candidate cheating.  For instance, Migicovsky \etal \citeltex{MDR+14} have recently shown how to outsmart invigilation using a smart watch. 
The difference between traditional testing and remote testing, namely between computer-assisted exams and Internet-based exams,  finds its analogue in voting: the former is comparable to paper-based electronic voting systems, while the latter is the analogous of Internet-voting systems.

In this chapter, we focus on a family of computer-assisted exam protocols called \emph{WATA}. The original versions of WATA progressively introduce more computer assistance in their design still keeping traditional testing and assuming TTP. We then propose a novel protocol, WATA IV, which includes the lightweight participation of a TTP, opens up also for computer-based exams, and ensures more security requirements despite a stronger threat model. WATA IV is further reconceived to completely remove the TTP. The underlying idea is to combine oblivious transfer and visual cryptography to allow candidate and examiner to jointly generate
a pseudonym that anonymises the candidate's test. The pseudonym is revealed only to the candidate at the beginning of testing. We analyse the protocol formally in ProVerif and prove that it satisfies all the stated security requirements.

\paragraph*{Outline of the chapter.}
Section~\ref{sec:relatedwata} discusses the related work about secure protocols for computer-assisted exams. \index{computer-assisted exams}
Section~\ref{sec:wataexam} describes the original WATA II \& III protocols and their informal analyses.
Section~\ref{sec:wataiv} introduces WATA IV according the four phases of an exam and points up the novelties respect to the original WATA schemes.
Section~\ref{sec:watav} contains a new exam protocol that redesigns parts of  WATA IV and removes the TTP. This section also introduces the formalisation of dispute resolution,  and provides the formal analysis of the new exam protocol in ProVerif.
Section~\ref{sec:conclusionremarkwata} discusses future work and ends the chapter.

\section{Related Work}\label{sec:relatedwata}
Nowadays, most of the exams employed in public competitions are computer-assisted or even computer-based. 
ETS and Pearson Vue develop various Com\-puter-assisted exams for skill and professional certifications \citeweb{ets, pearsonurl}.
The European Union adopts computer-based exams for the selection of EU personnel \citeweb{euexam}. 
The specification of such  exams is not publicly available, and their security fully relies on the developers, who have the prominent role of TTP during the exam execution. This choice has not prevented frauds on the administered exams \citeweb{cheatingets}.

Different exam protocols have been proposed to ensure anonymous marking.
INFOSAFE \citeweb{INFOSAFE} is an anonymous marking system for computer-assisted exam with traditional testing, and is adopted in university exams.
Candidates write down their personal details on top of a tamper-evident paper, and hide them with a flap which is bent and glued over. After marking, the personal details are
disclosed by tearing off the flap. Systems following a ``double envelope'' strategy, often used in public tenders, make use of two envelops to separate the identification details from the offers. The personal details are assumed to be read after marking. Many European universities, such as Dublin City University and University of Sheffield, use their own anonymous marking systems \citeweb{DCU, SHEFFIELD}. So do top USA academies, such as Stanford and Harvard Law School \citeweb{STANFORD, HLS}. The latter relies on the Blind Grading Number system which assigns candidates with numerical pseudonyms until the marking period ends. Nemo Scan \citeweb{NEMOSCAN}  uses a patented anonymity paper cover \citeltex{patent} consisting of two parts: one with the covered candidate details, the other with a section where to type the marks. At notification, a scanner with a proprietary software reads the paper with the candidate details and assigns her the mark.

All the systems outlined above assume a trusted authority to ensure Anonymous Marking. Moreover, it is not clear how such systems scale up to other security requirements. 
In this chapter, we discuss how to progressively remove trusted authorities from the design of the protocols, and consider the exam authority  corrupted to various extents.


\section{The WATA Exam Protocols }\label{sec:wataexam}
The acronym WATA stands for Written Authenticated Though Anonymous exams, and refer to a family of exams originally developed at the University of Catania. 
Historically, WATA exams have two main goals. The first goal is to mechanise the double envelope technique in a software. The second goal is to ensure authentication and anonymity despite a corrupted examiner.

The first two versions of the system are conceptually identical and only differ in the implementations: WATA I was written in Visual Basic and was only available for Microsoft Windows; WATA II \citeltex{BCR10} was implemented in Java, hence more efficient and portable. In this chapter, we only consider the second version.

 WATA III \citeltex{BCC+11} redesigns completely the exam system to offer remote management and remote notification, features not available in the previous versions. However, the new design    introduces a TTP that participates in all the phases of the exam. We first propose WATA IV, a new version  that  minimises the involvement of such TTP, and then we show how to remove it completely from the design.
Every version considers candidates free of long-term public keys, and contemplates either traditional testing or computer-based exams, but not remote testing. \index{remote testing} \index{remote notification} 

The WATA exams were originally incepted as software rather than protocols. In the first two versions, the software ran locally into the examiner's computer, while in WATA III the software ran into a remote machine.
We provide a different prospective of the WATA exams by originally describing them as protocols.

\subsubsection*{Notification Request Authentication}
The WATA exam was originally conceived for university exams,  and in some universities nowadays candidates can take the exam up to a fixed number of times. However, if the candidate withdraws, it is not counted towards the number of attempts. Other universities have a policy that does not allow the candidate to resit a failed exam the next session, unless the candidate withdraws from the exam before notification. Thus, WATA exams consider the additional requirement of \emph{Notification Request Authentication}. It says that a mark should be associated with the candidate only if she requests to learn her mark.

To formalise this requirement in the applied $\pi$-calculus, we need to define two new events that extend the list proposed in chapter \ref{chap:formal}.
\begin{itemize}
\item \texttt{requested}$\langle \idcand, \idform \rangle$ means that the candidate $\idcand$ accepts to learn the mark associated to the test $\idform$. The event is inserted into the process of the candidate at the location where the request is sent to the notifier.
\item \texttt{stored}$\langle \idcand, \mrk \rangle$ means that the authority officially considers the candidate $\idcand$ associated with $\mrk$. The event is inserted into the process of the authority at the location where it registers the mark to the candidate.
\end{itemize}

The requirement can be specified as follows:

\begin{prop}[\bf Notification Request Authentication] 
An exam protocol ensures \emph{Notification Request Authentication} if for every exam process $\EP$
\begin{center}
 $\storedtt{\idcand}{\mrk}~\leadsto~\mathit{inj}\requestedtt{\idcand}{\idform}$ 
\end{center}
on every execution trace.
\end{prop}

We use the specification above to formally analyse the protocol described in Section \ref{sec:watav}.

\subsection{WATA II} 
\WTWO considers the roles of examiner, invigilator, and candidate. 
The examiner, in addition to the usual tasks assigned to its role, runs tasks normally ascribed to other authorities, such as the question committee, the recorder and the notifier. The invigilator distributes the tests to the candidates, and collects them at the end of testing.
Every phase of the exam is executed locally, and testing takes place traditionally by pen and paper.

The examiner maintains data in three tables:
the history table $\mathit{DB}_h$ records the performances of the candidate over the past exam;
the mark table $\mathit{DB}_m$ stores the mark assigned to each test;
The question table $\mathit{DB}_q$ stores the questions.

\setlength{\abovecaptionskip}{25pt}
\setlength{\belowcaptionskip}{0pt}
\begin{figure}
\centering
\setlength{\unitlength}{3.1cm}
\begin{picture}(3,4)


\put(0,4.9){\line(0,-1){.25}}
\put(0,3.91){\line(0,-1){2.83}}
\put(0,0.88){\line(0,-1){.39}}
\put(0,0.17){\line(0,-1){.39}}

\put(1.5,4.9){\line(0,-1){.18}}
\put(1.5,3.82){\line(0,-1){.54}}
\put(1.5,2.94){\line(0,-1){.25}}
\put(1.5,2.23){\line(0,-1){.22}}
\put(1.5,1.58){\line(0,-1){.7}}

\put(3,4.9){\line(0,-1){1.675}}
\put(3,3.03){\line(0,-1){1.062}}
\put(3,1.65){\line(0,-1){1.87}}

\put(0,5){\makebox(0,0){Examiner}}
\put(1.5,5){\makebox(0,0){Invigilator}}
\put(3,5){\makebox(0,0){Candidate}}

\thicklines
\multiput(-0.5,4.8)(0.05,0){80}{\circle*{0.02}}
\put(-0.5,4.06){\rotatebox[origin=l]{90}{\makebox{\texttt{Preparation}}}}

\multiput(-0.5,3.57)(0.05,0){80}{\circle*{0.02}}
\put(-0.5,3.1){\rotatebox[origin=l]{90}{\makebox{\texttt{Testing}}}}
\put(-0.35,4.25){\framebox{
\parbox{9.9cm}{
$\mathit{quest} \leftarrow_{\mathcal{R}} \mathit{DB}_q $\\
$\mathit{id\_test}\leftarrow_{\mathcal{R}} \Sigma^{n}$\\
$\mathit{DB}_m  \leftarrow \langle \mathit{id\_test}, $\textvisiblespace$ \rangle$ \\
$\texttt{test} \leftarrow $ print\_aside$(\mathit{id\_test}, \mathit{auth\_form}, \mathit{id\_test}, \mathit{quest}, \mathit{answ\_form})$ \\
$\texttt{test}_{\scriptsize\texttt{signed}} \leftarrow $ bio\_sign$(M, \texttt{test})$
}}}

\put(0,3.72){\vector(1,0){1.5}}
\put(0.53,3.78){\makebox(0,0){1: pile\_of$(\texttt{test}_{\scriptsize\texttt{signed}})$}}

\put(1.5,3.35){\vector(1,0){0.6}}
\put(2.5,3.35){\vector(1,0){0.5}}
\put(1.85,3.43){\makebox(0,0){2: $\texttt{test}_{\texttt{signed}}$}}
\put(2.3,3.36){\makebox(0,0){random}}

\put(1.15,3.1){\framebox{
\parbox{6.8cm}{
$\texttt{test}_{\scriptsize\texttt{auth\_filled}} \leftarrow  $ fill\_auth\_form$(\texttt{test}_{\scriptsize\texttt{signed}} )$ 
}}}

\put(3,2.8){\vector(-1,0){1.5}}
\put(2.27,2.85){\makebox(0,0){\tiny 3:combine$(\texttt{id\_doc},$ hide\_id\_test$( \texttt{test}_{\scriptsize\texttt{auth\_filled}}))$}}

\put(0.4,2.43){\framebox{
\parbox{6.7cm}{
check\_validity$(\texttt{id\_doc})$ 
\\
check$(\texttt{id\_doc},reg\_list)$ 
\\
check$(\texttt{id\_doc},$ hide\_id\_test$( \texttt{test}_{\scriptsize\texttt{auth\_filled}})))$ 
}}}

\put(1.5,2.05){\vector(1,0){1.5}}
\put(2.27,2.11){\makebox(0,0){\tiny 4:combine$(\texttt{id\_doc},$ hide\_id\_test$( \texttt{test}_{\scriptsize\texttt{auth\_filled}}))$}}

\put(1.22,1.78){\framebox{
\parbox{6.5cm}{
$\texttt{test}_{\scriptsize\texttt{filled}} \leftarrow $fill\_answ\_form$(\texttt{test}_{\scriptsize\texttt{auth\_filled}})$ \\
$[\texttt{token}, \texttt{a\_test}] \leftarrow $ split $(\texttt{test}_{\scriptsize\texttt{filled}})$
}}}

\put(2.13,1.48){\vector(-1,0){0.63}}
\put(3,1.48){\vector(-1,0){0.5}}
\put(2.75,1.55){\makebox(0,0){5: $\texttt{a\_test}$}}
\put(2.32,1.5){\makebox(0,0){random}}

\multiput(-0.5,1.38)(0.05,0){80}{\circle*{0.02}}
\put(-0.5,0.9){\rotatebox[origin=l]{90}{\makebox{\texttt{Marking}}}}

\put(1.5,1.2){\vector(-1,0){1.5}}
\put(1.05,1.25){\makebox(0,0){6: pile\_of$(\texttt{a\_test})$}}

\put(-0.35,0.95){\framebox{
\parbox{4cm}{
$\mathit{DB}_m \leftarrow_{\mathit{id\_test}} \langle \Ldot,  mark \rangle$
}}}

\put(-0.5,-0.05){\rotatebox[origin=l]{90}{\makebox{\texttt{Notification}}}}
\multiput(-0.5,.75)(0.05,0){80}{\circle*{0.02}}

\put(3,0.6){\vector(-1,0){3}}
\put(2.77,0.65){\makebox(0,0){7: $\texttt{token}$}}

\put(-0.35,.3){\framebox{
\parbox{4.5cm}{
$\langle \Ldot,  mark \rangle \leftarrow_{\mathit{id\_test}} \mathit{DB}_m$ \\
$\mathit{DB}_h \leftarrow_{id} \langle \Ldot, \mathit{course}, \mathit{mark} \rangle$
}}}

\put(0,-0.06){\vector(1,0){3}}
\put(.22,0){\makebox(0,0){8: $\mathit{mark}$}}

\end{picture}
\caption{The \WTWO exam protocol}\label{fig:wata}
\end{figure}
\setlength{\abovecaptionskip}{10pt}
\setlength{\belowcaptionskip}{0pt}

In the following, we describe the protocol and refer to the message sequence chart depicted in Figure \ref{fig:wata}.

\begin{figure}[t]
\centering\includegraphics[scale=.25]{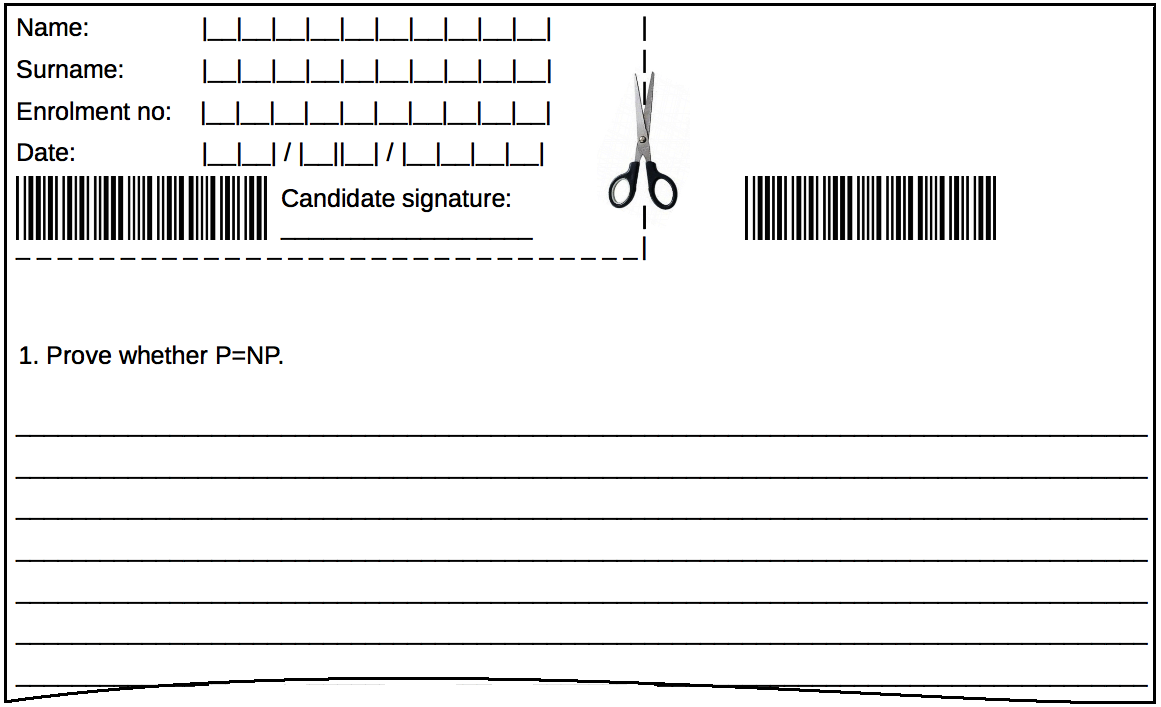}
\caption{Fragment of a test in WATA II}
\label{fig:test}
\end{figure}

\subsubsection{Preparation}
The examiner randomly extracts a list of questions $\mathit{quest}$ from the question table, and generates a random test identifier $\mathit{id\_test}$ of predetermined size $n$ using the alphabet $\Sigma$.  The test identifier is stored in the mark table next to an empty mark. The examiner then prints out the $\mathtt{test}$, which contains the following information: the test identifier \emph{id\_test}, an authentication form  $\mathit{\textit{auth\_form}}$, another occurrence of the test identifier, the questions, and a form for the answers $\mathit{\textit{answ\_form}}$. To facilitate the mechanical reading, both occurrences of the test identifier are encoded as a barcode. The test has a precise layout, notably with test identifier and authentication form framed at the top-left corner of the sheet through a dotted line; this can be seen in Figure \ref{fig:test}.
The examiner signs inside this frame across test identifier and authentication form and possibly reinforces the signature with the stamp of the exam organisation. It is assumed that this association is tamper-proof. This produces $\texttt{test}_{\scriptsize\texttt{signed}}$, which the examiner hands to the invigilator (step 1). 
This phase is repeated as many times as the number of registered candidates, so that the invigilator gets a pile of tests pre-signed by the examiner.

\subsubsection{Testing}
The invigilator leaves the pile of tests on a desk at the exam venue. Then, the candidate picks a test randomly (step~2), and is assigned a seat.
At her seat, the candidate fills out the authentication form with her personal details,  and temporarily  hides the test identifier prior to hand the test combined with her identity document to the invigilator for authentication (step~3). 
The hiding could be achieved, for example, by folding the top corners in. In so doing, the invigilator cannot learn the test identifier while authenticating the candidate.
 The invigilator checks whether the identity document is the candidate's valid one, whether the candidate identity matches an entry in the list of registered candidates and the details written on the authentication form. If so, the invigilator hands identity document and test back to the candidate (step~4), who can fill out the answer section. 

When the testing time is over, the candidate tears the test in two pieces of papers of different sizes. The smaller one, which we term $\mathit{token}$,  contains the filled authentication form and the test identifier. The larger piece of paper, which we term $\mathit{a\_test}$,  contains  questions, test identifier, and answers. The candidate keeps the $\mathit{token}$, and leaves the anonymous test in a random position through the current pile of tests (step~5).

\subsubsection{Marking}
The invigilator collects the pile of anonymous tests and distributes them to the examiners (step~6). It also removes the records of the mark table that refer to undistributed tests. 

The examiner evaluates the answers and assigns a mark to the anonymous test. The examiner then scans the barcode to get the corresponding test identifier, and enters the mark in the mark table, precisely in the record identified by the test identifier.

\subsubsection{Notification} 
The candidate who wants to know her mark brings her token at the venue announced to host the notification.
There the candidate hands the token to the examiner (step~7), who checks signature and personal details, and scans the barcode to get the corresponding test identifier. 
The examiner finds the record identified by the test identifier on the mark table, and obtains the corresponding mark. Finally, the examiner stores the mark into the history table, and notifies the mark to the candidate (step~8).

\subsubsection*{Discussion}
Except for the preparation of the tests, the presence of computers in \WTWO is minimal. Most of the tasks are run by humans, and the security of the protocol mostly relies on the physical properties of paper.

We consider the requirements proposed in chapter \ref{chap:formal}. It can be observed that \WTWO 
trivially ensures Test Authenticity  but not Anonymous Examiner because the protocol considers only one examiner. \index{Test Authenticity}
Candidate Authorisation is met because the invigilator authorises the candidate to take the exam only if the personal details reported on the identity document match an entry in the list of eligible candidates. \index{eligible candidates} \index{Candidate Authorisation} \index{Anonymous Examiner}
Also Answer Authenticity is met because the invigilator verifies that the candidate wrote the personal details on the authentication form. Moreover, the examiner's signature on the tests ensures their authenticity. It follows that \WTWO also ensures Test Origin Authentication. \index{Answer Authenticity} \index{Test Origin Authentication}
Mark Authenticity is met because the examiner inserts the mark into the mark table, exactly in the record identified by the random test identifier reported into the test. At notification, the examiner notifies the candidate with this mark, which is also stored in the history table. However, a malicious examiner may tell a different mark after the candidate hands him the token.
The novel requirement of Notification Request Authentication is met because only the candidate holds the token. A malicious examiner cannot generate a forged token because it would need to be signed by the candidate. In fact, the candidate is the sole entity who can establish the link between her identity and her test.

Concerning privacy requirements, \WTWO guarantees Question Indistinguishability provided that the candidate does not collude either with the examiner or the invigilator. In fact, the examiner hands the tests to the invigilator prior testing. However, if the questions that appear into a test are randomly chosen, it becomes harder for a candidate to learn which questions she will be assigned.
Anonymous Marking is met because only the candidate knows the test identifier associated to her. Moreover, the candidate submits the test in a random position of the pile of anonymous tests. \index{Mark Privacy}
\WTWO ensures Mark Privacy because the notification is face-to-face. The examiner notifies the mark to the corresponding candidate after a successful authentication, and only if she hands a valid token. Since Mark Privacy is met, it follows that also Mark Anonymity is met. \index{Anonymous Marking} \index{Mark Anonymity}

Although \WTWO ensures authentication and privacy without TTP, it has the major limitation that notification requires candidates to meet in person the examiner.

\subsection{WATA III}

\WWW allows remote notification and involves a major level of computer assistance than the previous version. The protocol considers the participation of the \emph{WATA Server} in addition to candidate, examiner, and invigilator roles. The WATA Server runs most of the tasks of preparation and notification, while the tasks of the examiner are now limited to the marking phase.  \index{remote notification}

The WATA Server maintains data in four tables.  The history, mark, and question tables have the same functionalities of the previous protocol. \WWW introduces  the \emph{sharec} table $DB_{c}$ that stores the partial information about the test identifier, which now is called \emph{pseudonym}.

The pseudonym is associated also to the candidate rather than only to the test. The idea is to split the pseudonym in two shares, print them on the test, and give one to the candidate and the other to the examiner. The latter can associate a test to its author only if the candidate reveals its share.


\setlength{\abovecaptionskip}{25pt}
\setlength{\belowcaptionskip}{0pt}
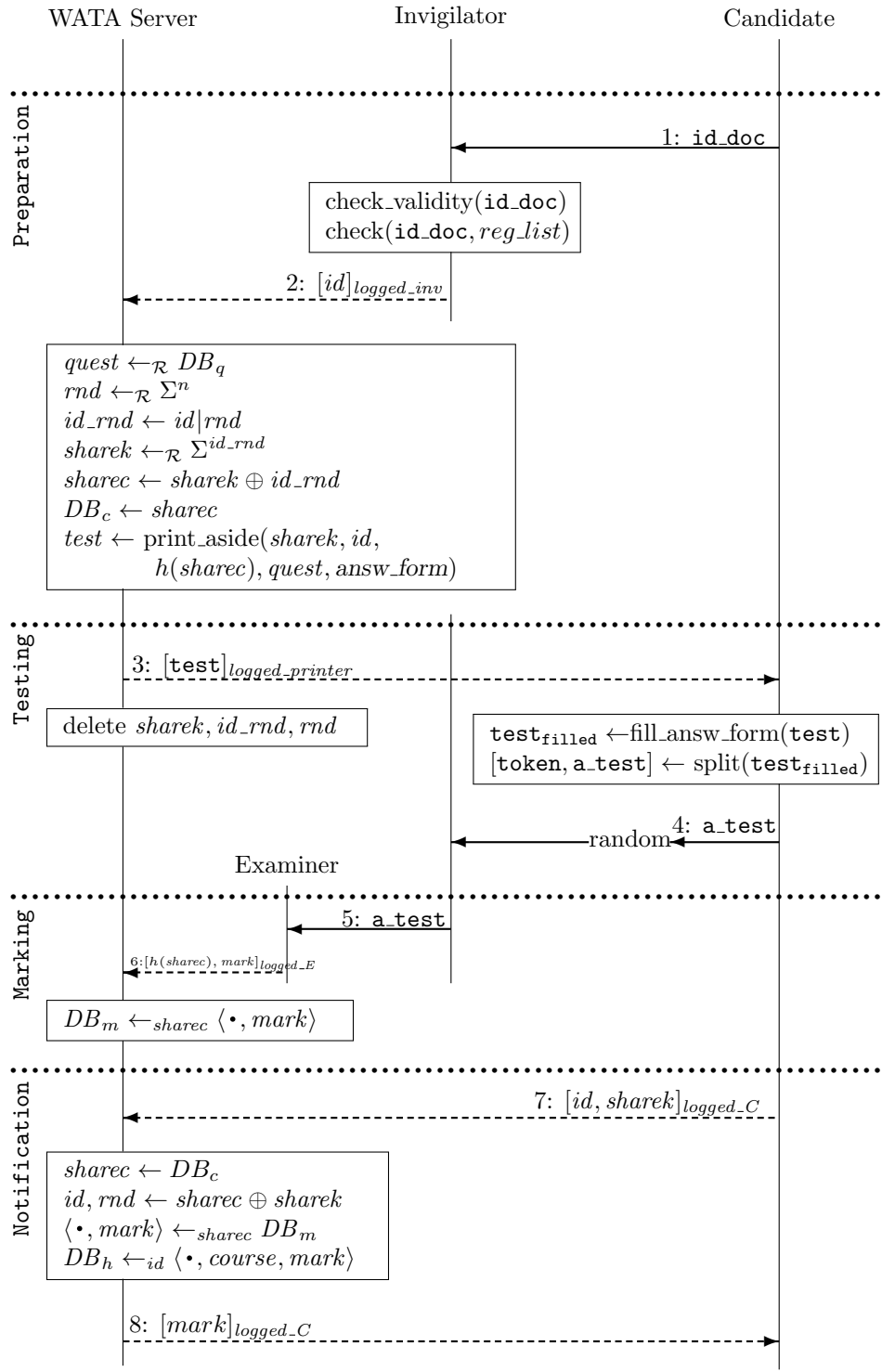
\begin{figure}
\centering
\setlength{\unitlength}{3.1cm}
\begin{picture}(3,6)


\put(0,5.9){\line(0,-1){1.41}}
\put(0,3.369){\line(0,-1){.55}}
\put(0,2.64){\line(0,-1){1.17}}
\put(0,1.285){\line(0,-1){0.51}}
\put(0,.188){\line(0,-1){0.4}}

\put(0.75,2){\line(0,-1){0.45}}

\put(1.5,5.9){\line(0,-1){0.66}}
\put(1.5,4.92){\line(0,-1){0.32}}
\put(1.5,3.25){\line(0,-1){1.7}}

\put(3,5.9){\line(0,-1){3.11}}
\put(3,2.47){\line(0,-1){2.7}}

\put(0,6){\makebox(0,0){WATA Server}}
\put(1.5,6){\makebox(0,0){Invigilator}}
\put(0.75,2.1){\makebox(0,0){Examiner}}
\put(3,6){\makebox(0,0){Candidate}}

\thicklines
\multiput(-0.5,5.65)(0.05,0){80}{\circle*{0.02}}
\put(-0.5,4.92){\rotatebox[origin=l]{90}{\makebox{\texttt{Preparation}}}}
\multiput(-0.5,3.2)(0.05,0){80}{\circle*{0.02}}
\put(-0.5,2.72){\rotatebox[origin=l]{90}{\makebox{\texttt{Testing}}}}

\put(3,5.4){\vector(-1,0){1.5}}
\put(2.7,5.45){\makebox(0,0){1: $\texttt{id\_doc}$}}

\put(0.85,5.05){\framebox{
\parbox{3.5cm}{
check\_validity$(\texttt{id\_doc})$ 
\\
check$(\texttt{id\_doc},reg\_list)$ 
}}}

\multiput(0,4.7)(0.05,0){30}{\line(1,0){0.03}}
\put(0.05,4.7){\vector(-1,0){0.05}}
\put(1.09,4.77){\makebox(0,0){ 2: $[\mathit{id}]_{\mathit{logged\_inv}}$}}

\put(-0.35,3.9){\framebox{
\parbox{6.3cm}{
$\mathit{quest} \leftarrow_{\mathcal{R}} \mathit{DB}_q $\\
$\mathit{rnd}\leftarrow_{\mathcal{R}} \Sigma^{n}$\\
$\mathit{id\_rnd}\leftarrow \mathit{id}|\mathit{rnd}$\\
$\mathit{sharek}\leftarrow_{\mathcal{R}} \Sigma^{\mathit{id\_rnd}}$\\
$\mathit{sharec}\leftarrow \mathit{sharek} \oplus \mathit{id\_rnd}$\\
$\mathit{DB}_{c}  \leftarrow \mathit{sharec}$ \\
$\mathit{test} \leftarrow $ print\_aside$(\mathit{sharek, id}, $\\
\hspace*{1.3cm}$h(\mathit{sharec}), \mathit{quest}, \textsl{answ\_form})$ 
}}}

\multiput(0,2.95)(0.05,0){60}{\line(1,0){0.03}}
\put(2.95,2.95){\vector(1,0){0.05}}
\put(.55,3.01){\makebox(0,0){3: $[\texttt{test}]_\mathit{logged\_printer}$}}

\put(-0.35,2.7){\framebox{
\parbox{4.2cm}{
delete $\mathit{sharek, id\_rnd, rnd}$
}}}




\put(1.6,2.6){\framebox{
\parbox{5.4cm}{
$\texttt{test}_{\scriptsize\texttt{filled}} \leftarrow $fill\_answ\_form$(\texttt{test})$ \\
$[\texttt{token}, \texttt{a\_test}] \leftarrow $ split$(\texttt{test}_{\scriptsize\texttt{filled}})$ 
}}}

\put(2.13,2.2){\vector(-1,0){0.63}}
\put(3,2.2){\vector(-1,0){0.5}}
\put(2.75,2.28){\makebox(0,0){4: $\texttt{a\_test}$}}
\put(2.32,2.22){\makebox(0,0){random}}

\multiput(-0.5,1.95)(0.05,0){80}{\circle*{0.02}}
\put(-0.5,1.45){\rotatebox[origin=l]{90}{\makebox{\texttt{Marking}}}}

\put(1.5,1.8){\vector(-1,0){0.75}}
\put(1.22,1.85){\makebox(0,0){ 5: $\texttt{a\_test}$}}

\multiput(0,1.6)(0.05,0){15}{\line(1,0){0.03}}
\put(0.05,1.6){\vector(-1,0){0.05}}
\put(0.46,1.64){\makebox(0,0){\tiny 6:$[h(\mathit{sharec}), \mathit{mark}]_{\mathit{logged\_E}}$}}


\put(-0.35,1.35){\framebox{
\parbox{4cm}{
$\mathit{DB}_m \leftarrow_{\mathit{sharec}} \langle \Ldot,  \mathit{mark} \rangle$
}}}

\put(-0.5,0.35){\rotatebox[origin=l]{90}{\makebox{\texttt{Notification}}}}
\multiput(-0.5,1.15)(0.05,0){80}{\circle*{0.02}}

\multiput(0,0.93)(0.05,0){60}{\line(1,0){0.03}}
\put(0.05,0.93){\vector(-1,0){0.05}}
\put(2.4,1){\makebox(0,0){7: $[\mathit{id, sharek}]_{\mathit{logged\_C}}$}}

\put(-0.35,0.45){\framebox{
\parbox{4.5cm}{
$\mathit{sharec} \leftarrow DB_{c}$ \\
$\mathit{id,rnd} \leftarrow \mathit{sharec} \oplus \mathit{sharek}$ \\
$\langle \Ldot,  \mathit{mark} \rangle \leftarrow_{\mathit{sharec}} \mathit{DB}_m$ \\
$\mathit{DB}_h \leftarrow_{\mathit{id}} \langle \Ldot, \mathit{course, mark} \rangle$
}}}

\multiput(0,-0.1)(0.05,0){60}{\line(1,0){0.03}}
\put(2.95,-0.1){\vector(1,0){0.05}}
\put(.45,-0.03){\makebox(0,0){8: $[mark]_\mathit{logged\_C}$}}

\end{picture}
\caption{The WATA III exam protocol}\label{fig:wata3}
\end{figure}
\setlength{\abovecaptionskip}{10pt}
\setlength{\belowcaptionskip}{0pt}

\WWW assumes that a list of registered candidates for the exam is available to the invigilator, and secure  TLS communications between the WATA Server and the other principals. The WATA Server authenticates invigilator and candidate via login and password.
Every communication between invigilator and candidate is face-to-face, while communications to the WATA Server are always remote. Remote communications are highlighted with dashed lines in the message sequence chart in Figure \ref{fig:wata3}.

\paragraph{Remark. } We found a security issue in the original specification of WATA III \citeltex{BCC+11}.
In a nutshell, the specification allowed a corrupted candidate to be assigned with the mark of the test submitted by another candidate, hence violating Mark Authenticity. The corrupted candidate could generate a fake pseudonym such that at notification she could convince the examiner that the pseudonym should associated with her identity. \index{Mark Authenticity}
This was possible because the original specification contemplates a weak generation of the pseudonym that reveals key information. 
We fix this issue by hide such information via hashing.
In the remainder, we only describe the fixed version.

\subsubsection{Preparation}
At exam venue, the candidate approaches the invigilator's desk and hands her identity document \texttt{id\_doc} (step~1). The invigilator checks whether the personal details of the candidate \emph{id} appears in the registered candidate list, and if so, logs in the WATA Server and sends  \emph{id} via secure channel (step~2). 
The WATA Server randomly  extracts the questions from the question table. Then, it generates a random value \emph{sharek}, whose length matches the \emph{id} augmented with some randomness \emph{id\_rnd}. Thus, it generates (\emph{sharec}) that is the result of the one-time pad of \emph{sharek} with \emph{id\_rnd}. The WATA Server stores \emph{sharec} in the database and finally generates the layout of the test.  The \emph{sharek} is placed on the top left of the printout, while the hashed version of \emph{sharec} is placed on the top right of the  printout, close to the answer section \emph{answ\_form}. Both \emph{sharek} and  hashed \emph{sharec} are represented in the form of barcode.
The test thus consists of two parts: the \emph{token}, which contains \emph{sharek} and the candidate's personal details \emph{id}; the anonymous test \emph{a\_test}, which contains the hash of \emph{sharec}, questions, and answer section.

\subsubsection{Testing}
The WATA Server sends the test via a secure channel to a printer, which is available at the exam venue.
The candidate approaches the printer and takes its \texttt{test}  (step~3).
Then, the WATA Server deletes all the data it used but the one stored in the tables, namely it removes \emph{sharek}, \emph{id\_rnd}, and the randomness. 

At her sit, the candidate fills out the answer section with the answers. When the time is over, the candidate splits the test, and takes the \texttt{token} at home while inserts the anonymous test \texttt{a\_test} into a random position of the pile of anonymous tests (step~4).

\subsubsection{Marking}
The invigilator collects the pile of anonymous tests and hands them to the examiner  (step~5), who
evaluates the answers and assigns a mark to each anonymous test. 
For each test, the examiner scans the barcode and gets the corresponding hash of \emph{sharec}. The examiner then logs into WATA Server and uploads the pair of hashed \emph{sharec} and mark via a secure channel (step~6).
The WATA Server can find the corresponding \emph{sharec} by hashing each entry of $DB_{c}$. In so doing,  it can store the mark in the entry identified by \emph{sharec} in the mark table.

\subsubsection{Notification} 
The candidate who wants to know her mark scans the barcode printed in the token and gets  \emph{sharek}, which she sends to the WATA Server via a secure channel, after she logged in (step~7).
The WATA Server XOR-es \emph{sharek} with each \emph{sharec} stored in the database until it decrypts a valid \emph{id} concatenated with some randomness. It then retrieves the record identified by the \emph{sharec} from the mark table, and obtains the corresponding mark. Finally, the WATA Server stores the mark into the history table, and notifies it to the candidate (step~8).

\subsubsection*{Discussion} 


It is clear that \WWW cannot be deemed secure assuming a corrupted WATA Server. This role is  ubiquitous in the design and is in charge of the critical steps.
Therefore, the WATA Server should be considered as an honest-but-curious role, which follows the protocol honestly but tries to learn as much as possible.

A critical part of the protocol is the deletion of data performed by the WATA Server.
Although this practice is found in other protocols \citeltex{EKO+14}, it may be impractical to force a party to delete data. However, it can be still possible to verify that the party actually deletes the data \citeltex{HCZ15}.

Concerning the authentication requirements, \WWW ensures Candidate Authorisation since the invigilator authorises the candidate to take the exam only if the candidate's personal  details reported on the valid identity document match an entry in the list of eligible candidates. \index{eligible candidates} The invigilator has to ensure that the correct candidate takes the test generated for her from the printer. This  avoids that corrupted candidates swap their tests before sitting for the exam.
In fact, the protocol does not require that the candidate writes her personal details down into the test as provided for the previous version. Thus, the invigilator does not need to check the test once the candidate sits for the exam.
Answer Authenticity is met because candidates are invigilated. If a corrupted candidate introduces an illegal test, she will not be able to receive a mark because the examiner would upload a forged hash of \emph{sharec} that the WATA Server could not retrieve in the database.
It follows that \WWW ensures Test Origin Authentication.
Similarly to WATA II, Test Authenticity is met but not Anonymous Examiner since the protocol considers only one examiner. 
Mark Authenticity is met because both candidate and examiner know only the hash of \emph{sharec} until notification. \WWW ensures Notification request authentication because only the candidate knows the \emph{sharek} after testing.

Concerning privacy requirements, Question Indistinguishability is met because the WATA Server generates the test with the questions. Each test is printed at testing and taken by the candidate directly, so the invigilator learns the question only when testing concludes. \index{Question Indistinguishability}
Although the WATA Server deletes some data at testing, we observe that it can violate Anonymous Marking. At preparation, a corrupted WATA Server can associate the candidate \emph{id} 
with the corresponding \emph{sharec}. At testing, the WATA Server receives from the examiner the mark associated with the hash of the \emph{sharec}. Therefore, the WATA Server can learn the author of a test without the knowledge of \emph{sharek}. Note that this attack is possible even considering an honest-but-curious threat model because it does not require that the WATA Server deviates from the protocol, but resorts solely on the knowledge of the WATA Server.
Finally, Mark Privacy is met because the WATA Server notifies the mark to the candidate only, after  receiving  a correct \emph{sharek}. It follows that \WWW ensures  Mark Anonymity as well.

WATA III allows for remote notification but assigns most of its critical tasks to a TTP. It turns out that considering an honest-but-curious WATA Server, WATA III fails to ensure Anonymous Marking.
\index{remote notification}


\section{WATA IV}\label{sec:wataiv}
In this section, we advance WATA IV, a new exam protocol that overcomes the limitations of WATA III. We design the protocol without the ubiquitous WATA Server and introduce the \emph{anonymiser} role, whose participation is confined to preparation only. Similar to the WATA Server, the anonymiser is honest-but-curious but its duties are drastically reduced. In WATA IV most of the critical tasks are run by a possibly corrupted examiner. Moreover, WATA IV opens for remote preparation and requires no printers at testing.
We anticipate that WATA IV meets the same security requirements as WATA III does, though augmented with Anonymous Marking and the two individual verifiability requirements of Mark Integrity and Mark Notification Integrity.

The main novelty of the design of  WATA IV is the use visual cryptography.
The idea consists of encoding the pseudonym into two visual cryptographic shares: one share is given to the candidate, and the other is given to the examiner. Neither the candidate nor the examiner knows the pseudonym until they meet at testing, when the candidate learns it by overlapping the examiner's share with hers.

Thanks to visual cryptography, the candidate can do a cryptographic operation at testing without the assistance of any computer device. In the following, we briefly discuss visual cryptography and commitment scheme, namely the  cryptographic primitives used in WATA IV.









\begin{figure}
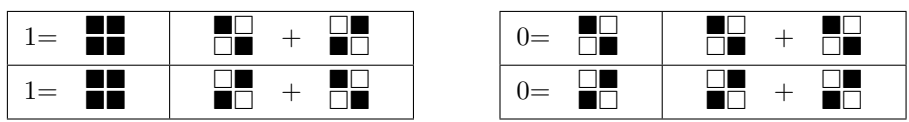

\begin{center}
{
\begin{tabular}{c c}
\begin{tabular}{|c|c|}
\hline 1=$\onev$ & $\zerova$+$\zerovb$\\ \hline
1=$\onev$ & $\zerovb$+$\zerova$\\ \hline
\end{tabular}
&~~~~~
\begin{tabular}{|c|c|}
\hline 0=$\zerova$ & $\zerova$+$\zerova$\\ \hline
0=$\zerovb$ & $\zerovb$+$\zerovb$\\ \hline
\end{tabular}
\end{tabular}
}
\caption{Representation of bits $0$ and $1$ using visual cryptography}
\label{fig:visualcrypto}
\end{center}
\end{figure}

\subsubsection{Visual Cryptography}
It is a secret sharing scheme, devised by Naor and Shamir \citeltex{Naor94}, that allows a visual decryption of a ciphertext.  A secret image is ``encrypted'' by splitting it into a number of image \emph{shares}.
The basic version of the scheme is the 2-out-of-2 secret sharing system, in which a secret image is split into two shares $\share_{A}$ and $\share_{B}$. The shares are printed on transparency sheets, which reveal the secret image when the shares are overlapped. This scheme is information-theoretic secure, namely each share leaks no information about the secret image.
In fact, it emulates the XOR operation though the visual decryption is actually equivalent to the OR operation. The scheme is information-theoretic secure because either a black or a white pixel, mapped respectively to 0 and 1, can originate by any of the sub-pixels shown in Figure \ref{fig:visualcrypto}.

Many schemes for visual cryptography have been proposed over the years. Although we consider the
Naor and Shamir scheme for WATA IV, we conjecture that other visual scheme can be used as well, but with different security guarantees.

\subsubsection{Commitment Scheme} A commitment scheme is used to bind a committer to a
secret value. The committer publishes a commitment that hides the value, which remains secret until he reveals it. Should the committer reveal a different value, this would be noticed because it cannot be mapped to the published commitment.
WATA IV uses the Pedersen commitment scheme \citeltex{Pedersen91}, which guarantees
unconditional hiding, namely the value remains secret despite a computational unbounded attacker.
The scheme consists of the algorithms of \emph{commitment}, in which the value is chosen, hidden, and bound to the committer, and of \emph{disclosure}, in which the value is publicly revealed. 
The commitment algorithm takes in two given public generators $g,h\in \mathbb{G}_q$, the secret value $v$, and a random value $r\in_{\mathcal{R}} \mathbb{Z}^*_{q}$. The algorithm outputs the commitment $g^vh^r$ denoted with $C_r(v)$. 
The disclosure algorithm takes in the commitment $C_r(v)$, the values $v$ and $r$, and outputs
\texttt{true} if the commitment is correct or \texttt{false} otherwise. 

WATA IV adopts the Pedersen commitment scheme at notification.
The examiner generates a commitment of the mark of the candidate. Once the candidate reveals her identity to know the mark, she can verify the examiner notifies her the committed mark. This deters the examiner to notify the candidate with a mark that is different than the one assigned to candidate's test.


\begin{figure}
\centering
\setlength{\unitlength}{3.5cm}
\begin{picture}(3,6)
\small

\put(0,5.7){\line(0,-1){1.91}}
\put(0,3.64){\line(0,-1){3.07}}
\put(0,3.64){\line(0,-1){3.07}}
\put(0,.425){\line(0,-1){.425}}

\put(1,5.7){\line(0,-1){2.745}}
\put(1,2.69){\line(0,-1){1.3}}

\put(2,5.7){\line(0,-1){0.42}}
\put(2,4.365){\line(0,-1){.5}}

\put(3,5.7){\line(0,-1){1.91}}
\put(3,3.64){\line(0,-1){1.25}}
\put(3,2.015){\line(0,-1){2.015}}

\put(0,5.8){\makebox(0,0){Examiner}}
\put(1,5.8){\makebox(0,0){Invigilator}}
\put(3,5.8){\makebox(0,0){Candidate}}
\put(2,5.8){\makebox(0,0){Anonymiser}}

\thicklines
\multiput(-0.22,5.63)(0.05,0){69}{\circle*{0.02}}
\put(-0.22,5.05){\rotatebox[origin=l]{90}{\makebox{\texttt{Preparation}}}}


\multiput(0,5.45)(0.05,0){40}{\line(1,0){0.03}}
\put(1.95,5.45){\vector(1,0){0.05}}
\put(0.35,5.52){\makebox(0,0){$1:[\mathit{id,ex}]_{\mathit{logged}\_E}$}}

\put(1.22,4.8){\framebox{
\parbox{5cm}{
$\mathit{pid} \leftarrow_{\mathcal{R}} \Sigma^{t \times u}$ \\
$\mathit{share_{A}} \leftarrow_{\mathcal{R}} \Sigma^{t \times u} $  \\
$\mathit{share_{B}} \leftarrow \mathit{pid} \oplus \mathit{share_{A}}$ \\
$\mathit{data_{A}}\leftarrow \mathit{id, ex, share_A}$\\
$\mathit{data_{B}}\leftarrow \mathit{id, ex, share_B}$\\
$\mathit{transp}\leftarrow \mathit{data_A, \sign_{\mathit{SSK_{An}}}(data_A)} $\\
$\mathit{paper}\leftarrow \mathit{data_B, \sign_{\mathit{SSK_{An}}}(data_B)} $ \\
}}}

\multiput(0.05,4.15)(0.05,0){40}{\line(-1,0){0.03}}
\put(0.05,4.15){\vector(-1,0){0.05}}
\put(1.8,4.2){\makebox(0,0){$2:\mathit{transp}$}}

\multiput(2,3.95)(0.05,0){20}{\line(1,0){0.03}}
\put(2.95,3.95){\vector(1,0){0.05}}
\put(2.2,4.0){\makebox(0,0){$3:\mathit{paper}$}}

\put(2.55,3.7){\framebox{
print $\texttt{paper}$
}}

\put(-0.1,3.7){\framebox{
print $\texttt{transp}$
}}

\multiput(-0.22,3.5)(0.05,0){69}{\circle*{0.02}}
\put(-0.22,3.1){\rotatebox[origin=l]{90}{\makebox{\texttt{Testing}}}}


\put(0.2,3.3){\makebox(0,0){$4:\texttt{transp}$}}
\put(0,3.25){\vector(1,0){1}}

\put(3,3.05){\vector(-1,0){2}}
\put(2.8,3.1){\makebox(0,0){$5:\texttt{id\_doc}$}}

\put(0.45,2.8){\framebox{
\parbox{3.5cm}{
check\_validity$(\texttt{id\_doc})$ 
\\
check$(\texttt{id\_doc},reg\_list)$ 
}}}

\put(1,2.55){\vector(1,0){2}}
\put(1.32,2.6){\makebox(0,0){$6:\texttt{transp}, \texttt{test}$}}

\put(1.55,2.18){\framebox{
\parbox{5.3cm}{
$\mathit{pid}\leftarrow$overlap$(\texttt{transp}, \texttt{paper})$ \\
$\texttt{test}_{\scriptsize\texttt{pid}} \leftarrow $fill\_pid\_form$(\texttt{test})$ \\
$\texttt{test}_{\scriptsize\texttt{filled}} \leftarrow $fill\_answ\_form$(\texttt{test}_{\scriptsize\texttt{pid}})$
}}}

\put(3,1.85){\vector(-1,0){2}}
\put(2.73,1.9){\makebox(0,0){$7:\texttt{test}_{\scriptsize\texttt{filled}}$}}

\put(0.73,1.55){\makebox(0,0){$8:\texttt{test}_{\scriptsize\texttt{filled}}$}}
\put(1,1.5){\vector(-1,0){1}}

\multiput(-0.22,1.67)(0.05,0){69}{\circle*{0.02}}
\put(-0.22,1.27){\rotatebox[origin=l]{90}{\makebox{\texttt{Marking}}}}

\multiput(0,1.20)(0.05,0){40}{\line(1,0){0.03}}
\put(1.96,1.20){\vector(1,0){0.05}}
\put(0.65,1.26){\makebox(0,0){$9:\sign_{\mathit{SSK_E}}(pid,C_r(\mrk))$}}

\put(2.02,1.17){\framebox{$\mathcal{BB}$}}

\multiput(-0.22,1)(0.05,0){69}{\circle*{0.02}}
\put(-0.22,0.35){\rotatebox[origin=l]{90}{\makebox{\texttt{Notification}}}}

\multiput(0,.78)(0.05,0){60}{\line(1,0){0.03}}
\put(0.05,.78){\vector(-1,0){0.05}}
\put(2.38,.85){\makebox(0,0){$10:[\sign_{\mathit{SSK_{An}}}(data_B)]_{\mathit{logged_C}}$}}

\put(-0.1,0.475){\framebox{
$\mathit{pid} \leftarrow \mathit{share_{A}} \oplus \mathit{share_{B}}$
}}

\multiput(0,.18)(0.05,0){60}{\line(1,0){0.03}}
\put(2.95,.18){\vector(1,0){0.05}}

\put(1.01,0.25){\makebox(0,0){$11:[\sign_{\mathit{SSK_E}}(pid,mark,C_r(\mrk),r)]_{\mathit{logged_C}}$}}
\end{picture}

\caption{The \WW exam protocol}\label{fig:wataIV}
\end{figure}

\subsection{Description}\label{sec:wataIV}
WATA IV relies on a append-only bulletin board in which the examiner publishes the pseudonyms and 
the commitment of the marks. Each session is identified with a unique exam code $\mathit{ex}$.
In the following we describe the protocol and refer to the message sequence chart depicted in Figure \ref{fig:wataIV}.

\begin{figure}
\centering
\begin{minipage}{1\textwidth}
\centering \fbox{\includegraphics[width=0.97\textwidth]{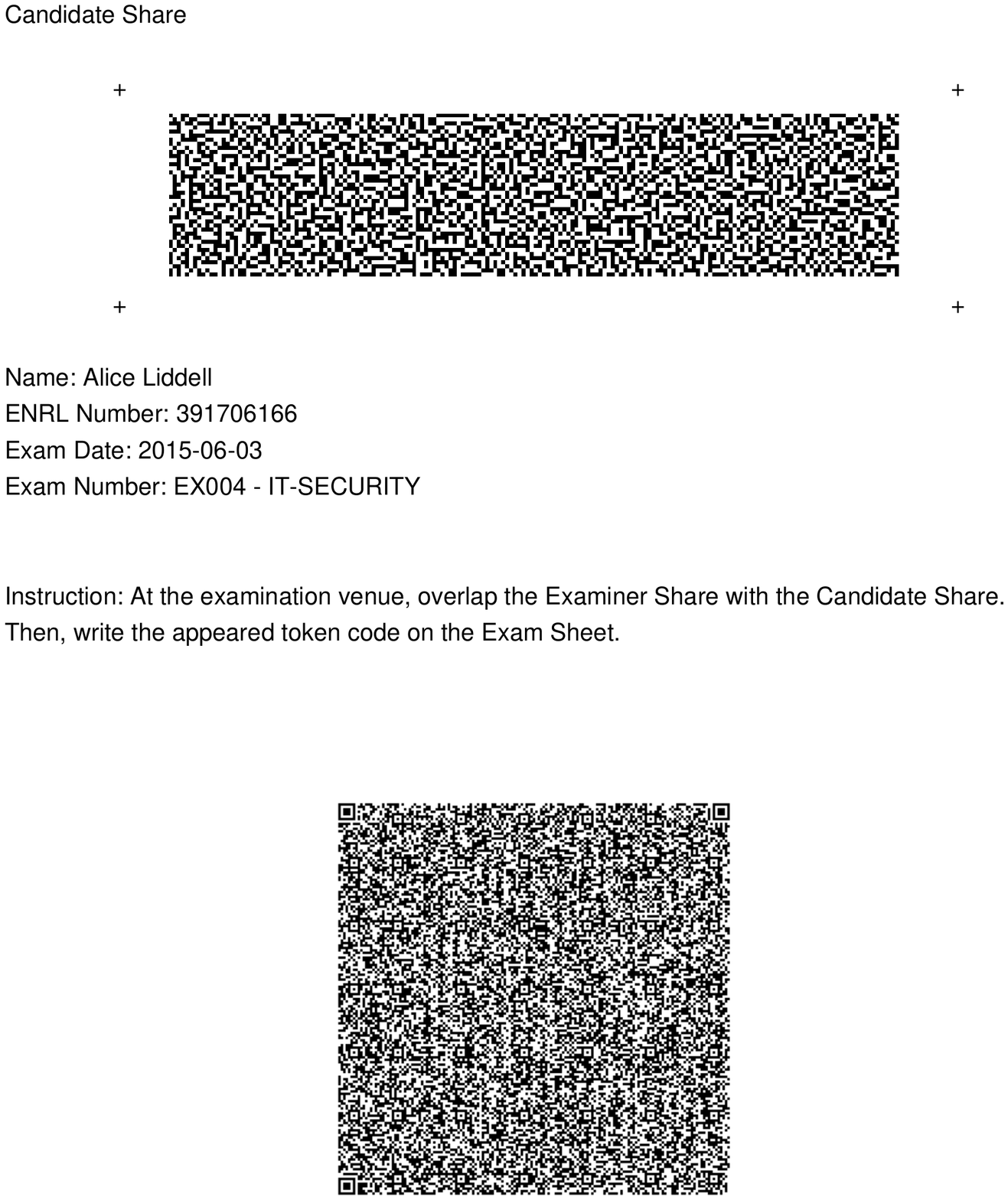}}
\caption{The candidate paper sheet in WATA IV}\label{fig:qrc}
\end{minipage}
\end{figure}


\subsubsection{Preparation}
The examiner checks the candidate eligible for the exam \emph{ex} and, if so, enters the candidate details \emph{id} in the dedicated list \emph{reg\_list}. After that, the examiner sends the candidate's details and the exam code to the anonymiser via a secure channel (step~1).

The anonymiser generates the pseudonym $\mathit{pid}$ that consists of a visual representation of a random alphanumeric string, and  a random visual cryptographic image, $\share_{A}$.
Then, the anonymiser generates the second visual cryptographic image, $\mathit{share}_{B}$, such that overlapping  $\share_{A}$ and $\share_{B}$ results in the image representing of the pseudonym.
Let $\data_{A}$ denote the triplet of \emph{id}, \emph{ex} and $\mathit{share_{A}}$.
The anonymiser signs $\data_{A}$, and generates the corresponding signature as follows. First, the plaintexts   $\data_{A}$ and  $\data_{B}$ are encoded in \emph{Base64} and signed with the signing key of the anonymiser $\mathit{SSK_{An}}$. Then, the signatures are encoded in \emph{Base64} again, and included in two QR codes with the corresponding encoded plaintext. 
To facilitate the printing, the anonymiser  includes such information in the digital versions of an A4 paper sheet, respectively $\mathit{transp}$ and $\mathit{paper}$, which layout is outlined in Figure \ref{fig:qrc}. The signatures printed on the bottom of the sheets self contain the data reported on each sheet plus the corresponding signature.
The anonymiser emails $\transp$ to the examiner (step 2) and $\paper$  to the candidate (step 3). We assume that the attacker is out of control of the email infrastructure, hence he cannot learn the contents. However secure emailing techniques, such as S/MIME \citeltex{SMIME} or MIME over OpenPGP \citeltex{mimePGP}, can be used to ensure confidentiality of the content of the emails.

For each candidate, the examiner stores the signed $\data_{A}$ into the database, and prints each $\texttt{transp}$ on a transparency sheet.
Similarly, each candidate prints her $\texttt{paper}$ on a common A4 paper sheet.
 
\subsubsection{Testing} 
The examiner hands the transparency sheet to the invigilator (step 4).
The candidate takes a seat at exam venue, and hands a valid identity document \texttt{id\_doc} to the invigilator for authentication (step 5). The invigilator checks that the  candidate is in the list of those registered for the exam. Then, the invigilator finds the transparency sheet $\texttt{transp}$ that reports the candidate's details, and hands it to her along with a \texttt{test} (step 6). 
If some registered candidates fail to show up, the invigilator puts the corresponding transparency sheets and the excess tests aside.

Once the invigilator delivers the transparency sheets to all candidates, the candidate can overlap her paper sheet with the corresponding transparency sheet and can read the pseudonym. The candidate writes down the pseudonym into the test and begins to answer the questions. When the testing time is over, the candidate submits her test (step 7), and takes the paper and transparency sheets back with her. 
The candidate can  place her test anywhere in the pile of already submitted tests. The invigilator collects the pile of tests when all candidates have submitted their tests.

\subsubsection{Marking}
The invigilator hands all the tests (step 8) and the remaining transparency sheets to the  examiner. 
The examiner evaluates the  test, generates a commitment of the \emph{mark}, signs the pair of pseudonym and mark (\emph{pid}, \emph{mark}) and publishes the signature on the public append-only bulletin board (step 9).
This allows the candidate to verify whether her test has been marked though ignoring the mark.

\subsubsection{Notification}
The examiner runs notification for a fixed time frame. 
The candidate who wants to know her mark sends  the signed $\data_{B}$
to the examiner via a secure channel (step 10).
The examiner verifies the signature, and overlaps $\share_{B}$ with its $\share_{A}$
to get the pseudonym \emph{pid}. Notably, this procedure can be implemented, hence requires no human involvement. The examiner thus retrieves the mark associated with the pseudonym, and finally sends to the candidate the signature of  pseudonym, mark, commitment, and commitment random value (step 11).

\subsection{Discussion}\label{sec:analysis}
WATA IV ensures all the security requirements as WATA III does plus a few more.
Concerning authentication, Test Authenticity is trivially met while $ $ Anonymous Examiner is not because the protocol assumes only one examiner. \index{Anonymous Examiner}
Candidate Authorisation  is met because the invigilator checks that the candidate is in the list of those registered for the exam. 
Answer Authenticity  is met because the invigilator gives to the candidate the transparency sheet that has her details. 
If a corrupted candidate prints a different visual crypto image on her paper sheet, she cannot read any intelligible pseudonym by overlapping the paper sheet with the transparency sheet. The same applies if any two corrupted candidates swap their paper sheets before testing.
As we shall see later, a dispute resolution procedure guarantees that a corrupted candidate cannot even claim that no pseudonym appears because the examiner misprinted the transparency sheet.
Still, a corrupted candidate could write a random pseudonym into the test, but  at notification the candidate would not be able to send a valid $\mathit{data}_{B}$. 
Since WATA IV ensures Candidate Authorisation and Answer Authenticity, it also ensures Test Origin Authentication. \index{Test Origin Authentication}
Mark Authenticity is met because the  examiner posts on the bulletin board the signature of the pseudonym associated with a commitment of the mark.
Notification request authentication  is also met because only the candidate holds her $\share_{B}$.

Concerning privacy, Question Indistinguishability is met by the assumptions on the origin of tests, in which the examiner generates the tests. \index{Question Indistinguishability}
WATA IV guarantees Anonymous Marking  because the examiner cannot associate a test with a candidate until notification. Anonymous Marking can last forever, even if examiner and the invigilator collude, provided that the candidate chooses not to get her mark. \index{Anonymous Marking}
Mark Privacy \index{Mark Privacy} is met because the examiner notifies the mark to the candidate only if the latter sends a valid $\share_{B}$. Thus each candidate can get only their corresponding mark. 
Notably, the anonymiser cannot associate a candidate with a mark, because the examiner only publishes the commitment of the mark on the bulletin board.
It follows that also Mark Anonymity is met. \index{Mark Anonymity}

Concerning the individual verifiability requirements, Mark Integrity is met because the candidates can verify that the examiner registered the same mark whose commitment was published on the bulletin board. Similarly, Mark Notification Integrity is met because the candidate can verify that the mark notified by the examiner is the same committed on the bulletin board.

\subsubsection*{Dispute resolution}
An interesting feature of \WW is the support for dispute resolution during testing. In fact, the combination of signatures and visual cryptography guarantees an easy procedure to find the culprit if the candidate or the examiner misbehave. Therefore, Dispute Resolution is an accountability requirement as it allow a judge to blame the principal who misbehave in the execution of the protocol. 

In WATA IV the judge is the invigilator, and the dispute originates if no intelligible pseudonym can be read when the candidate overlaps the paper sheet with the transparency sheet. Should such a dispute arise, the invigilator could then quickly resolve it as follows. First, he scans the QR code printed on the candidate's paper sheet and checks the correctness of the signature. Then, he checks if the candidate's details revealed by the signature match the ones written on the candidate's paper sheet. If so, the invigilator overlaps the visual crypto image revealed by the signature with the transparency sheet provided by the examiner. If this reveals no intelligible pseudonym then the examiner misprinted the corresponding transparency sheet. Otherwise the candidate misprinted hers. The outcome of the dispute can be double checked by repeating the procedure with the QR code printed on the transparency sheet of the examiner.

\subsubsection*{Comparison with WATA III}\label{ssec:cp}
\WW brings along significant improvements compared to WATA III. In particular:  
\begin{itemize}
\item \WW meets the same security requirements of WATA III augmented with Anonymous Marking and the two individual verifiability requirements of Mark Integrity and Mark Notification Integrity.
\item \WW meets the security requirements despite a more realistic threat model as it drastically limits the tasks assigned to honest-but-curious roles. 
\item \WW can support both computer-based and computer-assisted exams, while WATA III supports only computer-assisted exams with traditional testing. This is possible because test and pseudonym are \index{computer-assisted exams} generated independently in \WW. At testing, the candidate chooses one of the computer devices provided at exam venue, and enters the pseudonym retrieved from the paper sheets. Of course, the use of computers at testing raises similar security of Internet-based exam, as discussed in chapter \ref{chap:remark} for Remark!.   
\item In \WW the candidate receives part of the pseudonym by mail rather than at the exam venue.
\item In \WW the examiner cannot register a different mark to the candidate after he learns the corresponding test, because he commits the mark at marking.
\item In \WW any dispute between the candidate and the examiner can be solved with no efforts at testing: if an intelligible pseudonym appears, the candidate received the correct transparency sheet. If not, the signatures on the sheets reveal who misbehaved.
\end{itemize}

In the next section we propose an enhancement that removes the need of the honest-but-curios anonymiser in the design of the protocol.
Moreover, we corroborate the results of the informal security analysis outlined above, with the automated verification of the enhanced protocol in ProVerif.

\section{Removing the Need of Trusted Parties}\label{sec:watav} 

The major limitation of WATA IV is that it requires an honest-but-curious anonymiser. Although its lightweight participation, relying on a trusted third party in the design of a security protocol introduces obvious risks. The risks can be mitigated by distributing the trust across several parties, as we did for \remark~in chapter \ref{chap:remark}, but it still requires at least one party to be trustworthy. In the domain of exams this is critical because parties typically have conflicting interests, and it may be hard to find an entity who can play the role of a TTP, as recent exam scandals confirm. \index{exam scandal}

In this section, we propose a new protocol that guarantees several security properties without the need of a TTP. The protocol combines oblivious transfer and visual cryptography to allow candidate and examiner to jointly generate a pseudonym that anonymises the candidate’s test without the need of the anonymiser. The pseudonym is revealed only to the candidate at testing. We analyse the
protocol formally in ProVerif and prove that it satisfies the same security requirements stated for WATA IV.

We minimise the roles used in this protocol to candidate and examiner. The latter also runs the tasks associated to the roles of anonymiser and invigilator in WATA IV.  As we consider a  corrupted examiner who can deviate from any of its assigned tasks, it follows that any of its sub-roles can be corrupted as well.    

\subsubsection{Oblivious transfer} This protocol uses oblivious transfer to avoid the participation of the honest-but-curious anonymiser.
Oblivious transfer schemes allow a chooser to pick some pieces of information from a set that a sender offers him, in such a way that (a) the sender does not learn which pieces of information the choosers picks, and (b) the chooser learns no more than the pieces of information he picks. Our enhanced protocol adopts Tzeng's oblivious transfer scheme~\citeltex{T04}. In Tzeng's scheme, the chooser commits to some elements from a set, and sends the commitments to the sender. This, in turn, obfuscates all the set's elements, and the chooser will be able to de-obfuscate only the elements he has committed to. Tzeng's scheme guarantees unconditional security for the receiver's choice, and it is efficient since it
works with the sender and receiver's exchanging only two messages.

\begin{figure}
    \centering 
\FramedBox{20.3cm}{12.1cm}
{
\begin{enumerate}
\item \emph{C} calculates $y_i=g^{x_i}h^{\gamma_i}$ where: 
\begin{itemize}
\item[-] $x_i\in_{\mathcal{R}} \mathbb{Z}_q^*$.
\item[-] $\gamma_i\in_{\mathcal{R}} [1,k]$.
\item[-] $i=1,2,\ldots,l$ with $l>n$.
\end{itemize} 
\item \emph{C}$\rightarrow$\emph{E}: $y_1,y_2,\ldots,y_l$.
\item \emph{E} calculates $\beta_{ij}\leftarrow_{\pi_{\mathcal{R}}} \left(\alpha_i\oplus c_{j}\right)$,
$\omega_{ij}=\langle a_{ij}, b_{ij}\rangle \leftarrow \langle g^{r_{ij}}, \beta_{ij}\left(\frac{y_i}{h^j}\right)^{r_{ij}}\rangle$, \\
$\mi{com}=h^s\prod\limits_{i=1}^l {g_i}^{\alpha_i}$, and $\mathit{sign1}=\sign_{\mi{SSK}_E} ( \mi{idC}, \mi{ex}, \mi{com})$
where:
\begin{itemize}
\item[-] $\alpha_i\in_{\mathcal{R}} [0,1]^{t \times u}$.
\item[-] $s,r_{ij}\in_{\mathcal{R}}\mathbb{Z}_q^*$.
\item[-] $g_i\in_{\mathcal{R}}\mathbb{G}_q$.
\item[-] $i=1,2,\ldots,l$.
\item[-] $j=1,2,\ldots,k$.
\end{itemize}
or runs the challenge procedure against $y_1,y_2,\ldots,y_l$.
\item \emph{E}$\rightarrow$\emph{C}: $(\omega_{11}, \ldots, \omega_{1k}),
\ldots (\omega_{l1}, \ldots, \omega_{lk})$ and $\mathit{sign1}$.
\item \emph{C} calculates $\chi_i \in [1,l]$ and $\sigma_j \in [1,l]$ where:
\begin{itemize}
\item[-] $i=1,2,\ldots,m$.
\end{itemize}
\item \emph{C}$\rightarrow$\emph{E}: $\chi_1,\chi_2,\ldots,\chi_m$ and $\sigma_1,\sigma_2,\ldots,\sigma_n$.
\item \emph{E} calculates $ev_{\chi_i}= \langle\alpha_{\chi_i}, (\beta_{\chi_i1},\beta_{\chi_i2},\ldots, \beta_{\chi_ik}), (r_{\chi_i1}, r_{\chi_i2}, \ldots, r_{\chi_ik})             \rangle$ and \\ $\mathit{sign2}=\sign_{\mi{SSK}_E} ( \mi{idC}, \mi{ex}, (\sigma_1,\sigma_2,\ldots, \sigma_n)  )$ where
\begin{itemize}
\item[-] $i=1,2,\ldots,m$.
\item[-] $j=1,2,\ldots,k$.
\end{itemize}
and prints $\texttt{transp}=\left\langle(\alpha_{\sigma_1},\alpha_{\sigma_2}, \ldots, \alpha_{\sigma_n}), \mi{idC}, \mi{ex}, QR3 \right\rangle $ where
\begin{itemize}
\item[-] $QR3=\mi{idC}, \mi{ex}, (\alpha_1,\alpha_2, \ldots, \alpha_l,s)$.
\end{itemize}
\item \emph{E}$\rightarrow$\emph{C}: $ev_{\chi_1}, ev_{\chi_2}, \ldots, ev_{\chi_m}$ and $\mathit{sign2}$.
\item \emph{C} checks $ev_{\chi_i}$, calculates $\beta_{\sigma_j}=\frac{b_{\sigma_j\gamma_j}}{{(a_{\sigma_j\gamma_j})}^{x_{\sigma_j}}}$ where
\begin{itemize}
\item[-] $i=1,2,\ldots,m$.
\item[-] $j=1,2,\ldots,n$.
\end{itemize}
and prints $\texttt{paper}= \left\langle(\beta_{\sigma_1},\beta_{\sigma_2}, \ldots, \beta_{\sigma_n}), \mi{idC},\mi{ex}, QR1, QR2  \right\rangle$ where
\begin{itemize}
\item[-] $QR1=\mi{idC}, \mi{ex}, \mathit{sign1}$.
\item[-] $QR2=\mi{idC}, \mi{ex}, \mathit{sign2}$.
\end{itemize}

\end{enumerate}
}
   \caption{Preparation phase}
    \label{fig:preparation}
\end{figure}

\begin{figure}
    \centering 
\FramedBox{6cm}{12.1cm}
{
\begin{enumerate}
\setcounter{enumi}{9}
\item  \emph{C}$\xrightarrow{hands}$\emph{E}: $\texttt{id\_doc}$
\item  \emph{E} checks $\texttt{id\_doc}$
\item  \emph{E}$\xrightarrow{hands}$\emph{C}: $\texttt{transp}, \texttt{test}$
\item  \emph{C} calculates $\mathit{pid}= (\alpha_1,\alpha_2, \ldots, \alpha_n) \oplus (\beta_1,\beta_2, \ldots, \beta_n)$
and writes $\texttt{test}_{\texttt{filled}}=(\mathit{answers}, \mathit{pid})$\\
or runs the \dr algorithm if no pseudonym appears.
\item  \emph{C}$\xrightarrow{hands}$\emph{E}: $\texttt{test}_{\texttt{filled}}$
\end{enumerate}
}
   \caption{Testing phase}
    \label{fig:testing}
\end{figure}

\begin{figure}
    \centering 
\FramedBox{5.8cm}{12.1cm}
{
\begin{enumerate}
\setcounter{enumi}{14}
\item \emph{E} calculates $c=g^{v}h^{\mathit{mark}}$ and 
$\mathit{sign3}=\sign_{\mi{SSK}_E} ( \mathit{pid}, c)$
where: 
\begin{itemize}
\item[-] $v\in_{\mathcal{R}} \mathbb{Z}_q^*$.
\item[-] $mark\in M$.
\end{itemize} 
\item  \emph{E}$\rightarrow\mathcal{BB}$: $\mathit{sign3}$

\item  \emph{C}$\rightarrow$\emph{E}:$ (\beta_1,\beta_2, \ldots, \beta_n), \mathit{sign1}, \mathit{sign2}, \mathit{sign3} $
\item \emph{E} calculates $\mathit{sign4}=\sign_{\mi{SSK}_E} ( \mi{idC}, \mi{ex}, pid, \mathit{mark}, v)$
\item  \emph{E}$\rightarrow$\emph{C}: $ \mathit{sign4}$
\end{enumerate}
}
    \caption{Marking and Notification phases}
    \label{fig:notification}
\end{figure}

\subsection{Description}\label{sec:descriptionw5} 
In this protocol, we mainly revise the preparation phase, while the design of the other phases are similar to the design of WATA IV.
Candidate and examiner jointly generate the pseudonym \emph{pid}  as a pair of visual cryptography shares, by means of an oblivious transfer scheme rather than via the anonymiser. Notably, also the procedure for dispute resolution requires some modification since it cannot rely on the anonymiser's signatures.

We describe the protocol in reference to the four exam phases. We assume that all remote communications are via a secure channel. Figure \ref{fig:preparation} illustrates preparation in the Alice-Bob notation; Figure \ref{fig:testing} describes testing; Figure \ref{fig:notification} describes the protocol's steps that concern both marking and notification.   \index{exam phase}

In the description we assume the following public parameters:
\begin{center}
\begin{tabular}{p{4.2cm}|p{7cm}}
	$n$ &  length of the candidate's pseudonym \\\hline
	$\Sigma = \{s_1,\ldots,s_k\}$ &  alphabet of pseudonym's characters\\\hline
	$c_j\in \{0,1\}^{t \times u},~j=1,\ldots,k$ & 
		 ($t \times u$)-pixel representation of a character\\\hline
	$\mi{idC}$ &  candidate ID\\\hline
	$\mi{ex}$ &  exam code\\\hline
	$\mathit{SPK}_E$ &  signing key of the examiner\\\hline
	$M$ &  set of possible marks\\\hline
	$g,h\in_{\mathcal{R}}\mathbb{G}_q$ &  generators for bit-commitments \\\hline
\end{tabular}
\end{center}
\normalsize

\subsubsection{Preparation}
The goal of preparation is to generate a candidate's pseudonym, which is a string of $n$ characters taken from the alphabet $\Sigma$, and to encode it into two visual cryptographic shares. 
Both candidate and examiner cannot know the pseudonym until they meet at testing, when the candidate learns her pseudonym by overlapping the examiner's share with hers.
The underlying idea is that the candidate provides a commitment to an index into an array. The examiner fills the array with a secret permutation of the characters, and only when the two secrets are brought together is the selection of a character determined.

Part of this phase is inspired by one of the schemes used to print a secret, proposed by Essex \etal~\citeltex{ECHA09}. We tailor the scheme  in such a way to be able to generate a pseudonym and to support the dispute resolution algorithm. More precisely, the main technical differences between  preparation and their scheme are: (a) a modified oblivious transfer protocol that copes with several secret messages in only one protocol run; (b) the generation of signatures that will be used for accountability in the resolution of disputes. 

In the following we refer to the steps outlined in Figure \ref{fig:preparation}, \ref{fig:testing}, and \ref{fig:notification}.
The protocol begins with the candidate providing a sequence of $l$ commitments $y_i$ to an index into an array of length $k$. (steps 1-2).
More precisely, the parameter $l$, is chosen so that the $l-n$ elements can be later used for a cut-and-choose audit. The examiner can challenge the candidate to check whether the committed choices are in fact in the interval $[1,k]$. 
Otherwise, the examiner generates a sequence of randomly chosen 
${t \times u}$ images, indicated as $\alpha_1,  \ldots, \alpha_l$ in Figure \ref{fig:preparation}.
A sequence of $k$ images, ($\beta_{i1}, \ldots,  \beta_{ik}$), are generated from  $\alpha_i$ and each possible character $c_j$. 
The sequence is randomly permuted and repeated for all $i$, resulting in $l$ sequences of
($\beta_{11}, \ldots,  \beta_{1k}$), \ldots,
($\beta_{l1}, \ldots, \beta_{lk}$). The secret permutation and the commitment allow that the selection of character is determined only when the two secrets are brought together.

The examiner then generates the obfuscation $\omega_{ij}$ from each $\beta_{ij}$
and a commitment on each $\alpha_i$, indicated as $\mathit{com}$ (step 3),  which is signed and sent with the 
sequences of obfuscations $(\omega_{11}, \ldots, \omega_{1k})$,$ \ldots,$ $(\omega_{l1}, \ldots, \omega_{lk})$ to the candidate (step 4). 
The obfuscation allows the candidate to retrieve only the elements whose indexes 
correspond to the choices she committed in step 1 ($y_i$). 

The candidate performs a cut-and-choose audit, selecting a random set of $l-n$ sequences amongst the $\omega$. In so doing, she can check whether the examiner generated the sequence of images correctly.
%
The remaining  substitutions $\sigma_1, \sigma_2, \ldots, \sigma_n$ select the indexes 
of the images that make the pseudonym. Thus, the visual share of the examiner  
consists of the concatenated images $(\alpha_{\sigma_1},\ldots,\alpha_{\sigma_n})$ (step 5-6).

The examiner then generates the proofs for the cut-and-choose audit, and prints the visual share and the candidate's details in the transparency printout \texttt{transp}.
This also include all the elements $\alpha_1,\ldots,\alpha_l$ and the value used for their commitment (step 7), which are stored in the form of QR code.
The examiner then sends the proofs and the signed substitutions $\sigma$ to the candidate (step 8).
In turn, the candidate checks the proofs, de-obfuscates the elements $\omega$, 
and retrieves the visual share consisting of the concatenated image ($\beta_{\sigma_1},\beta_{\sigma_2}, \ldots, \beta_{\sigma_n}$). She finally prints the share, together with the two signatures, on a \texttt{paper} printout (step 9). At this point, both candidate and examiner have a visual share, which once overlapped reveal an intelligible sequence of characters that serves as pseudonym.

The candidate's paper printout includes two QR codes (\emph{QR1}, and \emph{QR2}) while the 
examiner's transparency only one (\emph{QR3}).
All the three QR codes share the same candidate identity $\mi{idC}$ and exam identifier $\mi{ex}$.
The QR codes \emph{QR1} and \emph{QR2} encode the two signatures of the examiner, respectively on commitment of the elements $\alpha$ 
and  on the substitutions $\sigma$, while \emph{QR3} encodes the  elements $\alpha$. 

\subsubsection{Testing}
The candidate brings the paper printout at exam venue, while the examiner brings the transparencies.
The examiner authenticates the candidate by checking her identity document (step 10-11).
He then gives the candidate her corresponding transparency and the test (step 12).
The candidate overlaps her paper printout with the transparency and learns her pseudonym, which she writes on the test (step 13). If no pseudonym appears, then this may happen only if the candidate or the examiner misprinted their printouts, and
the \dr outlined in Algorithm \ref{dispute} reveals the party that is accountable for the misbehaviour.
At the end of the phase, the candidate returns the filled test anywhere in the pile of tests (step 14), and takes both transparency and  paper printouts home.


\subsubsection{Marking and Notification}
At marking, the examiner evaluates the answers and generates a commitment on the assigned mark (step 15). Then, he signs both mark and  pseudonym found in the answer sheet, and publishes the signature on the bulletin board (step 16).

Notification opens for a fixed time, during which the candidate can remotely request to learn and register her mark. She has to send the ordered sequences of $\beta_1, \ldots, \beta_n$ and all the signatures so far she collected  to the examiner (step 17). The examiner checks the signatures, overlaps the given sequence with the corresponding sequences of $\alpha_1, \ldots, \alpha_n$, and learns the pseudonym. Again, if no registered pseudonym appears, Dispute Resolution can reveal the party who misbehaved.
The examiner signs the mark and the secret parameter used to commit the mark (step 18), and sends the signature to the candidate (step 19). In so doing, the candidate can verify the correctness of the mark by looking at the bulletin board. 

\begin{algorithm}
{
\SetAlgoLined
\KwData {Public parameters: ($C, n, g_i, h, \mi{idC}, \mi{SPK}_E )$}
\begin{itemize}
\item [-] $\mathit{paper}=( (\beta_{\sigma_1},\beta_{\sigma_2}, \ldots, \beta_{\sigma_n}), \mi{idC}',\mi{ex}', \mathit{sign1}, \mathit{sign2} )$ where:
\begin{itemize}
\item [-] $\mathit{sign1}=\sign_{\mi{SSK}_E} ( \mi{idC}'', \mi{ex}'', \mi{com})$
\item [-] $\mathit{sign2}=\sign_{\mi{SSK}_E} ( \mi{idC}''', \mi{ex}''', (\sigma'_1,\sigma'_2,\ldots, \sigma'_n)  )$
\end{itemize}
\item [-] $\mathit{transp}=(\alpha_{\sigma_1''},\alpha_{\sigma_2''}, \ldots, \alpha_{\sigma_n''}), \mi{idC}', \mi{ex}', (\alpha'_1,\alpha'_2, \ldots, \alpha_l',s) $.
\end{itemize}
\KwResult {Corrupted principal}

\eIf  {$\mathit{sign1}$ is verifiable with $\mi{SPK}_E$
\textbf{and} $\mathit{sign2}$ is verifiable with $\mi{SPK}_E$
\textbf{and} $\mi{idC}=\mi{idC}'=\mi{idC}''=\mi{idC}'''$
\textbf{and} $\mi{ex}=\mi{ex}'=\mi{ex}''=\mi{ex}'''$}
{
	\eIf {$\mi{com}\neq h^s\prod\limits_{i=1}^l {g_i}^{\alpha'_i}$
\textbf{or} $pid$=$(\alpha'_{\sigma_1'},\alpha'_{\sigma_2'}, \ldots, \alpha'_{\sigma_n'}) \oplus (\beta_{\sigma_1},\beta_{\sigma_2}, \ldots, \beta_{\sigma_n})$ }
{\KwRet {\emph{Examiner}}}
{  \KwRet {\emph{Candidate}}}
}
{ \KwRet {\emph{Candidate}}}
}
\caption{Dispute resolution}\label{dispute}
\end{algorithm}

\subsubsection*{Dispute resolution}
The Algorithm \ref{dispute} provides an efficient way to resolve a dispute if the candidate retrieves no intelligible pseudonyms when she overlaps the visual shares.
We assume that at exam venue it is available an electronic device with a camera, such as a smart phone or tablet, which stores the public key of the examiner. The input of the Algorithm are the two QR codes printed on the paper printout (QR1 and QR2) and the QR code printed on the transparency (QR3), all scanned with the camera of the electronic device.

First, the algorithm checks the correctness of the signatures encoded in QR1 and QR2. It also checks whether the candidate identity and the exam identifier reported on the paper printout match the ones in QR1 and QR2. If any one of the checks fails then the candidate misprinted her paper printout thus she is the culprit. Otherwise, the algorithm uses the data in QR3 to check the correctness of the examiner's commitment and that no pseudonym appears using the $\alpha$ elements  indexed with the $\sigma$ substitutions encoded in QR3. If any one of these checks fails then the examiner misprinted the transparency and thus he is guilty, otherwise the candidate is the culprit.

\subsection{Formal Analysis}\label{analysis}
We analyse the protocol in ProVerif.
In the remainder, we first presents the formal model, and then the results of the analysis of authentication, privacy, verifiability, and accountability requirements. 

\subsubsection*{Model choices} We model TLS and face-to-face communications between the roles using the cryptographic primitive of probabilistic symmetric encryption rather than using ProVerif's private channels. 
This choice is motivated because the attacker cannot monitor communications via ProVerif's private channels, and cannot even know if any communication happens. We think this is a too strong assumption that may miss attacks. By renouncing to private channels, we achieve stronger security guarantees for the analysis of the protocol. 
Moreover, our choice has a triple advantage: it allows the attacker to learn when a candidate registers for the exam or is notified with a mark; it allows modelling either corrupted candidate or examiner by just sharing the private key with the attacker; 
it increases the chance that  verification in ProVerif terminates.  Also, the attacker has more discretional power because he can observe when a candidate is given the test and when she submits the answers.

\begin{table}
\begin{center} 
\begin{tabular}{|c|c|}
\hline
{\bf Primitive} & {\bf Equation}
\\ \hline
Probabilistic symmetric enc.  &
$\begin{aligned}
\phantom{\mi{checksign}(\mi{sign}}\mi{sdec}(\mi{senc}(m,k,r),k) & = m 
\end{aligned}$
\\ \hline
Signature &
$\begin{aligned}
\mi{getmess}(\mi{sign}(m,ssk)) & = m \\
\mi{checksign}(\mi{sign}(m,ssk), \mi{spk}(ssk)) & = m
\end{aligned}$ 
\\ \hline
Visual cryptography & 
$\begin{aligned}
\mi{overlap}(share,gen\_share(m,share)) & = m \\
\mi{overlap}(share,share) & = share \\
\end{aligned}$
\\ \hline
Oblivious transfer & 
$\begin{aligned}
\mi{deobf}(\mi{obf}(r,m,\texttt{sel},\mi{commit}(r',\texttt{sel})), r') & = m \\
\end{aligned}$
\\ \hline
\end{tabular}

\caption{Equational theory to model the enhanced protocol}
\label{tab:eqtwata5}
\end{center}
\end{table}
\normalsize

We use the  equational theory illustrated in Table \ref{tab:eqtwata5} to model the cryptographic primitives of the  protocol. 
The theory for probabilistic symmetric key consists of two functions $\mi{senc}$ and $\mi{sdec}$. A message encrypted with a private key
can only be decrypted using the same private key. Note that the randomness $r$ on the encryption algorithm causes that the same message encrypted several times outputs different ciphertexts. 
The equational theory for signature is the same used in the other protocols considered in this dissertation.

We introduce a novel theory in ProVerif to model oblivious transfer and visual cryptography. The function $\mi{obf}$ allows the examiner to obfuscate 
the elements $\beta_{1}, \ldots,  \beta_{i}$, while the function $\mi{deobf}$ returns the correct element $\beta_{sel}$ to the candidate, depending on the choice she committed. We also provide the theory for the Pedersen commitment scheme with the function $\mi{commit}$. Finally, we model the generation of a  visual cryptography share with $\mi{gen\_share}$, and their overlapping with the function $\mi{overlap}$. 

\setlength{\abovecaptionskip}{-10pt}
\setlength{\belowcaptionskip}{0pt}
\begin{figure}
\begin{center}
\begin{lstlisting}
let C (idC: host, k: key, dvk: key, answer: bitstring, 
       choiceC: bitstring, spkE: pkey) =
(*Preparation*)
 out(ch, idC);
 (* commitment on choices *)
 new x: rand;
 let commitC= commit(x, choiceC) in
 out(ch, senc((commitC, idC),k));
 (* sign1 *)
 in(ch, encomega: bitstring); 
 let (omega1: bitstring, omega2: bitstring, sign1: bitstring)=
     sdec(encomega,k) in
 let (=idC, =spkE, com: bitstring)=checksign(sign1, spkE) in
 (* cut-and-choose & retrieve beta *)
 in(ch, encsign2: bitstring);
 let (sign2: bitstring)=sdec(encsign2,k) in
 let (=idC,=spkE,alpha:bitstring,beta1:bitstring,beta2:bitstring)=
     checksign(sign2, spkE) in
  if ((code1=overlap(c0,alpha,beta1) && 
       code2=overlap(c0,alpha,beta2)) || 
      (code2=overlap(c0,alpha,beta1) && 
       code1=overlap(c0,alpha,beta2))) then
   if beta1= deobf(omega1,x) then
    Ctesting(dvk, k, answer,  beta1, idC,spkE,sign2,sign1)
   else if beta2= deobf(omega2,x) then
    Ctesting(dvk, k, answer,  beta2, idC,spkE,sign2,sign1).

let Ctesting(dvk: key, k:key, answer: bitstring, beta: bitstring, 
             idC: host, SPKe: pkey, sign2: bitstring, 
             sign1: bitstring) =
(*Testing*)
 in(ch, dvtransp: bitstring); 
 let (question: bitstring, alpha': bitstring, =idC, =SPKe) =
     sdec(dvtransp, dvk) in
 let pid=overlap(c0,alpha',beta) in
 if pid=code1 || pid=code2 then (*otherwise dispute resolution*)
  event submitted(idC, SPKe , question, answer,pid);
  out(ch, senc((answer, pid, question),dvk));

(*Marking*) 
  in(ch, sign3: bitstring);
  let (=pid, commitM: commitment)=checksign(sign3, SPKe) in

(*Notification*) 
  (* request to learn her mark *)
  event requested(idC, pid);
  out(ch, senc( (beta, sign1, sign2, sign3, pid),k)); 
  in(ch, encsign4: bitstring);
  let sign4=sdec(encsign4,k) in  
  let (=idC, =pid, =SPKe, mark: bitstring, v: rand) =
      checksign(sign4, SPKe) in 
  if commitM=commit(v, mark) then 
   event notified(idC, mark,pid).
\end{lstlisting}
\end{center}
\caption{The process of the candidate}
\label{fig:candidatewatav}
\end{figure}
\setlength{\abovecaptionskip}{10pt}
\setlength{\belowcaptionskip}{0pt}

\setlength{\abovecaptionskip}{-10pt}
\setlength{\belowcaptionskip}{0pt}
\begin{figure}
\begin{center}
\begin{lstlisting}
let E (sskE:skey, question: bitstring, idX: host) =
 new perm_ch: channel;
 (
  (
(*Preparation*) 
   get sslkey(=idX,k) in (* check if the host is eligible *)
   in(ch, enccommit: bitstring);
   let (commitX: commitment, =idX)=sdec(enccommit,k) in
   new alpha: bitstring;
   in(ch, (c': bitstring, c'':bitstring)); (* calculate beta *)
   let beta1=share(c',alpha) in 
   let beta2=share(c'',alpha) in 
   new r1: rand; new r2: rand; (* obfuscate the betas *)
   let omega1=obf(r1, beta1, s1, commitX) in
   let omega2=obf(r2, beta2, s2, commitX) in
   new s: rand;    (*commitment on alphas *)
   let com=commit(s, alpha) in
   let sign1=sign( (idX, spk(sskE), com), sskE) in
   out(ch, senc( (omega1, omega2, sign1), k));(*cut-and-choose*) 
   let sign2=sign( (idX, spk(sskE), alpha, beta1, beta2), sskE) in
   out(ch, senc(sign2, k) );(* Print the transparency *)
   let dvtransp=(question, alpha, idX, spk(sskE)) in
   event registered(idX);

(*Testing*) 
   (* de visu authentication *)
   (* get the key to emulate de visu scenario at testing *)
   get dvkey(=idX, dvk) in 
   out(ch, senc(dvtransp,dvk)); (* hand the transparency *)  
   in(ch, enctest: bitstring); (* get the filled test *)
   let (answer: bitstring, pid: bitstring, =question) = 
       sdec(enctest, dvk) in 
   if (pid=code1 || pid=code2) then 
    event collected(idX, spk(sskE), question, answer,pid);
    event distributed(idX,question, answer,pid);
   (*otherwise dispute resolution*)

(*Marking*)
    new v: rand;  new mark: bitstring;
    let commitM= commit(v, mark) in 
    let sign3= sign( (pid, commitM), sskE) in 
    (* publish the commitment of the mark *)
    event marked(question,answer,mark,pid);
    out(ch, sign3); 

(*Notification*)
    in(ch, encnotif: bitstring); 
    let (beta: bitstring, =sign1, =sign2, =sign3, =pid)=
        sdec(encnotif, k) in 
    if pid=overlap(c0,alpha,beta) then (* is the correct beta? *) 
     event stored(idX, mark,pid);  (* notify the candidate *) 
     out(ch, senc(sign((idX, pid, spk(sskE), mark, v), sskE), k))
  )  | 
  (out(perm_ch, (code1,code2)) | out(perm_ch, (code2,code1)))
 ).
\end{lstlisting}
\end{center}
\caption{The process of the examiner}
\label{fig:examinerwatav}
\end{figure}
\setlength{\abovecaptionskip}{10pt}
\setlength{\belowcaptionskip}{0pt}

\begin{figure}
\begin{center}
\begin{lstlisting}
process
 !(	
   new sskE: skey; let spkE = spk(sskE) in out (ch, spkE); 
   
   (!in(ch, idhost: host); new questions:bitstring; 
   E(sskE, questions, idhost))  |
   (!processA) |
   (! (new answer: bitstring; new k: key; insert sslkey(CA, k); 
      new dvk:key; insert dvkey(CA, dvk); 
      C(CA, k, dvk,  answer, s1, spkE)) 
   ) 
  )
\end{lstlisting}
\caption{The exam process}
\label{fig:processwatav}
\end{center}
\end{figure}

\begin{figure}
\begin{center}
\begin{lstlisting}
let processA =
 in(ch, (h: host, kX: key, dvkX: key));
 if h<>CA then
  insert sslkey(h,kX).
  insert dvkey(h,dvkX).
\end{lstlisting}
\caption{The process of corrupted candidate}
\label{fig:corruptedwatav}
\end{center}

\end{figure}

\begin{figure}
\begin{center}
\begin{lstlisting}
process
 !(	
   new sskE: skey; let spkE = spk(sskE) in out (ch,(sskE, spkE)); 
   new coll_key: key;

   (!processA) | (!collector(coll_key)) |  
   (new idC: host; new k: key; insert sslkey(idC, k); 
    (* The examiner is corrupted so the attacker knows the keys *)
    new dvk:key; insert dvkey(idC, k); out(ch, k); out(ch, dvk); 
    C(idC, k, dvk,  choice[ansA,ansB], coll_key, s1, spkE) 
   ) |
   (new idC: host; new k: key; insert sslkey(idC, k); 
    (* The examiner is corrupted so the attacker knows the keys *)
    new dvk:key; insert dvkey(idC, k); out(ch, k); out(ch, dvk);
    C(idC, k, dvk, choice[ansB,ansA], coll_key, s2, spkE) )
  )
\end{lstlisting}
\caption{The exam process to analyse Anonymous Marking}
\label{fig:amwatav}
\end{center}
\end{figure}

\begin{figure}
\begin{center}
\begin{lstlisting}
let collector (coll_key: key) =
 in(ch, (encA: bitstring, idcA: bitstring));
 in(ch, (encB: bitstring, idcB: bitstring));
 let (answerA: bitstring, codeA: bitstring, =idcA) =
     sdec(encA, coll_key) in
 let (answerB: bitstring, codeB: bitstring, =idcB) =
     sdec(encB, coll_key) in
 if idcA<>idcB && codeA<>codeB then
  (* A candidate cannot submit a test twice *)
  let tA= (answerA,codeA) in
  let tB= (answerB,codeB) in
  out(ch,choice[tA,tB]);
  out(ch,choice[tB,tA]).
\end{lstlisting}
\caption{The collector process}
\label{fig:collectorwatav}
\end{center}
\end{figure}

The process of the candidate is modelled in Figure \ref{fig:candidatewatav}; the process of the examiner is in Figure \ref{fig:examinerwatav};
the exam process is depicted in Figure  \ref{fig:processwatav}. In addition, we also model an unbounded number of corrupted candidates who can register for the exam as in Figure \ref{fig:corruptedwatav}. All the processes are augmented with the events that allow the verification of authentication requirements. 
We verify \am in presence of a corrupted examiner and corrupted co-candidates. The corresponding ProVerif process is depicted in Figure \ref{fig:amwatav}. We add the process \emph{collector} that simulates the desk where candidates leave their tests (Figure \ref{fig:collectorwatav}). 
To verify Question Indistinguishability, we consider corrupted candidates, while for both \mpr and \man we consider corrupted eligible candidates \index{eligible candidates} who can register for the exam but cannot participate at testing. 

\begin{algorithm}
\label{verifiabilityalg}
{
\SetAlgoLined
\KwData {Public parameters: ($g,h,\mi{SPK}_E )$}
\begin{itemize}
\item [-] $\mathit{sign3}=\sign_{\mi{SSK}_E} ( \mi{pid}, c  )$
\item [-] $\mi{idC}, \mi{pid'}, \mi{mark}, v$.
\end{itemize}
\KwResult {Whether the candidate was notified with the mark assigned to her test.}
{
\eIf  {$\mathit{pid}=\mi{pid'}$
\textbf{and} $c=g^{v}h^{\mathit{mark}}$}
{\KwRet {\texttt{true}}}
 {\KwRet {\texttt{false}}}
 }
}
\caption{The verifiability-test for Mark Integrity I.V.}\label{verifiability}
\end{algorithm}

\begin{figure}
\begin{center}
\begin{lstlisting}
let testMI(spkE: pkey,  priv_ch: channel, bbpk: pkey) =
 in(ch, signsign_bb:bitstring);
 in(priv_ch, (idX:host, pid':bitstring, mark:bitstring, v:rand));

 let (sign_bb:bitstring)=checksign(signsign_bb, bbpk) in 
 let (pid: bitstring, commitM: commitment) =
     checksign(sign_bb, spkE) in 

 if pid=pid' && commitM=commit(v, mark) then 
  event OK
 else KO.
\end{lstlisting}
\caption{The ProVerif process of the verifiability-test \emph{testMI}}
\label{fig:vermarkv}
\end{center}
\end{figure}

\paragraph{Modelling Verifiability.} The individual verifiability definition of Mark Integrity can be modelled with Algorithm \ref{verifiabilityalg}, which defines the verifiability-test \emph{testMI}. The corresponding ProVerif model is depicted in Figure \ref{fig:vermarkv}.

The verifiability-test \emph{testMI}  takes in the pseudonym \emph{pid}, the randomness value of the commitment, and the mark notified to the candidate via a private channel. It also takes as input the notification \emph{sign3} signed by the examiner from the bulletin board. 
The verifiability-test checks if the examiner's signature is correct and the disclosure of the commitment contained on the notification reveals the mark provided by the candidate. 
Since this requirement is interesting in presence of corrupted co-candidates and a corrupted examiner, we model the bulletin board as a ProVerif process to check soundness. In so doing, we can annotate the process of the bulletin board with the event \verb+marked+ in the location where it publishes
the notification. We also annotate the process of the candidate with the event \verb+assigned+ in the location where the process receives the notification.  We then use the following correspondence assertion to check soundness:
\begin{center}
 $\OKb{\id}{\pid}{\mathit{mark}}~\leadsto~\markingc{\pid} \cup \assignedb{\id}{\pid}{\mathit{mark}}$ 
\end{center}
To prove completeness, ProVerif checks that the verifiability-test process does not emit the event \texttt{KO} when the input data is correct.

It can be observed that  Mark Notification Integrity I.V. can be modelled with the same test used to check of Mark Integrity I.V.
In fact, the candidate checks if the commitment associated to her pseudonym  contains the same mark that the examiner notified to her.  

\begin{figure}
\begin{center}
\begin{lstlisting}
let dispute(spkE: pkey, dvk: key) =
 in(ch, transppaper: bitstring);  

 let ((q: bitstring, alpha: bitstring, idhost: host, =spkE, 
       alpha': bitstring, s: rand), 
      (beta: bitstring, idhost: host, =spkE, sign1: bitstring, 
       sign2: bitstring)) = sdec(transppaper, dvk) in 
 let (idX: host, ex': pkey, com: commitment) = 
     checksign(sign1, spkE) in
 let (idX': host, ex'': pkey, alpha'': bitstring, 
      beta1: bitstring, beta2: bitstring) =
     checksign(sign2, spkE) in

 if idX=idX' && idX=idhost && ex'=ex'' && ex'=spkE then  
  if com<>commit(s, alpha') || code1=overlap(c0,alpha',beta) || 
     code2=overlap(c0,alpha',beta) || alpha'<>alpha'' then
   event Eguilty
  else event Cguilty
 else event Cguilty.
\end{lstlisting}
\caption{The ProVerif process of \emph{dispute}}
\label{fig:dispute}
\end{center}
\end{figure}

\paragraph{Modelling Dispute Resolution.} We introduce the first formalisation of accountability requirement for exams, namely \emph{Dispute Resolution}. Accountability allows us to identify which principal is responsible for a protocol failure. In the case of exam, a candidate should be able to submit a test and receive the corresponding mark. If she fails\ in any of these, Dispute Resolution prescribes that the participant who caused such failure can be identified.
We formally model Dispute Resolution with a similar approach advanced for individual verifiability,  with two differences: first, we resort to unreachability of an event to prove soundness; second, we check that the exam process does not execute the algorithm of \dr to prove completeness. 

The process \emph{dispute} is illustrated in Figure \ref{fig:dispute}. The process takes in the examiner's transparency and the candidate's paper, and outputs the corrupted principal according to Algorithm \ref{dispute} specified in section \ref{sec:descriptionw5}. It is annotated with the event
 $\texttt{Cguilty}$, which the process emits when the candidate is considered to be the culprit. The process is also annotated with the event $\texttt{Eguilty}$, which is emitted when the examiner is considered to be the culprit.

If the protocol executes the process \texttt{dispute}, then either the examiner or the candidate is corrupted. Thus, regarding soundness, the idea is to check that \texttt{dispute} cannot return an honest principal instead of the corrupted one. This is captured by the following definitions.
\begin{prop}[\bf Soundness (corrupted examiner)] 
The process   \emph{dispute} is sound with respect to a corrupted examiner and honest candidate if the event 
$\mathtt{Cguilty}$ is emitted in no execution trace of the exam protocol.
\end{prop}
\begin{prop}[\bf Soundness (corrupted candidate)] 
The process   \emph{dispute} is sound with respect to a corrupted candidate and honest examiner if the event 
$\mathtt{Eguilty}$ is emitted in no execution trace of the exam protocol.
\end{prop}

To prove completeness, we check that the exam protocol never runs the process \emph{dispute}, hence both events \texttt{Cguilty} and \texttt{Eguilty} are not emitted. This is captured by the following definition.
\begin{prop}[\bf Completeness (exam process)] 
The exam protocol is complete respect to \dr if  neither the event
$\mathtt{Eguilty}$ nor $\mathtt{Cguilty}$ are emitted in any execution trace of the protocol with honest roles.
\end{prop}

We thus can say that the exam protocol ensures \dr if \texttt{dispute} is sound and the exam protocol complete.


\subsubsection*{Limitations} A limitation of the formal model is the specification of the cut-and-choose audit due to the powerful ProVerif's attacker model. In fact, if the attacker plays the cutter's role, he might cut the set of elements such that the subset audited by the chooser is correct, while the other subset not. Although in reality the probability of success of this attack for a large set of elements is small, it is a valid attack in ProVerif irrespective of the number of elements. In our case, the chooser is the candidate and the cutter the examiner. We thus have a false attack when the examiner is corrupted, namely controlled by the attacker. In this case, we avoid this situation by allowing the candidate to check all the elements of the set. This is sound because the candidate plays the role of the chooser, thus she is honest and follows the protocol although she knows the extra information.

\begin{table}
\begin{center}
\begin{tabular}{|c|c|c|}
\hline
{\bf Requirement} & {\bf Result} & {\bf Time} \\ \hline
  {\cautho} &  $\have$  &  8 s \\ \hline
  {\aau} &  $\have$  & 7 s  \\ \hline 
  {\CAu} &  $\have$  & 7 s  \\ \hline 
      {\ta} &  $\have$ & 8 s \\ \hline
    {Mark Authenticity} &  $\have$  &  8 s \\ \hline
Notification Request Auth.  & $\checkmark$ & 8s  \\ \hline
     {Question Indistinguishability} &  $\have$  &  $<$1 s\\ \hline
      {Anonymous Marking} &  $\have$ & 27 s\\ \hline
       {Mark Privacy} &  $\have$ & 28 m 41 s  \\ \hline       
       {Mark Anonymity} &  $\have$ & 52 m 12 s  \\ \hline       
Mark Integrity I.V.  & $\have$ & $<$1s   \\ \hline
Dispute Resolution & $\have$ & $<$1s   \\ \hline
\end{tabular}
\caption{Summary of the analysis of the enhanced protocol}
\label{tab:results_watav}
\end{center}
\end{table}%

\subsubsection*{Results}
Table~\ref{tab:results_watav} outlines the results of our analysis. 
ProVerif confirms that the protocol guarantees all the authentication requirements
despite allowing an unbounded number of corrupted eligible co-candidates. \index{eligible candidates} Thus, the exam protocol ensures authentication although the attacker can register to the exam. 

Concerning privacy properties, ProVerif proves that the exam protocol guarantees all the privacy requirements. We do not consider Anonymous Examiner since we assume only one examiner, hence the protocol trivially fails to meet the requirement.   

ProVerif confirms the verifiability-test \emph{testMI} is sound and complete, thus the exam protocol is Mark Integrity I.V. and Mark Notification Integrity I.V. verifiable. 

Finally, the exam protocol ensures Dispute Resolution: ProVerif shows that the protocol does not blame honest principals when the \texttt{dispute} algorithm is executed (soundness), and that does not run the algorithm when both examiner and candidate are honest (completeness).

\section{Conclusion}\label{sec:conclusionremarkwata}
This chapter draws its motivation by observing that exam security has a major role in the widespread acceptance of computer-assisted exams. \index{computer-assisted exams}
It focuses on a family of computer-assisted exams called WATA, and shows how to gradually remove the need of trusted third party in their design, though ensuring more security requirements.

This chapter provides a new outlook of WATA II and WATA III, which originally were conceived as software running into the examiner's machine, by re-engineering them as exam protocols.
It discusses their security properties, trust assumptions, and both security and functional limitations.  
Then, it advances WATA IV, a novel exam protocol that allows both remote registration and remote notification. \index{remote registration} \index{remote notification}
WATA IV reduces significantly the participation of TTP respect to the previous versions, while  guaranteeing more security requirements.
Moreover, WATA IV supports both computer-based exam and traditional testing. 

WATA IV is further enhanced resulting in a new protocol that meets the same security requirements of WATA IV but without the need of a TTP.
The underlying idea is to combine oblivious transfer and visual cryptography to generate a pseudonym that anonymises the test for the marking.
A formal analysis in ProVerif confirms that the enhanced protocol ensures all the stated requirements.

Finally, this chapter advances the accountability requirement of Dispute Resolution and its formal specification. Using ProVerif, we prove that the enhanced protocol also meets such requirement without the need of a TTP.


\chapter{Formal Analysis of Certificate Validation in SEB and  Modern Browsers}\label{chap:socio}

Computer-assisted exams often include some remote tasks, such as remote registration or remote notification of candidates, which normally take place by means of a browser. \index{remote registration} \index{remote notification} \index{computer-assisted exams}
Internet-based exams also require that candidates take the exam via software, possibly a browser \citeltex{SBV+12} \citeweb{questionmark, questbase}. Although these browsers are customised to prevent unauthorised access to resources during the exam, they still appeal on TLS to meet authentication and privacy. In the design of secure exam protocols, as well as in many security protocols, it is often assumed that an implementation of TLS channel is available to secure the communications among the principals. \index{secure exam protocol} \index{Internet-based exams}

While we can reasonably assume that \tls provides privacy and integrity as it uses robust cryptographic schemes, 
we should be more careful in assuming the same for authentication.
The authentication of a website depends, to various degrees, on trust. One element of trust comes either from the web-of-trust concept \citeltex{Zimmerman92} or from the public-key infrastructure (PKI) \citeltex{rfc5280}. 
Either way, the authentication of a website works through the emission of certificates. A
certificate binds an identity with a public key, and contains other pieces
of information that the verifier, also known as authenticator or trustor, needs to check to accept the certificate.
The verifier is a software, normally the browser of the user who accesses a website.
The browser verifies that the identity on the  certificate corresponds to the identity on the website, and that the certificate is signed by a trusted authority.
Thus, an authentication that succeeds  seems to  depend mostly on the browser than on the user.
But when the validation fails, browsers usually resort on the choices of users. In this case, the user is the ultimate responsible for the website's authentication.

Invalid certificates are not rare. For example, exam authorities and institutions often self-issue their own certificates rather than purchase them from accredited certification authorities. Self-issuing a certificate is in fact cheaper. But, even if institutions purchase their certificates from a 
recognised authority, they may still use the certificate beyond its expiration date or abuse it to certify different domains and sub-domains that are not actually provided for the certificate.
With a security take, an invalid certificate may originate from a network attacker who attempts a man-in-the-middle attack, namely the attacker replaces the server's certificate with his own. 

We do not intend to contribute to the long-established debate on the interpretation of the technical meaning of authentication \citeltex{Gollmann}.
We rather observe that browser's security is critical for exams, and we expect to substantiate our observation that server authentication goes beyond the technical \emph{certification path validation} algorithm as described in the standard X.509 \citeltex{rfc5280}.
The validation of a certificate is in fact \emph{socio-technical} as modern browsers consider user's choices and support the validation with novel technologies, such as HSTS.
Thus, with  \emph{certificate validation} we refer to the extended protocol that browsers  customise with user's involvement and additional security mechanisms.

We note that the way the certificate validation is accomplished varies considerably among browsers.
This variety motivates a number of research questions:
\begin{itemize}
\item
What are the differences in terms of user involvement in how modern browsers implement server authentication?
\item
Which browsers reduce the security risks for users when a certificate is invalid?
\item
Can browsers improve their security by involving the user more profitably than they do at present?
\end{itemize}
This list of questions, purposely truncated to length three here, arises when server authentication, and certificate validation in particular, is assessed from a socio-technical standpoint.

This work complements traditional human-computer interaction studies by advancing what seems to be the first formal analysis of browser's certificate validation that is not only logically conditioned on the technology but also on user actions.

In this chapter, we consider six browsers. One is Secure Exam Browser (SEB), which specifically addresses the security of remote testing of Internet-based exams. \index{remote testing} The others are Firefox, Chrome, Safari, Internet Explorer, and Opera Mini, which are the most popular browsers and may support the remote tasks of an exam. We also consider the analysis of private browsing mode, and the interleaving of classic and private browsing for the browsers that support it. We introduce five socio-technical requirements each binding elements like TLS session, certificates and, notably, user choices.
The mix of browsers, modes, and requirements allows for the analysis of 60 different scenarios.
We anticipate that the results of the analysis turn out to be interesting, from which we are able to state four recommendations.
Notably, the results of our analysis include two bugs that concern Safari. We reported the bugs to Apple, and were replied that a fix would be available in the upcoming versions of Safari for iOS and OS X. 

The contribution of this chapter exceeds the formal analysis of the 60 scenarios. This formal analysis was not carried out using a known approach. By contrast, it was not obvious how to represent (portions of) the functioning of a browser in order for the analyser to quickly get to grasps with its properties without reading long prose. Various graphical notations were tried out, and finally we found \ac{uml} activity diagrams \citeweb{OMG2011} to bear the necessary flexibility. Building these diagrams is a major hallmark in our understanding of the technicalities of the browsers. However, they are only semi-formal and not directly executable; a formal model is needed for a fully automatic analysis. We therefore translate our UML diagram models to models in the \ac{csp} process algebra  \citeltex{Hoare}. The model is then extended with an \ac{ltl} specification of the requirements of interest. Each extended model forms the input to the Process Analysis Toolkit (PAT) model checker \citeltex{SunLDP09}.

\paragraph{Outline of the chapter.} Section \ref{sec:relatedsocio} discusses the related work about formal approaches for the security analysis of browsers and certificate validation, and related studies of human aspects. 
Section \ref{sec:positioning} details the different aspects of certificate validation.
Section \ref{sec:seb} describes the SEB kiosk exam browser. \index{kiosk browser}
Section \ref{sec:browser} describes the five most popular general-purpose browsers.
Section \ref{sec:modellingsocio} details the browsers' certificate validation ceremonies using UML Activity Diagrams.
Section \ref{sec:analysisocio} concerns the analysis of the five socio-technical requirements on the browser's ceremonies with the PAT model checker.
Section \ref{sec:findings} discusses the results of the analysis.
Section \ref{sec:concsocio} explores some implications and concludes the chapter.


\section{Related Work}\label{sec:relatedsocio}

A few works have developed formal verification techniques to model and analyse web browsers.
Akhawe \etal \citeltex{Akhawe2010} introduce a formal model of web security. They define the main components of the web, namely Non-Linear Time, Browser, Servers, and the Network as \emph{Web Concepts}. They consider a spectrum of threats that span from a malicious web server to a more advanced attacker who is able to inject contents into an honest web server. Finally, they analyse two security requirements, \ie security invariants and session integrity, in five web security \emph{mechanisms}.
The considered mechanisms include neither TLS nor certificate validation, and the formal model assumes the user correctly interprets the browser's security indicators. Our work focuses on certificate validation and considers users who may not correctly understand security indicators.
Gro\ss~\etal \citeltex{GroB:2005:BMS:2156732.2156760} propose a formal framework to model a web browser and  the behaviour of a user who interacts with the browser. They validate their framework by analysing the security of password-based user authentication.
Contrary to modelling an ideal browser, we analyse the actual implementation of modern browsers. It would be interesting to merge our approach with their model of web browser.
Formal verification techniques have been used to analyse Human-Automation Interaction \citeltex{BBS13}. They mostly focus on cognitive aspects of users rather than on the technical systems they interact with. We consider user choices, which do depend on human cognitive factors, but our focus is more on the technical part, namely on the web browsers.

There are many studies that conduct a security analysis of certificate validation.
Georgiev \etal \citeltex{Georgiev:2012:MDC:2382196.2382204} analyse certificate validation in the context of TLS for different web applications, such as shopping carts, cloud storage, and payment gateways.
They find that several APIs perform certificate validations that do not follow any standard. The authors show how such customisation leads to security vulnerabilities.
Differently to our automated approach, they detect attacks by visual inspection  of the source code; moreover, the analysed applications involve no browsers.
Kaminsky \etal \citeltex{Kaminsky:2010:PLC:2163571.2163593} discuss different attacks against the certificate infrastructure. They point out that certificate issuers and browsers may differently interpret 
the subject name in an X.509 certificate. Such a lack of standardisation makes the subject name vulnerable to injection attacks. A recent update to the X.509 standard \citeltex{rfc6818} aims to fix the issue.
Clark and Van Oorschot \citeltex{CV13} provide a comparative evaluation of enhancements implemented into browsers for certificate validation. They argue that it is becoming more common that attackers own valid certificates for a website. Attackers' attention focuses on certificate infrastructure because of its reliance on human factors. We share their view that certificate validation goes beyond the mere binding between a domain name and a public key.

Different solutions have been proposed to modify the structural flaws of the certificate infrastructure \citeweb{convergence,Marlinspike2013} \citeltex{dane}, but they also require dedicated  modifications to the  protocols that use certificates to achieve authentication.
For example, the Certificate Transparency project \citeltex{certtrasn} suggests an improvement to the current TLS certificate system by providing supplemental monitoring and auditing services via certificate logs. The technique requires a different server implementation to accommodate a TLS extension. Structural solutions such as TLS extensions may take a long time before being implemented extensively, because both browser and server need to support the extension.
We also end up with a similar proposition, but it requires a modification of the browser only, with no change required on the server and on the TLS protocol.

The human aspect of certificate validation has recently been analysed via different empirical studies. 
Akhawe and Porter Felt \citeltex{APF13} make an empiric analysis to assess the effectiveness of browser security warnings. In particular, they measure the users' click-through rates on certificate and malware warnings. They find that users tend to ignore warnings as they click through the Chrome certificate warnings. Such finding led Google to redesign Chrome's certificate warnings. 
Similarly, Flinn and Lumsden \citeltex{Flinn05} conduct a survey to assess whether users are aware of the security risk they face online.
They find that users have different interpretations of the term ``secure website'' and are generally unaware that TLS provides server authentication. By contrast, our formal analysis does not pertain to user perception, but investigates the various ways in which user's choices influence server authentication.
J{\o}sang \etal \citeltex{Josang2012} point out that web browsers can only do syntactic server authentication, as TLS
cannot provide semantic server authentication. Thus, the attacker can exploit semantic attacks to trick the user. They advocate the need of a framework to determine the assurance level of server authentication. Our approach aims at analysing how web browsers help users to avoid such server authentication attacks.
Gajek \etal \citeltex{Gajek2008} propose a new authentication protocol that consists of a mix between the Password Authentication Key Exchange (PAKE) and TLS protocols without relying on a PKI. They formalise a user as a probabilistic machine. The user's behaviour can recognise the so-called human-perceptible indicators like pictures and sounds. In contrast, we are not interested in the cognitive aspects, and make minimal assumptions about the user capabilities.
Akhawe \etal \citeltex{AAVS13} produce a taxonomy of certificate validation warnings and collect data over more 10 billion TLS connections that are not under MITM attacks. Thus, they calculate the false positive rate of showing warnings and present a number of recommendations to improve browser design decisions. Conversely, our approach considers MITM attacks.     

The security of browsers has been studied variously, for example to avoid the user's oversight of warning messages \citeltex{Sunshine09}, or to improve the readability of their contents \citeltex{Biddle}. These works are positioned over the cognitive aspects of human-computer interaction with the browsers. To position our work, it is useful to note that we see the socio-technical system consisting of a web server, a computer network, a browser, a user and possibly an intruder as a \emph{ceremony} in the sense of Ellison \citeltex{Ellison2007}. The various technical and social layers of a ceremony have been recently identified \citeltex{Bella}, with a practically useful suggestion that certain layers can be neglected, namely virtually compressed, during the analysis to sharpen the analyser's focus on other layers.



\section{Basics}\label{sec:positioning}
This section clarifies a few technical notions and sets the terminology used throughout the chapter. First it details the constituents of a web certificate and explains how their validation works or can fail. Then, it discusses the regulations that specifies the \emph{path validation} of a web certificate, and observes how the standard eventually leaves the  interpretation of certificate validation to browser manufacturers. In consequence, certificate validation becomes a mix of user's choices and technical security mechanisms, from which it emerges that certificate validation in modern browsers is a socio-technical process. The sequel of this Section outlines a few basic concepts needed later.

\subsection*{Web certificates}
A web certificate binds an identity to a public key. The X.509 standard \citeltex{rfc5280} specifies the structure of a certificate by listing a set of mandatory and optional fields.  
Among the mandatory fields, four are fundamental to understand the authentication purpose of a certificate: \emph{subject}, which  specifies the certificate owner's identity;  \emph{subject public key}, which specifies the public key associated to the subject;   \emph{issuer}, which specifies the entity who verified that the public key belongs to the owner described in the subject;  \emph{certificate signature}, which specifies the digital signature generated by the issuer on subject and public key. 

Other typical mandatory  fields are the following: \emph{version}, which specifies whether optional fields are expected to be used;  \emph{serial number}, which is unique among the certificates generated by the issuer;  and  \emph{validity}, which specifies the time interval during which the issuer maintains information about the status of the  certificate.

\subsection*{Certificate validation}

The algorithm for the path validation of a certificate checks whether the certificate is valid. It consists in verifying that the signature is correct provided that the verifier trusts the issuer. In this case the public key can be used to communicate with its owner. 
However, even if the signature is correct, the verifier may not trust the issuer. In this case the issuer, which is known as \emph{intermediate authority}, needs itself to be certified. 
This forms a chain of intermediate authorities called  \emph{certification path}. The certification path always chains up to a root called  \emph{certification authority (CA)}, whose certificate is self-signed, namely the issuer and the subject coincide. The verifier is assumed to trust the public key of the CA. 
Thus, the validation of a certificate path is to check the fields of each certificate up to a trustful root certificate.
The standard  X.509 details a certification path validation algorithm, but verifiers are free to implement their own algorithms, provided they offer equivalent functionality \citeltex{rfc5280}.

\subsection*{Invalid certificates}\label{sec:cv}
If the path validation of the certificate succeeds then the certificate is valid, namely the certificate binds correctly the identity with the public key. However, the validation may fail due to a number of errors:
\begin{description} 
\item
\textbf{Unknown or untrusted certificate issuer}.
The certificate path chains up to a
certification authority that is not in the list of CAs the verifier trusts.
Verifiers may trust different certification authorities, since there is no
universally trusted list of CAs.

\emph{Possible reasons.} A certificate may be invalid  because
entities, such as web servers, may prefer to self issue a certificate rather
than purchase expensive certificates by commercial CAs. This choice may be even
necessary when a single entity owns many different domains that need to be
certified. Self-issuing a certificate is a quick procedure and has no costs.
Self-issued certificates are widespread through a large number of public
institutions, such as universities, or the US Army \citeweb{U.S.Army}. 
\end{description}

\begin{description} 
\item
\textbf{Expired certificate}.
A certificate expires after a specific time interval.
Specifically this happens because the \emph{validity} field contains a past
date. 

\emph{Possible reasons.} Entities may forget to renew their certificates before they expire.
According to a recent survey \citeltex{AAVS13}, expired certificates are the most
common form of benign (\ie false positives) certification path validation
failures. 
\end{description}

\begin{description} 
\item
\textbf{Revoked certificate}.
A certification authority may revoke the
certificate due to either administrative or security reasons. For example, if
an entity believes that an attacker has learned the private key, it may ask the
CA to revoke the certificate. 

\emph{Possible reasons.} Again, verifiers may trust different certificate
authorities, thus revocation also depends on the specific CA store used by the
verifier. Moreover, certificate revocation is a protocol by itself: CRL, OCSP, and
CRLSets are three common protocols to revoke and check certificates
\citeltex{rfc6818,OCSP} \citeweb{CRLSets}, and verifiers may use any of these.
\end{description}

\begin{description} 
\item
\textbf{Mismatched certificate subject.}
The \emph{subject}  expected by
the verifier mismatches the one shown in the certificate.  According to a
large-scale survey on certificates \citeltex{Vratonjic2013}, mismatched certificate
subject is a frequent case of certificate path validation failure.

\emph{Possible reasons.} False positives may occur because an entity
needs to secure its own sub-domain (\eg \url{www.sub1.entity.com}), but, to
save costs, the entity does not purchase a certificate for each sub-domain.
\end{description}

The X.509 standard says that if any one
of the checks of the certification path validation fail, the algorithm
terminates, returning a failure indication to the concerned protocol.  The \tls
protocol uses X.509 certificates to support authentication, and browsers
implement \tls over HTTP to provide confidentiality, integrity, and
authentication on the communications with web servers.  Since authentication is
an optional \tls requirement, the corresponding RFC standard \citeltex{rfc5246}
outsources the certificate validation to the browsers: \emph{``How to interpret
	the authentication certificates exchanged is left to the judgement of
the designers and implementors of protocols that run on top of TLS''}.
Moreover, the HTTP over TLS standard \citeltex{rfc2818} advocates the involvement
of the user when the certification path validation fails: \emph{``User oriented
clients MUST either notify the user (clients MAY give the user the opportunity
to continue with the connection in any case) or terminate the connection.''}.
Ultimately, browsers can implement differently the certificate validation,
which becomes socio-technical when the technical approach, namely the
certification path validation, fails. 

\subsection*{Socio-Technical aspects of certificate validation}

Browsers communicate with the user through  different ways, such as text warnings, pop-up windows, open or closed padlocks, and coloured address bars. The main component that browsers use to interact with the user is the viewport, which is depicted in Figure \ref{fig:viewport}

The user can choose any of the options proposed in the browser's viewport: a cautious user may close the browsing session, while a curious one may click through a warning.
Users who interact with a browser may be variously skilled and educated. They are influenced by a huge variety of local or global cultural values. 
Malicious websites nowadays can use scripts to gain major control on browser's viewport and deceive user's \citeltex{CV13} on security warnings.
Although the design of browser's security indicators has improved over the years \citeltex{YSA05}, a number of studies have shown that users tend to click through a warning without paying attention \citeweb{sslpass12} \citeltex{AAVS13,Sunshine09}.
These social factors make the problem of certificate validation harder than it seems, namely a security problem that cannot be solved by purely technical means.

\begin{figure}[h]
\centering\includegraphics[scale=.35]{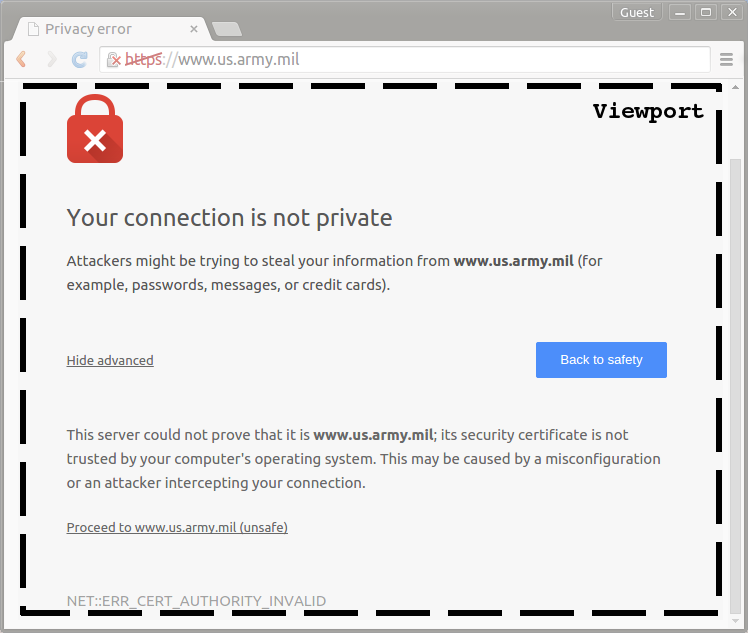}
\caption{The viewport component of a browser}
\label{fig:viewport}
\end{figure}

A few proposals have been recently advanced to minimise the 
participation of users in certificate validation and 
make security less reliant on user's choice \citeweb{TheChromiumBlog2011} \citeltex{rfc6797}.
Different browsers have adopted HSTS \citeltex{rfc6797}, a security mechanism originally conceived to thwart TLS stripping attacks \citeltex{marlinspike2009new,marlinspike2009more}. In a TLS stripping attack, the attacker forces the user to communicate via an HTTP connection although the web server supports HTTPS connections. HSTS-compliant browsers prevent ``unsecured'' HTTP connections to HSTS-compliant web servers. To do so, the web server sends the browser an HSTS header during a secured TLS session. Optionally, the HSTS header may include a list of the only certification authorities allowed to issue  the web server's certificate.  Then, the browser adds an internal policy stating that the concerned web server must be accessed via HTTPS only, and with a valid certificate. Thus, once the authentication of the web server succeeded, it shall not fail in the future. 

HSTS avoids user's participation in favour of a purely technical enforcement of security: if the certificate validation fails for a browser-known HSTS-complaint web servers, the browser shows the user an error message (not a warning) and aborts the connection.
Some browsers implement a pre-loaded whitelist of HSTS-complaint web servers to mitigate bootstrap attacks.
However, similar to the certification path validation, browsers implement HSTS differently from each other, as we shall see later.


\section{Safe Exam Browser} \label{sec:seb}
Safe Exam Browser (SEB) is a kiosk browser developed at ETH Zurich \citeltex{SBV+12}.  \index{kiosk browser}
Exam authorities can employ SEB at testing to thwart candidate cheating in computer-based and Internet-based exams. \index{Internet-based exams}
The main functionalities of SEB allow the candidate (a) to connect to a remote web-based exam system and (b) to access to a selection of third party applications and web sites during testing.
SEB prevents the candidate from accessing resources and utilities of the operating system, switching to undesired applications, quitting the browser any time, and opening a website not allowed by the authority. 

SEB turns the host computer device in a kiosk browser for exams by removing any other components of a browser but the viewport. It can be installed in a computer device either provided by the exam authority or personally owned by the candidate. In the latter scenario, the browser is secured by a signed and encrypted exam configuration. Moreover, SEB supports remote testing \index{remote testing} by means of a \emph{SEB Server} that provides more security on unmanaged computer devices thanks to screen recording and logging of all kinds of activities on the host computer. 
However, we stress that the threat model considered in this chapter does not concern malicious candidates or users. Rather, we focus on the security risks that a malicious observer can pose to the human principal via the browser. Similarly, screen recording and logs do cause concern about privacy, but this is a requirement not considered in this chapter.

Over the 10\% of all written exams \index{written exam} at ETH Zurich have been carried out using SEB in 2014. Besides, SEB has already been used for remote testing \citeltex{SBV+12}. The particular purpose of this browser of minimising the participation of the user makes interesting to check how this can affect certificate validation. \index{remote testing}

Since SEB is open source, we studied it by looking at its official documentation and source code. We consider the version 2.0 of the browser.


\section{Modern Browsers} \label{sec:browser}
We also consider the most popular browsers available nowadays.
According to StatCounter \citeweb{StatCounter}, the most used browsers are Firefox, Chrome, Internet Explorer, and Safari. Opera Mini is the most popular platform-independent browser \citeweb{OperaMini2014} and is available for many mobile devices. Its analysis is motivated because nowadays more and more users prefer to browse the Web with touchscreen mobile devices, and in particular Opera Mini dramatically reduces the amount of data transferred. In doing so, we aim to evaluate how such restrictions affect certificate validation.

\paragraph{Firefox.}
The inception of Mozilla Firefox originates from Netscape Navigator. According to StatCounter  \citeweb{StatCounter}, it is the third most popular browser over desktop, mobile, tablet, and console devices. Among the browsers we consider, Firefox seems to be the most complete: it supports HSTS, distinguishes two different certificate stores, and allows users to store server certificates either permanently or temporarily. Since Firefox is open source, we studied it by looking at its official documentation and source code. We consider Firefox version 36.0.4.
\paragraph{Chrome.}
Although Google Chrome is the youngest  browser we consider, it is the most popular. It was the first browser to support HSTS policies, and adopts different certificate stores depending on the operating system underlying the browser. Chrome is based on the Chromium open source code with minor differences. We analysed Chrome inspecting the Chromium source code and using empirical tests.
This work considers Chrome version 41.0.
\paragraph{Internet Explorer.}
Microsoft Internet Explorer was the most popular browser for years, and has been overtaken by Chrome only recently. Currently, it does not support HSTS, which is however planned to be implemented soon \citeweb{iehsts14}. Internet Explorer is available only for Windows operating systems, and uses their certificate stores. Since Internet Explorer is closed source, we relied on empirical tests, also supported by network analysers. The version of Internet Explorer analysed is 11.0.16.
\paragraph{Safari.}
Safari is the browser developed by Apple and is popular on the company devices. It supports a  HSTS and, similarly to Firefox, distinguishes two different certificate stores allowing users to store server certificates. Safari is available only on Apple's operating systems and is closed source. We thus analysed it empirically and assisted by network analysers. We studied Safari 8.0.3.
\paragraph{Opera Mini.}
Opera Mini is used by more than 244 million people per month and is particularly popular in emerging countries, according to the company data \citeweb{OperaMini2014}. It aims at being the most lightweight  browser for any java-capable device. To do so, the browser uses Opera proxy servers and compression technologies to reduce traffic and speed up page display. Thus, although  communications to the  proxy server are encrypted, there is no end-to-end TLS encryption between the browser and the web server. We analysed the (closed-source) browser empirically, using a java emulator with network analysers. 
In this work we consider the version 7.6.4.

\subsection{Private Browsing}
All the browsers we consider in this chapter support \emph{private browsing}\footnote{SEB supports private browsing sessions only}, which is a privacy protection mode that disables browser's history and cache. 
Private browsing gives the user no guarantee about Internet privacy as an eavesdropper can still learn the websites visited by the user. Rather, private browsing protects user's privacy only over the data stored in the local machine.
Each browser implements private browsing differently, and this is usually done by inhibiting different features  \citeltex{ABJ+10}. 
Private browsing is becoming increasingly popular among users \citeweb{Elie12}, so we consider it in our analysis.
In particular, it is interesting to study how browsers balance security (e.g., HSTS, user's approved certificate) with privacy technologies.
Concerning certificate validation, one would expect no differences between private and classic browsing. However, as we shall see later, this is not true.

\paragraph{Technical notes.}
We tested the certificate validation in SEB, Firefox, Chrome, and Internet Explorer using an Intel Core i7 3.0 GHz with  8 GB RAM running Windows 8.1 on a virtual machine.
We tested certificate validation on Opera Mini inside MicroEmulator, a Java implementation of JavaME, and analysed certificate validation on Safari on an Apple MacBook Pro Intel Core i5 2.5 GHz with 8 GB RAM running OS X Yosemite 10.10.
The network analysers we used to understand how certificate validation works, especially on closed-source browsers, are ``Wireshark'' \citeweb{wireshark}, ``mitmproxy'' \citeweb{mitmproxy}, and ``Charles'' \citeweb{charles}. They ran on a second virtual machine with Linux Ubuntu 14.04 and intercepted any traffic between a server and the target browser on the main virtual machine.



\section{Modelling Certificate Validation} \label{sec:modellingsocio}
As seen in Section~\ref{sec:positioning}, the certificate validation is a protocol rather than an algorithm,  which may involve users. Therefore, we refer to is as a ceremony.

The certificate validation  ceremony includes a user, the browser, and the web server.
In this section, we provide a formalisation of the ceremony for each browser. However, finding the right formalism for the socio-technical analysis is not easy.
The standard notation to describe security protocols is the Alice-and-Bob notation \citeweb{LSV}. Although this notation provides a simple and clear description, it comes with some limitations: it cannot capture fork, join, and  branching, which are essential, for example, to model  multiple user's choices in the certificate validation.
Flowcharts offer a graphical description and look suitable to describe browsers and algorithms in general, but are less appropriate for the description of  protocols: we have the three roles of  browser, user,  and server, and we need to detail the messages that the roles exchange. Message sequence charts extend flowcharts to the domain of protocols, emphasising the interaction among the roles. 
Although message sequence charts have been extended with formal semantics \citeltex{MR94}, they still have the same limitation of the Alice-and-Bob notation.

Thus, we choose the semi-formal and graphical description of UML Activity Diagram \citeweb{OMG2011}. An UML activity diagram is a graphical scheme that defines the activities needed to meet a given functionality. UML activity diagrams are made of shapes  that model choices, interactions, and concurrency. Table \ref{table:shape} recalls the main UML activity diagram's shapes.

The contribution of activity diagrams is threefold. First, they give an intuitive representation of a protocol session, highlighting the mechanisms used on each role. Second, they can represent parallel actions (fork/join) and multiple choices (branching).
Third, they can be easily translated in a fully formal language, thanks to their semi-formal semantics \citeltex{Abdelhalim2012}. In particular, we translate activity diagrams to CSP\# \citeltex{SunLDP09}, a modelling language that enriches the high-level operators of \ac{csp} (e.g., choices, interleaving, hiding, etc.) with low-level programming constructs (e.g., arrays, while, etc.).
The  \acs{csp}\# code is then fed to an automatic tool that checks whether the input model guarantees a set of properties. The code can be found in Appendix \ref{app:csp} and in  \citeweb{thesiscode}. 

\subsection{UML Activity Diagrams for Certificate Validation}\label{sec:ad}
\begin{table}[!h] 
\begin{center}
{\small
\begin{tabular}{|>{\centering\arraybackslash}m{3cm}|>{\centering\arraybackslash}m{7cm}|}
\hline
{\bf Shape}  & {\bf Description} 
\\\hline

\vspace{1ex}
\begin{tikzpicture} 
\draw[fill=-red!80!green,rounded corners=2pt] (0.1,0.1) rectangle (1.5,0.5);
\end{tikzpicture}
& Activity node  
\\\hline

\vspace{1ex}
\begin{tikzpicture} 
	\draw[fill=-red!80!green] (0.1,0.1) rectangle (1.5,0.5);
\end{tikzpicture} 
& Object (datastore) node\\\hline

\vspace{1ex}\begin{tikzpicture} 
\tiny{\umlNarynode[x=0,y=0,name=final, fill=-red!80!green]{}}
\end{tikzpicture} & 
Decision or merge point
\\\hline

\vspace{1ex}\begin{tikzpicture}
\node(G)[rectangle, fill=black, text width=1cm]{};
\end{tikzpicture}
& Input and output objects of activities
\\\hline

\vspace{1ex}\begin{tikzpicture}
\node(G)[rectangle,draw=black,text width=1cm, rotate=90]{};
\end{tikzpicture}
& Distinguishing the activities by role
 \\\hline
 

\vspace{1ex}\begin{tikzpicture} 
\tiny{\umlstateinitial[]}
\end{tikzpicture} & 
Initial node\\\hline

\vspace{1ex}\begin{tikzpicture} 
\tiny{\umlstatefinal[]}
\end{tikzpicture} & 
Activity final node\\\hline

\vspace{1ex}\begin{tikzpicture} 
\tiny{\umlstateexit[]}
\end{tikzpicture} & 
Activity final node within a role
\\\hline

\begin{tikzpicture}
\draw[->,line width=0.5pt] (0,0) to (1,0);
\end{tikzpicture} & 
Control flow within a role 
\\\hline

\begin{tikzpicture}
\draw[->,line width=1pt] (0,0) to (1,0);
\end{tikzpicture} & 
Control flow among different roles
\\\hline

\begin{tikzpicture}
\draw[dashed, ->,line width=0.5pt] (0,0) to (1,0);
\end{tikzpicture} &
Object flow from or to an object node
\\\hline
\end{tabular}
}
\caption{Description of the shapes defined in UML Activity Diagram.}
\label{table:shape}
\end{center}
\end{table}

We build nine activity diagrams that model the certificate validation ceremonies for the browsers both in classic and private browsing.
Although we consider six browsers, we do not model the classic browsing for SEB since it
supports only private browsing, and we do not model the private
browsing of Internet Explorer and Opera Mini because they are identical to the corresponding classic browsing modes.  

The UML Activity Diagram for SEB is in Figure \ref{fig:seb}; the diagrams modelling Firefox in classic and private browsing are respectively
in Figure \ref{fig:ff} and in Figure \ref{fig:ffpb}; the diagrams modelling Chrome in classic and private browsing are respectively
in Figure \ref{fig:ch} and \ref{fig:chpb}; the  diagrams for Safari are in Figure \ref{fig:sa} and \ref{fig:sapb}; the diagrams for Internet Explorer and Opera Mini are respectively in Figure \ref{fig:ie} and \ref{fig:om}.

The UML activity diagrams include the functionalities limited to describe how
each browser achieves certificate validation. Each  activity diagram has
four columns, each representing a communicating role. From left to right we have
the \emph{user}, the \emph{browser user interface}, the \emph{browser
engine}, and the \emph{server}. Each role begins with a filled dot that points to their first activity. We assume that the browser bootstraps with the start web page, which is displayed in the browser user interface. Next, the browser user interface can load a web page because the user types a URL or clicks on an active link in the currently displayed web page. 
A label close to a thick arrow defines the object exchanged between activities.
Activities may need to get access to datastores, which are represented as
object nodes.

\begin{figure}[h!]
  \centering\includegraphics[scale=.69]{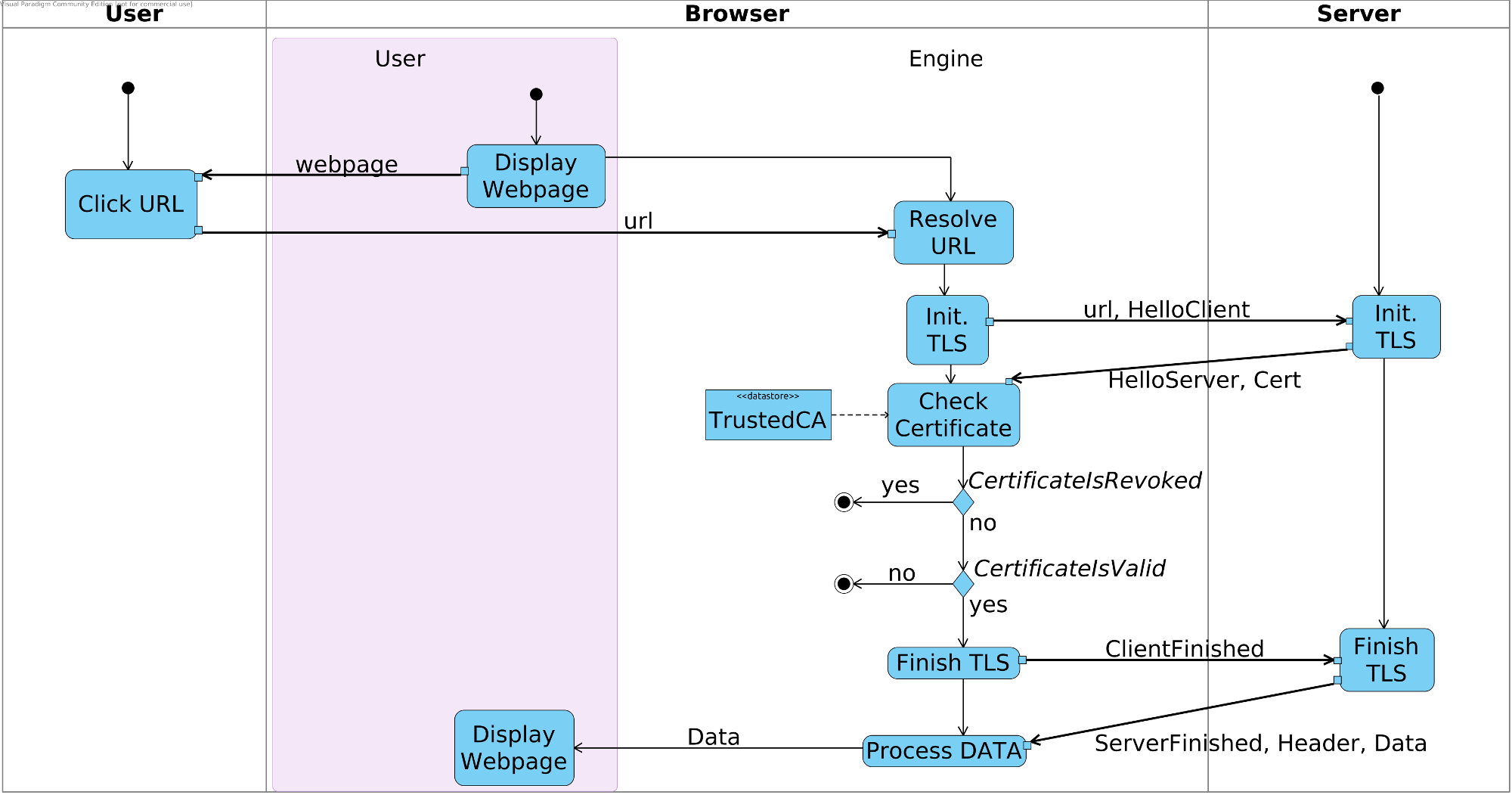}
\caption{Activity diagram for certificate validation in SEB}
\label{fig:seb}
\end{figure}

\begin{figure}[h!]
  \centering\includegraphics[scale=.69]{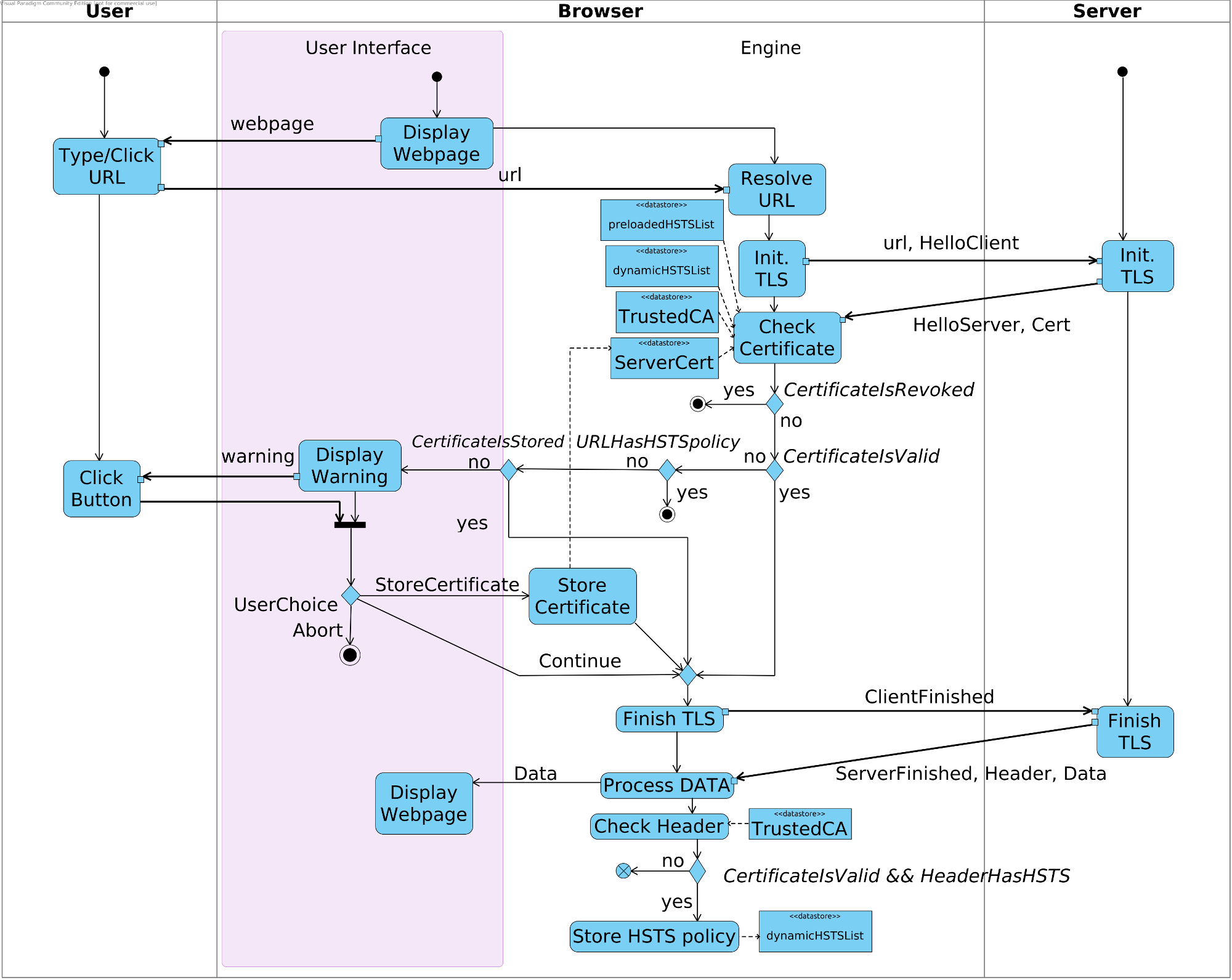}
\caption{Activity diagram for certificate validation in Firefox}
\label{fig:ff}
\vspace*{13mm}
  \centering\includegraphics[scale=.69]{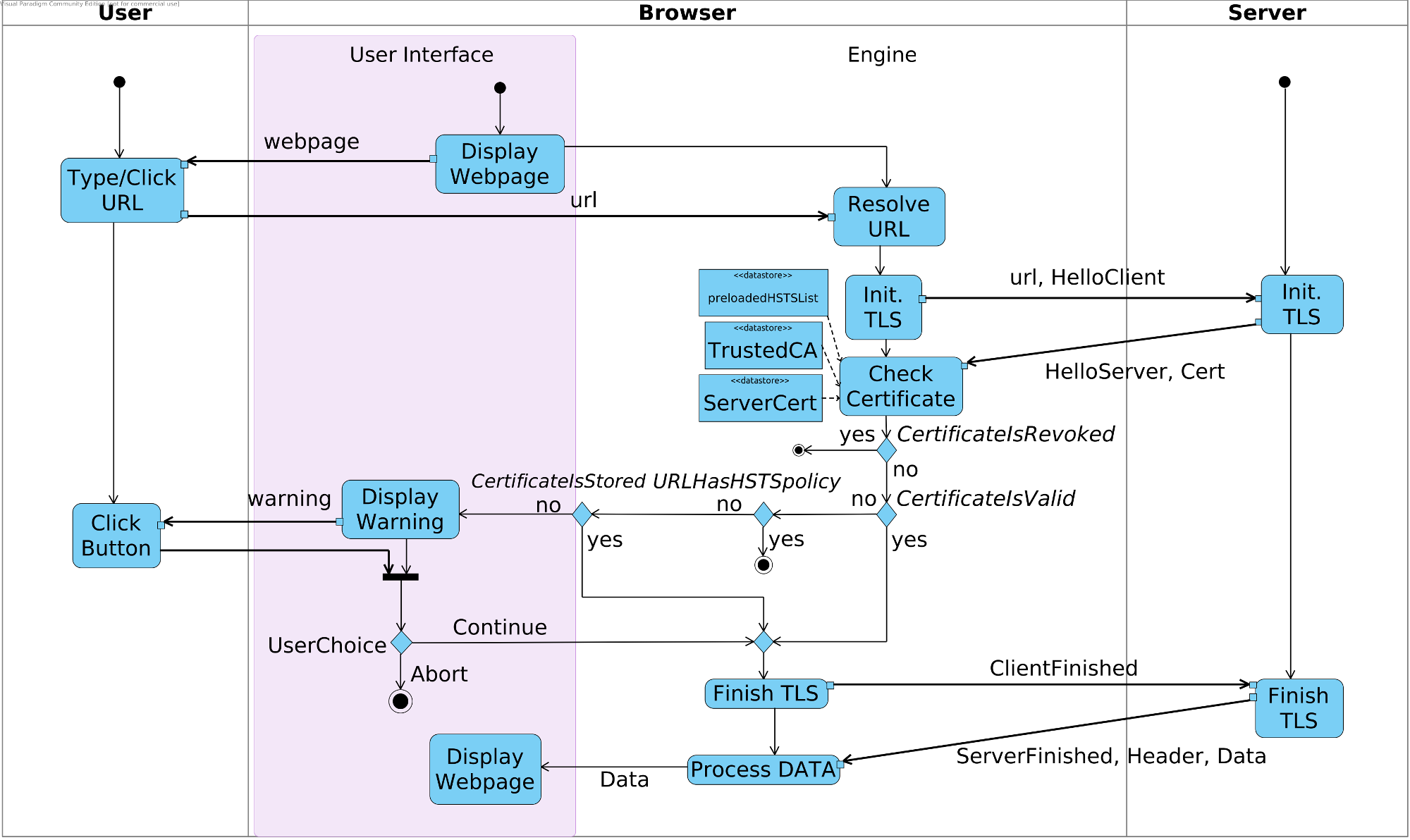}
\caption{Activity diagram for certificate validation in Firefox in private browsing}
\label{fig:ffpb}
\end{figure}

\begin{figure}[h!]
  \centering\includegraphics[scale=.69]{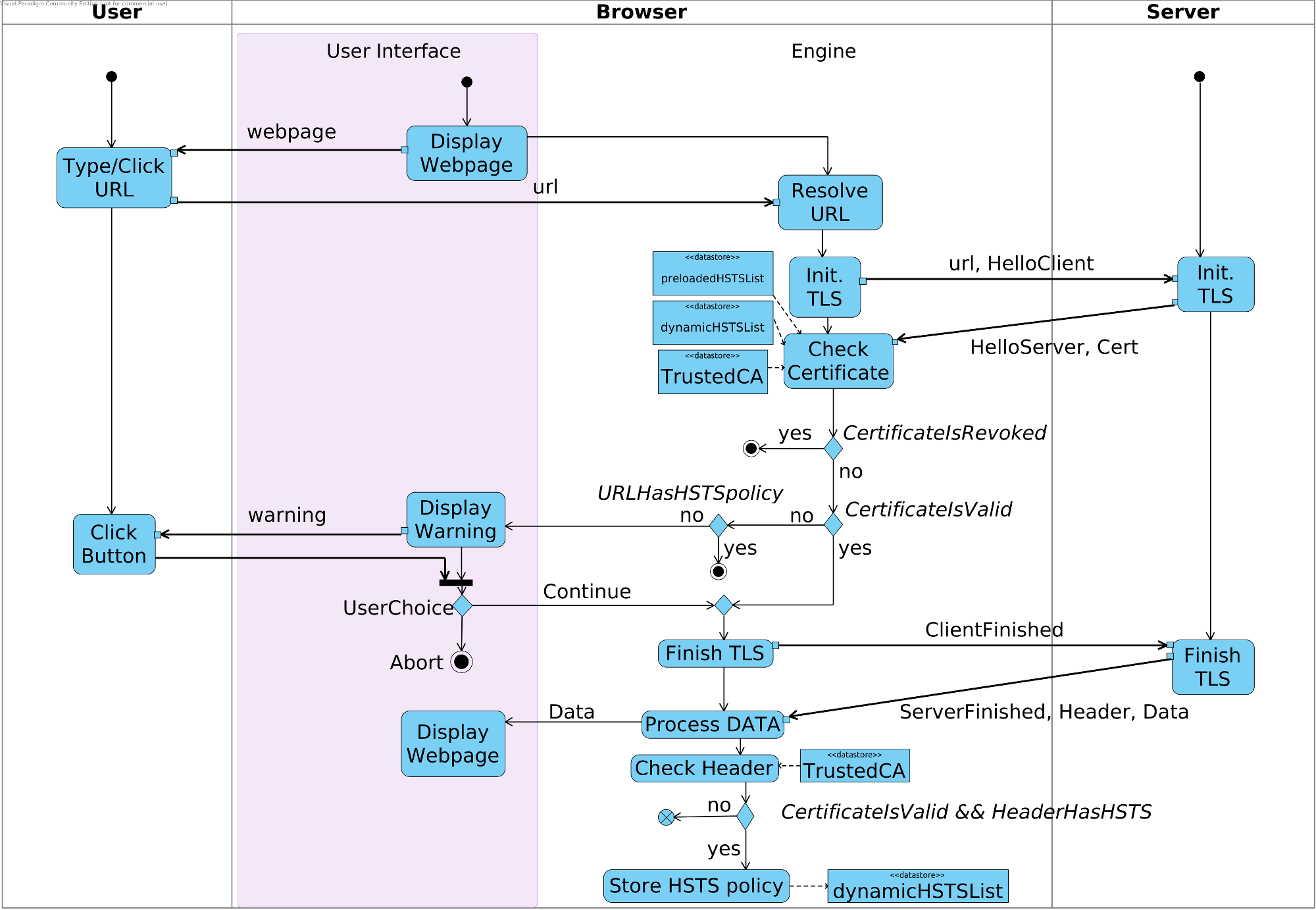}
\caption{Activity diagram for certificate validation in Chrome}
\label{fig:ch}
\vspace*{25mm}
  \centering\includegraphics[scale=.69]{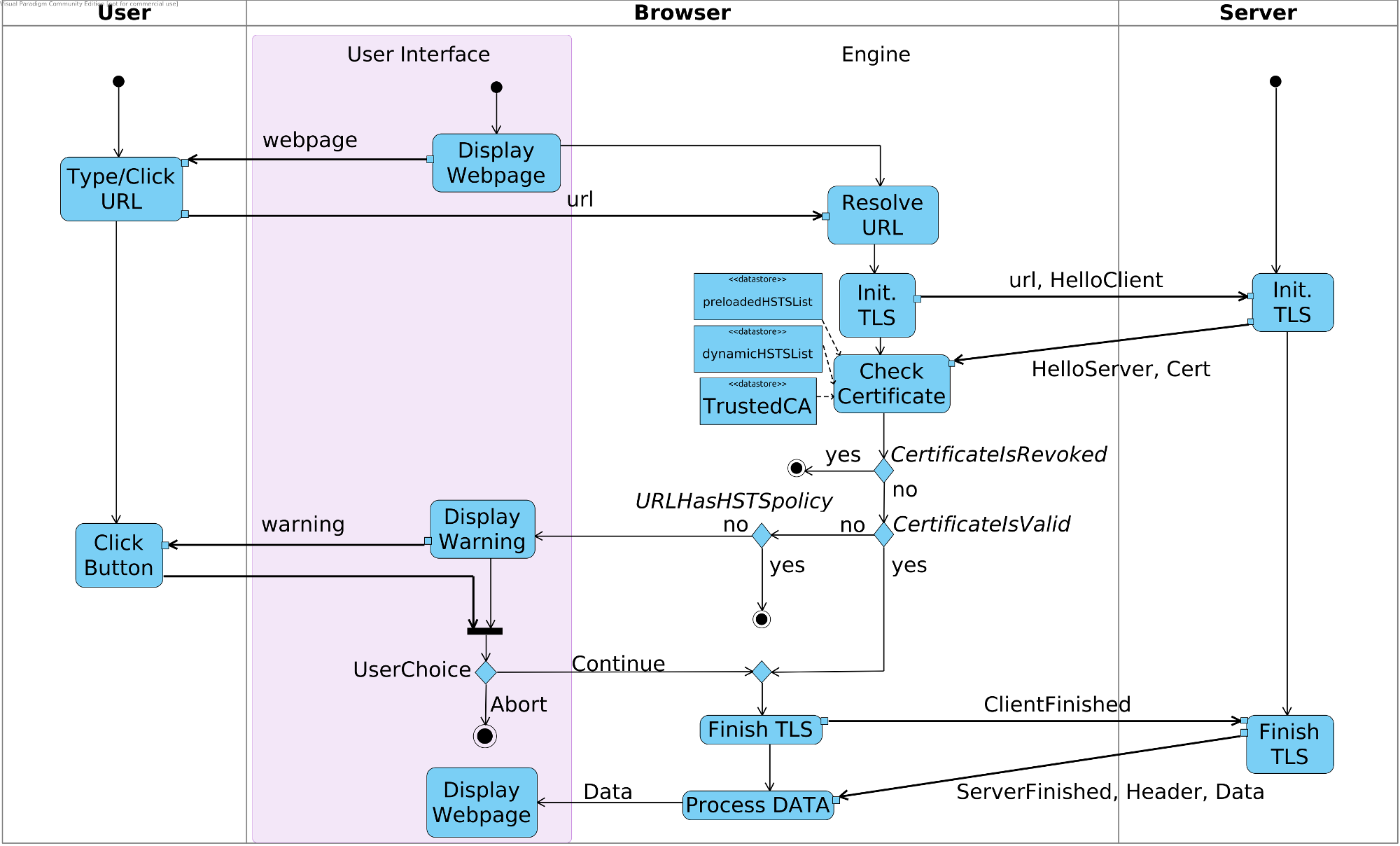}
\caption{Activity diagram for certificate validation in Chrome in private browsing}
\label{fig:chpb}
\end{figure}


\begin{figure}[h!]
  \centering\includegraphics[scale=.693]{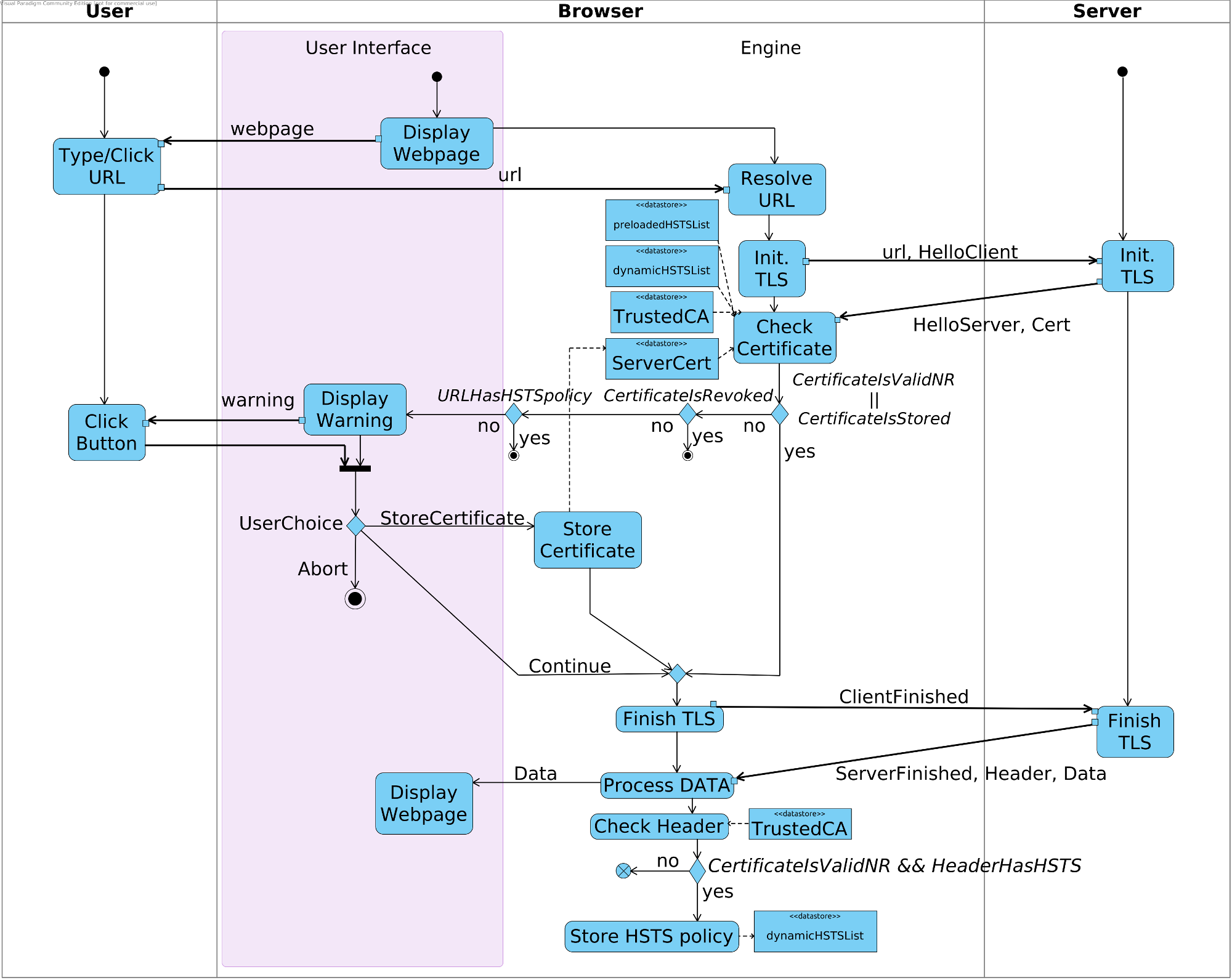}
\caption{Activity diagram for certificate validation in Safari}
\label{fig:sa}
\vspace*{5mm}
  \centering\includegraphics[scale=.693]{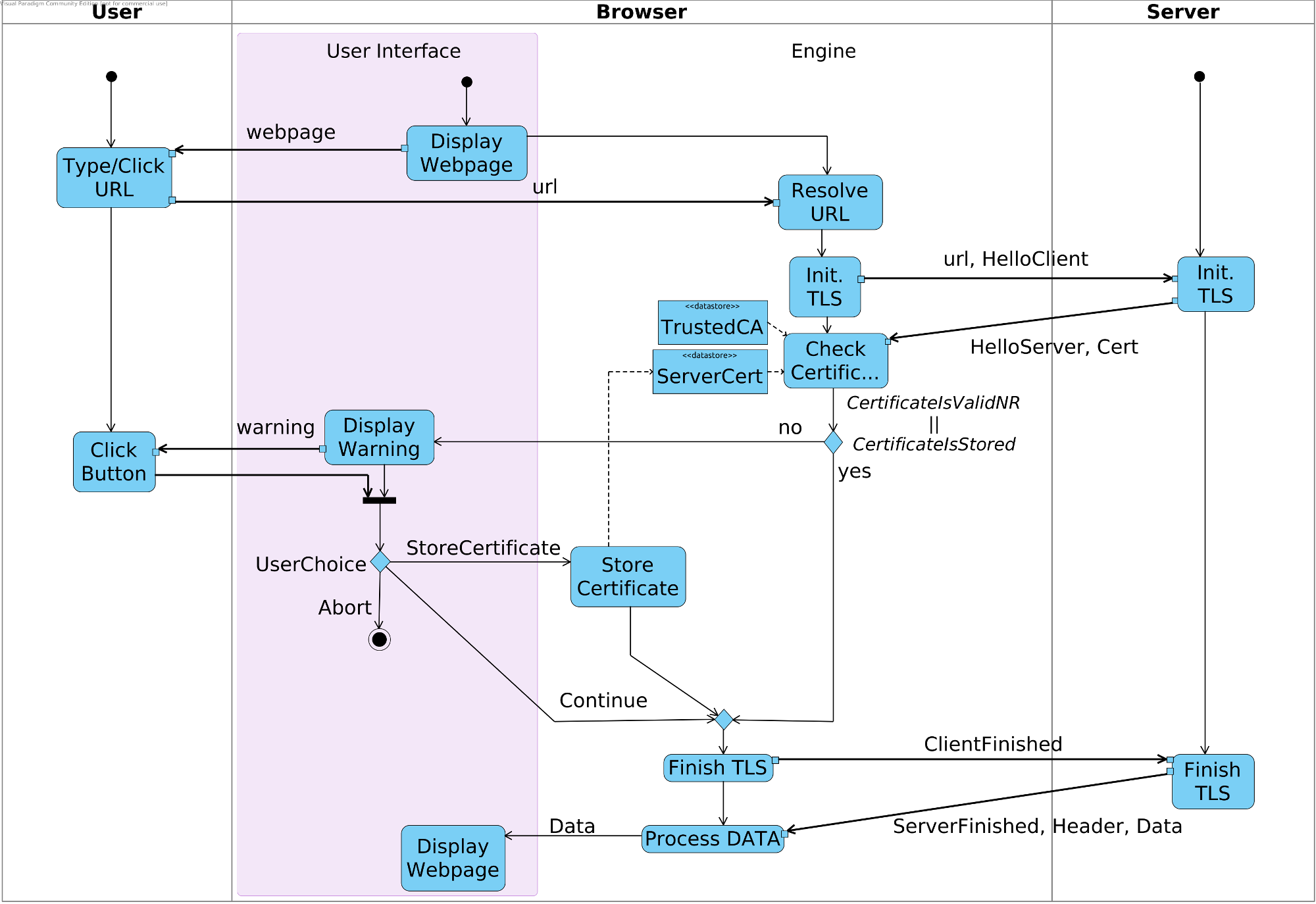}
\caption{Activity diagram for certificate validation in Safari in private browsing}
\label{fig:sapb}
\end{figure}

\begin{figure}[h!]
  \centering\includegraphics[scale=.695]{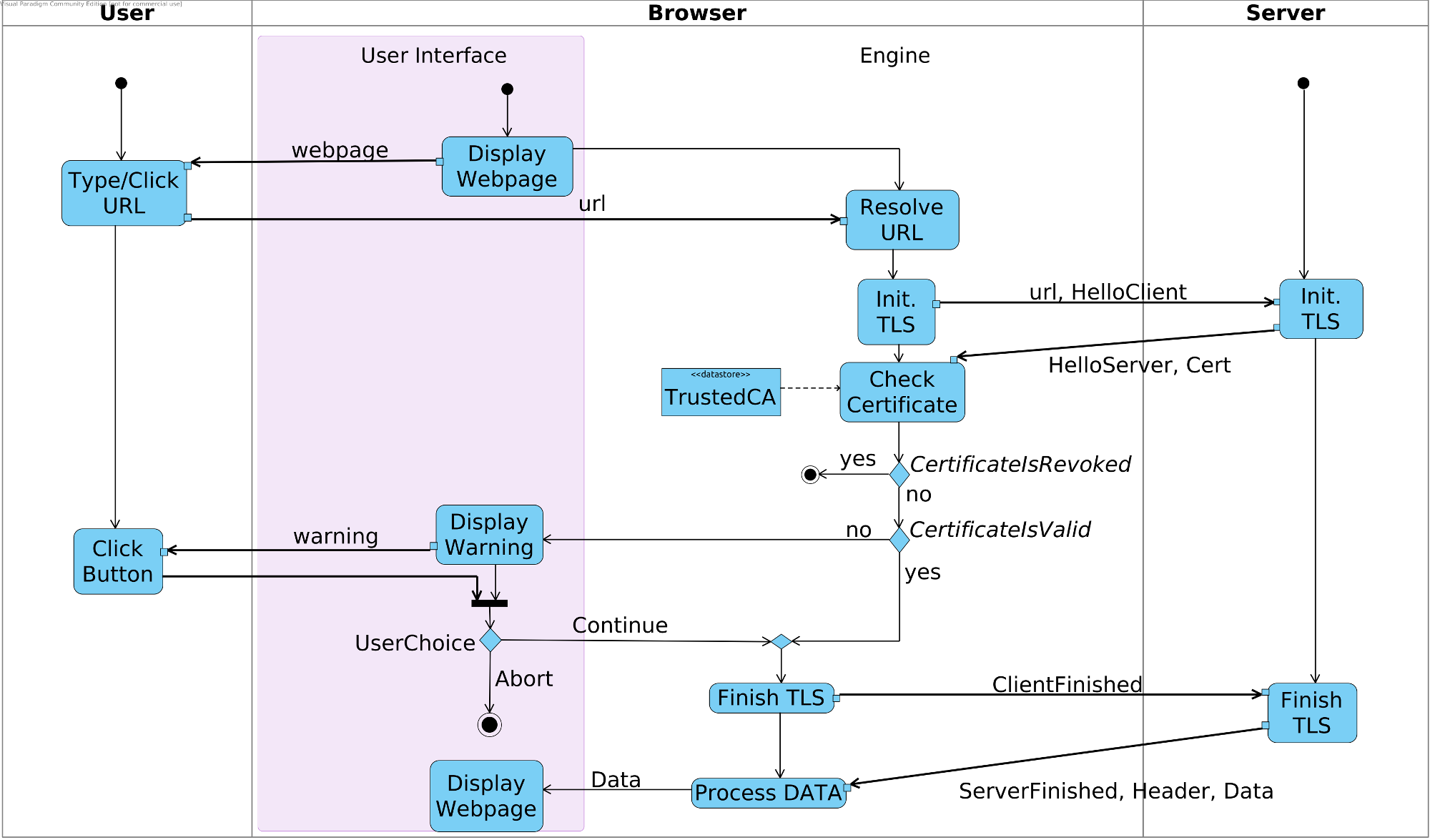}
\caption{Activity diagram for certificate validation in Internet Explorer}
\label{fig:ie}
\end{figure}
\vspace*{5mm}
\begin{figure}[h!]
  \centering\includegraphics[scale=.695]{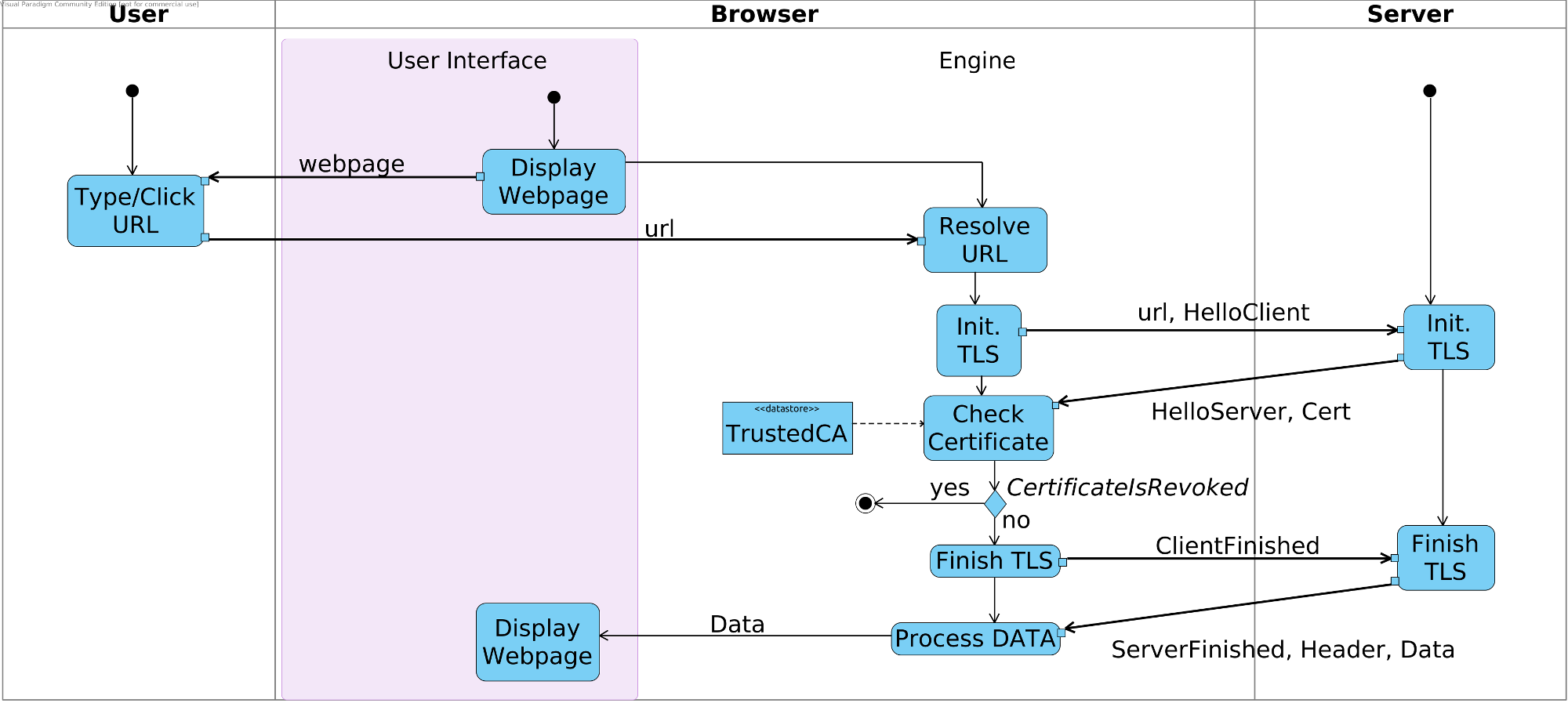}
\caption{Activity diagram for certificate validation in Opera Mini}
\label{fig:om}
\end{figure}

\subsection{Description of the Main UML  Activities}
For each role involved in the certificate validation, we describe the principal activities and checks that concern their UML activity diagrams.
In the remainder, we denote UML activities in $\UMLactivity{serif}$ and UML decisions in $\UMLcheck{italics}$.

\paragraph{User.} 
A user is modelled as a non-deterministic entity. This means that the
user may potentially choose any of the paths of interaction that the browser
offers, namely $\UMLactivity{Type/Click~URL}$ or $\UMLactivity{Click~Button}$. It logically follows that this is the weaker assumption about the user
skills: a ceremony that is secure for a non-deterministic user will be secure
for any user. Modelling the user as a non-deterministic entity is the best
approximation that we can envisage at present. A more elaborate formal model that captures the
complexities of user behaviour is currently an open issue. Moreover, the feasibility of such a model is often questioned \citeltex{LMM99}. 

\paragraph{Browser.} The representation of the browser is split into user interface and engine. The former has the activities of $\UMLactivity{Display~Webpage}$ and $\UMLactivity{Display~Warning}$. The engine normally begins with the activity $\UMLactivity{Resolve~URL}$ and then starts the TLS handshake with the activity $\UMLactivity{Init.TLS}$.
The engine has the fundamental check that concerns certificate validation, namely $\UMLcheck{CertificateIsValid}$. The check follows the activity $\UMLactivity{Check~Certificate}$, which verifies if the issuer is known, the certificate is not expired, and the certificate subject matches the intended one. Moreover, the activity may verify additional information with the assistance of datastores, such as $\UMLactivity{preloadedHSTSList}$, which stores the HSTS policy whitelist, or $\UMLactivity{ServerCert}$, which stores the user's approved certificates.
Note that the check $\UMLcheck{CertificateIsRevoked}$ is subject to the revocation protocol implemented by the browser.
If the flow of certificate validation has not been aborted, the engine of the browser concludes a successful TLS handshake with the activity $\UMLactivity{Finish~TLS}$. Then, it runs the activity $\UMLactivity{Process~DATA}$ to get the data encrypted by the server. It possibly verifies if the data contains new information about HSTS with the activity $\UMLactivity{Check~Header}$, which may lead the browser to store a new HSTS policy with the activity $\UMLactivity{Store~HSTS~policy}$.

\paragraph{Server.} We purposely consider only two activities of the server. We aim in fact to focus more on the model of the browser rather than that of the web server. The server starts the TLS handshake on its side with the activity $\UMLactivity{Init.TLS}$, and concludes it with the activity $\UMLactivity{Finish~TLS}$.  As we shall see below, a
server may be corrupted by the attacker and may deviate from the supposed activity flow.

Before  formally analysing the socio-technical properties on each browser, we note that the activity diagrams already offer some interesting insights.
This may support the case that graphical models may be more insightful than formal ones.

As expected, SEB minimises the participation of the user also in certificate validation. In fact,
Firefox, Chrome, Internet Explorer, and Safari involve the user more than SEB and Opera
Mini. In particular, Firefox and Safari allow the user to store a
server certificate either permanently or temporarily. However there is a
fundamental difference between Firefox and Safari: the first prioritises the
HSTS policy check (\ie $\UMLcheck{URLHasHSTSpolicy}$) over user's choices, the latter prioritises the user's stored server certificate (\ie $\UMLcheck{CertificateIsStored}$) over the HSTS policy.
Also the activity diagrams of Opera Mini and SEB show a fundamental difference,
although they look generally similar: in Opera Mini, unless the certificate is revoked,
the certificate validation  always leads to a successful termination of the TLS
handshake; in SEB the certificate validation may instead lead to aborting the
handshake according to the check $\UMLcheck{CertificateIsValid}$.

Some browsers show also differences between their classic and private
browsing.  Firefox involves the user and implements HSTS differently in private
browsing: the user cannot store a server certificate, and HSTS policies stored
in earlier sessions are not considered.  Also Chrome has a different
implementation of HSTS in private browsing: no new HSTS policies can be
permanently stored while the ones stored in previous classic sessions are
considered.  More surprisingly, Safari neither permanently stores HSTS policies
in private browsing nor considers the ones previously stored in classic
sessions.

This brief informal analysis corroborates the statement that certificate validation differs among browsers. In the next section, we see in depth by means of a formal approach how these differences affect the socio-technical security aspect of browser certificate validation.



\section{Socio-Technical Formal Analysis} \label{sec:analysisocio}
We use model checking to formally analyse certificate validation. We provide a systematic method to translate the UML activity diagrams to a formalisation in  \ac{csp}\# that is amenable to automatic validation by means of \acs{pat}.

From a security perspective, we define a threat model and specify the socio-technical properties that concern certificate validation in \ac{ltl}.

\subsubsection*{Threat model}
We consider a \ac{mitm} attacker who wants to violate server authentication. He partially controls the network and can divert the browser's $\UMLactivity{Init.TLS}$ request to a corrupted server that the attacker owns. The attacker can generate a self-issued certificate, namely a new certificate signed by himself.
He may alternatively have a valid certificate, namely signed by a certification authority, for a server that he controls. The attacker can interpose between the browser and the honest server that the user requests, and can replace the server certificate with one of his own. The sole limitation is that the attacker cannot sign a certificate on behalf of a certification authority.

\subsection{Socio-Technical Security Requirements}\label{sec:informal}
We select five different socio-technical requirements that we deem relevant. They bind elements that span from TLS session identifiers to user choices. They  aim to demonstrate how the technical mechanisms implemented in browsers interrelate the  user choices with the overall system security.
We first give informal and intuitive descriptions of the requirements and then express them formally.
\begin{prop}[Warning Users]\label{prop1}
A user whose browser receives an invalid certificate is warned before the browser completes the session.
\end{prop}

This requirement is about a browser's warning the user that the certificate of the required server is invalid.
As explained in Section \ref{sec:cv}, a certificate can be invalid for different reasons, each of them being more or less risky for the user. For example, some circumstances observe a server 
that self-issues its certificate, others conceal an attacker who attempts a MITM by injecting a fake certificate of his own.

\begin{prop}[Storing Server Certificates]
A~user who approves a server certificate via a browser is protected from man-in-the-middle attacks on future sessions with the same server  via the same browser.
\end{prop}

This requirement is about how storing server certificate relates with MITM protection.
When browsers receive an invalid certificate, they may still allow the user to store it. If the user assumes that a certificate in fact is trustworthy, one would expect that future sessions with the same server will be protected from MITM attacks. We shall see in the discussion that follows that this is not true for all browsers.

\begin{prop}[Applying HSTS User Security]
A user who accesses a $ $ server via a browser that receives a valid certificate and an HSTS header is protected from man-in-the-middle attacks on future sessions with the same server via the same browser.
\end{prop}
This requirement stands on a different scenario from that of the previous one, although their conclusions are equal. This scenario sees an HSTS-compliant server who sends a valid certificate to the browser  the user is using. However, it remains to be checked whether the browser is HSTS-compliant too, in which case it whitelists the server because its certificate is validated once.

\begin{prop}[Applying HSTS Bootstrap]
A user who accesses a server that is pre-loaded on the browser's HSTS list is protected from man-in-the-middle attacks on future sessions with the same server via the same browser.
\end{prop}
This requirement concerns the relation between pre-loaded whitelist and MITM protection.
In particular, we check whether HSTS-complaint browsers correctly implement the pre-loaded server whitelist to mitigate bootstrap attacks, namely MITM attacks at the first server visit.

\begin{prop}[Learning from Server Certificate History]
A~user who $ $ completes a  TLS session with a server via a browser receiving an invalid certificate, and then completes another session with the same server via the same browser receiving a valid certificate is warned by the browser about the risk of man-in-the-middle attack.
\end{prop}

This last requirement aims at checking whether the browser informs the user that a MITM attack may have occurred in a \emph{previous} TLS session. For example, if one considers a session where the browser receives an invalid certificate, then the browser may warn the user about this (according to Definition \ref{prop1}). If in a subsequent session the browser receives a valid certificate for the same web page, 
it may be the case that the former session experienced a MITM attack,  hence the browser warns the user.

\subsection{Automated Verification}
Our automatic analysis relies on PAT, a model checker for the analysis of concurrent and real-time systems. Its layered design separates modelling languages from model checking algorithms, thus supporting different languages via different algorithms. It supports a range of application domains that span from bio-systems to security protocols.

PAT supports an enriched version of CSP, called CSP\#. The extensions of CSP\# include low-level constructs that offer a connection between data states and executable operations. Moreover, PAT supports user defined C\# functions and data types that can be used directly in  CSP\# code as external libraries. We take advantage of this by defining an advanced data structure to model the certificate stores of browsers.
PAT can model safety (i.e., bad things do not happen) and liveness (i.e., good things eventually happen) properties.  In PAT, a requirement can be specified in the same language used to specify the system model, i.e., CSP,  or in a temporal logic language. In our case, we use Linear Temporal Logic to specify our requirements, and resort on PAT's model checking techniques for their verification.
PAT supports symbolic and explicit model checking. Since we specify our requirements in LTL, we rely on the dedicated temporal-logic model checker of PAT. In particular, we use  depth-first-search as searching strategy algorithm to check whether a property is valid. If the property turns out not to be valid, we use  the breadth-first-search algorithm to find the shortest  witness trace that falsifies the property.
We choose to use PAT over the more popular refinement checker FDR \citeltex{Roscoe}, because PAT supports low-level constructs written in CSP\# and the specification of our requirements in LTL. However, we expect the same analysis results using another tool such as FDR.    
\begin{table}[h]
\begin{center}
{
\begin{tabular}{|c|l|}
\hline
{\bf \ac{csp}\#}  & {\bf Description} \\ \hline
$\mathit{Stop}$ & deadlock  \\ \hline
$\mathit{Skip}$ & termination \\ \hline
$a \rightarrow P$ &  event prefixing \\ \hline
$c?[b]x \rightarrow P$ &  input communication \\ \hline
$c!e \rightarrow P$ &  output communication \\ \hline
$P[*]Q$ &  external choice \\ \hline
$P; Q$ &  sequential composition \\ \hline
$P$ \textbackslash $~A$ &  hiding \\ \hline
$x:=e$ &  assignment \\ \hline
$\mathit{if}~b~\mathit{then}~P~\mathit{else}~Q$ &  conditional choice  \\ \hline
$P~||~Q$ &  parallel composition \\ \hline
$P~|||~Q$ &  interleaving \\ \hline
$P~\mathit{interrupt}~Q$ &  hiding \\ \hline
\end{tabular}
}
\caption{\ac{csp}\# syntax.}
\end{center}
\label{tab:csp}
\end{table}

\paragraph{Mapping UML activity diagrams to \ac{csp}\#.} 
We systematically generate the \ac{csp}\# code from UML activity diagrams, and then validate the code with PAT.
Such generation is quite straightforward because we define a map between the shapes of UML activity diagrams and the \ac{csp}\# syntax, which is outlined in Table~\ref{tab:csp}.
More precisely: 
\begin{itemize}
\item the activity node maps to the \ac{csp}\# event; 
\item the object (datastore) node maps to the \ac{csp}\# array; 
\item the decision point maps to \ac{csp}\# conditional choice; 
\item input and output objects of activities map to the \ac{csp}\# values of input and output communications; 
\item activities roles are distinguished within a  \ac{csp}\# process; 
\item the beginning of the activity flow is a \ac{csp}\# event;
\item the ending of the activity flow maps to \ac{csp}\#  termination;  
\item the flow of activities within a role maps to the \ac{csp}\# event prefixing; 
\item the flow of activities among different roles maps to \ac{csp}\# input and output communications; 
\item the flow of data of an object node maps to the \ac{csp}\# assignment.
\end{itemize}


\subsubsection*{Socio-technical requirements in LTL}
PAT fully supports the LTL syntax to define a requirement about the system behaviour. An LTL formula is defined by \emph{events}, \emph{predefined propositions},  logical operators, and modal operators. An LTL formula can be evaluated over an infinite sequence of truth evaluations and paths. Thus, the assertion is true if every execution of the system  satisfies the formula.

Although LTL defines five different modal operators, our requirements can be expressed using the combination of two operators only: $\square$, whose semantics is that the formula holds on the current state and the entire subsequent path; $\bigcirc$, whose semantics is that the formula holds at the next state on the path. The combination  $\bigcirc\square$ expresses that the formula holds on the entire subsequent path (not necessarily in the current state).

The propositions of our LTL formulas refer to ``choices'' (\eg $\UMLcheck{CertificateIs}$\-$\UMLcheck{Valid}$) and  to the activities (\eg $\UMLactivity{Init.TLS}$) of our UML activity diagrams. 
In the following definitions, we employ the same font styles used in the activity diagrams consistently.  Those predicates evaluate true in states, respectively,
where that choice has been selected and where that activity is executed with
success.  Also, our requirements include three additional \ac{ltl} predicates,
which we code in \ac{csp}\# as macros, and one more event, namely:

\begin{itemize}
\item $\LTLprop{UserWantsS}$ is true when a user wants to visit an
honest server, namely when she types a URL or clicks a link that points to a server
not corrupted by the attacker; 

\item $\LTLprop{AuthFail}$ is true when a 
\ac{mitm} attack succeeds, namely when the browser completes the TLS session with the attacker; 

\item $\LTLprop{Preloaded}$ is true when an honest server is in the
preloaded HSTS list of the browser; 

\item $\UMLactivity{ServerFinished.HSTS.Data}$ is true when the server sends
that message to the browser at the end of the TLS handshake.  
\end{itemize}

\begin{asse}[Warning Users]
\begin{equation*}
\square((\UMLactivity{Finish TLS} \wedge \lnot~\UMLactivity{DisplayWarning})
\implies \UMLcheck{CertificateIsValid})
\end{equation*}
\end{asse}

This requirement says that it is always the case that, when the browser concludes the TLS session without warnings, the certificate must have been valid.

\begin{asse}[Storing Server Certificates]
\begin{multline*}
\square((\UMLcheck{CertificateIsStored} \wedge 
\UMLcheck{UserWantsS} \wedge \UMLactivity{DisplayWebpage} 
\wedge \lnot~\LTLprop{AuthFail})\implies\\
\bigcirc \square(\LTLprop{UserWantsS}\implies
\lnot~\LTLprop{AuthFail}))
\end{multline*}
\end{asse}

This requirement signifies that it is always the case that if the user visited a web page whose authentic certificate was stored, then the user can safely visit the web page in next sessions as the browser precludes MITM attacks.

\begin{asse}[Applying HSTS User Security]
\begin{multline*} 
\square((\UMLcheck{CertificateIsValid} \wedge 
\UMLactivity{ServerFinished.HSTS.Data} \wedge \LTLprop{UserWantsS})\implies\\
\bigcirc \square(\LTLprop{UserWantsS} \implies \lnot~\LTLprop{AuthFail}))
\end{multline*}
\end{asse}

This requirement is structured as the previous requirement, and can be interpreted
similarly, but it is about the \acs{hsts} policy. It says that it is always the case that if the
web page is HSTS complaint, then its certificate is valid and the user can safely visit the web page in the next sessions as the browser precludes MITM attacks.
               
\begin{asse}[Applying HSTS Bootstrap] $$
\square(\emph{Preloaded} \implies(\LTLprop{UserWantsS} \implies
\lnot~\LTLprop{AuthFail}))
$$ \end{asse}

This requirement also relates to the  \ac{hsts} policy. It expresses that it is always the case that if the web page is preloaded in browser's  \ac{hsts} whitelist, then the user can safely visit the web page as the browser precludes MITM attacks.

\begin{asse}[Learning from Server Certificate History]\noindent
\begin{multline*} 
\square((\UMLactivity{Finish TLS} \wedge \lnot~\UMLcheck{CertificateIsValid} \wedge \LTLprop{UserWantsS})\implies\\
\bigcirc \square((\UMLactivity{Finish TLS} \wedge \UMLcheck{CertificateIsValid} \wedge \LTLprop{UserWantsS})\implies\\
\UMLactivity{DisplayWarning}))
\end{multline*}
\end{asse}

Finally, this requirement signifies that it is always the case that if the user visited a web page whose certificate was invalid, and later she visits again the same web page but with an associated valid certificate, then the browser warns the user about the potential past MITM attack.


\section{Findings} \label{sec:findings}

We studied our five requirements on the six browsers by checking the satisfiability of our formulas on the CSP\# models of the certificate validation. As said above, we considered Firefox, Chrome, and Safari in three different modes: classical browsing, private browsing, and their interleaving. In total, the analysis has considered 60 different scenarios due to the mix of browsers, modes, and properties.

It was possible to encode most of the scenarios without incurring into state explosion. We needed to assume no expired certificates to avoid non termination in four scenarios that turn out to ensure the requirements: \emph{Applying HSTS User Security} on classic browsing of Firefox, and \emph{Applying HSTS Bootstrap} on classic browsing of Safari, Firefox, and their interleaving.

Over an Intel I7 processor running Ubuntu Linux 14.04 with 8 GB RAM, PAT 3.5.1 generally terminates  the verification of each scenario in just a few seconds.
However, the HSTS-related requirements take minutes on Firefox and Safari.
The longest runtime is for \emph{Applying HSTS Bootstrap} over Firefox,  which takes 448 seconds. These runtimes show that our approach to socio-technical requirements is practical and can be usefully boosted further.

Interpreting the output of the tool required some effort. Table~\ref{tab:results} summarises the findings.  At first glance, it can be seen that the browsers that verified the highest number of requirements are SEB and Chrome, then come Internet Explorer and Firefox, and lastly Safari and Opera Mini. As a practical outcome, it turns out that an exam system that uses Safari or Opera Mini increases the chances that a user might think they are taking a valid exam when they are not.

As expected, the results of interleaving summarises the failing results of classical  and private browsing: it suffices that a property fails in one of the modes (\ie classic or private) to fail also on their interleaving. 
However, a more interesting result is that a session in one mode may influence a later session in a different browsing mode. This is the case of Safari for \emph{Applying HSTS Bootstrap}.
In the remainder, we comment on each requirement in detail.

\subsubsection*{Warning Users}
The first requirement  is found valid over SEB, Chrome, and Internet Explorer.
It is also valid in Firefox in private browsing.
 By contrast, the model checker shows traces that falsify it over the other modes for Firefox, Safari, and Opera Mini.
With Firefox and Safari the traces are similar. They report a sequence of two TLS sessions both with a MITM attack. In the first session, the browser connects to the corrupted server and warns the user, who chooses to store the certificate of the attacker anyway. In the second session, the user tries to get access to the same server, but this time the browser has the attacker server certificate stored, and completes the session without warning the user. This is due
to the drawbacks of storing server certificates, which Firefox and Safari allow their users to do.
Notably, Firefox in private browsing forbids the user to store server certificates, hence the requirement turns out to be valid.
The trace that falsifies the property with Opera Mini is rather trivial because the browser does not involve the user at all. Opera Mini in fact shows a padlock when the certificate is valid, but even if the certificate is invalid, the browser completes the TLS session anyway, without informing the user.

\subsubsection*{Storing Server Certificate}The second property turns out to be the most tricky. It is found that all browsers verify the property except Firefox and Safari, although the latter are the only browsers that allow a user to store server certificates. This is because the property is a logical implication whose precondition is trivially falsified by the browsers that do no store server certificate. Surprisingly, the property does not hold on Firefox in classic browsing and on Safari in all modes, as they do not falsify the precondition. The user can in fact replace a server certificate as many times as she wishes to, while the browser does not inform the user that a server certificate was already stored. In support of this, the tool exhibits the following counterexample. In one session, the user engages with an honest server that transmits a self-issued certificate; the browser warns the user about the invalid certificate, but she chooses to store the certificate, thus the browser successfully concludes the TLS session. In a subsequent session, the user wants to connect again with the same server, but this time the attacker interposes and sends a self-issued certificate pretending to be the honest server; the browser warns the user about this second invalid certificate, regardless the fact that another certificate was already stored for the same server; the user decides to store also this certificate and the browser concludes the TLS session.

\subsubsection*{Applying HSTS User Security} The third requirement  is valid on SEB and in classic browsing on Firefox and Chrome. The requirement does not hold in private browsing because Firefox and Chrome trade off privacy and security: they prefer to remove HSTS policies stored during private browsing sessions to protect user's privacy. In fact an inspection of the stored HSTS policies would reveal the  HSTS-complaint website the user visited in private browsing.
Surprisingly, the property is not valid on Safari in any mode. The model checker shows a trace as follows: in the first session the attacker interposes in the communication, and the user chooses to store the attacker's certificate. In a subsequent session, the browser communicates with the honest server, from which it receives the HSTS header. Then, in a new session, the attacker interposes again in the communication, and the browser does not abort but concludes with no warnings. This is because a user's approved certificate bypasses the HSTS policy in Safari\footnote{After we filed a bug report to Apple, we received this reply:``The information you've provided will be valuable in our efforts to determine the cause of the issue you reported."}. As expected, the property is not valid on browsers that do not support HSTS. However, it holds on SEB although it does not support HSTS, because the browser aborts when the certificate is invalid.

\subsubsection*{Applying HSTS Bootstrap} The fourth property is checked over the browsers that support HSTS. It is valid on all browsers that support HSTS except on private browsing of Safari, which does not consider the HSTS pre-loaded whitelist in private browsing\footnote{We received this reply from Apple:``Your reported issue will be addressed in upcoming releases.  If you are a member of our developer program, you can test our fix in the current beta release of iOS 9 and OS X 10.11 El Capitan''}. Moreover, the interleaving of Safari modes also does not guarantee the property:  since Safari allows the user to permanently store a certificate even in private browsing, and such storing supersedes the HSTS policy, future sessions with HSTS-complaint websites are compromised also in classic browsing.

\subsubsection*{Learning from Server Certificate History} Finally, the fifth requirement holds only on SEB, in which the precondition $(\textsc{FinishTLS} \wedge \lnot~\mathit{CertificateIsValid})$ is always falsified since SEB aborts the session when the certificate is invalid. However, the property is not valid in the other browsers.
 This denounces the stateless philosophy whereby browsers do not record  warnings they issued in the past, hence browsers cannot leverage upon them at present.

\begin{table}[h]
\begin{center}
{\begin{tabular}{cc|L|L|L|L|L|}
\cline{3-7}
 & &Warning Users&  Storing Server Certificate& Applying HSTS User Sec.& Applying HSTS Bootstrap& Learning from Cert. History\\
\hline
\multicolumn{1}{ |c|  }{SEB} & \emph{co} &  $\checkmark$ & $\checkmark$ & $\checkmark$ & $-$ & $\checkmark$  \\\hline\hline
\multicolumn{1}{ |c|  }{\multirow{3}{*}{Firefox} }  &
\multicolumn{1}{ c|  }{\emph{cb}} &  $\times$ & $\times$ & $\checkmark$ & $\checkmark$ & $\times$  \\\cline{2-7} 
\multicolumn{1}{ |c|  }{} &
\multicolumn{1}{ c|  }{\emph{pb}} &  $\checkmark$ & $\checkmark$ & $\times$ & $\checkmark$ & $\times$  \\\cline{2-7}
\multicolumn{1}{ |c|  }{}&
\multicolumn{1}{ c|  }{\emph{in}} &  $\times$ & $\times$ & $\times$ & $\checkmark$ & $\times$  \\\hline\hline

\multicolumn{1}{ |c|  }{\multirow{3}{*}{Chrome}}  &
\multicolumn{1}{ c|  }{\emph{cb}} &  $\checkmark$ & $\checkmark$ & $\checkmark$ & $\checkmark$ & $\times$  \\\cline{2-7}
\multicolumn{1}{ |c|  }{}&
\multicolumn{1}{ c|  }{\emph{pb}} &  $\checkmark$ & $\checkmark$ & $\times$ & $\checkmark$ & $\times$  \\\cline{2-7}
\multicolumn{1}{ |c|  }{}&
\multicolumn{1}{ c|  }{\emph{in}} &  $\checkmark$ & $\checkmark$ & $\times$ & $\checkmark$ & $\times$  \\\hline\hline

\multicolumn{1}{ |c|  }{IE} & \emph{co} &  $\checkmark$ & $\checkmark$ & $\times$ & $-$ & $\times$  \\\hline\hline

\multicolumn{1}{ |c|  }{\multirow{3}{*}{Safari}} &
\multicolumn{1}{ c|  }{\emph{cb}} &  $\times$ & $\times$ & $\times$ & $\checkmark$ & $\times$  \\\cline{2-7}
\multicolumn{1}{ |c|  }{}&
\multicolumn{1}{ c|  }{\emph{pb}} &  $\times$ & $\times$ & $\times$ & $\times$ & $\times$  \\\cline{2-7}
\multicolumn{1}{ |c|  }{}&
\multicolumn{1}{ c|  }{\emph{in}} &  $\times$ & $\times$ & $\times$ & $\times$ & $\times$  \\\hline\hline

\multicolumn{1}{ |c|  }{OM} & \emph{co} &  $\times$ & $\checkmark$ & $\times$ & $-$ & $\times$ \\\hline
\end{tabular}}
\begin{tablenotes}%
\item{Note:}{The term \emph{cb} indicates classic browsing, \emph{pb} is for private browsing, and \emph{in} is  the interleaving of  classic and private browsing sessions. The term \emph{co} indicates that  classic and private browsing activity diagrams coincide.
The symbol $\checkmark$ indicates that 
the property holds; the symbol  $\times$ indicates that it does not; the symbol $-$ indicates that the property cannot be checked because the corresponding browsers do not support HSTS.}
\end{tablenotes}%
\caption{The five socio-technical requirements studied over six browsers.}
\end{center}
\label{tab:results}
\end{table}%

\subsection{Recommendations}
Upon the basis of our findings, we formulate the four recommendations outlined below.

\begin{rec} Browsers should trigger users when something wrong $ $ \mbox{happens} rather than when something good happens. \end{rec}
Browsers used to inform users when a TLS connection was going to be employed. 
Nowadays, most of the browsers tend to trigger and interrupt the user when some problems with TLS occur.  Of course, browsers also provide positive feedback such as a 
locked padlock in the chrome bar, but this does not require user's participation.
Opera Mini, for the sake of a lightweight implementation, shows the locked padlock as a positive feedback but does not warn users in case of problems with TLS.

\begin{rec} Browsers should consider the preloaded whitelist of \phantom{a} \mbox{HSTS-complaint} servers also in private browsing. \end{rec} 
Firefox and Chrome show that it is possible to protect the user and mitigate bootstrap attacks with HSTS without breaking user's privacy in private browsing.
The choice of Safari not to consider the preloaded whitelist leads to some weaknesses that when mixed with other features (i.e., user approved server certificates) may become serious vulnerabilities.

\begin{rec} HSTS policies should have priority over user's past \phantom{a} choices. \end{rec}
Also this recommendation comes from the findings on Safari, which implements a customised HSTS mechanism. HSTS has been conceived to avoid user's participation on security choices. A server that chooses to be HSTS-complaint cannot self-issue a certificate. Thus, any check on user's approved certificates should be superseded by the HSTS policy stored in the browser.

\begin{rec} Browsers should keep track of invalid certificate history to warn users more appropriately.  \end{rec}
Users may forget past security warnings, and browsers may help. For example, browsers could maintain a cache of invalid certificate hashes. In doing so, it would be possible for browsers to  warn users when a different invalid certificate is presented by a server with which the browser communicated in the past. It is worth noting that looking at past interactions is the strategy that the Session Description Protocol~\citeltex{rfc4572} advances to strengthen the management of self-issued certificates. 
Surprisingly, it has not been used in HTTPS.


\section{Conclusion} \label{sec:concsocio}
Since humans play critical roles in an exam, the security of such protocol should consider their participation.
We observe that browsers are the main component that nowadays interfaces the user with the network, not only in exam protocols, but also in many other security protocols.

The socio-technical analysis of the security of browsers is yet to be considered innovative at present. It combines traditional analysis of the technologies underlying browsers on the one hand, with elements of user participation on the other. By doing so, the socio-technical approach  is oriented at characterising security requirements also in terms of what the user may accomplish, with the ultimate aim of building browsers that are secure \emph{in} the presence of humans.

This chapter describes our work in this area. It focuses on server authentication with the user via the browser. More specifically, it studies the socio-technical ceremony of certificate validation in the various circumstances where this validation can fail, including MITM attacks. 

The security analysis of the ceremony of certificate validation from a socio-technical standpoint inspires a number of research questions, and we concentrated on three:
the first addresses the differences in terms of user participation in server authentication; the second concerns the strategies that browsers use to reduce the security risks for users in the presence of an invalid certificate; the third relates to how browsers improve the security by involving the users more profitably than they do at present.

To address these questions  we formulated five requirements that tackle how users are involved in the ceremony of certificate validation.
The outcome of our analysis demonstrates that each browser implements the ceremony of certificate validation differently, and that this is the origin of a few security problems. In particular, our analysis shows that HSTS fails on its goal when implemented in the wide customised process of certificate validation. Microsoft announced that the next version of Internet Explorer will support HSTS \citeweb{ieannouncment}. We argue that its usefulness will depend on how HSTS is implemented in the browser's certificate validation.

A major hallmark throughout our work is the adoption of UML activity diagrams as a semi-formal language to represent portions of browser functioning compactly, so that the human analyser could quickly get to grasps with their niceties. However, rigorous security analysis requires a formal approach. Thus,  the diagrams that represent the ceremonies are systematically mapped into CSP\# and validated in the PAT model checker against the LTL specification of our five socio-technical requirements. These are the main steps of our approach to the socio-technical formal analysis of the security of browsers. The current findings encourage us to develop this approach further, for example by automating it fully, and by trying it out on additional socio-technical requirements.

It is worth to stress that our current choices of formal languages and supporting tools are not meant to be binding; rather, they aim at demonstrating our approach. Also, we are currently working on reproducing the experiments described above on different tools. 
We advocate alternative verification methods to check other requirements such as privacy, a requirement that cannot be modelled in PAT. While looking at the interleaving of sessions in different browser modes, we noted that a session in private browsing should not interfere the following sessions in classic browsing.
A different approach,  possibly a different model checker such as FDR \citeltex{Roscoe}, is required to understand whether this interference may leak information, since PAT cannot verify privacy requirements. We leave this further development as future work.




\chapter{Conclusions}\label{chap:conclusion}
The main argument of this dissertation is that an exam must be designed and analysed as carefully as security protocols are. Thus, a rigorous understanding of the security requirements of exams is fundamental to design protocols that withstand threats coming from malicious candidates and authorities. Moreover, it is necessary to develop formal approaches that allow one to prove that an exam protocol meets the stated security requirements.
We address these needs using formal methods for the security analysis of exams, and cryptographic techniques for their design.

This dissertation has found that secure exam should meet ten fundamental requirements, five concerning authentication and five concerning privacy. Authentication requirements aim to preserve the association between candidate identity, mark, and test throughout the entire examination. Privacy requirements aim to provide anonymity to both candidates and examiners. This dissertation has also shown that exam should provide verifiability. Similarly to the end-to-end verifiability requirement in voting, having verifiability in any exam is fundamental to raise the credibility of the exam process.
Thus, we have identified six individual and five universal verifiability requirements. Exam protocols should provide enough information to allow candidates and auditors to verify the correctness of the exam using private and public available information. In summary, this dissertation has achieved Objective 1, as specified in chapter \ref{chap:introduction}, by advancing 21 security requirements for exams.

This research work has also provided novel approaches that allow the study of security requirements for exams. The formal framework for the analysis of authentication and privacy enables the evaluation of exam protocols, which can be specified in the \appi-calculus. The framework is flexible as it supports various types of exams, namely traditional, computer-assisted, and Internet-based exams.
The formal framework for the analysis of verifiability consists of an abstract model \index{abstract model} that supports a wide choice of verification methods based either on symbolic or on computational models.
Both frameworks can be extended with additional requirements. In fact, we formulate the extra requirements of Notification Request Authentication and Dispute Resolution.
Overall, this dissertation has introduced two formal frameworks for the analysis of exam protocols, hence it has met Objective 2.

The scarcity of secure exam protocols \index{secure exam protocol} in the literature, has led us to propose three new protocols for traditional, computer-assist\-ed, and Internet-based exam. The protocols exploit different cryptographic techniques that span from exponentiation mixnet to visual cryptography, but share the same design principle of minimising the reliance on the trusted third parties. In this vein, we have proposed two protocol versions of existing exam software (WATA).  \index{exponentiation mixnet}
Using our formal frameworks, it has been possible to analyse the security of existing and proposed exam protocols. Some security issues have been found and modification have been suggested.
The three novel protocols and the two protocol versions of exam software contribute to the achievement of Objective 3.

It has been observed that the human should be also considered for an end-to-end security analysis of exams. Moreover, browsers are often the main components of an exam protocol and are critical to exam security as they interface users and machines. Thus, this dissertation has advanced a socio-technical analysis of certificate validation as carried out in a browser for secure exams (SEB). It has proposed a method that allows us to expand the analysis to the most popular browsers. 
The method consists of modelling the certificate validation ceremonies using UML Activity Diagram, and of the systematic translation of such diagrams to CSP processes. We have validated the method with the model checking of five socio-technical requirements that binds TLS session, certificates and user choices. 
Thus, this dissertation has achieved Objective 4 by exploring a socio-technical understanding of the certificate validation in exam and modern browsers.

This research sees interesting outcomes, and we recall them. One is about the formal analysis of the existing exam protocols: it is observed that authentication can fail because of the presence of logic flaws in the design of the protocols, while privacy can fail because the protocols rely upon inadequate cryptographic primitives. A second insight is about the relation between exams and similar domains, such as voting. Although the approaches to design and analysis  are similar, we find that several requirements are different. Another interesting outcome is the design of cryptographic exam protocols that meet contrasting requirements  without the need of a TTP and in presence of remote and face-to-face phases. Exams should not rely on TTP and should guarantee some forms of accountability. One last notable result derives from the socio-technical analysis of certificate validation ceremonies. It turns out that the security of such ceremonies varies considerably when implementors customise the ceremonies, possibly including users in security decisions.

As any research, this work required us to make some choices. One is about the classes of security requirements. We considered the formalisation of authentication, privacy, and verifiability, but definitions of non-repudiation and accountability are also relevant for exams. Our formalisation of Dispute Resolution goes in that direction.
Another choice is about the list of considered requirements. We deemed the elements of this list highly desirable according to our experience and discussions with colleagues. However, the list is not meant to be exhaustive: some exams may demand additional requirements, as is the case of WATA protocols.

Our analysis inherits the limitations of ProVerif. Although the analysis is sound, ProVerif implements safe abstractions that might lead to non termination or false attacks. We overcome some limitations with manual proofs, as in the case of the analysis of \remark for universal verifiability.

In summary, this work has identified the security requirements for exams, developed frameworks for their analysis, and designed new protocols. The results of this research offer a promising foundation for the design and analysis of secure and practical protocols that accommodate any type of exam. The next section outlines a number of possible research directions for this area.




\section{Future Work}
Future work that continues the work presented in this dissertation can be envisaged over different research directions.

Concerning the formal frameworks, it might be possible to extend them with the specification of new security requirements and to study formally the relation between the proposed requirements.
Also, the analysis of more exam protocols would help to corroborate the flexibility of the proposed frameworks. In particular, it is interesting to analyse the verifiability of exams in the computational model, possibly with the assistance of the  CryptoVerif~\citeltex{Blanchet08} automatic tool. Another interesting research direction is to study whether our approach in defining verifiability can be applied to e-voting.

Regarding Remark!, future work includes the extension of open-source platforms like Moodle with our protocol. Another interesting research direction is to expand \remark~with techniques to detect plagiarism and candidate cheating at testing. This is a significant point since \remark~is designed to allow candidates to take the exam from home. We envisage that misbehaviour detection strategies such as data mining used to derive patterns described by Pieczul and Foley \citeltex{PF14} can be useful for this purpose. Another research direction includes the support for collaborative marking, in which the questions are categorised by subject, and examiners evaluate only the answers that pertain to the examiner subject area. 

Future work can be envisaged also for computer-assisted exam protocols. One is to extend the proposed protocols to accommodate different exam scenarios. For instance, some scenarios may not require the participation of the candidate at notification. To achieve this, we envisage a temporal deanonymization solution similar to the one specified in chapter \ref{chap:remark} for Remark!. 
We note that Dispute Resolution may conflict with the anonymity of the candidate's test. Thus it would be interesting to study how we can get both of them in the  design of exam protocols, and in general how to ensure both accountability and privacy requirements in the same system.

Another line of research concerns the formal analysis. It might be possible to study compositional proofs that integrate computational proofs of the cryptographic primitives used in our protocol with the symbolic ones obtained in ProVerif. A practical research direction is the implementation of a prototype of the protocol and the verification of whether different visual cryptography schemes can be used to increase the perceptual security of an exam.

The socio-technical analysis of certification validation shows that SEB is a promising browser, so it would be interesting to integrate it with \remark~to achieve secure testing. 
Finally, the socio-technical analysis approach can be developed further at least in two directions: one is to fully automate the translation from UML Activity Diagram to CSP\#; the other is to model additional socio-technical requirements, perhaps considering privacy.


\bibliographystyleltex{alpha}
\bibliographyltex{emarks,sarobib,sociotech,new,sarobibJ,sociotechJ}

\bibliographystyleweb{alpha}
\bibliographyweb{emarks,sarobib,sociotech,new,sarobibJ,sociotechJ}

\renewcommand\bibname{Publications}
\bibliographystyle{alpha}
\bibliography{emarks,sarobib,sociotech,new,sarobibJ,sociotechJ}

\newpage

\appendix


\chapter{CSP\# Code}\label{app:csp}

\section{Specification of Common Parts}
\begin{verbatim}
#import "PAT.Lib.Set";
#import "PAT.Lib.SetMine";

//UML activities' objects and certificate's fields
enum {HelloClient, HelloServer, ClientFinished, ServerFinished, 
      Data, Warning, Webpage, Continue, Abort, StoreCertificate, 
      Pk, HSTS, No_HSTS, S, I, SignCA, SignS, SignI, expi, 
      noexpi, revo, norevo};
		
channel ui 0;
channel network 0;

//UML datastores, certificate, and typed/clicked url
var<Set> dynamicHSTSList;
var<Set> preloadedHSTSList;
var<SetArray> ServerCert;
var cert[3];
var extendedcert[6]; 
var typed_url: {S..I}=S;

//UML decision points
#define CertificateIsValid cert[0]==typed_url && 
                           cert[2]==SignCA && 
                           extendedcert[4]==noexpi;

#define CertificateIsValidNR cert[0]==typed_url && 
                             cert[2]==SignCA && 
                             extendedcert[4]==noexpi && 
                             extendedcert[5]==norevo; 
                             //This used by Safari
#define URLhasHSTSpolicy dynamicHSTSList.Contains(typed_url) || 
                         preloadedHSTSList.Contains(typed_url);
#define CertificateIsStored ServerCert.Contains(extendedcert);

//Variables to keep track of some session's event
var intruder_server=false;
var user_warned=false;
var finishTLS=false;
var preload=false;

//Intruder process chooses which server plays session by session//
Intruder()= ServerI() [] ServerH();

//-----Intruder server process-----//
ServerI() = []header:{HSTS, No_HSTS}@ []url:{S,I}@
            []sk:{SignI,SignCA}@
            Init_TLS ->
            network?urlx.HelloClient ->
            //Intruder cannot sign certificate on behalf of CA
            ifa (url==S && sk==SignCA) {
               network!HelloServer.url.Pk.SignI -> Skip}
            else {network!HelloServer.url.Pk.sk -> Skip};
            Finish_TLS ->
            network?m  ->
            ifa (m==ClientFinished) {
               INTRUDER_IN{intruder_server=true} ->
               network!ServerFinished.header.Data ->Skip
            };					
            Intruder();

//-----Honest server process-----//
ServerH() = []header:{HSTS, No_HSTS}@  []sk:{SignS,SignCA}@
            Init_TLS ->
            network?urlx.HelloClient ->
            network!HelloServer.S.Pk.sk ->
            Finish_TLS ->
            network?m ->
            ifa (m==ClientFinished) {
               network!ServerFinished.header.Data ->Skip};
            Intruder();

//-----User process-----//
User() =    ui?webpage ->
            case {
            //The user can type or click on either honest's or 
            //intruder's url.
            webpage == Webpage: ui!S{typed_url=S} -> User() [] 
                                ui!I{typed_url=I} -> User()
            webpage == Warning: ui!StoreCertificate -> User()[]
                                ui!Continue -> User() []
                                ui!Abort -> User()
            default: User()};

//------Model process-----//
Model = Preloading() [] Begin();
Preloading = PreloadHSTSpolicy->{preloadedHSTSList.Add(S); 
                                 preload=true} -> Begin;
Begin = Intruder() ||| User() ||| Browser();

//User who wants to visit the honest server
#define UserwantS typed_url==S;
//Successful MITM attack: User wants to visit the honest server,
//but browser completed with the intruder
#define AuthFail intruder_server && UserwantS;
#define User_warned user_warned;
#define CompleteTLS finishTLS;
#define Preload preload;

///---------Properties--------///
#assert Model deadlockfree; 
//Property 1
#assert Model |=[] ((CompleteTLS && !User_warned) -> 
                    CertificateIsValid);

//Property 2
#assert Model |=[]((CertificateIsStored && UserwantS && 
                   ui.Data && !AuthFail)->
                   X([](UserwantS -> !AuthFail)));

//Property 3
#assert Model |=[]((CertificateIsValid && 
                   network.ServerFinished.HSTS.Data && UserwantS)->
                   X([](UserwantS -> !AuthFail)));

//Property 4
#assert Model |=[] (Preload-> (UserwantS -> !AuthFail));


//Property 5
#assert Model |=[]((CompleteTLS && !CertificateIsValid && 
                   UserwantS)->
                  X([]((CompleteTLS && CertificateIsValid && 
                        UserwantS)-> 
                       User_warned)));

\end{verbatim}

\newpage

\section{Specification of Web Browsers}

\subsection*{SEB}
\begin{verbatim}
Browser() = []rev:{revo, norevo}@[]exp:{expi,noexpi}@
            Display_Webpage ->
            //New session, variables used in macros are reset
            ui!Webpage{finishTLS=false; intruder_server=false; 
                       user_warned=false;} ->
            ui?url ->
            Resolve_URL ->
            Init_TLS ->
            network!url.HelloClient ->
            network?HelloServer.id.pk.sk{cert[0]=id;cert[1]=pk;
                                         cert[2]=sk} ->
            Check_Certificate ->
            if (rev==revo) {{finishTLS=false} -> Skip}
            else {
                if (CertificateIsValid) {{finishTLS=true}->Skip}
                else {  {finishTLS=false} -> Skip }
            };
            if (!finishTLS)
             //The browser informs the server about Abort (sync)
             {network!Abort -> Skip}
            else {
                Finish_TLS ->
                network!ClientFinished ->
                Process_DATA ->
                network?ServerFinished.header.Data ->
                Display_Webpage ->
                ui!Data -> Skip
            };
            Browser();
\end{verbatim}

\subsection*{Firefox}
\begin{verbatim}
Browser() = []rev:{revo, norevo}@[]exp:{expi,noexpi}@
            Display_Webpage ->
            //New session, variables used in macros are reset
            ui!Webpage{finishTLS=false; intruder_server=false; 
                       user_warned=false;} ->
            ui?url ->
            Resolve_URL ->
            Init_TLS ->
            network!url.HelloClient ->
            network?HelloServer.id.pk.sk{extendedcert[0]=cert[0]=id;
                                         extendedcert[1]=cert[1]=pk;
                                         extendedcert[2]=cert[2]=sk;
                                         extendedcert[3]=url;
                                         extendedcert[4]=exp} -> 
            Check_Certificate ->
            ifa (CertificateIsValid ) {{finishTLS=true} -> Skip}
            else {
                ifa (URLhasHSTSpolicy || rev==revo) 
                   {{finishTLS=false} -> Skip}
                else {
                    ifa (CertificateIsStored) 
                       {{finishTLS=true} -> Skip}								
                    else {
                        DisplayWarning ->
                        ui!Warning{user_warned=true} ->
                        ui?userchoice ->
                        tau{
                            if (userchoice == Abort) 
                              {finishTLS=false}
                            else {
                                finishTLS=true;
                                if (userchoice == StoreCertificate) 
                                  {ServerCert.Add(extendedcert);} 
                                  //associates a url to 
                                  //the server certificate
                            }
                        } -> Skip
                    }
                }
            };
            ifa (!finishTLS)
             //The browser informs the server about Abort (sync)
              {network!Abort -> Skip}
            else {
                Finish_TLS ->
                network!ClientFinished ->
                Process_DATA ->
                network?ServerFinished.header.Data ->
                Display_Webpage ->
                ui!Data ->
                Check_Header ->
                ifa (header==HSTS && CertificateIsValid) 
                   {StoreHSTSpolicy->
                    {dynamicHSTSList.Add(cert[0])} -> Skip}	 
            };
            Browser();
\end{verbatim}

\subsection*{Firefox - Private browsing}
\begin{verbatim}
Browser() = []rev:{revo, norevo}@[]exp:{expi,noexpi}@
            Display_Webpage ->
            //New session, variables used in macros are reset
            ui!Webpage{finishTLS=false; intruder_server=false; 
                       user_warned=false;} ->
            ui?url ->
            Resolve_URL ->
            Init_TLS ->
            network!url.HelloClient ->
            network?HelloServer.id.pk.sk{extendedcert[0]=cert[0]=id;
                                         extendedcert[1]=cert[1]=pk;
                                         extendedcert[2]=cert[2]=sk;
                                         extendedcert[3]=url;
                                         extendedcert[4]=exp} ->
            Check_Certificate ->
            if (CertificateIsValid) {{finishTLS=true} -> Skip}
            else {
                if (URLhasHSTSpolicy || rev==revo) 
                  {{finishTLS=false} -> Skip}
                else {
                    if (CertificateIsStored) 
                      {{finishTLS=true} -> Skip}								
                    else {
                        DisplayWarning ->
                        ui!Warning{user_warned=true} ->
                        ui?userchoice ->
                        tau{
                            if (userchoice == Abort) 
                              {finishTLS=false}
                            else {
                                finishTLS=true;                                
                            }
                        } -> Skip
                    }
                }
            };
            if (!finishTLS)
            //The browser informs the server about Abort (sync)
            {network!Abort -> Skip}
            else {
                Finish_TLS ->
                network!ClientFinished ->
                Process_DATA ->
                network?ServerFinished.header.Data ->
                Display_Webpage ->
                ui!Data -> Skip
            };
            Browser();
\end{verbatim}

\subsection*{Chrome}
\begin{verbatim}
Browser() = []rev:{revo, norevo}@[]exp:{expi,noexpi}@
            Display_Webpage ->
            //New session, variables used in macros are reset
            ui!Webpage{finishTLS=false; intruder_server=false; 
                       user_warned=false;expc=exp} ->
            ui?url ->
            Resolve_URL ->
            Init_TLS ->
            network!url.HelloClient ->
            network?HelloServer.id.pk.sk{cert[0]=id;cert[1]=pk;
                                         cert[2]=sk} ->
            Check_Certificate ->
            if (CertificateIsValid) {{finishTLS=true} -> Skip}
            else {
                if (URLhasHSTSpolicy || rev==revo) 
                  {{finishTLS=false} -> Skip}
                else {
                    DisplayWarning ->
                    ui!Warning{user_warned=true} ->
                    ui?userchoice ->
                    tau{
                        if (userchoice == Abort) 
                          {finishTLS=false}
                        else {	finishTLS=true; }
                    } -> Skip
                }
            };
            if (!finishTLS)
             //The browser informs the server about Abort (sync)
              {network!Abort -> Skip}
            else {
                Finish_TLS ->
                network!ClientFinished ->
                Process_DATA ->
                network?ServerFinished.header.Data ->
                Display_Webpage ->
                ui!Data ->
                Check_Header ->
                if (header==HSTS && CertificateIsValid)
                  {StoreHSTSpolicy->
                   {dynamicHSTSList.Add(cert[0])} -> Skip}	
            };
            Browser();
\end{verbatim}

\subsection*{Chrome - Private browsing}
\begin{verbatim}
Browser() = []rev:{revo, norevo}@[]exp:{expi,noexpi}@
            Display_Webpage ->
            //New session, variables used in macros are reset
            ui!Webpage{finishTLS=false; intruder_server=false; 
                       user_warned=false;expc=exp} ->
            ui?url ->
            Resolve_URL ->
            Init_TLS ->
            network!url.HelloClient ->
            network?HelloServer.id.pk.sk{cert[0]=id;cert[1]=pk;
                                         cert[2]=sk} ->
            Check_Certificate ->
            if (CertificateIsValid) {{finishTLS=true} -> Skip}
            else {
                if (URLhasHSTSpolicy || rev==revo) 
                  {{finishTLS=false} -> Skip}
                else {
                    DisplayWarning ->
                    ui!Warning{user_warned=true} ->
                    ui?userchoice ->
                    tau{
                        if (userchoice == Abort) 
                          {finishTLS=false}
                        else {	finishTLS=true; }
                    } -> Skip
                }
            };
            if (!finishTLS)
             //The browser informs the server about Abort (sync)
              {network!Abort -> Skip}
            else {
                Finish_TLS ->
                network!ClientFinished ->
                Process_DATA ->
                network?ServerFinished.header.Data ->
                Display_Webpage ->
                ui!Data -> Skip
            };
            Browser();
\end{verbatim}

\subsection*{Safari}
\begin{verbatim}
Browser() = []rev:{revo, norevo}@[]exp:{expi,noexpi}@
            Display_Webpage ->
            //New session, variables used in macros are reset
            ui!Webpage{finishTLS=false; intruder_server=false; 
            user_warned=false;} ->
            ui?url ->
            Resolve_URL ->
            Init_TLS ->
            network!url.HelloClient ->
            network?HelloServer.id.pk.sk{extendedcert[0]=cert[0]=id;
                                         extendedcert[1]=cert[1]=pk;
                                         extendedcert[2]=cert[2]=sk;
                                         extendedcert[3]=url;
                                         extendedcert[4]=exp;
                                         extendedcert[5]=rev} ->
            Check_Certificate ->
            if (CertificateIsValidNR || CertificateIsStored) 
              {{finishTLS=true} -> Skip}
            else {
                if (URLhasHSTSpolicy || rev==revo) 
                  {{finishTLS=false} -> Skip}
                else {
                    DisplayWarning ->
                    ui!Warning{user_warned=true} ->
                    ui?userchoice ->
                    tau{
                        if (userchoice == Abort) 
                        {finishTLS=false}
                        else {
                            finishTLS=true;
                            if (userchoice == StoreCertificate) 
                              {ServerCert.Add(extendedcert);} 
                              //associates a url to 
                             //the server certificate
                        }
                    } -> Skip                    
                }
            };
            if (!finishTLS)
            //The browser informs the server about Abort (sync)
            {network!Abort -> Skip}
            else {
                Finish_TLS ->
                network!ClientFinished ->
                Process_DATA ->
                network?ServerFinished.header.Data ->
                Display_Webpage ->
                ui!Data ->
                Check_Header ->
                if (header==HSTS && CertificateIsValidNR) 
                  {StoreHSTSpolicy->
                   {HSTSList.Add(cert[0])} -> Skip}	 
            };
            Browser();
\end{verbatim}

\subsection*{Safari - Private browsing}
\begin{verbatim}
Browser() = []rev:{revo, norevo}@[]exp:{expi,noexpi}@
            Display_Webpage ->
            //New session, variables used in macros are reset
            ui!Webpage{finishTLS=false; intruder_server=false; 
                       user_warned=false;} ->
            ui?url ->
            Resolve_URL ->
            Init_TLS ->
            network!url.HelloClient ->
            network?HelloServer.id.pk.sk{extendedcert[0]=cert[0]=id;
                                         extendedcert[1]=cert[1]=pk;
                                         extendedcert[2]=cert[2]=sk;
                                         extendedcert[3]=url;
                                         extendedcert[4]=exp;
                                         extendedcert[5]=rev} ->
            Check_Certificate ->
            if (CertificateIsValidNR || CertificateIsStored) 
              {{finishTLS=true} -> Skip}
            else {                       
                DisplayWarning ->
                ui!Warning{user_warned=true} ->
                ui?userchoice ->
                tau{
                    if (userchoice == Abort) {finishTLS=false}
                    else {
                        finishTLS=true;
                        if (userchoice == StoreCertificate) 
                          {ServerCert.Add(extendedcert);} 
                        //associates a url to the server certificate
                    }
                } -> Skip                        
            };
            if (!finishTLS)
            //The browser informs the server about Abort for syncing
            {network!Abort -> Skip}
            else {
                Finish_TLS ->
                network!ClientFinished ->
                Process_DATA ->
                network?ServerFinished.header.Data ->
                Display_Webpage ->
                ui!Data -> Skip
            };
            Browser();
\end{verbatim}

\subsection*{Internet Explorer}
\begin{verbatim}
Browser() = []rev:{revo, norevo}@[]exp:{expi,noexpi}@
            Display_Webpage ->
            //New session, variables used in macros are reset
            ui!Webpage{finishTLS=false; intruder_server=false; 
                       user_warned=false;expc=exp} ->
            ui?url ->
            Resolve_URL ->
            Init_TLS ->
            network!url.HelloClient ->
            network?HelloServer.id.pk.sk{cert[0]=id;cert[1]=pk;
                                         cert[2]=sk} ->
            Check_Certificate ->
            if (rev==revo) {{finishTLS=false} -> Skip}
            else {
                if (CertificateIsValid) {{finishTLS=true} -> Skip}
                else {
                   DisplayWarning ->
                   ui!Warning{user_warned=true} ->
                   ui?userchoice ->
                   tau{
                       if (userchoice == Abort) {finishTLS=false}
                       else {finishTLS=true; }
                   } -> Skip
                }
            };
            if (!finishTLS)
            //The browser informs the server about Abort (sync)
            {network!Abort -> Skip}
            else {
                Finish_TLS ->
                network!ClientFinished ->
                Process_DATA ->
                network?ServerFinished.header.Data ->
                Display_Webpage ->
                ui!Data -> Skip
            };
            Browser();
\end{verbatim}

\subsection*{Opera Mini}
\begin{verbatim}
Browser() = []rev:{revo, norevo}@[]exp:{expi,noexpi}@
            Display_Webpage ->
            //New session, variables used in macros are reset
            ui!Webpage{finishTLS=false; intruder_server=false;
                       user_warned=false;expc=exp} ->
            ui?url ->
            Resolve_URL ->
            Init_TLS ->
            network!url.HelloClient ->
            network?HelloServer.id.pk.sk{cert[0]=id;cert[1]=pk;
                                         cert[2]=sk} ->
            Check_Certificate ->
            if (rev==revo) {{finishTLS=false} -> Skip}
            else {{finishTLS=true} -> Skip};
            if (!finishTLS)
            //The browser informs the server about Abort (sync)
            {network!Abort -> Skip}
            else {
                Finish_TLS ->
                network!ClientFinished ->
                Process_DATA ->
                network?ServerFinished.header.Data ->
                Display_Webpage ->
                ui!Data -> Skip
            };
            Browser();
\end{verbatim}

\normalsize



\end{document}